\documentclass[11pt, a4paper]{article}
\usepackage{a4wide, url}
\usepackage[authoryear,round]{natbib}
\usepackage{graphicx}
\usepackage{amsfonts}
\usepackage{amsmath}
\usepackage{amssymb}
\usepackage{bm}
\usepackage{times}
\usepackage{color}
\usepackage{setspace}
\usepackage{rotating}
\usepackage{subfigure}
\usepackage{amsfonts,amsmath,amssymb,amsthm}
\usepackage{paralist}
\usepackage[linesnumbered,ruled,vlined]{algorithm2e}
\SetKwInput{KwInput}{Input}                
\SetKwInput{KwOutput}{Output}              

\usepackage[top=1in,bottom=1in,left=.7in,right=.7in]{geometry}

\usepackage[T1]{fontenc}
\usepackage{a4wide, url}  
\usepackage{latexsym}       
\usepackage{multirow} 
\usepackage[english]{babel}
\usepackage[latin1]{inputenc}
\usepackage{subfigure} 

\usepackage{pgf} 
\usepackage{endnotes}
\usepackage{bm} 
\usepackage{eurosym}   
\usepackage[normalem]{ulem}
\usepackage[textfont=it,labelfont=sc]{caption} 
\newcommand{\mbf}[1]{\mathbf{#1}} 	

\newcommand{\beq}{\begin{equation}}
\newcommand{\eeq}{\end{equation}}
\newcommand{\bea}{\begin{eqnarray}}
\newcommand{\eea}{\end{eqnarray}}
\newcommand{\ba}{\begin{array}}  
\newcommand{\ea}{\end{array}}      
\newcommand{\bi}{\begin{itemize}}
\newcommand{\ei}{\end{itemize}}
\newcommand{\ben}{\begin{enumerate}} 
\newcommand{\een}{\end{enumerate}}
\newcommand{\nn}{\nonumber}

\renewcommand{\r}{\right}
\renewcommand{\l}{\left}
\newcommand{\E}{\mathrm E}
\newcommand{\Var}{\text{\rm Var}}
\newcommand{\Cov}{\text{\rm Cov}}

\def\Black{} 

\renewcommand{\theequation}{\thesection.\arabic{equation}}

\long\def\symbolfootnote[#1]#2{\begingroup\def\thefootnote{\fnsymbol{footnote}}\footnote[#1]{#2}\endgroup}
    
\newtheoremstyle{thmstyle}
  {}
  {}
  {\it}
  {}
  {\sc}
  {.}
  {1mm}
  {}
\theoremstyle{thmstyle} 

\newtheorem{lemma}{Lemma}

\newtheorem{proposition}{Proposition}

\newtheorem*{assumption}{Assumption}

\newtheorem{remark}{Remark}

\begin{document}
\title {\textbf{\Large{Generalized Dynamic Factor Models and Volatilities:\\ \Large{Consistency, Rates, and Prediction Intervals}} }  }
\author {{\sc Matteo Barigozzi$^{\dag}$ \qquad   Marc Hallin$^{\ddag}$}\vspace{3mm} \\  \normalsize 
$^{\dag}${\it  LSE, Department of Statistics, Houghton Street, London WC2A 2AE, UK.} \\ \normalsize  
E-Mail:  m.barigozzi@lse.ac.uk\vspace{2mm} \\ \normalsize  
$^{\ddag}$ {\it ECARES, Universit\'e libre de Bruxelles CP114/4 B-1050 Bruxelles, Belgium. }\\ \normalsize  
E-Mail:  mhallin@ulb.ac.be
}
\date{\small{\today}}

\maketitle

\begin{abstract}
Volatilities, in high-dimensional panels of economic time series with a dynamic factor structure on the levels or returns, 
typically also admit a dynamic factor decomposition. We consider a two-stage dynamic factor model method recovering the common and idiosyncratic components of both levels and log-volatilities. Specifically, in  a  first estimation step, we extract the common and idiosyncratic shocks for the levels, from which a log-volatility proxy is computed. In a second step, we estimate a dynamic factor model, which is equivalent to a multiplicative factor structure for volatilities, for the log-volatility panel. By exploiting this two-stage factor approach, we build one-step-ahead conditional prediction intervals for  large $n\times T$ panels of returns. Those intervals are based on empirical quantiles, not  on conditional variances; they can be either equal- or unequal-tailed.  We provide  uniform consistency and consistency rates results for the proposed estimators as both $n$ and~$T$ tend to infinity.  We study the finite-sample properties of our estimators by means of Monte Carlo simulations. Finally, we apply our methodology to a panel of asset returns belonging to the S\&P100 index in order to compute  one-step-ahead conditional prediction intervals for the period 2006-2013.  A comparison with the componentwise GARCH benchmark (which does not take advantage of cross-sectional information) demonstrates the superiority of  our approach, which is genuinely multivariate (and high-dimensional),   nonparametric, and model-free.   \\
\\
\noindent {\itshape JEL Classification}: C32, C38, C58.\\
\noindent
{\itshape Keywords}: Volatility, Dynamic Factor Models, Prediction intervals, GARCH.
\end{abstract}

\symbolfootnote[0]{\\ We thank Christian Brownlees, Christian Francq, and Haeran Cho for helpful comments.  This paper was also presented at: ``Panel Data Forecasting Conference'', University of Southern California, Dornsife, Los Angeles, April, 2019; the ``6th Rimini Centre for Economic Analysis (RCEA) Time Series Econometrics Workshop'', University of Cyprus, Larnaca, June 2019,   the ``International Association for Applied Econometrics (IAAE) 2019 Annual Conference'', University of Cyprus, Nicosia, June 2019, and the ``Workshop on High-Dimensional Data Analysis", Durham University, June
 2019.
\noindent
}
            
\section{Introduction}

Data in high dimension unquestionably constitute one of the main challenges of contemporary statistics/econo\-metrics, and have become pervasive in most domains related with data sciences. 
Time series have not escaped that evolution, and the analysis of  high-dimensional time series---equivalently, large cross-sections of univariate time series or panels---today ranks among the most active topics in theoretical and applied econometrics. 

  The most successful methods so far in the analysis and prediction of high-dimensional time
series are based on the so-called factor model
approach. That approach, under its various forms, is based on a (non-observed) decomposition
of the observation (a large cross-section of time series with complex interrelations)  into the sum of two mutually orthogonal (all
leads, all lags) components: the {\it common component}, driven by a small number of {\it factors} or
{\it common shocks}, and an {\it idiosyncratic component}, with some variations in the definitions of
``common'' and ``idiosyncratic,'' and    the assumptions made. Regardless of the  definition adopted, the common and idiosyncratic components typically are disentangled by means of adequate cross-sectional and/or temporal aggregation of the observed time series. 

Those aggregation and factor model  approaches are strongly  rooted in the multivariate time-series methods developed in the eighties and nineties, of which George Tiao and his collaborators have been most  influential  and unremittable pioneers: see, for instance,   \citet{tiao1972asymptotic}, \citet{tiao1978some}, \citet{tiao1980forecasting}, \citet{tiao1981modeling}, \citet{tsay1985use}, \citet{pena1987identifying}, and \citet{tiao1989model}.

The type of factor model we are considering here is the General or Generalized Dynamic Factor Model (GDFM) introduced by \citet{FHLR00}, which, by taking into account all leading and lagging linear dependencies among the data,  encompasses most other models, as e.g. the static factor approaches by \citet{baing02}, \cite{stockwatson2002}, and \citet{FLM13}.  Moreover, as emphasised in \citet{fornilippi01} and \citet{hallinlippi13}, beyond the usual assumptions of second-order stationarity and existence of spectral densities, the GDFM decomposition into a common and an idiosyncratic component basically does not place any structural constraints on the data-generating process. In this sense,  contrary to   static factor approaches, it is  canonical, nonparametric and model-free. In this paper, we consider the one-sided  GDFM estimation method   recently   described in \citet{FHLZ15,FHLZ17}.
 
Prediction,  in  classical univariate and moderately multivariate  time series analysis,  is an obvious and natural objective; it is certainly no less crucial in high dimension. Efficient prediction, however, should exploit the amount of information available, due to the complex cross-dependencies among the many cross-sectional components, in the present and lagged values of the whole cross-section; the larger the cross-section  (i.e., the higher the dimension), the more crucial the role of that information, and the more delicate its recovering. Factor models naturally  have been used in the construction of {\it point-predictors}, and quite successfully so:  see, e.g., \citet{stockwatson2002}, \citet{baing08JoE}, \citet{FGLS18}, to quote only a very few.  Those authors,  however, are  dealing, mostly,  with macroeconomic data, while less attention has been given to factor model methods in the analysis and prediction of financial returns: see, e.g. \citet{chamberlainrotshild83},  \citet{CK93}, or \citet{ait2017}.  
In particular, when dealing with returns, due to the  presence of conditional distribution heterogeneity (of which conditional heteroskedasticity is only a very particular case), conditional volatility phenomenons are essential, and definitely should be taken into account when building conditional prediction limits or conditional prediction  intervals. 

Most multivariate methods available in the literature for the analysis of conditional heterogeneity are restricted to the study of conditional heteroskedasticity, and rely on  parametrisations of the ARCH-GARCH or Stochastic Volatility type: see, for instance, the reviews by \citet{BLR06} and \citet{AMY06}. Because of the curse of dimensionality, however, only the very simplest models can be considered in high-dimensional panels,  possibly inducing a nonnegligible  loss of efficiency.  Among those, the factor GARCH approach is the most popular, see e.g. \citet{DN89}, \citet{ENR92}, \citet{HRS92}, and \citet{SCF08}. {Static factor models directly based on volatilities have also been considered, but these fail to exploit the information contained in the idiosyncratic components of returns, see e.g. \citet{CKL06} and \citet{fan15}.}  For these reasons, \citet{barigozzihallin15a}   introduce a two-step GDFM approach by which the nonparametric and model-free virtues of factor models are used in a joint analysis of returns and volatilities. In \citet{barigozzihallin15b}, that two-step GDFM is combined with a GARCH strategy in order to produce point-forecasts for volatilities (see  also  \citealp{Trucios19} for a recent example), while  \citet{barigozzihallin15c} and \citet{BHS18} apply the same methodology in a study of the dynamic interdependencies of US and international financial markets. A two-stage factor approach similar to ours but in a static factor model setting is proposed in \citet{CB15}.

The objective of this paper is to combine the same two-step GDFM approach with a quantile-based construction of conditional confidence limits producing conditional interval predictions rather than point-forecasts for returns. That objective requires nontrivial consistency results on the two-step GDFM estimation method, which are not provided in \citet{barigozzihallin15a,barigozzihallin15b,barigozzihallin15c}. The first part of this paper, therefore, is devoted to a careful asymptotic analysis of the two-step GDFM. We then describe the quantile-based construction of conditional confidence limits, which we apply to a dataset of S\&P100 daily returns.

The paper is organised as follows. In Section \ref{sec:mod}, we present the GDFM model for the stochastic processes of returns (levels) and   log-volatilities, and  give sufficient conditions for its existence and identification.  Section~\ref{sec:est_summary}    describes the estimation of the model, and  Section \ref{sec:ap}   establishes the  consistency properties (with rates) of the proposed estimators. In Section \ref{eq:int}, we define the one-step-ahead conditional prediction confidence limits and  intervals.   In Section \ref{sec:sim}, we study the finite-sample properties of our estimators via simulations. 
Section \ref{sec:emp}   applies our methodology to a panel of daily returns of stocks listed in the S\&P100 index and investigates the resulting coverage performance. In Section \ref{sec:conc}, we conclude. Proofs  are  postponed to an Appendix. 

\subsection*{Notation}
The sub-exponential norm of a scalar random variable~$X$ is defined as $\Vert X\Vert_{\psi_1}:=\sup_{p\ge 1} p^{-1}\E[|X|^p]^{1/p}$ (see e.g.~Definition 5.13 in \citealp{vershynin12}). The transposed complex conjugate of a complex vector $\bf p$ is denoted as ${\bf p}^\dag$ and $\Vert \bf p\Vert=\bf p^\dag\bf p$. For an hermitian complex $n\times n$ matrix $\mbf A$ with generic $(i,j)$ entry $a_{ij}$ and largest (in modulus) eigenvalue~$\mu^{\mbf A}_1$, let~$\Vert \mbf A\Vert_1:=\max_{j=1,\ldots,n} \sum_{i=1}^n |a_{ij}|$ and $\Vert \mbf A\Vert:={\mu_1^{\mbf A}}$. 
As usual, $L$ stands for the lag operator, such that, given a stochastic vector process $\{\mbf Y_t | t\in\mathbb Z\}$,  $L^k\mbf Y_t:=\mbf Y_{t-k}$ for any integer $k$ and any~$t\in\mathbb Z$. Last, we denote by $\mathbb I(\mathcal A)$ the indicator function of an event $\mathcal A$.

\section{A General Dynamic Factor Model for levels and volatilities}\label{sec:mod}

We throughout assume that all stochastic variables in this
paper belong to the Hilbert space $L_2(\Omega, \mathcal F , \mathrm P)$, \linebreak where~$(\Omega, \mathcal F , \mathrm P)$ is
some common probability space. We study   double-indexed stochastic processes of the\linebreak form~$\mbf Y\!:= \{Y_{it} \vert i\in\mathbb{N} ,  \  t\in\mathbb{Z}\}$, with $n$-dimensional sub-processes $\mbf Y_n\!:= \{Y_{it} \vert i=~\!1,\ldots,n,~t\in~\!\mathbb{Z}\}$, $n\in\mathbb{N}$. In practice, we deal with the  finite observed $n\times T$ realisation 
$${\bf Y}_{n,T}:=\left(\begin{array}{cccc}
Y_{11}, & Y_{12}, & \ldots ,
& Y_{1T} \\
\vdots&\vdots&&\vdots \\
Y_{n1}, & Y_{n2}, & \ldots ,
& Y_{nT} 
\end{array}\right)
$$
of $\mbf Y$. In the empirical application of Section \ref{sec:emp}, the $Y_{it}$'s are observed values of daily stock returns, and we therefore  call  $\mbf Y$  the ``levels'' process. The assumptions in Section \ref{sec:mod_level} are mainly taken from \citet{FHLZ17}, with some modifications, mostly concerning the idiosyncratic components. 
On the other hand, the assumptions in Section \ref{sec:mod_vol} are new and are related to the log-volatility proxies originally introduced in \citet{barigozzihallin15a,barigozzihallin15b}. 

\subsection{Model and assumptions for levels}\label{sec:mod_level}

The Generalized Dynamic Factor Model (GDFM) for the   levels  process $\mbf Y$    is a decomposition of $Y_{it}$ into
\begin{equation} 
Y_{it}-\E[Y_{it}]= X_{it}+ Z_{it}, \quad i\in\mathbb{N} ,  \  t\in\mathbb{Z}\label{GDFM}  
\end{equation} 
with
\begin{equation} 
X_{it} = \sum_{j=1}^q \sum_{k=0}^{\infty} b_{ijk} u_{jt-k}=\mbf b_i'(L)\mbf u_t\quad\text{and}\quad Z_{it}=\sum_{k=0}^{\infty} d_{ik} v_{it-k} = d_i(L) v_{it},\label{eq:idio_level}
\end{equation} 
where $\E[Y_{it}]$ stands for the   expected value of $Y_{it}$ and the processes   $\mbf u:= \{u_{jt} \vert j=1,\ldots,q,  \  t\in\mathbb{Z}\}$ \linebreak and~$\mbf v_{n}:= \{v_{it} \vert i=1,\ldots,n, \   t\in\mathbb{Z}\}$ are mutually orthogonal (at all leads and lags) $q$- and $n$-dimensional white noises, respectively. Call~$\mbf u$ the process of {\it common factors} or {\it common shocks} and~$\mbf v_n$ the process of {\it idiosyncratic shocks};  $X_{it}$ and~$Z_{it}$ are~$Y_{it}$'s  {\it common} and {\it idiosyncratic  components}, 
respectively. 
 
 Letting  $\mbf X_n:= \{X_{it} \vert i=1,\ldots,n,  \  t\in\mathbb{Z}\}$ and $\mbf Z_n:= \{Z_{it} \vert i=1,\ldots,n,  \  t\in\mathbb{Z}\}$, equations~\eqref{eq:idio_level} in vector notation  takes the form 
 \begin{align}\label{eq:common_lev_idio_vec}
\mbf X_{nt}=\mbf B_n(L)\mbf u_t,\qquad\mbf Z_{nt}=\mbf D_n(L)\mbf v_{nt}, \quad n\in\mathbb{N} ,  \  t\in\mathbb{Z}.
\end{align}
with  $\mbf B_n(L):=(\mbf b_1(L)\ldots \mbf b_n(L))'$, and $\mbf D_n(L):=\text{\rm diag}(d_1(L)\ldots d_n(L))$. \medskip

 More precisely, we assume that \eqref{GDFM}-\eqref{eq:idio_level} hold and satisfy the following assumptions:\vspace{-1mm}
\begin{assumption}[L1] $\,$
\begin{compactenum}[(i)] 
\item the dimension $q$ of  $\mbf u_t$ does not depend on $n$;  the process $\mbf u :=\{\mbf u_t \vert t\in\mathbb{Z}\}$ is   second-order white noise, with mean $\mbf 0_q$ and  diagonal  positive definite  covariance $\bm{\Gamma}^{\rm u}$;
\item writing ${\mbf b}_{ik}:=(b_{i1k}\ldots b_{iqk})^\prime$ for  the $q\times 1$ coefficient of $L^k$ in ${\mbf b}_i(L)$, there exists a constant~$M_1>0$ such that   $\sum_{k=0}^{\infty}\Vert \mbf b_{ik}\Vert\, \vert k\vert\le  M_1$ for all $i\in\mathbb{N}$;
\item the process $\mbf v := \{\mbf v_{nt} \vert t\in\mathbb{Z}\}$ is    second-order white noise, with mean $\mbf 0_n$ and   positive definite covariance~$\bm\Gamma_n^{\rm v}$;  moreover, $\E[v_{it}|v_{is}]=0$ for all $i\in\mathbb N$ and $t,s\in\mathbb Z$ such that $t>s$;  
\item there exists a constant  $C_{\rm v}>0$ such that $\Vert \bm\Gamma_n^{\rm v}\Vert_1\le C_{\rm v}$ for all~$n\in\mathbb{N}$;
\item there exists a constant   $M_2>0$ such that  $\sum_{k=0}^{\infty}\vert  d_{ik}\vert\, \vert k \vert\le M_2$ for all $i\in\mathbb{N}$; 
\item $\Cov(u_{jt},v_{is})=0$ for all $i\in\mathbb N$,  $j=1,\ldots, q$, and $t,s\in\mathbb Z$;
\item there exists a constant   $M_3>0$ such that  $\sum_{k_1,k_2,k_3\in\mathbb Z} 
 \big\vert\E[u_{j_1t}u_{j_2,t-k_1}u_{j_3,t-k_2}u_{j_4,t-k_3}]\big\vert\le M_3$ for all $j_1,j_2,j_3,j_4=1,\ldots,q$;  
\item  there exists a constant   $M_4>0$ such that  $\sum_{k_1,k_2,k_3\in\mathbb Z}
 \big\vert\E[v_{i_1t}v_{i_2,t-k_1}v_{i_3,t-k_2}v_{i_4,t-k_3}]\big\vert \le M_4$  for all\linebreak $i_1,i_2,i_3,i_4\in\mathbb N$.\end{compactenum}
\end{assumption}

These assumptions are standard in the literature with the exception of part {\it (iv)} which imposes a mild form of sparsity on the covariance matrix of the idiosyncratic innovations. A similar condition can be found   in \citet{FLM13} and is empirically verified by \citet{boivinng06} and \citet{baing08JoE} for US macroeconomic data, and by \citet{barigozzihallin15c} for stock returns. As a consequence of parts {\it (iv)} and {\it (v)}, the idiosyncratic components are allowed to be serially autocorrelated and   mildly cross-correlated (see also Lemma \ref{lem:dyn_eval} below). Moreover, it is easy to check that such assumption is nesting other typical conditions on the cross-sectional dependence of idiosyncratic components (see e.g. \citealp{baing02}, and \citealp{stockwatson2002}, in the static factor model case). Parts {\it (ii)} and {\it (v)} imply absolute summability of the autocovariances and therefore the existence of a purely continuous spectral density. 
 Moreover these assumptions and existence of fourth-order moments  in parts (vii) and (viii) are classical requirements for consistent estimation of the autocovariances and  the spectral density (see e.g. Chapter IV,  Theorem 6, in \citealp{hannan1970}, for the autocovariances, and the results in Section 6.2 in \citealp{priestley01}, and Theorem 5A in \citealp{parzen57}, for the spectral density). Last, in part {\it (iii)} we also make the typical assumption of martingale difference innovations used  in the GARCH literature (see e.g. Definition 2.1 in \citealp{FZ11}).

It should be insisted, however, that the GDFM  is not a {\it statistical model} in the usual sense, inasmuch as, beyond the requirement of second-order stationarity, the existence of a finite (but unspecified) $q$, and the existence of a spectrum,  it does not  really impose any restrictions on the data-generating process: as argued by \citet{fornilippi01} and \citet{hallinlippi13}, \eqref{GDFM}-\eqref{eq:idio_level} indeed constitute a representation result rather than a model equation. 

On the filters $\mbf b_i(L)$ and $d_i(L)$ we furthermore impose   the following assumptions:
\begin{assumption}[L2] $\,$
\begin{compactenum}[(i)] 
\item $\mbf b_i(L)$ has rational entries, i.e.  $b_{ij}(L)=\theta_{ij}(L)\phi^{-1}_{ij}(L)$, where $\phi_{ij}(z)$ and $\theta_{ij}(z)$, for all $i\in\mathbb N$ and \linebreak$j=1,\ldots, q$, are finite-order polynomials;
\item there exists a constant $\bar \phi>1$ such that $\phi_{ij}(z)\neq 0$  for all $i\in\mathbb N$, all $j=1,\ldots, q$,  and  all~$z\in\mathbb C$ such that $|z|\le \bar \phi$;
\item the coefficients $\theta_{ijk}$ of $\theta_{ij}(L)$ are such that $|\theta_{ijk}|\le B^X$ for some positive  constant $B^X$, \linebreak all $k\in\mathbb N\cup \{0\}$, all~$i\in\mathbb N$, and $j=1,\ldots, q$;
\item $d_i(L)$ is of the form $c_i^{-1}(L)$ where $c_i(z)$, for all $i\in\mathbb N$, is a finite-order polynomial,~$c_{i}(0)=1$ and $c_{i}(z)\neq 0$ for all $z\in\mathbb C$ such that $|z|\le 1$.
\end{compactenum}
\end{assumption}
This latter assumption is not strictly needed and could be easily relaxed to allow for infinite order autoregressive dynamics---at the expense, however, of heavier notation and longer proofs;  see also Section \ref{sec:ap} for a short discussion. 
 This assumption implies that both the common and idiosyncratic components have a rational spectral density. Rational filters for the common component are also assumed in \citet{FHLZ17}, while here we also assume that the idiosyncratic component admits a finite autoregressive representation. In particular, using part~{\it (iv)}, we can rewrite the second equation in  \eqref{eq:idio_level} as 
\beq
c_i(L) Z_{it} = v_{it}.\label{eq:var_lev_idio}
\eeq

Let $\bm\Sigma_n^Y(\theta)$, $\bm\Sigma_n^X(\theta)$ and $\bm\Sigma_n^Z(\theta)$, $\theta\in[-\pi,\pi]$, be the $n\times n$ spectral density matrices of the observed panel, the  common, and the idiosyncratic components, respectively;  the existence of those spectral densities is guaranteed by Assumption (L1).  Denote by $\lambda_{nj}^Y(\theta)$, $\lambda_{nj}^X(\theta)$, and $\lambda_{nj}^Z(\theta)$ their respective~$j$-th largest eigenvalues---the {panel}, {common}, and idiosyncratic {\it dynamic eigenvalues}, on which we assume  the following. {Hereafter, ``for\linebreak  all~$\theta\in[-\pi,\pi]$'' or ``$\theta-a.e.$'' is to be understood as  ``for all~$\theta$ but over a subset of values included in a set with Lebesgue measure zero.'' Similarly, $\sup_{\theta\in[-\pi,\pi]}$ in the sequel is an {\it essential} $\sup$, etc. \Black

\begin{assumption}[L3] There exist a positive integer $\bar n$ and continuous functions $\alpha_{j}$ and $\beta_{j-1}$  from~$[-\pi,\pi]$ to $\mathbb R\,$, $j=1,\ldots,q$, independent of $n$, and such that 
$$0< \beta_{j-1}(\theta) < \alpha_{j}(\theta)\le {\lambda_{nj}^X(\theta)}/{n}\le \beta_j(\theta)<\infty\quad\!\!\text{ $\theta$-a.e. in $[-\pi,\pi]$, all $j=1,\ldots, q$, and all $n>\bar n$. }$$
\end{assumption}

Under this assumption, the first $q$ common dynamic eigenvalues, irrespective of the frequency~$\theta$ (except possibly over a set of measure zero), are diverging linearly as~$n\to~\!\infty$. The following results then hold for the  idiosyncratic dynamic eigenvalues 
 and  those of the panel. 

\begin{lemma} \label{lem:dyn_eval} Under Assumptions (L1) and (L3),
\begin{compactenum}[(i)] 
\item there exists a constant $C^Z>0$ such that $\sup_{\theta\in[-\pi,\pi]} \lambda_{n1}^Z(\theta)\le C^Z$ for all $n\in\mathbb N$;
\item there exist a positive integer $\bar n$ and continuous functions $\alpha_{j}^Y$ and $\beta_{j-1}^Y$  from~$[-\pi,\pi]$ to $\mathbb R\,$,\linebreak $j=1,\ldots,q$, independent of $n$ and such that $0< \beta_{j-1}^Y(\theta) < \alpha_{j}^Y(\theta)\le {\lambda_{nj}^Y(\theta)}/{n}\le \beta_j^Y(\theta)\!<\infty$, $\theta$-a.e. in $[-\pi,\pi]$, all $j=1,\ldots, q$, and all $n>\bar n$; 
\item there exists a constant $C^Y>0$ such that $\sup_{\theta\in[-\pi,\pi]} \lambda_{n,q+1}^Y(\theta)\le C^Y$ for all $n\in\mathbb N$.
\end{compactenum}
\end{lemma}
As a consequence of Lemma \ref{lem:dyn_eval}, identification of the model, i.e., consistently disentangling  the unobserved common and idiosyncratic components, is possible, under the assumptions made   in the limit, as $n\to\infty$, thanks to the behaviour of the dynamic eigenvalues.

 Based on  results by \citet{andersondeistler08} for singular vector processes with a rational spectrum, \citet{fornilippi11} and \citet{FHLZ15} prove that, for generic values of the coefficients of the filters $\mbf b_i(L)$ as defined in Assumption (L2), the space spanned by $u_{j,t-k}$ for~$j=1,\ldots, q$ and $k\ge 0$ is the same as the space spanned by any $(q+1)$-dimensional subvector of $\mbf X_{t}$ and its lags;  moreover, those subvectors admit an autoregressive representation driven by the common shocks ${\bf u}_t$. 
   
   More precisely, any  $(q+1)$-dimensional subvector $\mbf X^\ddag_t$    of ${\mbf X}_{nt}$  admits an  autoregressive  representation of the form
\begin{equation}\label{VAReq:lev}
\mbf A^\ddag(L)\mbf X^\ddag_t= \mbf H^\ddag\mbf u_t,
\end{equation}
where $\mbf A^\ddag(L)$ is a finite-order VAR operator such that $\mbf A^\ddag(0)=\mbf I_{q+1}$, $\mbf u_t$ is  the vector of common shocks in \eqref{eq:common_lev_idio_vec},  and~$\mbf H^\ddag$ an appropriate~$(q+1)\times q$ matrix. On that representation, we make the following assumptions.
\begin{assumption}[L4] 
Let $\mbf X^\ddag_t$ be an arbitrary $(q+1)$-dimensional subvector of ${\mbf X}_{nt}$: the  autoregressive  representation \eqref{VAReq:lev} is such that 
\begin{compactenum}[(i)] 
\item $\mbf A^\ddag(L)$ is uniquely defined;
\item the degree $S^\ddag$ of $\mbf A^\ddag(z)$ is uniformly bounded, that is,  $S^\ddag\le S$ for some    integer $S>0$   independent of $n$ and the choice of the subvector ${\mbf X}^\ddag_t$;
\item $\text{\rm det}[\mbf A^\ddag(z)]\neq 0$ for all $z\in\mathbb C$ such that $|z|\le 1$;
\item $\mbf H^\ddag$ is $(q+1)\times q$, with full rank $q$;
\item denoting by $\bm\Gamma_h^{X^\ddag}$ the lag-$h$ autocovariances of $\mbf X^\ddag:=\{ \mbf X^\ddag_t \vert t\in\mathbb{Z} \}$  and defining 
\[
\bm{\mathcal C}^\ddag:=\l[
\ba{cccc}
\bm\Gamma_0^{X^\ddag}& \bm\Gamma_1^{X^\ddag}& \cdots & \bm\Gamma_{S-1}^{X^\ddag}\\
\bm\Gamma_{-1}^{X^\ddag}&\bm\Gamma_0^{X^\ddag}& \cdots & \bm\Gamma_{S-2}^{X^\ddag}\\
\vdots&\vdots&\ddots&\vdots\\
\bm\Gamma_{-S+1}^{X^\ddag}& \bm\Gamma_{-S+2}^{X^\ddag}& \cdots & \bm\Gamma_{0}^{X^\ddag}\\
\ea
\r],
\]
$\text{\rm det}(\bm{\mathcal C}^\ddag) >d>0$, where $d$ is independent of the choice of the subvector~${\mbf X}^\ddag_t$.
\end{compactenum}
\end{assumption}

This assumption allows us to derive an alternative representation of the GDFM \eqref{eq:idio_level} which is particularly useful for estimation and for the construction,  in Section \ref{sec:mod_vol} below, of a further GDFM for log-volatilities. Without loss of generality, let $n$ factorise into $n=m(q+1)$ for some positive integer~$m$, so that we can partition~${\mbf X}_n$ into~$m$ subprocesses, each of dimension $(q+1)$, of the form~${\mbf X}^{(k)}_t:=(X_{(k-1)(q+1),t}\ldots X_{k(q+1)-1,t})^\prime$, $k=1,\ldots,m$, with superscript $^{(k)}$ substituted for~$^\ddag$. Each ${\mbf X}^{(k)}$ satisfies \eqref{VAReq:lev} and Assumption~(L4). Defining the $n\times q$~ma\-trix~$\mbf H_n:=(\mbf H^{(k)'}\cdots  \mbf H^{(m)'})'$, we thus have the VAR representation  
\beq\label{comVAReq}
\mbf A_n(L) \mbf X_{nt} = \mbf H_n\mbf u_t,
\eeq
where $\mbf A_n(L)$ is $n\times n$ block-diagonal with diagonal blocks $\mbf A^{(1)}(L),\ldots, \mbf A^{(m)}(L)$. Moreover, in view of~\eqref{eq:common_lev_idio_vec}, we have $\l[\mbf A_n(L)\r]^{-1} \mbf H_n= \mbf B_n(L)$ (see Proposition 3 in \citealp{FHLZ17}).
Then,   the following alternative and equivalent representation of the GDFM holds:
\beq\label{eq:gdfm_lev_static}
\mbf A_n(L)\l\{\mbf Y_{nt}-\E[\mbf Y_{nt}]\r\} = \mbf H_n\mbf u_t+\mbf A_n(L)\mbf Z_n.
\eeq
The advantage of this  representation is that it is ``static'' in the sense that the common shocks $\mbf u$  now are loaded only contemporaneously and not via filters as  in \eqref{eq:common_lev_idio_vec}.

To conclude with, note that the  Yule-Walker equations 
\begin{align}\label{eq:yw_pop}
\big(\mbf A_1^\ddag\cdots\mbf A_S^\ddag\big)=\big(\bm\Gamma_1^{X^\ddag}\cdots \bm\Gamma_S^{X^\ddag} \big)\big[\bm{\mathcal C}^\ddag\big]^{-1},
\end{align}
characterising the $S$ matrix coefficients of~$\mbf A^\ddag(L)$ in \eqref{VAReq} 
 are well defined in view of part~{\it (v)} of Assumption~(L4); the same conclusion holds, blockwise, for the $n$ -dimensional VAR \eqref{comVAReq}. 

For ease of notation, define  the filtered processes 
\[
{\mbf Y}_n^*:=\mbf A_n(L)\l\{\mbf Y_n-\E[\mbf Y_n]\r\},\quad{\mbf X}^*_n:=\mbf A_n(L)\mbf X_n,\quad\text{and}\quad {\mbf Z}^*_n:=\mbf A_n(L)\mbf Z_n
\]
with traditional (static) covariance eigenvalues $\mu_{nj}^{Y^*}$, $\mu_{nj}^{X^*}$, and $\mu_{nj}^{Z^*}$, respectively. 
Since~\eqref{eq:gdfm_lev_static} is a static factor model, it is  natural to make the following assumption on the eigenvalues of the covariance of~${\mbf X}_n^*$ (see Assumption~4 in \citealp{FGLR09} or Assumption~6 in \citealp{FHLZ17}). Unless $q=1$, indeed, it  does not even follow from Assumption~(L3) that $\E ({\mbf X}_n^*{\mbf X}_n^{*\prime})$  has rank $q$.
 
\begin{assumption}[L5] There exist a positive integer $\bar n$ and constants  $a_{j}>b_{j-1}$, $j=1,\ldots, q$, independent of $n$ such that $0<a_j \le {\mu_{nj}^{X^*}}/{n}\le b_j<\infty$ for all $j=1,\ldots, q$ and all $n>\bar n$. 
\end{assumption}
The following results then hold for the eigenvalues $\mu_{nj}^{Z^*}$ and $\mu_{nj}^{Y^*}$  of the covariance matrices of  ${\mbf Z}^*_n$ and~${\mbf Y}^*_n$, respectively.
\begin{lemma}\label{lem:stat_eval}
Under Assumptions (L1), (L3), (L4), and (L5), 
\begin{compactenum}[(i)] 
\item there exists a constant $C^{Z^*}>0$ such that $\mu_{n1}^{Z^*}\le C^{Z^*}$ for all  $n\in\mathbb{N}$;
\item  there exist a positive integer $\bar n$ and constants $a_{j}^{Y^*}>b_{j-1}^{Y^*}$, $j=1,\ldots, q$, independent of $n$ such that  \linebreak
$0<a_j^{Y^*}\le {\mu_{nj}^{Y^*}}/{n}\le b_j^{Y^*}<\infty$ for all $j=1,\ldots, q$ and all $n>\bar n$;
\item there exists a constant $C^{Y^*}>0$ such that $\mu_{n,q+1}^{Y^*}\le C^{Y^*}$  for all  $n\in\mathbb{N}$.
\end{compactenum}
\end{lemma}

\subsection{Model and assumptions for volatilities}\label{sec:mod_vol}
We define the vector of common innovations (at time $t$) as the  $n$-dimensional vector$$\mbf e_{nt}:=(e_{1t},\ldots,e_{nt})^\prime:=\mbf H_n\mbf u _t;$$
for $n>q$,   the processes ${\mbf e}_n:=\{\mbf e_{nt}\vert t\in\mathbb{Z}\}$, $n\in\mathbb{N}$ clearly are singular. Then, letting $s_{it}:= e_{it}+ v_{it}$,
our log-volatility proxy is  
\beq
h_{it} :=\log s_{it}^2= \log(e_{it}+ v_{it})^2,
\eeq
yielding  the double-indexed stochastic process $\mbf h:= \{h_{it} \vert i\in\mathbb{N} ,  \  t\in\mathbb{Z}\}$, with $n$-dimensional sub-process\-es~$\mbf h_n:= \{h_{it} \vert i=1,\ldots,n,  \  t\in\mathbb{Z}\}$. We call $\mbf h$  the ``log-volatilities'' process. Similar definitions are used  in \citet{EM06} and our   previous work  (\citealp{barigozzihallin15a,barigozzihallin15b,barigozzihallin15c}, and \citealp{BHS18}). In order for such processes to be well defined we make the following assumption.  
\begin{assumption}[V0] For all $i\in\mathbb N$ and $t\in\mathbb Z$, $\vert s_{it}\vert >0$ almost surely.
\end{assumption}

This assumption makes sure that no cancellation can happen between common and idiosyncratic innovations; it is required, since    $e_i$ and $v_i$, although mutually orthogonal by  Assumption~(L1.vi), need not be mutually independent (assuming, for instance, that $e_i$ and $v_i$ are absolutely continuous is not sufficient).

Assuming a GDFM with $Q$ factors for the log-volatilities, we obtain  
\begin{align}
& h_{it}-\E[ h_{it}]= \chi_{it}+\xi_{it}\quad i\in\mathbb N ,\ t\in\mathbb Z\label{GDFMVol}\\
\text{with }\ & \chi_{it} = \sum_{j=1}^Q \sum_{k=0}^{\infty} f_{ijk} \varepsilon_{jt-k}=\mbf f_i'(L)\bm\varepsilon_t\quad\text{and}\quad \xi_{it}=\sum_{k=0}^{\infty} g_{ik} \nu_{it-k} = g_i(L) \nu_{it},\label{eq:idio_vol}
\end{align}
where $\E[h_{it}]$ is $h_{it}$'s expected value, $\chi_{it}$ and~$\xi_{it}$ are~$h_{it}$'s  {\it common} and {\it idiosyncratic} components,   and the pro\-cess\-es~$\bm\varepsilon:= \{\varepsilon_{jt} \vert j=1,\ldots,Q,  \  t\in\mathbb{Z}\}$ and $\bm\nu_{n}:= \{\nu_{it} \vert i=1,\ldots,n,  \  t\in\mathbb{Z}\}$, $n\in \mathbb N$ are  mutually orthogonal (at all leads and lags) $Q$-  and $n$-dimensional white noise, respectively.  Note that a GDFM for   log-volatilities implies a multiplicative GDFM representation \[
s_{it}^2=\exp(h_{it}) = \exp(\chi_{it}) \exp(\xi_{it}) \exp(\E[h_{it}]).
\]
 for the volatilities themselves. Letting  
 $$\bm\chi_n:= \{\chi_{it} \vert i=1,\ldots,n,  \  t\in\mathbb{Z}\}\quad\text{and}\quad\bm\xi_n:= \{\xi_{it} \vert i=1,\ldots,n,  \  t\in\mathbb{Z}\},$$
equations  \eqref{eq:idio_vol} in vector notation  take the form  
\beq\label{eq:common_vol_idio_vec}
\bm\chi_{nt}=\mbf F_n(L)\bm\varepsilon_t,\qquad \bm\xi_{nt}=\mbf G_n(L)\bm\nu_{nt}
\eeq
with $\mbf F_n(L):=(\mbf f_1(L)\ldots \mbf f_n(L))'$ and $\mbf G_n(L):=\text{\rm diag}(g_1(L)\ldots g_n(L))$.\medskip

 The following assumptions then are   the analogues, for  log-volatilities and \eqref{GDFMVol}-\eqref{eq:idio_vol} , of Assumption~(L1). 
\begin{assumption}[V1]$\,$
\begin{compactenum}[(i)] 
\item  The dimension $Q$ of  $\bm\varepsilon_t$ does not depend on $n$; the process  $\bm\varepsilon:=\{\bm\varepsilon_t \vert t\in\mathbb{Z}\}$ is    second-order white noise, with mean $\mbf 0_Q$ and diagonal  positive definite covariance~$\bm\Gamma^\varepsilon$;  
\item writing ${\mbf f}_{ik}:=(f_{i1k}\ldots f_{iqk})^\prime$ for  the $Q\times 1$ coefficient of $L^k$ in ${\mbf f}_i(L)$, there exists a constant $M_5>0$ such that   $\sum_{k=0}^{\infty}\Vert \mbf f_{ik}\Vert\, \vert k\vert\leq M_5$  for all $i\in\mathbb{N}$; 
\item the process $\{\bm\nu_{nt} \vert t\in\mathbb{Z}\}$ is     second-order white noise, with mean $\mbf 0_n$ and   positive definite covariance~$\bm\Gamma^\nu_n$;  moreover, $\E[\nu_{it}|\nu_{is}]=0$ for all $i\in\mathbb N$ and and $t,s\in\mathbb Z$ such that $t>s$;  
\item there exists a constant  $C_{\nu}>0$ such that $\Vert \bm\Gamma_n^{\nu}\Vert_1 \leq C_{\nu}$ for all $n\in\mathbb{N}$; 
 \item there exists a constant   $M_6>0$ such that  $\sum_{k=0}^{\infty}\vert  g_{ik}\vert\, \vert k \vert\le M_6$ for all $i\in\mathbb{N}$; 
\item $\Cov(\varepsilon_{jt},\nu_{is})=0$  for all $i\in\mathbb N$,  $j=1,\ldots, q$, and $t,s\in\mathbb Z$;
\item there exists a constant   $M_7>0$ such that $\sum_{k_1,k_2,k_3\in\mathbb Z} \vert\E[\varepsilon_{j_1t-k_1}\varepsilon_{j_2t-k_2}\varepsilon_{j_3t-k_3}\varepsilon_{j_4t}]\vert\le M_7$ for all \linebreak $j_1,j_2,j_3,j_4=1,\ldots,Q$; 
\item   there exists a constant   $M_8>0$ such that $\sum_{k_1,k_2,k_3\in\mathbb Z} \vert\E[\nu_{i_1t-k_1}\nu_{i_2t-k_2}\nu_{i_3t-k_3}\nu_{i_4t}]\vert\le M_8$ for all \linebreak$i_1,i_2,i_3,i_4\in\mathbb N$.
\end{compactenum}
\end{assumption}

The same comments made for Assumption (L1) apply here. Moreover, note that all moments of log-transforms of heavy-tailed variables  exist and are finite, even for stable distributions (see e.g. Theorem 5.8.1 in \citealp{UZ11}). Pursuing with assumptions, the following one is the log-volatility counterpart of (L2). 

\begin{assumption}[V2] $\,$
\begin{compactenum}[(i)] 
\item $\mbf f_i(L)$ has rational entries $f_{ij}(L)=\tilde\theta_{ij}(L)\tilde\phi_{ij}^{-1}(L)$, where $\tilde\phi_{ij}(z)$ and $\tilde\theta_{ij}(z)$, for all $i\in\mathbb N$ and $j=1,\ldots, Q$, are finite-order polynomials;
\item there exists a constant $\underline \phi>1$ such that $\tilde\phi_{ij}(z)\neq 0$  for all $i\in\mathbb N$, all $j=1,\ldots, Q$,  and  all~$z\in\mathbb C$ such that~$|z|\le \underline \phi$;
\item the coefficients $\tilde\theta_{ijk}$ of  $\tilde\theta_{ij}(L)$ are such that   $|\tilde\theta_{ijk}|\le B^\chi$ for some constant $B^\chi>0$    and all $i\in\mathbb N$,\linebreak $j=1,\ldots, Q$, and~$k\in\mathbb N\cup \{0\}$;
\item $g_i(L)$  is of the form  $p_i^{-1}(L)$  where  $p_i(z)$, for all $i\in\mathbb N$, is a finite-order polynomial,~$p_{i}(0)=1$ and $p_{i}(z)\neq 0$ for all $z\in\mathbb C$ such that $|z|\le 1$.
\end{compactenum}
\end{assumption}
Assumptions (V2.iv) implies that we can rewrite \eqref{eq:idio_vol} also as
\beq
p_i(L) Z_{it} = \nu_{it}.\label{eq:var_vol_idio}
\eeq
As in the case of levels, this assumption could be relaxed to allow for an infinite autoregressive order.

Let $\bm\Sigma_n^h(\theta)$, $\bm\Sigma_n^\chi(\theta)$, and $\bm\Sigma_n^\xi(\theta)$, $\theta\in[-\pi,\pi]$ denote the $n\times n$ spectral density matrices of $\mbf h_n$, its common and its idiosyncratic components, with $j$-th largest   eigenvalues  $\lambda_{nj}^h(\theta)$, $\lambda_{nj}^\chi(\theta)$ and $\lambda_{nj}^\xi(\theta)$,   respectively. As in~(L3), we assume the following.

\begin{assumption}[V3] There exist a positive integer $\bar n$ and continuous functions $\tilde\alpha_{j}(\theta)$ and $\tilde\beta_{j-1}(\theta)$ from~$[-\pi,\pi]$ to~$\mathbb R\,$, $j=1,\ldots,Q$,  such that  
$0<\tilde\beta_{j-1}(\theta)<\tilde\alpha_j(\theta)\le {\lambda_{nj}^\chi(\theta)}/{n}\le \tilde\beta_j(\theta)<\infty$, $\theta$-a.e. in $[-\pi,\pi]$, \linebreak all~$j=1,\ldots, Q$, and all $n>\bar n$. 
\end{assumption}

Finally,  the analogue (V4)  of (L4)   again is based on the representation results in \citet{FHLZ15}:  \linebreak any~$(Q+1)$- dimensional subvector $\bm\chi^\ddag_t$    of $\bm\chi_{nt}$  admits an  autoregressive  representation of the form
\begin{equation}\label{VAReq}
\mbf M^\ddag(L)\bm\chi^\ddag_t= \mbf R^\ddag\bm\varepsilon_t,
\end{equation}
where $\mbf M^\ddag(L)$ is a finite-order VAR operator such that $\mbf M^\ddag(0)=\mbf I_{Q+1}$, $\bm\varepsilon_t$ is  the vector of common shocks in~\eqref{eq:common_vol_idio_vec},  and~$\mbf R^\ddag$ an appropriate~$(Q+1)\times Q$ matrix. On that representation, we make the following assumptions: 
\begin{assumption}[V4] $\,$
\begin{compactenum}[(i)] 
\item $\mbf M^\ddag(L)$ is uniquely defined;
\item the degree $\tilde S^\ddag$ of $\mbf M^\ddag(z)$ is uniformly bounded, that is,  $\tilde S^\ddag\le \tilde S$ for some    integer $\tilde S>0$   independent of $n$ and the choice of the subvector ${\bm \chi}^\ddag_t$;
\item $\text{\rm det}[\mbf M^\ddag(z)]\neq 0$ for all $z\in\mathbb C$ such that $|z|\le 1$.
\item the $(Q+1)\times Q$ matrix  $\mbf R^\ddag$ has  full rank $Q$;
\item denoting by $\bm\Gamma_h^{\chi^\ddag}$ the lag-$h$ autocovariances of $\bm\chi^\ddag:=\{ \bm\chi^\ddag_t ,t\in\mathbb{Z} \}$  and defining $\bm{\mathcal V}^\ddag$ analogously to $\bm{\mathcal C}^\ddag$ in (L4),
$\text{\rm det}(\bm{\mathcal V}^\ddag) >\tilde d >0$, where $\tilde d$ is independent of the choice of the subvector~$\bm\chi^\ddag_t$.
\end{compactenum}
\end{assumption}

Now,  Assumption (V4) implies $\l[\mbf M_n(L)\r]^{-1} \mbf R_n= \mbf F_n(L)$,  
so that, assuming without loss of genera\-lity that~$n=\bar m(Q+1)$  (with $\bar m\neq m$ if $Q\ne q$)  and defining  a block-diagonal autoregressive operator $\mbf M_n(L)$ the way we defined $\mbf A_n(L)$ in the previous section, we can rewrite the GDFM for log-volatilities under the static form
\beq\label{eq:gdfm_vol_static}
\mbf M_n(L) \l\{\mbf h_{nt}-\E[\mbf h_{nt}]\r\} = \mbf R_n\bm\varepsilon_t+\mbf M_n(L)\bm\xi_{nt}.
\eeq
After defining, with obvious notation, the filtered processes $\vspace{1mm}{\mbf h}^*_n:=\mbf M_n(L)\l[\mbf h_n-\E[\mbf h_n]\r]$,  ${\bm\chi}^*_n:=\mbf M_n(L)\bm\chi_n$, and~${\bm\xi}^*_n:=\mbf M_n(L)\bm\xi_n$, with (static) spectral eigenvalues $\mu_{nj}^{h^*}$, $\mu_{nj}^{\chi^*}$, and $\mu_{nj}^{\xi^*}$, we conclude with the analogues of~(L5) and  Lemmas~\ref{lem:dyn_eval} and~\ref{lem:stat_eval} for the log-volatility panels.

\begin{assumption}[V5] There exist a positive integer $\bar n$ and constants  $\tilde a_{j}>\tilde b_{j-1}>0$, $j=1,\ldots, Q$, independent of~$n$ such that $0< \tilde a_j \le {\mu_{nj}^{\chi^*}}/{n}\le \tilde b_j<\infty$ for all $j=1,\ldots, Q$ and all $n>\bar n$.
\end{assumption}
We then have the following. 

\begin{lemma}\label{lem:eval_vol}
Under Assumptions (V0), (V1), (V3), (V4), and (V5),
\begin{compactenum}[(i)] 
\item there exists a constant $C^\xi>0$ such that $\sup_{\theta\in[-\pi,\pi]} \lambda_{n1}^\xi(\theta)\le C^\xi$ for all $n\in\mathbb N$;
\item  there exist a positive integer $\bar n$ and continuous functions $\alpha_{j}^h(\theta)$ and $\beta_{j-1}^h(\theta)$ from~$[-\pi,\pi]$ to $\mathbb R\,$, \linebreak $j=1,\ldots, Q$,~independent of~$n$ and such that~$0<\beta_{j-1}^h(\theta) <\alpha_j^h(\theta)\le {\lambda_{nj}^h(\theta)}/{n}\le \beta_j^h(\theta)\!<\infty$, $\theta$-a.e. in~$[-\pi,\pi]$, all $j=1,\ldots, Q$, and all $n>\bar n$; 
\item there exists a constant $C^h>0$ such that $\sup_{\theta\in[-\pi,\pi]} \lambda_{n,Q+1}^h(\theta)\le C^h$ for all~$n\in\mathbb N$;
\item there exists a constant $C^{\xi^*}>0$  such that $\mu_{n1}^{\xi^*}\le C^{\xi^*}$ for all  $n\in\mathbb{N}$;
\item  there exist a positive integer $\bar n$ and constants $a_{j}^{h^*}>b_{j-1}^{h^*}$, $j=1,\ldots, Q$, independent of $n$ such  \linebreak that~$0<a_j^{h^*}\le {\mu_{nj}^{h^*}}/{n}\le b_j^{h^*}<\infty$, for all~$j=1,\ldots, Q$ and all $n>\bar n$;
\item there exists a constant $C^{h^*}>0$ such that $\mu_{n,Q+1}^{h^*}\le C^{h^*}$  for all  $n\in\mathbb{N}$.
\end{compactenum}
\end{lemma}


\setcounter{equation}{0}

\section{Estimation, consistency, and rates}\label{sec:est}

Hereafter, the terminology ``estimation'', ``estimator'', etc.\ is used, in an orthodox way, for data-driven quantities attempting at evaluating  parameters (covariances, spectra, loadings, \ldots) but also, with a slight abuse, for data-driven quantities attempting at reconstructing unobserved variables (such as common factors, common and idiosyncratic components, \ldots ). All those ``estimators'', which are~${\bf Y}_{n,T}$-measurable random variables (hence  depend both on $n$ and $T$) are carrying  ``hats''.

\subsection{Summary of estimation}\label{sec:est_summary}
Estimation proceeds in two parts. The first part deals with the observed $n\times T$ panel ${\bf Y}_{n,T}$ of levels, 
and   follows  along similar lines as in  \citet{FHLZ17}, yielding estimated log-volatility proxies; the second part   consists in repeating   the same estimation steps, now based on those estimated  log-volatility quantities. 
Global consistency of the   procedure is discussed in the next section, along with further necessary conditions. 

To start with, we assume that  $q$ and $Q$ are known---an assumption we are relaxing later on. 
For simplicity of notation, we also  assume $\mbf Y_n$ and $\mbf h_n$ to be  centred, i.e., to have zero mean;  in practice, sample means are to be subtracted in order to obtain centred variables---which has no impact on consistency nor consistency rates.

Here is a detailed list of the steps required for estimation. Further comments on the choice of the quantities needed for estimation and a schematic description of the procedure are given at the end of this section (see also Algorithms 1 and  2). 
\ben
\item [(\textit {L.i})] To start with, compute the lag-window estimator 
\[
\widehat{\bm\Sigma}_{n}^Y(\theta_h):=\frac 1{2\pi} \sum_{k=-T+1}^{T-1}\mathrm K\l(\frac{k}{B_T}\r)e^{-ik\theta_h}\widehat{\bm\Gamma}_{nk}^Y,\quad \theta_h=\frac{\pi h}{B_T}, \quad \vert h\vert \le B_T,
\]
of the spectral density matrix of  returns, where $\widehat{\bm\Gamma}_{nk}^Y:=T^{-1}\sum_{t=|k|+1}^T {\mbf Y}_{nt}{\mbf Y}_{nt-|k|}'$ is the usual  lag-$k$ sample autocovariance matrix of levels and $\mathrm K$ is a suitable kernel with bandwidth $B_T$. 
We here adopt the common choice of a Bartlett kernel 
\[
\mathrm K\l(x\r)=\l\{\ba{cl}
1-|x|& \mbox{if } |x|\le 1\\
0& \mbox{otherwise},
\ea
\r. 
\]
but other classical kernels are also possible. 
  
\item [(\textit{L.ii})]  Collect the $q$ normalised column eigenvectors associated with $\widehat{\bm\Sigma}^Y_{n}(\theta_h)$'s $q$ largest 
  eigenvalues into the $n\times q$ matrix $\widehat{\bf P}^Y_{n} (\theta_h)$, and collect the corresponding eigenvalues into the~$q\times q$ diagonal matrix $\widehat{\bm \Lambda}^Y_{n} (\theta_h)$. Take 
  \[
\widehat{\bm\Sigma}_{n}^X(\theta_h):=\widehat{\bf P}^Y_{n} (\theta_h) \widehat{\bm \Lambda}^Y_{n} (\theta_h) \widehat{\bf P}^{Y\dag}_{n} (\theta_h),
\]
as an estimate of the spectral density matrix of the level-common component process~${\bf X}_n$. 
\item [(\textit{L.iii})] By inverse Fourier transform of $\widehat{\bm\Sigma}_{n}^X(\theta_h)$, estimate the autocovariance matrices of $\mbf X_n$: 
\[
\widehat{\bm\Gamma}^X_{nk}:= \frac{\pi}{B_T}\sum_{h=-B_T}^{B_T} e^{ik\theta_h} \widehat{\bm\Sigma}_{n}^X(\theta_h), \qquad k\in\mathbb Z.
\]
\item [(\textit{L.iv})] Assuming, for simplicity,
\footnote{{In practice, the last $n-\lfloor n/(q+1) \rfloor (q+1)$ cross-sectional items can be added to the last block in the analysis which will then have size larger than $(q+1)$. Since the arguments in \citet{FHLZ17} used in the next section apply to any partition of blocks of size $(q+1)$ or larger, nothing changes in what follows. }}
that $n=m(q+1)$, consider the $m$    diagonal  $(q+1)\times (q+1)$ blocks of the 
$\widehat{\bm\Gamma}^X_{nk}$'s. For each block, estimate, via Yule-Walker methods,  the coefficients of a~$(q+1)$-dimensional VAR  model (order determined via AIC or BIC). In other words, compute the sample analogue of \eqref{eq:yw_pop}. This yields, for the $\ell$-th diagonal  block, an estimator~$\widehat{\mbf A}^{(\ell)}(L)$ of the autoregressive filter $\mbf A^{(\ell)}(L)$ appearing in Assumption~(L4), hence an estimator $\widehat{\mbf A}_n(L)$ of the VAR filter $\mbf A_n(L)$. The resulting estimated  filtered process and its estimated covariance matrix are~$
\widehat{{\mbf Y}}^*_{nt}:=\widehat{\mbf A}_n(L)\mbf Y_{nt}$ and 
$\widehat{\bm\Gamma}^{\widehat{{Y}}^*}_n:=T^{-1}\sum_{t=1}^T\widehat{{\mbf Y}}^*_{nt}\widehat{{\mbf Y}}^{*'}_{nt}$, 
respectively.
\item [(\textit{L.v})] Collect the $q$ normalised (column) eigenvectors corresponding to $\widehat{\bm\Gamma}^{\widehat{{Y}}^*}_n$'s $q$ largest
  eigenvalues into the~$n\times q$ matrix $\widehat{\bf Q}^{\widehat{{Y}}^*}_{n}$. Projecting $\widehat{{\mbf Y}}^*_{nt}$ onto the space spanned by the columns of $\widehat{\bf Q}^{\widehat{{Y}}^*}_{n}$ provides an estimate $\widehat{\mbf e}_n$ of the innovation process~$\mbf e_n$. Taking into account the set of identifying restrictions described in Assumption~(I) below, we obtain the estimators
\[
\widehat{\mbf H}_n:= \sqrt n \widehat{\bf Q}^{\widehat{{Y}}^*}_{n},\qquad \widehat{\mbf u}_t:=\frac{1}{n}{\widehat{\mbf H}_n' \widehat{{\mbf Y}}^*_{nt}},\quad\text{and}\quad  \widehat{\mbf e}_{nt}:=\widehat{\mbf H}_n\widehat{\mbf u}_t=\widehat{\bf Q}^{\widehat{{Y}}^*}_{n}\widehat{\bf Q}^{\widehat{{Y}}^{*'}}_{n}\widehat{{\mbf Y}}^*_{nt}.
\]
Our estimator of the dynamic loadings  then is 
 $
\widehat{\mbf B}_n(L):= \widehat{\mbf A}_n^{-1}(L)\widehat{\mbf H}_n$, 
where we truncate the filter $\widehat{\mbf A}_n^{-1}(L)$ at some finite lag $\bar k_1$. From this we obtain an estimator 
$
\widehat{\mbf X}_{nt} := \widehat{\mbf B}_n(L)\widehat{\mbf u}_t$ 
of the common component. 
\item [(\textit{L.vi})] The resulting estimator of the idiosyncratic component is   $\widehat{\mbf Z}_{nt} := \mbf Y_{nt}-\widehat{\mbf X}_{nt}$. Fitting a univariate AR model (order determined via AIC or BIC), either by least squares or via Yule-Walker methods, to each of the $n$ components of~$\widehat{\mbf Z}_{nt}$ yields   estimators $\widehat{\mbf v}_n$ of the residuals and $\widehat{\mbf C}_n(L)$ of the diagonal matrix of coefficients from which we also obtain $\widehat{\mbf D}_n(L)\!:=\!\widehat{\mbf C}_n^{-1}(L)$ with $\widehat{\mbf C}_n^{-1}(L)$ truncated at some  finite lag~$\bar k_2$.
\item [(\textit{R})] For all $i=1,\ldots, n$ and $t=1,\ldots,T$, let $\widehat s_{it}:= \widehat e_{it}+\widehat v_{it}$ and define the estimated log-volatility proxies as  capped values of $\log (\widehat s_{it}^{\ 2})$:
\[
\widehat h_{it}:=\log (\widehat s_{it}^{\ 2})\ \mathbb I(\vert \widehat s_{it}\vert\ge \kappa_T) +\log (\kappa_T^2)\ \mathbb I(\vert \widehat s_{it}\vert< \kappa_T),
\]
where $\kappa_T> 0$ is a sequence of constants to be chosen in order to make our proxy robust to the log-transform. Note that consistency of our estimation procedure requires an adaptive choice of~$\kappa_T$, depending on the sample size as explained in Assumption (R) below. 
In particular, $\kappa_T$ must be strictly positive for consistency to hold.

\item [(\textit{V.i})]   Denote by $\widehat{\mbf h}_{nt}:=\big(\widehat h_{1t} \ldots \widehat h_{nt}\big)^\prime$, $t=1,\ldots,T$  the $n$-dimensional vector of log-volatility proxies   and compute  the lag-window estimator 
\[
\widehat{\bm\Sigma}_{n}^{\widehat{h}}(\theta_\ell):=\frac 1{2\pi} \sum_{k=-T+1}^{T-1}\mathrm K\l(\frac{k}{M_T}\r)e^{-ik\theta_\ell}\widehat{\bm\Gamma}_{nk}^{\widehat h},\quad \theta_\ell=\frac{\pi \ell}{M_T}, \quad \vert \ell\vert \le M_T,
\]
of its spectral density matrix,  where $\widehat{\bm\Gamma}_{nk}^{\widehat h}:=T^{-1}\sum_{t= |k|+1}^T {\widehat{\mbf h}}_{nt}{\widehat{\mbf h}}_{n,t-|k|}'$ is the  lag-$k$ sample autocovariance matrix of estimated log-volatilities. Here again we adopt the Bartlett kernel,   with bandwidth $M_T$, which could be different from $B_T$ in step (\textit{L.i}). 
\item [(\textit{V.ii})-(\textit{V.vi})] Repeat steps (\textit{L.ii})-(\textit{L.vi}) for $\widehat{\mbf h}_n$. In particular, steps (\textit{V.ii})-(\textit{V.v}) yield the estimators~$\widehat{\mbf M}_n(L)$ and $\widehat{\mbf R}_n$, from which we compute
\begin{align}
\widehat{\mbf h}_{nt}^* := \widehat{\mbf M}_n(L)\widehat{\mbf h}_{nt},\qquad
\widehat{\bm\varepsilon}_t:=\frac{1}{n}{\widehat{\mbf R}_n'\widehat{\mbf h}_{nt}^*},\quad \text{and}\quad 
\widehat{\mbf F}_n(L):= \widehat{\mbf M}_n^{-1}(L)\widehat{\mbf R}_n,\nn
\end{align}
while from (\textit{V.vi}) we obtain   $\widehat{\bm\nu}_{n}$ and $\widehat{\mbf P}_n(L)$, hence  $\widehat{\mbf G}_n(L):=\widehat{\mbf P}_n^{-1}(L)$. As before, $\widehat{\mbf M}_n^{-1}(L)$ and~$\widehat{\mbf P}_n^{-1}(L)$ are truncated at finite lags $\bar k_1^{*}$ and $\bar k_2^{*}$.
\een

\vskip .2cm
\begin{algorithm}\label{tab:alg1}
\DontPrintSemicolon
\footnotesize{  
  \KwInput{data in levels $\mathbf Y$ of dimension $n\times T$, number of factors $q$, bandwidth for estimating spectral density $B_T$, number of lags for impulse responses  $\bar k_1$ and $\bar k_2$, number of permutations for estimating the common component $nrep$}
  \KwOutput{common component $\widehat{\mathbf X}$, idiosyncratic component $\widehat{\mathbf Z}$, common shocks $\widehat {\mathbf u}$ and $\widehat{\mathbf e}$, common impulse responses $\widehat{\mathbf B}(L)$, idiosyncratic shocks $\widehat{\mathbf v}$, idiosyncratic impulse responses $\widehat{\mathbf D}(L)$}
\medskip
Compute autocovariance matrices of data $\widehat{\bm \Gamma}_{k}^Y$ for $|k|\le B_T$\smallskip

Compute the lag-window estimator of the spectral density matrix of data $\widehat{\bm\Sigma}^Y(\theta_h)$ for $\theta_h=\pi h/B_T$ and $|h|\le B_T$, using $\widehat{\bm \Gamma}_{k}^Y$ and the Bartlett kernel \smallskip
 
 \For{$h\leftarrow -B_T$ \KwTo $B_T$}
 {Compute the $q$ largest eigenvalues $\widehat \lambda_{1}^Y(\theta_h),\ldots,\widehat \lambda_{q}^Y(\theta_h)$ of $\widehat{\bm\Sigma}^Y(\theta_h)$ and collect the corresponding eigenvectors into the columns of 
 $\widehat{\mathbf P}^Y(\theta_h)$. Let $\widehat{\bm\Sigma}^X(\theta_h)=\widehat{\mathbf P}^Y(\theta_h)\mbox{diag}(\widehat \lambda_{1}^Y(\theta_h),\ldots,\widehat \lambda_{q}^Y(\theta_h))\widehat{\mathbf P}^{Y\dag}(\theta_h)$}\smallskip
 
Compute the autocovariance matrices of the common component $\widehat{\bm \Gamma}_{nk}^X$ for $|k|\le B_T$ by inverse Fourier transform of $\widehat{\bm\Sigma}^X(\theta_h)$\smallskip

\For{$\mathcal P\leftarrow 1$ \KwTo $nrep$}
{Choose a random partition $\mathcal P(1),\ldots, \mathcal P(m(q+1))$ of the $n$ series into $m=\lfloor n/(q + 1)\rfloor(q + 1)$ blocks 
such that the first $q$ series are always included 
and let $\mathbf Y_{\mathcal P}=(Y_{\mathcal P(1)},\ldots ,Y_{\mathcal P(m(q+1))})^\prime$\,\smallskip

\If({Add the last $n-m$ series to the last block}\,){$m(q+1)<n$}\smallskip

\For{$\ell \leftarrow 1$ \KwTo $m$}{Obtain the coefficients $\widehat{\mathbf A}^{(\ell)}_{\mathcal P}(L)$ fitting a VAR($p_1^{(\ell)}$) on $\mathbf Y^{(\ell)} := (Y_{\mathcal P((\ell-1)(q+1))}\ldots Y_{\mathcal P(\ell(q+1)-1)})^\prime$ via Yule Walker equations using  $\widehat{\bm \Gamma}_{k}^X$ for $k=0,\ldots, \ell$, with $p_1^{(\ell)}\le B_T$ and determined via BIC}\,\smallskip

Let $\widehat{\mathbf A}_{\mathcal P}(L)=\mbox{diag}(\widehat{\mathbf A}^{(1)}_{\mathcal P}(L),\ldots, \widehat{\mathbf A}^{(m)}_{\mathcal P}(L))$ and let $\widehat{\mathbf Y}_{t,\mathcal P}^{*}=\widehat{\mathbf A}_{\mathcal P}(L)\mathbf Y_{t,\mathcal P}$ for $t=1,\ldots, T$\,\smallskip

Compute $\widehat{\mathbf H}_{\mathcal P}$ as $\sqrt n$ times the $q$ leading eigenvectors of the sample covariance matrix of $\widehat{\mathbf Y}_{\mathcal P}^{*}$ \,\smallskip

Compute $\widetilde{\mathbf B}_{\mathcal P}(L)=\widehat{\mathbf A}^{-1}_{\mathcal P}(L)\widehat{\mathbf H}_{\mathcal P}$ truncating at lag $\bar k_1$\,\smallskip

Compute $\widehat{\mathbf B}_{\mathcal P}(L)=\widetilde{\mathbf B}_{\mathcal P}(L)\bm{\mathcal R}_{\mathcal P}$ with $\bm{\mathcal R}_{\mathcal P}$ is $q\times q$ orthogonal and such that the $q\times q$ block of $\widehat{\mathbf B}_{\mathcal P}(0)$ obtained by isolating  the rows corresponding to the first $q$ series in $\mathbf Y$ is lower triangular\,\smallskip

Compute $\widehat{\mathbf u}_{t,\mathcal P}=n^{-1}\bm{\mathcal R}_{\mathcal P}^\prime\widehat{\mathbf H}_{\mathcal P}^\prime\widehat{\mathbf Y}_{t,\mathcal P}^{*}$ for $t=1,\ldots, T$\,\smallskip
}

Compute the common shocks as $\widehat{\mathbf u}_t=(nrep)^{-1}\sum_{\mathcal P=1}^{nrep}\widehat{\mathbf u}_{t,\mathcal P}$ for $t=1,\ldots, T$\,\smallskip

Compute $\widehat{\mathbf e}_t=(nrep)^{-1}\sum_{\mathcal P=1}^{nrep}\widehat{\mathbf H}_{\mathcal P}\bm{\mathcal R}_{\mathcal P}\widehat{\mathbf u}_{t,\mathcal P}$ \,\smallskip

Compute the impulse response functions $\widehat{\mathbf B}(L)=(nrep)^{-1}\sum_{\mathcal P=1}^{nrep}\widehat{\mathbf B}_{\mathcal P}(L)$\,\smallskip

Compute the common component as $\widehat{\mathbf X}_{t}=\widehat{\mathbf B}(L)\widehat{\mathbf u}_{t}$ for $t=1,\ldots, T$\,\smallskip

Compute the idiosyncratic component as $\widehat{\mathbf Z}={\mathbf Y}-\widehat{\mathbf X}$ such that $\widehat{\mathbf Z}=(\widehat Z_1\ldots \widehat Z_n)^\prime$\,\smallskip

\For{$i\leftarrow 1$ \KwTo $n$}{Obtain the coefficients $\widehat{c}_i(L)$ fitting a VAR($s_{1i}$) on $\widehat{Z}_i$ via least squares, with $s_{1i}$ determined via BIC\,\smallskip

Let $\widehat{v}_{it}=\widehat{c}_i(L)\widehat{Z}_{it}$ for $t=1,\ldots, T$}\,\smallskip

Compute the impulse response functions as $\widehat{\mathbf D}(L)=\mbox{diag}(\widehat c_1^{\,-1}(L),\ldots, \widehat c_n^{\, -1}(L))$ truncating at lag $\bar k_2$\,\smallskip

Let the idiosyncratic shocks be $\widehat{\mathbf v}_t=(\widehat{v}_{1t}\ldots \widehat{v}_{nt})^\prime$ for $t=1,\ldots, T$
\smallskip
}
\caption{\small Estimation of dynamic factor model for levels}
\end{algorithm}
\begin{algorithm}\label{tab:alg2}
\DontPrintSemicolon
\footnotesize{  
  \KwInput{from Algorithm 1: common and idiosyncratic shocks $\widehat{\mathbf e}$ and $\widehat{\mbf v}$ both of dimension $n\times T$\\ 
  number of factors $Q$, capping constant $\kappa_T$, bandwidth for estimating spectral density $M_T$, number of lags for impulse responses~$\bar k_1^*$ and $\bar k_2^*$, number of permutations for estimating the common component $nrep$}
  \KwOutput{common component $\widehat{\bm\chi}$, idiosyncratic component $\widehat{\bm \xi}$, common shocks $\widehat {\bm\varepsilon}$ and $\widehat{\bm\eta}$, common impulse responses $\widehat{\mathbf F}(L)$, idiosyncratic shocks $\widehat{\bm\nu}$, idiosyncratic impulse responses $\widehat{\mathbf G}(L)$}
\medskip
\For{$i\leftarrow 1$ \KwTo $n$}{
\For{$t\leftarrow 1$ \KwTo $T$}{Compute log-volatility proxy $\widehat h_{it}$\,\smallskip

\If({$\widehat h_{it}=\log(\widehat e_{it}+\widehat v_{it})^2$}\,){$|\widehat e_{it}+\widehat v_{it}|\ge \kappa_T$\,}\smallskip
\Else($\widehat h_{it}=\kappa_T$)\,
}}
 

Compute autocovariance matrices of log-volatility $\widehat{\bm \Gamma}_{k}^{\widehat h}$ for $|k|\le M_T$\smallskip

Compute the lag-window estimator of the spectral density matrix of log-volatility $\widehat{\bm\Sigma}^{\widehat h}(\theta_h)$ for $\theta_h=\pi h/M_T$ and $|h|\le M_T$, using $\widehat{\bm \Gamma}_{k}^{\widehat h}$ and the Bartlett kernel \smallskip
 
 \For{$h\leftarrow -M_T$ \KwTo $M_T$}
 {Compute the $Q$ largest eigenvalues $\widehat \lambda_{1}^{\widehat h}(\theta_h),\ldots,\widehat \lambda_{Q}^{\widehat h}(\theta_h)$ of $\widehat{\bm\Sigma}^{\widehat h}(\theta_h)$ and collect the corresponding eigenvectors into the columns of 
 $\widehat{\mathbf P}^{\widehat h}(\theta_h)$. Let $\widehat{\bm\Sigma}^{\widehat \chi}(\theta_h)=\widehat{\mathbf P}^{\widehat h}(\theta_h)\mbox{diag}(\widehat \lambda_{1}^{\widehat h}(\theta_h),\ldots,\widehat \lambda_{Q}^{\widehat h}(\theta_h))\widehat{\mathbf P}^{\widehat h \dag}(\theta_h)$}\smallskip
 
Compute the autocovariance matrices of the common component $\widehat{\bm \Gamma}_{nk}^{\widehat \chi}$ for $|k|\le M_T$ by inverse Fourier transform of $\widehat{\bm\Sigma}^{\widehat \chi}(\theta_h)$\smallskip

\For{$\mathcal P\leftarrow 1$ \KwTo $nrep$}
{Choose a random partition $\mathcal P(1),\ldots, \mathcal P(m(Q+1))$ of the $n$ series into $m=\lfloor n/(Q + 1)\rfloor(Q + 1)$ blocks 
such that the first $Q$ series are always included 
and let $\widehat {\mbf h}_{\mathcal P}=({\widehat h}_{\mathcal P(1)},\ldots ,{\widehat h}_{\mathcal P(m(Q+1))})^\prime$\,\smallskip

\If({Add the last $n-m$ series to the last block}\,){$m(Q+1)<n$}\smallskip

\For{$\ell \leftarrow 1$ \KwTo $m$}{Obtain the coefficients $\widehat{\mathbf M}^{(\ell)}_{\mathcal P}(L)$ fitting a VAR($p_2^{(\ell)}$) on $\widehat{\mathbf h}^{(\ell)} := ({\widehat h}_{\mathcal P((\ell-1)(Q+1))}\ldots {\widehat h}_{\mathcal P(\ell(Q+1)-1)})^\prime$ via Yule Walker equations using  $\widehat{\bm \Gamma}_{k}^{\widehat \chi}$ for $k=0,\ldots, \ell$, with $p_2^{(\ell)}\le M_T$ and determined via BIC}\,\smallskip

Let $\widehat{\mathbf M}_{\mathcal P}(L)=\mbox{diag}(\widehat{\mathbf M}^{(1)}_{\mathcal P}(L),\ldots, \widehat{\mathbf M}^{(m)}_{\mathcal P}(L))$ and let $\widehat{\mathbf h}_{t,\mathcal P}^{*}=\widehat{\mathbf M}_{\mathcal P}(L)\widehat{\mathbf h}_{t,\mathcal P}$ for $t=1,\ldots, T$\,\smallskip

Compute $\widehat{\mathbf R}_{\mathcal P}$ as $\sqrt n$ times the $q$ leading eigenvectors of the sample covariance matrix of $\widehat{\mathbf h}_{\mathcal P}^{*}$ \,\smallskip

Compute $\widetilde{\mathbf F}_{\mathcal P}(L)=\widehat{\mathbf M}^{-1}_{\mathcal P}(L)\widehat{\mathbf R}_{\mathcal P}$ truncating at lag $\bar k_1^*$\,\smallskip

Compute $\widehat{\mathbf F}_{\mathcal P}(L)=\widetilde{\mathbf F}_{\mathcal P}(L)\bm{\mathcal R}_{\mathcal P}$ with $\bm{\mathcal R}_{\mathcal P}$ is $Q\times Q$ orthogonal and such that the $Q\times Q$ block of $\widehat{\mathbf M}_{\mathcal P}(0)$ obtained by isolating  the rows corresponding to the first $Q$ series in $\widehat{\mathbf h}$ is lower triangular\,\smallskip

Compute $\widehat{\bm \varepsilon}_{t,\mathcal P}=n^{-1}\bm{\mathcal R}_{\mathcal P}^\prime\widehat{\mathbf R}_{\mathcal P}^\prime\widehat{\mathbf h}_{t,\mathcal P}^{*}$ for $t=1,\ldots, T$\,\smallskip
}

Compute the common shocks as $\widehat{\bm\varepsilon}_t=(nrep)^{-1}\sum_{\mathcal P=1}^{nrep}\widehat{\bm\varepsilon}_{t,\mathcal P}$ for $t=1,\ldots, T$\,\smallskip

Compute $\widehat{\bm\eta}_t=(nrep)^{-1}\sum_{\mathcal P=1}^{nrep}\widehat{\mathbf R}_{\mathcal P}\bm{\mathcal R}_{\mathcal P}\widehat{\bm\varepsilon}_{t,\mathcal P}$ \,\smallskip

Compute the impulse response functions $\widehat{\mathbf F}(L)=(nrep)^{-1}\sum_{\mathcal P=1}^{nrep}\widehat{\mathbf F}_{\mathcal P}(L)$\,\smallskip

Compute the common component as $\widehat{\bm\chi}_{t}=\widehat{\mathbf F}(L)\widehat{\bm\varepsilon}_{t}$ for $t=1,\ldots, T$\,\smallskip

Compute the idiosyncratic component as $\widehat{\bm\xi}=\widehat{\mathbf h}-\widehat{\bm\chi}$ such that $\widehat{\bm\xi}=(\widehat {\xi}_1\ldots \widehat {\xi}_n)^\prime$\,\smallskip

\For{$i\leftarrow 1$ \KwTo $n$}{Obtain the coefficients $\widehat{p}_i(L)$ fitting a VAR($s_{2i}$) on $\widehat{\xi}_i$ via least squares, with $s_{2i}$ determined via BIC\,\smallskip

Let $\widehat{\nu}_{it}=\widehat{p}_i(L)\widehat{\xi}_{it}$ for $t=1,\ldots, T$}\,\smallskip

Compute the impulse response functions as $\widehat{\mathbf G}(L)=\mbox{diag}(\widehat p_1^{\,-1}(L),\ldots, \widehat p_n^{\, -1}(L))$ truncating at lag $\bar k_2^*$\,\smallskip

Let the idiosyncratic shocks be $\widehat{\bm\nu}_t=(\widehat{\nu}_{1t}\ldots \widehat{\nu}_{nt})^\prime$ for $t=1,\ldots, T$
\smallskip
}
\caption{\small Estimation of dynamic factor model for log-volatilities}
\end{algorithm}
An important remark needs to be made here. The cross-sectional  ordering of the panel has an impact on the selection of the  diagonal blocks in steps (\textit{L.iv}) and (\textit{V.iv}). Each cross-sectional  permutation of the panel, thus,  would lead to distinct estimators---all sharing  the same asymptotic properties. A Rao-Blackwell argument (see \citealp{FHLZ17} for details) suggests aggregating these estimators   into a unique one by  simple averaging (after obvious reordering of  the cross-section) of the resulting estimated shocks.~Although averaging over all~$n!$ permutations is clearly unfeasible, as stressed by \citet{FHLZ17} and verified empirically also in \citet{FGLS18}, a few of them are enough, in practice,  to deliver stable averages (which therefore are  matching the infeasible average over all~$n!$ permutations). 

Implementation of the above estimation steps is described in Algorithms 1 and 2. Those algorithms  require   setting   bandwidths $B_T$ and $M_T$ for the estimation of the spectral densities, a capping constant $\kappa_T$, and the number of factors $q$ and $Q$. Concerning the bandwidths and the capping constant, we refer to Section \ref{sec:ap} for the required asymptotic properties (see Assumptions (K) and (R), respectively), while a numerical assessment of the impact of these quantities is provided in Section \ref{sec:sim} on simulated data (see also the results in Appendix \ref{app:sim}) and in Section~\ref{sec:emp} on real data. Overall, our numerical analysis shows that low   levels of capping or even no capping at all  are  preferable, as they avoid inducing too much bias in the log-volatility distributions. As for the bandwidths, large values of $T$ are required to construct reliable estimates, since they allow setting $M_T$ large enough to capture the high persistence of log-volatility series. Our results are quite insensitive to the choice of $B_T$, due to the fact that financial returns typically are only weakly autocorrelated.

Finally, we can determine the numbers  $q$ and $Q$ of common shocks by means of the information criteria proposed by \citet{hallinliska07} and applied on the panels $\mbf Y_n$ and $\widehat{\mbf h}_n$, respectively.  The resulting data-driven estimators $\widehat{q}$ and $\widehat{Q}$ converge in probability to $q$ and $Q$, respectively. Since $q$ and $Q$ are integers, this means that, for any $\epsilon >0$,   there exist 
$n(\epsilon)$ and $T(\epsilon)$ such that, for all $n>n(\epsilon)$ and~$T>T(\epsilon)$,  $\widehat{q}=q$ and $\widehat{Q}=Q$ with probability larger than $1-\epsilon$. Hence, in Section~3.2 below, we safely can assume that $q$ and~$Q$ are known.


\subsection{Consistency and rates}\label{sec:ap}
Consistency of the estimators of the GDFM model for  levels is proved in \citet{FHLZ17}. Some differences exist, though,  between their approach and ours.
First,   \citet{FHLZ17} make slightly weaker assumptions on  idiosyncratic serial dependence and, by exploiting  results in \citet{WZ18} on spectral density estimation, they derive their consistency results under the constraint that~${B_T\log B_T}/T\to 0$ as $T\to\infty$. A more classical approach is adopted here, based on Assumptions~(L1) and~(V1), which as a consequence requires mildly stronger constraints on the range of admissible values for the bandwidths $B_T$ and $M_T$. Specifically, we require the following.

\begin{assumption}[K] As $T\to\infty$, ${B_T}=o(\sqrt T)$ and ${M_T}=o(\sqrt T)$.
\end{assumption}
Note that for $T\simeq 1000$ as in our empirical study,   the range of admissible bandwidths is still such that most of the serial dependence in the data is captured when estimating the spectral density (see Section \ref{sec:emp} for more details on the choice of the bandwidths).

Second, the results in \citet{FHLZ17} hold pointwise in $t$, which is not sufficient for our needs  when it comes  to prove consistency in the second part of the estimation procedure. Indeed, we need uniform (over all~$t\in\{1,\ldots,T\}$) consistency of the estimators of the common and idiosyncratic components. For this reason, we make additional assumptions on the distribution of common and idiosyncratic components. 

\begin{assumption}[T] There exist constants $K_u>0$, $K_{\varepsilon}>0$, $K_Z>0$, and $K_\xi>0$, such that, for any~$t=1,\ldots ,T$,
\begin{compactenum}[(i)] 
\item $\max_{j=1,\ldots, q}\Vert u_{jt}\Vert_{\psi_1}\le K_u$;
\item $\max_{j=1,\ldots, Q}\Vert \varepsilon_{jt}\Vert_{\psi_1}\le K_\varepsilon$;
\item  $\sup_{\bm w_n: \Vert\bm w_n\Vert = 1}\Vert \bm w_n'\mbf Z_{nt} \Vert_{\psi_1}\le K_Z$, for all $n\in\mathbb N$; 
\item  $\sup_{\bm w_n: \Vert\bm w_n\Vert = 1}\Vert \bm w_n'\bm \xi_{nt} \Vert_{\psi_1}\le K_\xi$, for all $n\in\mathbb N$.
\end{compactenum}
\end{assumption}

This assumption is equivalent to an assumption of sub-exponential tails of the common factors and the normed linear combinations of idiosyncratic components.  Specifically, it can be shown that (T{\it i}) is equivalent to requiring for any $j=1,\ldots,q$, that $\mathrm{P}(|u_{jt}|>\epsilon)\le K_u^* \exp\l(- {\epsilon}/K_{u}^{**}\r)$ for any $\epsilon>0$ and some finite $K_u^*,K_u^{**}>0$ (see also \citealp{vershynin12}, and Appendix \ref{app:prop1} for details). The same holds also for (T{\it ii}), (T{\it iii}), and (T{\it iv}).  See Remark 1 at the end of this section for a  discussion of the implications and possible relaxations of this assumption.

Two remarks on (T{\it iii}) and (T{\it iv}) are in order here (see  Sections 5.2.4 and 5.2.5 in \citealp{vershynin12} for   details). First, note that by letting $\bm w_n =(0\ldots w_i \ldots 0)'$, with $w_i=1$ for a given $i$, those assumptions imply that each idiosyncratic component has marginal sub-exponential distribution. Second,   an implication of Lemmas \ref{lem:stat_eval} and \ref{lem:eval_vol} is that   vectors of the form~$\bm w_n'\mbf Z_n$ and $\bm w_n'\bm\xi_n$ have finite variance for all $n$, a necessary condition for pointwise consistency. However, 
(T{\it iii}) and (T{\it iv}) are stricter on idiosyncratic cross-sectional dependence, since they control all moments of  normed linear combinations of idiosyncratic components. Indeed, since the common components~$\mbf X_n$ and $\bm\chi_n$ are recovered by aggregation across the~$n$ elements of $\mbf Y_n$ and $\widehat{\mbf h}_n$, respectively,  uniform consistency requires  limiting the contribution of the tails of the distribution of cross-sectional averages of idiosyncratic components. 

Finally, since  factors and factor loadings are not separately identified, we can, without loss of generality,  impose the following assumptions, which are just  identification constraints (see \citealp{FGLR09} for  similar conditions).
\begin{assumption}[I]
\begin{compactenum}[(i)] 
\item Denoting by $\mbf P^{X^*}_n$ the $n\times q$ matrix of normalized column eigenvectors corresponding to the $q$ largest eigenvalues of the covariance matrix of $\mbf X_n^*$, put  $\mbf H_n := \sqrt n\mbf P^{X^*}_n$ and~$\mbf u_t:={{\mbf P^{X^*}_n}'\mbf X_n^*}/{\sqrt{n}}$;
\item  denoting by  $\mbf P^{\chi^*}_n$ the $n\times Q$ matrix of normalized eigenvectors corresponding to the $Q$ largest eigenvalues of the covariance matrix of $\bm\chi_n^*$,  put $\mbf R_n := \sqrt n\mbf P^{\chi^*}_n$ and $\bm\varepsilon_t:= {{\mbf P^{\chi^*}_n}'\bm\chi_n^*}/{\sqrt{n}}$.
\end{compactenum}
\end{assumption}
In other words, Assumption (I) requires  the common factors $\mbf u_t$ ($\bm\varepsilon_t$) to be the (non-normalised) principal components of $\mbf X_n^*$ ($\bm\chi_n^*$). Note that, under Assumption~(I), both the factors and their loadings depend on $n$; their product, however, does not, which is particularly convenient and simplifies the proofs. Other   identification constraints are commonly used in principal component analysis (see e.g. \citealp{FLM13}); they do not affect the   results below, but lead to much heavier notation.

The consistency properties of the estimated GDFM for the levels as described in steps (\textit{L.i})-(\textit{L.vi}) are as follows.

\begin{proposition} \label{prop:level} Let $\rho_{nT}:=\max\big({B_T}/{\sqrt T}, 1/{B_T},  1/{\sqrt n}\big)$. 
Then, under Assumptions (L1)-(L5), (K), (T), and (I), there exists a $q\times q$ diagonal matrix $\mbf J$ with entries~$\pm 1$ such that
\begin{compactenum}[(a)]
\item $\max_{i=1,\ldots, n}\Vert \widehat{\mbf b}_{ik}'- \mbf b_{ik}'\mbf J\Vert=O_{\rm P}(\rho_{nT})$, for all $k\le \bar k_1$;
\item $\max_{t=1,\ldots, T}\Vert \widehat{\mbf u}_t-\mbf J\mbf u_t\Vert=O_{\rm P}(\rho_{nT}\log T)$;
\item $\max_{i=1,\ldots, n}\vert \widehat{d}_{ik}-  d_{ik}\vert=O_{\rm P}(\rho_{nT}\log^2 T)$,  for all $k\le \bar k_2$;
\item $\max_{i=1,\ldots, n}\max_{t=1,\ldots, T}\vert \widehat{v}_{it}-v_{it}\vert=O_{\rm P}(\rho_{nT}\log^2 T)$.
\end{compactenum}
\end{proposition}

The proof of parts  {\it (a)} and {\it (b)} of Proposition \ref{prop:level} follows directly from \citet{FHLZ17} together with Assumptions (T{\it i}) and (T{\it iii}). However, parts  {\it (c)} and {\it (d)} concerning the idiosyncratic components are new results and provide uniform consistency over both time and the cross-section (see also Remark 1 below).  In particular, notice that parts  {\it (c)} and {\it (d)}  of Proposition \ref{prop:level} are proved under Assumption (L2{\it iv}) of a finite-order autoregressive representation for the idiosyncratic component. Relaxing that assumption into possibly  infinite-order autoregressive repressentations would require addressing, in the proofs of parts {\it (c)} and {\it (d)}, the issue  of truncation errors related to finite-order~AR fitting. Consistency   still could be proved, but with  rates  depending on  the rate of decay of the autocovariances of  idiosyncratic components, as shown, for example, in \citet{denhaan97}. For simplicity, we do not consider this   here. 

As for the global consistency properties (after the second estimation step), we need a final condition on the choice of the capping sequence $\kappa_T$ in step (\textit{R}). 
\begin{assumption}[R] The sequence $\kappa_T>0$ is such that the sets ${\mathcal T}_{i;nT} :=\big\{t \in\{1,\ldots ,T\}\, \big\vert \,  \vert \widehat s_{it}\vert <~\! \kappa_T\big\}$ satisfy $\max_{i=1,\ldots, n}\vert{\mathcal T}_{i;nT}\vert =o_{\rm P}(\sqrt T)$ uniformly in $n$ as $T\to\infty$. Moreover, there exist  a positive integer~$\bar T$ and    constants~$\varphi>1$ and  $0< \underline c \leq  \overline c$, independent of $n$, such that $\underline c\le\kappa_T \log^{\varphi} T\le \overline c$ for all~$T>\bar T$.
\end{assumption}
The intuition behind this assumption is as follows.~As shown in Appendix \ref{app:prop2}, an immediate consequence of Proposition \ref{prop:level} is that the volatility proxies are consistently estimated, namely, 
\[
\max_{i=1,\ldots, n} \max_{t=1,\ldots,T} \vert\widehat s_{it}- s_{it}\vert = O_{\rm P}(\rho_{nT}\log^2 T),\quad\text{ as } n,T\to\infty.
\]
Now,   setting $\kappa_T=0$  in step (\textit{R}), then, due to the log-transform, uniform consistency of $\widehat h_{it}$ becomes problematic when $\widehat s_{it}$ gets ``close to zero''. For this reason, we need $\kappa_T>0$. The set ${\mathcal T}_{i;nT}$ is that of all time points $\{1,\ldots, T\}$ at which~$\widehat s_{it}$ is close to zero, and uniform consistency of $\widehat h_{it}$ for $t\in{\mathcal T}_{i;nT}^c$   straightforwardly follows from uniform consistency of $\widehat s_{it}$. On the other hand,  the sets ${\mathcal T}_{i;nT}$ should not contain too many time points, and have cardinality going to zero at appropriate rate---whence Assumption~(R). In particular, we suggest to choose $\kappa_T$ of the order of~$\log ^{-\varphi} T$  for all $i$.  
Although we do not have  theoretical results  justifying this choice of $\kappa_T$ in practice, simulation-based results (see Appendix \ref{sec:app_R})  indicate that, the condition on the cardinality of the sets ${\mathcal T}_{i;nT}$ is indeed satisfied for $\kappa_T$ decreasing logarithmically in $T$.

Consistency of the estimated GDFM for   log-volatilities as described in steps (\textit{R}) and (\textit{V.i})-(\textit{V.vi}) then follows. 
\begin{proposition}\label{prop:vol} 
Let $\tau_{nT}:=\max\big( {B_TM_T}/{\sqrt T},  {M_T}/{\sqrt n}\big)$ and assume that $B_T\ge c T^{1/4}$ for some finite $c>0$. \linebreak Then, under Assumptions (L1)-(L5), (V1)-(V5), (K), (T), (I), and (R), there exists a $Q\times Q$ diagonal matrix $\mbf S$ with entries $\pm 1$ such that
\begin{compactenum}[(a)]
\item $\max_{i=1,\ldots, n}\Vert \widehat{\mbf f}_{ik}'- \mbf f_{ik}'\mbf S\Vert=O_{\rm P}(\tau_{nT} \log^{3+\varphi} T)$ for all $k\le \bar k_1^*$;
\item $\max_{t=1,\ldots, T}\Vert \widehat{\bm \varepsilon}_t-\mbf S\bm \varepsilon_t\Vert=O_{\rm P}(\tau_{nT}\log^{4+\varphi} T)$;
\item $\max_{i=1,\ldots, n}\vert \widehat{g}_{ik}-  g_{ik}\vert=O_{\rm P}(\tau_{nT}\log^{5+\varphi} T)$ for all $k\le \bar k_2^*$;
\item $\max_{i=1,\ldots, n}\max_{t=1,\ldots, T}\vert \widehat{\nu}_{it}-\nu_{it}\vert=O_{\rm P}(\tau_{nT}\log^{5+\varphi} T)$.
\end{compactenum}
\end{proposition}

This result, which  is new, provides the theoretical foundation for the consistency of the estimators used in  \citet{barigozzihallin15a,barigozzihallin15b,barigozzihallin15c} and in this paper. Note that parts {\it (c)} and {\it (d)}, just as  parts {\it (c)} and {\it (d)} of Proposition \ref{prop:level}, are proved under Assumption (V2{\it iv}) of a finite-order autoregressive representation for the idiosyncratic components; the same comments as for Proposition \ref{prop:level} apply.

Our results show that, up to logarithmic factors and the bandwidth-related ones, the rates of consistency of our estimators are of order $\min(\sqrt T,\sqrt n)$ as in classical one-step factor models. 
The following three technical remarks discuss how our assumptions, in particular Assumptions (T) and (K), affect the consistency rates, and how the effect of those logarithmic  and  bandwidth-related factors  could be controlled further if we were willing to make additional assumptions.

\begin{remark}[Serial dependence of idiosyncratic components]\upshape{Inspection of the proof of part {\it (c)} of Proposition \ref{prop:level}  shows that the extra (with respect to part {\it (b)})  $\log T$ factor there is due to terms of the type $T^{-1}\sum_{t=1}^T Z_{it}$. Now, while the  cross-sectional dependence of idiosyncratic components is controlled via   Assumption (T{\it iii}), we   do not impose  (beyond weak stationarity)  any specific assumption on their serial dependence. However, it is worth noting that, if we made some  mild additional mixing assumption controlling that serial dependence, then those terms could be bounded by a Bernstein-type inequality, as for example in Theorem 1 by \citet{MPR11}. Similar comments   apply to Proposition \ref{prop:vol} and bounds on the idiosyncratic sums $T^{-1}\sum_{t=1}^T \xi_{it}$. If  such additional assumptions were made,  the rates in Proposition \ref{prop:level} parts {\it (c)} and {\it (d)} would change to $O_{\rm P}(\rho_{nT} \log T)$, those in Proposition \ref{prop:vol} part {\it (a)}   to $O_{\rm P}(\tau_{nT} \log^{1+\varphi} T)$, those  in  part {\it (b)}   to $O_{\rm P}(\tau_{nT}\log^{2+\varphi} T)$, and those in parts {\it (c)} and {\it (d)}   to $O_{\rm P}(\tau_{nT}\log^{2+\varphi} T)$.}
\end{remark}

\begin{remark}[Tail behavior]\upshape{
In Section \ref{sec:emp}, we analyze a panel of stock returns, and it is therefore worth discussing how our assumptions relate to the distributional properties of financial data.  First, let us stress that  it is common, in the financial econometrics literature, to assume Gaussianity of log-volatility proxies \citep[see e.g.][]{ABD02}. This is in agreement with the tail Assumptions~(T{\it ii}) and~(T{\it iv}) since sub-Gaussians tails are lighter than sub-exponentials. 
In the Gaussian case, the rates in Proposition \ref{prop:vol} part {\it (a)} would change to $O_{\rm P}(\tau_{nT}\log^{5/2+\varphi} T)$, those in
part {\it (b)} to $O_{\rm P}(\tau_{nT}\log^{3+\varphi} T)$, and those in parts  {\it (c)}, and {\it (d)} to $O_{\rm P}(\tau_{nT}\log^{7/2+\varphi} T)$.

Second,     Assumption (T{\it i}) straightforwardly generalizes to  more general classes of distributions such that, for some finite constants $K_u^*>0$, $K_u^{**}>0$, and~$\vartheta>0$, $\mathrm{P}(|u_{jt}|>\epsilon)\le K_u^* \exp\l(- {\epsilon}^\vartheta/K_{u}^{**}\r)$ for any $\epsilon>0$\linebreak  and~$j=1,\ldots,q$; (T{\it iii})   can be generalized similarly  for  level idiosyncratic components. These distributions are studied in the literature under the name of {\it sub-Weibull distributions} (\citealp{KC18}, and \citealp{VA19}) or {\it semi-exponential} (\citealp{borovkov00}).\footnote{Note that the assumption of a sub-Weibull tail decay is equivalent to the moment condition $(\E[|u_{jt}|^k])^{1/k}\le Ck^{1/\vartheta}$ for all $k\ge 1$ and some finite $C>0$ (see \citealp[Theorem 2.1]{VA19});   fourth-order moments in that case always exist.} By letting $\vartheta<1$, we could allow for tails, which, although still  exponentially decaying, could be heavier than  assumed in Assumption~(T), thus accounting for moderately extreme events. 
Following the same steps as in Appendix \ref{app:prop1}, it is easily seen that in this case the rates in Proposition \ref{prop:level} part {\it (b)}
would change to~$O_{\rm P}(\rho_{nT}\log^{1/\vartheta} T)$ and  those  in parts~{\it (c)} and {\it (d)})   to~$O_{\rm P}(\rho_{nT}\log^{2/\vartheta} T)$. As for Proposition \ref{prop:vol}, would we assume a sub-Weibull distribution also in (T{\it ii}) and (T{\it iv}) (with the same value of $\vartheta$), then rates would change to $O_{\rm P}(\tau_{nT}\log^{3/\vartheta+\varphi} T)$  in part {\it (a)}, to $O_{\rm P}(\tau_{nT}\log^{4/\vartheta+\varphi} T)$ in part {\it (b)}, and to~$O_{\rm P}(\tau_{nT}\log^{5/\vartheta+\varphi} T)$ in parts {\it (c)} and {\it (d)}. To conclude,   assuming sub-Gaussian tails in (T{\it ii}) and~(T{\it iv}) modifies the rates  in part {\it (a)} of Proposition \ref{prop:vol} 
 into $O_{\rm P}(\tau_{nT}\log^{2/\vartheta+1/2+\varphi} T)$, those in part {\it (b)} in\-to~$O_{\rm P}(\tau_{nT}\log^{2/\vartheta+1+\varphi} T)$, and those in parts {\it (c)} and {\it (d)}   into $O_{\rm P}(\tau_{nT}\log^{2/\vartheta+3/2+\varphi} T)$.


Finally, in principle, we also could assume {\it power-law decay}---that is, the existence of finite constants $K_u^*>0$ and  $\beta>0$ such that $\mathrm{P}(|u_{jt}|>\epsilon)\le K_u^* \epsilon^{-\beta}$ for any $j=1,\ldots,q$ and   $\epsilon>0$; we similarly could generalize~(T{\it iii}) for level idiosyncratic  components. We do not explore this possibility in detail, but we notice that , in order to  have consistency under this setting, we would need at least $\beta>2$; moreover, the smaller $\beta$,  the smaller   the range of admissible choices for the bandwidths $B_T$ and $M_T$. Notice however that, in practice, determining the actual values of~$\vartheta$ and $\beta$ is very tricky,  and that small values of $\vartheta$ can generate a tail behavior which is comparable to the power-law behavior (see Figure \ref{fig:powerlaw}).}
\end{remark}

\begin{figure}[t!]\caption{\small Comparison of the tails of log-normal (black), power-law (red) and sub-Weibull (blue) probability density functions $f(x)$ (log-scales on both axis).}\label{fig:powerlaw}
\centering \smallskip\noindent 
\setlength{\tabcolsep}{.01\textwidth}
\begin{tabular}{@{}c}
 \includegraphics[width=.7\textwidth,trim=1cm 1cm 2cm 0cm,clip]{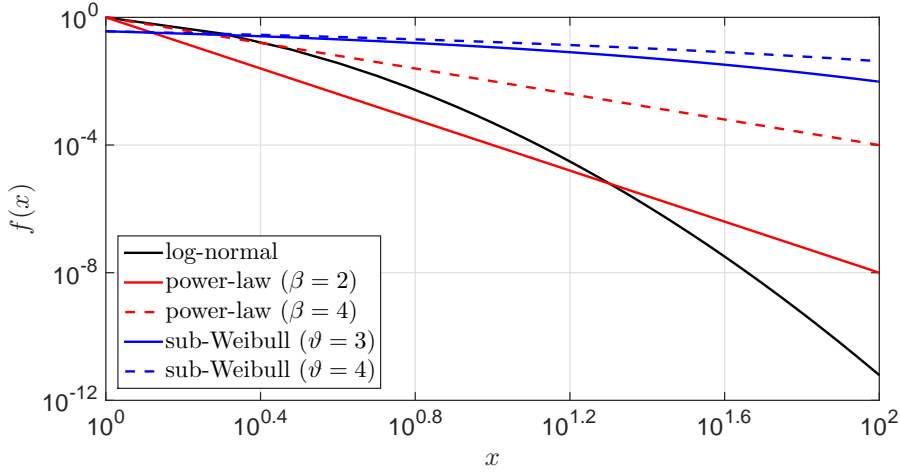} \\
\end{tabular}
\end{figure}

\begin{remark}[Bandwidths and estimation of spectral densities]\upshape{The results in Propositions \ref{prop:level} and \ref{prop:vol} require uniform consistency of the estimated spectral density over all frequencies. For this reason, we have stronger than usual asymptotic constraints on the bandwidths. These could be relaxed if we made stronger assumptions on the shocks. First, notice that in our setting the level shocks are just uncorrelated (see Assumptions (L1{\it i}) and~(L1{\it iii})), and are by no means independent. However, if we are willing to assume the existence, for level shocks,  of  moments of all orders,  then we could apply Theorem 7.7.4 in  \cite{brillinger2001}, which would allow us to replace~$B_{T}$ with~$B_T^\epsilon \sqrt {B_T}$ for any~$\epsilon>0$ in the definition of $\rho_{nT}$ in Proposition \ref{prop:level}. Second, we could, in principle, allow for independent shocks on log-volatilities (e.g. assuming Gaussianity, see Remark 1) and therefore make use of Theorem 4 and Section 4.2 in  \cite{WZ18}, which would allow us to replace $M_{T}$ with $\sqrt {M_T\log M_T}$ in the definition of $\tau_{nT}$ in Proposition \ref{prop:vol}.
}\end{remark}
}
\section{Conditional prediction intervals}\label{eq:int}
Before describing our prediction intervals,   let us summarise here the main notation developed  in the previous sections. Given an observed dataset of size $n\times T$, we have,  for the levels,
\begin{align}
Y_{it} &= X_{it} + Z_{it} +\E[Y_{it}],   \label{eq:summary1}\\
X_{it} &= \mbf b_{i0}' \mbf u_t + \sum_{k=1}^\infty \mbf b_{ik}' \mbf u_{t-k}:=e_{it} + X_{it|t-1},\quad Z_{it} = d_{i0} v_{it} + \sum_{k=1}^\infty d_{ik} v_{it-k} := v_{it} + Z_{it|t-1}, \nn\\
s_{it} &:= e_{it} + v_{it},\ \ i=1,\ldots, n,\   t=1,\ldots, T\nn
\end{align}
where  $d_{i0}=1$ because of \eqref{eq:var_lev_idio} and, for the log-volatilities,
\begin{align}
h_{it} &:= \log s_{it}^2 = \chi_{it}+ \xi_{it} + \E[h_{it}],  \label{eq:summary2}\\
\chi_{it}&= \mbf f_{i0}' \bm\varepsilon_t + \sum_{k=1}^\infty \mbf f_{ik}' \bm\varepsilon_{t-k}:=\eta_{it} + \chi_{it|t-1},\quad \xi_{it}= g_{i0} \nu_{it} + \sum_{k=1}^\infty g_{ik} \nu_{it-k} := \nu_{it} + \xi_{it|t-1}, \nn\\
\omega_{it} &:= \eta_{it} + \nu_{it},\ \   i=1,\ldots, n,\   t=1,\ldots, T\nn
\end{align}
where  $g_{i0}=1$ because of \eqref{eq:var_vol_idio}.

The optimal one-step-ahead linear predictors  of level $Y_{it}$ and log-volatility $h_{it}$ are thus
\begin{equation}\label{predpop}
Y_{it|t-1} :=X_{it|t-1} + Z_{it|t-1}+ \E[Y_{it}] \quad\text{and }\quad h_{it|t-1} :=\chi_{it|t-1} + \xi_{it|t-1}+ \E[h_{it}], 
\end{equation}
with innovations   $s_{it}$ and $\omega_{it}$, respectively. As a consequence, the level innovations are  
\[
s_{it} = \exp\big( {h_{it}}/{2}\big) \mbox{sign}(s_{it})= \exp\big({h_{it|t-1}}/{2}\big) \exp\big({\omega_{it}}/{2}\big) \mbox{sign}(s_{it}).
\]
We therefore define a one-step-ahead predictor of the volatilities as
\[
s_{it|t-1} := \exp\big({h_{it|t-1}}/2\big),
\]
with associated ``multiplicative innovations'' 
\[
w_{it} := \exp\big({\omega_{it}}/{2}\big) \mbox{sign}(s_{it}).
\]

Note, however, that, due to the nonlinear nature of the exponential transformation from $h_i$ to $s_i$,  this multiplicative decomposition of volatilities into a predictor and an ``innovation'' does not enjoy (in the space of volatilities) the traditional $L^2$  optimality properties, which only hold for their logarithms (in the space of log-volatilities). This, however, will not be a concern in the quantile-based construction we now describe, due to the fact that the coverage probabilities of  a interquantile interval are invariant under continuous monotone transformations: the quantile of $w_{it}$ .


Denoting by  $q(\alpha;w_{i})$  the (unconditional) $\alpha$-quantile of $w_{i}:=\{w_{it}\vert  t=1,\ldots, T\}$, $i=1,\ldots, n$ (which, by stationarity, does not depend on~$t$), theoretical 
 lower and upper prediction bounds with confidence level $(1-\alpha )$ and $\alpha\in(0,1)$ are
\begin{equation}\mathcal L_{it|t-1}(\alpha):=Y_{it|t-1} + s_{it|t-1}\, q(\alpha;w_{i}) \label{eq:LUtt1} 
\ \text{ and }\  
\mathcal U_{it|t-1}(\alpha):=Y_{it|t-1} + s_{it|t-1}\, q(1-\alpha;w_{i}), 
\end{equation}
respectively. Note that $Y_{it|t-1}$  lies above $\mathcal L_{it|t-1}(\alpha)$ for $\alpha <{\rm P}[w_{it}\leq 0]$ and lies  below $\mathcal U_{it|t-1}(\alpha)$ \linebreak  for~$\alpha <1-{\rm P}[w_{it}\leq 0]$. 
Prediction intervals with coverage probability $(1-\alpha )$ can be constructed as 
\beq\label{eq:Itt1}
\mathcal I_{it|t-1}(\alpha) :=\big[\,\mathcal L_{it|t-1}(\alpha^-) ,\, \mathcal U_{it|t-1}(\alpha^+)\; \big]
\eeq
with $\alpha^\pm<1/2$ \Black and  $\alpha^- + \alpha^+ =\alpha $, covering $Y_{it|t-1}$ (see \eqref{eq:Htt1}) provided~that 
\begin{equation}
\alpha^-<{\rm P}[w_{it}\leq 0]\quad\text{ and }\quad
\alpha^+ <1-{\rm P}[w_{it}\leq 0].\label{unqtails}
\end{equation}
Clearly, the lower bound $\mathcal L_{it|t-1}(\alpha)$ provides a measure of the Value-at-Risk of level $\alpha$ at time~$t$, which we denote as ${\rm VaR}_{it}(\alpha):=-\mathcal L_{it|t-1}(\alpha)$ (see Section 12.3.1 in \citealp{FZ11} for a review).\footnote{Usually, a Value-at-Risk is reported as a positive quantity. That will be the case with ${\rm VaR}_{it}(\alpha)$ for $\alpha$ small enough. Positive values of~$\mathcal L_{it|t-1}(\alpha)$ are possible, though: in such cases, ${\rm VaR}_{it}(\alpha)$ is defined to be zero by convention (see \citealp{FZ11}, Definition~12.1).}

The   advantage of quantile-based prediction intervals of the form \eqref{eq:Itt1} over their conditional heteroske\-dasticity-based competitors  stems from the fact that,  irrespective of the way $w_{i}$ has been obtained,  the conditional $\alpha$-quantiles of $Y_{it}$ (conditional on $Y_{i,t-1}, Y_{i,t-2}, \ldots$) {\it are}  of the form \eqref{eq:LUtt1}.  This quantile-based approach moreover allows for unequal tails ($\alpha^- \neq \alpha^+ $ in \eqref{unqtails}---hence, distinct attitudes towards losses and gains) and automatically takes into account  the typical  skewness of   financial data distributions. 
 
In practice, the model is estimated from a $n\times T$ observed panel; the empirical counterparts of~${Y}_{i,T+1|T} $ and~${h}_{i,T+1|T}$ for  $i=1,\ldots,n$   are\vspace{-3mm}
\[ 
\widehat{Y}_{i,T+1|T} = \widehat{X}_{i,T+1|T}+\widehat{Z}_{i,T+1|T}+ \frac 1 T\sum_{t=1}^T Y_{it}= \sum_{k=1}^{\bar k_1} \widehat{\mbf b}_{ik}'\widehat{\mbf u}_{T-k+1} + \sum_{k=1}^{\bar k_2} \widehat{d}_{ik}\widehat{v}_{i,T-k+1}+ \frac 1 T\sum_{t=1}^T Y_{it}\vspace{-3mm} \]
and\vspace{-3mm}
\[\widehat{h}_{i,T+1|T}=\widehat{\chi}_{i,T+1|T}+\widehat{\xi}_{i,T+1|T}+ \frac 1 T\sum_{t=1}^T \widehat h_{it}= \sum_{k=1}^{\bar k_1^*} \widehat{\mbf f}_{ik}'\widehat{\bm\varepsilon}_{T-k+1} + \sum_{k=1}^{\bar k_2^*} \widehat{g}_{ik}\widehat{\nu}_{i,T-k+1}+ \frac 1 T\sum_{t=1}^T \widehat h_{it},
\vspace{-1mm}\] 
and we accordingly define 
$
\widehat{s}_{i,T+1|T}:=\exp\big({\widehat{h}_{i,T+1|T}}/{2}\big)
$; 
  based on the estimates  $ 
\widehat s_{it}$ and~$ 
\widehat{\omega}_{it}$  of   $s_{it}$ and~$\omega_{it}$, let~$\widehat w_{it}:=\exp\big({\widehat {\omega}_{it}}/2\big)\text{sign}(\widehat s_{it})$. 

For any $i$, denote by $\widehat w_{i(1)},\ldots, \widehat w_{i(T)}$ the order statistic of $\widehat w_{i1},\ldots, \widehat w_{iT}$; the empirical quantile~$w_{i(\lceil T\alpha\rceil)}$ then can be used as an estimator of $q(\alpha;w_{i})$.  Empirical versions of the prediction limits and  intervals \eqref{eq:LUtt1} and \eqref{eq:Itt1} are 
\beq\nn
\widehat{\mathcal L}_{i,T+1|T}(\alpha):=\widehat Y_{i,T+1|T} + \widehat s_{i,T+1|T}\, \widehat w_{i(\lceil T\alpha\rceil)},\quad 
\widehat{\mathcal U}_{i,T+1|T}(\alpha):=\widehat Y_{i,T+1|T} + \widehat s_{i,T+1|T}\, \widehat w_{i(\lceil T(1-\alpha)\rceil)}
\eeq
and  
\beq\label{eq:Itt1hat}
\widehat{\mathcal I}_{i,T+1|T}(\alpha) :=\big[\widehat{\mathcal L}_{i,T+1|T}(\alpha^-),\widehat{\mathcal U}_{i,T+1|T}(\alpha^+)\big] 
\eeq
with  $\alpha^\pm<1/2$ and $\alpha^-+\alpha^+=\alpha\in(0,1)$. A schematic description of this procedure is given in Algorithm~3.

 If the $w_{it}$'s were i.i.d.\  instead of weak white noise, the convergence (for given $\alpha^-$ and $\alpha^+$, without rates) of~\eqref{eq:Itt1hat} to~\eqref{eq:Itt1} would follow from the fact that, as a consequence of the consistent estimation of the GDFMs for levels and volatilities, for any~$n_0$ and $T_0$,  $\max_{1\leq i\leq n_0}\max_{1\leq t\leq T_0}\vert\widehat w_{it}-w_{it}\vert$ converges to zero   as $n$ and $T$ tend to infinity.

 Then, the difference between the empirical quantile of order~$\alpha$ computed from $\{\widehat w_{1t},\ldots,\widehat w_{iT_0}\}$ and the empirical quantile of order~$\alpha$ computed from the unobservable $\{w_{i1},\ldots,w_{iT_0}\}$ is~$o_{\rm P}(1)$ for given $1\leq i\leq n_0$ as $n$ and~$T$ tend to infinity.   Now, for given $i$, were the $w_{it}$'s i.i.d.,  the empirical $\alpha$-quantile computed from $\{w_{i1},\ldots,w_{iT_0}\}$ is, for~$T_0$ large enough,    arbitrarily close to its theoretical counterpart $q(\alpha; w_i)$ with probability arbitrarily close to one. The same conclusion extends to the present case where the  $w_{it}$'s are stationary and uncorrelated provided that they satisfy some additional mild ergodicity or mixing assumption. The literature on Glivenko-Cantelli and quantile consistency under ergodicity and mixing is abundant, and we will not proceed with imposing any specific mixing conditions here which anyway hardly can be checked from the data. The reader may like to refer to Theorem 3.1 in \citet{FZ19} for details.

Once prediction regions have been constructed, it is important to evaluate their actual coverage performance.  For this, it is useful to define  the conditional coverage  indicators---namely, for prediction intervals $ \widehat{\mathcal I}_{i,T+1|T}(\alpha)$, 
\beq\label{eq:Htt1}
\widehat{\mathcal H}_{i,T+1|T}(\alpha):=\mathbb I\big(Y_{i,T+1}\in \widehat{\mathcal I}_{i,T+1|T}(\alpha)\big).
\eeq
For a given $i$, we say that $ \widehat{\mathcal I}_{i,T+1|T}(\alpha)$ provides the correct coverage if 
\beq\nn
\mathrm{P} (Y_{i,T+1}\in  \widehat{\mathcal I}_{i,T+1|T}(\alpha) | Y_{i,T},\ldots ,Y_{i1} )=\E[\widehat{\mathcal H}_{i,T+1|T}(\alpha) | Y_{i,T},\ldots ,Y_{i1}]=(1-\alpha), 
\eeq
which is   equivalent   (see e.g. Lemma 1 in \citealp{christoffersen1998}) to the hypothesis that 
	\beq\label{eq:null}
	\widehat{\mathcal H}_{i,T+1|T}(\alpha)\stackrel{iid}{\sim}\mbox{Bernoulli}(1-\alpha).
	\eeq 
	That hypothesis can   be tested  against alternatives of insufficient coverage probability values, against non-sharp prediction limits, or against alternatives of serial dependence. We refer to Section~\ref{backtestSec} for details and  implementation.  

\begin{algorithm}[t]
\DontPrintSemicolon
\footnotesize{  
  \KwInput{
  data in levels $\mbf Y$, $\alpha^-\in[0,1/2]$ and  $\alpha^+\in[0,1/2]$ such that the confidence level is $\alpha=\alpha^++\alpha^-\in(0,1)$\\
  from Algorithm 1: common level shocks $\widehat{\mathbf u}$ of size $q\times T$ and $\widehat{\mathbf e}$ of size $n\times T$, idiosyncratic level shocks $\widehat{\mbf v}$ of size $n\times T$, common level impulse responses $\widehat{\mbf B}(L)$ of size $n\times q\times \bar k_1$, idiosyncratic level impulse responses $\widehat{\mbf D}(L)$ of size $n\times n\times \bar k_2$\\
  from Algorithm 2: log-volatility proxy $\mbf h$ of size $n\times T$, common log-volatility shocks $\widehat{\bm\varepsilon}$ of size $Q\times T$ and $\widehat{\bm\eta}$ of size $n\times T$, idiosyncratic log-volatility shocks $\widehat{\bm \nu}$ of size $n\times T$, common level impulse responses $\widehat{\mbf F}(L)$ of size $n\times Q\times\bar k_1^*$, idiosyncratic level impulse responses $\widehat{\mbf G}(L)$ of size $n\times n\times \bar k_2^*$}
  \KwOutput{lower bounds of conditional prediction interval $\widehat{\mathcal L}_{1,T+1|T}(\alpha^-),\ldots,\widehat{\mathcal L}_{n,T+1|T}(\alpha^-)$\\
  upper bounds of conditional prediction interval $\widehat{\mathcal U}_{1,T+1|T}(\alpha^+),\ldots,\widehat{\mathcal U}_{n,T+1|T}(\alpha^+)$}
\medskip

Compute $\bar {\mbf Y}$ the sample mean of levels $\mbf Y$\,\smallskip

Compute the one-step-ahead prediction of common and idiosyncratic components of levels $\widehat{\mbf X}_{T+1|T}=\sum_{k=1}^{\bar k_1} \widehat{\mbf B}_k\widehat{\mathbf u}_{T-k+1}$\,\smallskip

Compute the one-step-ahead prediction of idiosyncratic component of levels $\widehat{\mbf Z}_{T+1|T}=\sum_{k=1}^{\bar k_2} \widehat{\mbf D}_k\widehat{\mathbf v}_{T-k+1}$\,\smallskip

Compute the one-step-ahead prediction of levels $\widehat{\mbf Y}_{T+1|T}=\widehat{\mbf X}_{T+1|T}+\widehat{\mbf Z}_{T+1|T}+\bar {\mbf Y}$ such that $\widehat{\mbf Y}_{T+1|T}=(\widehat{Y}_{1,T+1|T}\ldots \widehat{Y}_{n,T+1|T})^\prime$\,\smallskip

Compute $\bar {\widehat{\mbf h}}$ the sample mean of log-volatilities $\widehat{\mbf h}$\,\smallskip

Compute the one-step-ahead prediction of common component of log-volatilities $\widehat{\bm\chi}_{T+1|T}=\sum_{k=1}^{\bar k_1^*} \widehat{\mbf F}_k\widehat{\bm\varepsilon}_{T-k+1}$\,\smallskip

Compute the one-step-ahead prediction of idiosyncratic component of log-volatilities $\widehat{\bm\xi}_{T+1|T}=\sum_{k=1}^{\bar k_2^*} \widehat{\mbf G}_k\widehat{\bm\nu}_{T-k+1}$\,\smallskip

Compute the one-step-ahead prediction of log-volatilities $\widehat{\mbf h}_{T+1|T}=\widehat{\bm\chi}_{T+1|T}+\widehat{\bm\xi}_{T+1|T}+\bar {\widehat{\mbf h}}$\,\smallskip

Compute the one-step-ahead prediction of volatilities $\widehat{\mbf s}_{T+1|T}=\exp(\widehat{\mbf h}_{T+1|T}/2)$ such that $\widehat{\mbf s}_{T+1|T}=(\widehat{s}_{1,T+1|T}\ldots \widehat{s}_{n,T+1|T})^\prime$\,\smallskip\,\smallskip

Compute the log-volatility innovations $\widehat{\bm\omega}_t=\widehat{\bm\eta}_t+\widehat{\bm\nu}_t$ for $t=1,\ldots, T$\,\smallskip

Compute the volatility proxy $\widehat{\mbf s}_{t}=\exp(\mbf h_t/2)$ or equivalently $\widehat{\mbf s}_{t}=\widehat{\mbf e}_t+\widehat{\mbf v}_t$ for $t=1,\ldots, T$\,\smallskip

Compute the volatility innovations $\widehat{\mbf w}_{t}=\exp(\widehat{\bm\omega}_t/2)\mbox{sign}(\widehat{\mbf s}_t)$ 
such that $\widehat{\mbf w}_{t}=(\widehat{w}_{1t}\ldots \widehat{w}_{nt})^\prime$ for $t=1,\ldots, T$\,\smallskip
\,\smallskip

\For{$i\leftarrow 1$ \KwTo $n$}{Compute the order statistics $w_{i(\lceil T\alpha^-\rceil)}$ and $w_{i(\lceil T(1-\alpha^+)\rceil)}$ of $w_i$\,\smallskip

Compute the lower bound $\widehat{\mathcal L}_{i,T+1|T}(\alpha^-)=\widehat Y_{i,T+1|T} + \widehat s_{i,T+1|T}\, \widehat w_{i(\lceil T\alpha^-\rceil)}$\,\smallskip

Compute the upper bound $\widehat{\mathcal U}_{i,T+1|T}(\alpha^+)=\widehat Y_{i,T+1|T} + \widehat s_{i,T+1|T}\, \widehat w_{i(\lceil T(1-\alpha^+)\rceil)}$\,\smallskip
}
 
}
\caption{\small Estimation of conditional prediction intervals}
\end{algorithm}

\setcounter{equation}{0}
\section{Simulation study}\label{sec:sim}

\subsection{Setup}
To study the performance of our estimator on finite samples, we simulate data ($\mathcal M$ replications) according to the model described in \eqref{eq:summary1}-\eqref{eq:summary2}. 

For each Monte Carlo replication $m=1,\ldots, \mathcal M$ and for given values of $n,T,q$, and $Q$, we first simulate a multiplicative factor model for the volatilities which in turn implies a factor structure also for the levels. The common component of the log-volatilities is generated as 
\[
\bm\chi_{nt,m}:=(\mbf M_{n,m}(L))^{-1} \mbf R_{n,m}\bm\varepsilon_{t,m}, \quad t=1,\ldots, T,
\]
where $\bm\varepsilon_{t,m}\stackrel{iid}{\sim}  N(\mbf 0_Q,\mbf I_Q)$, $\mbf R_{n,m}$ is $n\times Q$ with entries $[\mbf R_{n,m}]_{ij}\stackrel{iid}{\sim} N(0,1)$ and rescaled such that $\mbf R_{n,m}^\prime \mbf R_{n,m}=n$,\linebreak and $\mbf M_{n,m}(L)=\mbf I_n-\sum_{k=1}^3\mbf M_{kn,m}L^k$ where the coefficients $\mbf M_{kn,m}$ are diagonal $n\times n$ matrices with entries~$[\mbf M_{kn,m}]_{ij}\stackrel{iid}{\sim} N(0,1)$ and rescaled in such a way that $\det(\mbf M_{n,m}(z))\ne 0$ for $|z|\le 1$.\footnote{In particular, when looking at simulated data 25\% of the total $3n^2$ roots are found to be in the range $(0.7,1)$, thus accounting for high persistence in log-volatilities, see also Table \ref{tab:dgp1} below.} Then, we generate the process
\[
\bm\xi_{nt,m}^*:=(\mbf P_{n,m}^*(L))^{-1} \bm\nu_{nt,m}^*, \quad t=1,\ldots, T,
\]
where $\bm\nu_{nt,m}^*\stackrel{iid}{\sim}  N(\mbf 0_n,\bm \Sigma_{n,m})$, with $\bm \Sigma_{n,m}$  a Toeplitz matrix with entries $[\bm \Sigma_{n,m}]_{ij}:=0.5^{|i-j|}$, if $|i-j|\le 2$ and zero otherwise, and $\mbf P_{n,m}^*(L)$  generated in the same way as $\mbf M_{n,m}(L)$. Denoting by~$\xi^*_{it,m}$  the $i$th element of~$\bm\xi^*_{it,m}$, we rescale it into $\xi^{**}_{it,m}:= \xi^*_{it,m} [\Var( \chi_{it,m})/\{2\Var( \xi^*_{it,m})\}]^{1/2}$ so that the signal-to-noise ratio is~2.

Define\vspace{-1mm} 
\begin{align}
e_{it,m}^*&:=\exp(\chi_{it,m}/2)\ \pi_{it,m},\;\mbox{ and }\; v_{it,m}^*:=\exp(\chi_{it,m}/2)\ \exp(\xi^{**}_{it,m}/2) \pi_{it,m},\quad t=1,\ldots, T,\  i=1,\ldots, n,\nn
\end{align}
where $\pi_{it,m}=\pm1$ with equal probabilities 0.5 and 
 $\chi_{it,m}$ is the $i$th element of~$\bm\chi_{t,m}$. The volatility and log-volatility proxies  then are
\begin{align}
s_{it,m}^2&:=(e_{it,m}^*+v_{it,m}^*)^2 = \exp(\chi_{it,m})\l[1+\exp(\xi^{**}_{it,m})+2\exp(\xi^{**}_{it,m}/2) \r],\nn\\
h_{it,m}&:=\log(s_{it,m}^2)= \chi_{it,m}+\log \l[1+\exp(\xi^{**}_{it,m})+2\exp(\xi^{**}_{it,m}/2) \r],\; t=1,\ldots, T,\  i=1,\ldots, n,\nn
\end{align}
from which we see that, since each $\chi_{i,m}$ is driven by the $Q$-dimensional vector of shocks $\bm\varepsilon_m$, it has the role of common log-volatility, while the $n$ shocks $\bm\nu^*_{n,m}$ have only an idiosyncratic role.

Letting $\mbf V$ be the $q$ normalized eigenvectors corresponding to the $q$ largest eigenvalues of the sample covariance of the vector $\mbf e_{nt,m}^*:=(e_{1t,m}^*\ldots e_{nt,m}^*)^\prime$, we build the level shocks as
\begin{align}
\mbf e_{nt,m}&:=\mbf V\mbf V^\prime\mbf e^*_{nt,m},\;\mbox{ and }\;\mbf v_{nt,m}:=\mbf V_\perp\mbf V_\perp^\prime\mbf e^*_{nt,m}+\mbf v_{nt,m}^*,\quad t=1,\ldots, T,\nn
\end{align}
where $\mbf V_\perp$ is $n\times (n-q)$ such that $\mbf V_\perp^\prime \mbf V=\mbf 0_{(n-q)\times q}$, and $\mbf v_{nt,m}^*:=(v_{1t,m}^*\ldots v_{nt,m}^*)^\prime$. Note that, by construction, the elements $e_{it,m}$ and $v_{it,m}$ of the vectors $\mbf e_{nt,m}$ and $\mbf v_{nt,m}$ are such that    $(e_{it,m}+v_{it,m})= (e_{it,m}^*+ v_{it,m}^*)$: therefore, we can also write $s_{it}^2=(e_{it,m}+v_{it,m})^2$.

Finally, we  generate the vectors of common and idiosyncratic components of the levels as
\begin{align}
\mbf X_{nt,m} &:= (\mbf I_n-\mbf A_{n,m}L)^{-1}\mbf e_{nt,m}, \;\mbox{ and }\;\mbf Z_{nt,m}:= (\mbf I_n-\mbf C_{n,m}L)^{-1}\mbf v_{nt,m}, \quad t=1,\ldots, T\nn
\end{align}
where $\mbf A_{n,m}$ is a diagonal $n\times n$ matrix with entries $[\mbf A_{n,m}]_{ij}\stackrel{iid}{\sim}  U[-0.3,0.7]$, and $\mbf C_{n,m}$ is generated in the same way but with entries from a uniform distribution over $[\mbf C_{n,m}]_{ij}\stackrel{iid}{\sim} U[-0.5,0.5]$; since these matrices are diagonal, the autoregressive models for $\mbf X_{n,m}$ and $\mbf Z_{n,m}$ are causal. The panel of
 levels 
  then is generated \linebreak as~$\mbf Y_{nt,m}:=\mbf X_{nt,m}+\mbf Z_{nt,m}$.

In our numerical study, we let $n\in\{100,200\}$, $T\in\{200,500,1000\}$, and either $q=1$ and $Q=1$,  $q=3$ and $Q=2$ (as in the empirical application of the next section), or $q=2$ and $Q=3$. For each configuration considered, we simulate and estimate the model $\mathcal M=200$ times.

It has to be noticed that the data-generating process we are considering is similar to a stochastic volatility model. To illustrate the properties of the generated data, we report in Table \ref{tab:dgp1} the autocorrelations up to lag 10 of $h_{i,m}$, $s_{i,m}$, $e_{i,m}$, $v_{i,m}$, $X_{i,m}$, $X_{i,m}^2$, $Z_{i,m}$, $Z_{i,m}^2$, $Y_{i,m}$, and $Y_{i,m}^2$, averaged over all $\mathcal M$ replications and over all~$n$ series, and when $n=200$, $T=1000$.
It can be seen that log-volatilities $h_{i,m}$ and volatilities $s_{i,m}$ have high persistence, while, due to the way they are generated, the shocks $e_{i,m}$ and $v_{i,m}$ display no linear serial dependence, i.e. are weak white noises. Turning to the kurtosis of the level shocks reported in the left panel of Table \ref{tab:kurt}, these display heavy tails (especially the common ones) for the case $q=1$ and $Q=1$, while the kurtosis tends to decrease when increasing $Q$, possibly due to the aggregation of shocks in generating the common components of the log-volatility~$\chi_{i,m}$. Similar comments apply to the absolute values of skewness reported in the right panel of Table~\ref{tab:kurt}: especially in the  case $q=1$ and $Q=1$, the common shocks display a high degree of asymmetry. Because of these features of the simulated data the case  $q=1$ and $Q=1$ is particularly interesting to study to assess the performance of our estimators when dealing with heavy-tailed and skewed data.

\begin{table}[t!]
\centering
\caption{\small Autocorrelations of simulated variables. Average values over all $n$ series and all $\mathcal M$ replications for  $n=200$, $T=1000$, and $\mathcal M=200$. Values outside the $[\pm 1.96/\sqrt T]=[\pm 0.0620]$ interval are starred.}\label{tab:dgp1}
\vskip .2cm
\footnotesize
\begin{tabular}{l | ccc ccc }
\hline\hline
$q=1$ & \multicolumn{6}{|c}{lag}\\
$Q=1$& 1&2&3&4&5&6\\
\hline\hline
$h_{i,m}$	&	0.2967*	&	0.2856*	&	0.1178*	&	0.1691*	&	0.0528	&	0.1142*	\\
$s_{i,m}$	&	0.3082*	&	0.2743*	&	0.1201*	&	0.1426*	&	0.0552	&	0.0758*	\\
\hline
$e_{i,m}$	&	-0.0378	&	0.0696*	&	-0.0565	&	0.0089	&	-0.0099	&	-0.0283	\\
$e^2_{i,m}$	&	0.0326	&	0.0495	&	0.0020	&	0.0035	&	0.0031	&	0.0042	\\
\hline
$v_{i,m}$	&	-0.0023	&	-0.0017	&	-0.0035	&	0.0005	&	-0.0015	&	-0.0028	\\
$v^2_{i,m}$	&	0.2961*	&	0.2674*	&	0.1196*	&	0.1445*	&	0.0565	&	0.0771*	\\
\hline
$X_{i,m}$	&	0.1753*	&	0.1770*	&	0.0096	&	0.0302	&	-0.0032	&	-0.0262	\\
$X^2_{i,m}$	&	0.0989*	&	0.0791*	&	0.0068	&	0.0023	&	0.0016	&	-0.0037	\\
\hline
$Z_{i,m}$	&	0.0020	&	0.0783*	&	-0.0033	&	0.0118	&	-0.0018	&	-0.0002	\\
$Z^2_{i,m}$	&	0.2877*	&	0.2260*	&	0.1054*	&	0.1200*	&	0.0522	&	0.0622*	\\
\hline
$Y_{i,m}$	&	0.1174*	&	0.1439*	&	0.0051	&	0.0219	&	-0.0031	&	-0.0172	\\
$Y^2_{i,m}$	&	0.0994*	&	0.0798*	&	0.0073	&	0.0023	&	0.0018	&	0.0029	\\
\hline
\hline
$q=3$ & \multicolumn{6}{|c}{lag}\\
$Q=2$& 1&2&3&4&5&6\\
\hline\hline
$h_{i,m}$	&	0.2654*	&	0.2826*	&	0.1183*	&	0.1611*	&	0.0508	&	0.1272*	\\
$s_{i,m}$	&	0.2757*	&	0.2607*	&	0.1237*	&	0.1305*	&	0.0511	&	0.0913*	\\
$e_{i,m}$	&	-0.0089	&	-0.0690*	&	-0.0028	&	0.0304	&	-0.0353	&	-0.0071	\\
$e^2_{i,m}$	&	0.1939*	&	0.0721*	&	0.0034	&	0.0166	&	0.0077	&	0.0113	\\
$v_{i,m}$	&	-0.0010	&	-0.0011	&	-0.0030	&	0.0002	&	-0.0038	&	-0.0030	\\
$v^2_{i,m}$	&	0.2635*	&	0.2515*	&	0.1212*	&	0.1283*	&	0.0503	&	0.0907*	\\
$X_{i,m}$	&	0.1977*	&	0.0592	&	0.0465	&	0.0492	&	-0.0142	&	0.0002	\\
$X^2_{i,m}$	&	0.2545*	&	0.0822*	&	0.0161	&	0.0222	&	0.0073	&	0.0101	\\
$Z_{i,m}$	&	-0.0172	&	0.0814*	&	-0.0052	&	0.0119	&	-0.0045	&	-0.0013	\\
$Z^2_{i,m}$	&	0.2708*	&	0.2133*	&	0.1076*	&	0.1059*	&	0.0502	&	0.0764*	\\
$Y_{i,m}$	&	0.1227*	&	0.0671*	&	0.0285	&	0.0374	&	-0.0123	&	-0.0003	\\
$Y^2_{i,m}$	&	0.2166*	&	0.0909*	&	0.0237	&	0.0242	&	0.0079	&	0.0009	\\
\hline
\hline
$q=2$ & \multicolumn{6}{|c}{lag}\\
$Q=3$& 1&2&3&4&5&6\\
\hline\hline
$h_{i,m}$	&	0.2730*	&	0.2692*	&	0.1254*	&	0.1717*	&	0.0494	&	0.1293*	\\
$s_{i,m}$	&	0.2375*	&	0.2348*	&	0.1036*	&	0.1481*	&	0.0238	&	0.0885*	\\
\hline
$e_{i,m}$	&	-0.0221	&	0.0127	&	-0.0204	&	-0.0001	&	-0.0071	&	-0.0067	\\
$e^2_{i,m}$	&	0.0025	&	0.0640*	&	0.0222	&	0.0616	&	0.0001	&	0.0134	\\
\hline
$v_{i,m}$	&	-0.0026	&	-0.0002	&	0.0000	&	0.0026	&	-0.0011	&	-0.0035	\\
$v^2_{i,m}$	&	0.2330*	&	0.2313*	&	0.1015*	&	0.1456*	&	0.0229	&	0.0879*	\\
\hline
$X_{i,m}$	&	0.1835*	&	0.1254*	&	0.0364	&	0.0277	&	0.0079	&	0.0008	\\
$X^2_{i,m}$	&	0.1159*	&	0.1020*	&	0.0388	&	0.0377	&	0.0131	&	0.0057	\\
\hline
$Z_{i,m}$	&	0.0317	&	0.0872*	&	0.0032	&	0.0156	&	-0.0008	&	-0.0007	\\
$Z^2_{i,m}$	&	0.2509*	&	0.1918*	&	0.0969*	&	0.1227*	&	0.0293	&	0.0711*	\\
\hline
$Y_{i,m}$	&	0.1313*	&	0.1121*	&	0.0253	&	0.0220	&	0.0040	&	0.0012	\\
$Y^2_{i,m}$	&	0.1047*	&	0.0869*	&	0.0426	&	0.0425	&	0.0043	&	0.0133	\\
\hline
\hline
\end{tabular}
\end{table}

\begin{table}[t!]
\centering
\caption{\small Kurtosis and absolute value of skewness of simulated common level shocks $e_{i,m}$ and idiosyncratic level shocks~$v_{i,m}$. Maximum and average values over all $n$ series and all $\mathcal M$ replications for $n=200$, $T=1000$, and~$\mathcal M=200$.}\label{tab:kurt}
\vskip .2cm
\footnotesize
\begin{tabular}{l | cc | cc | cc|| cc | cc | cc}
\hline
\hline
& \multicolumn{6}{c||}{kurtosis}& \multicolumn{6}{c}{skewness}\\
&\multicolumn{2}{c|}{$q=1$, $Q=1$}&\multicolumn{2}{c|}{$q=3$, $Q=2$}&\multicolumn{2}{c||}{$q=2$, $Q=3$}&\multicolumn{2}{|c|}{$q=1$, $Q=1$}&\multicolumn{2}{c|}{$q=3$, $Q=2$}&\multicolumn{2}{c}{$q=2$, $Q=3$}\\
& max.& aver.&max.& aver.&max.& aver.& max.& aver.&max.& aver.&max.& aver.\\
\hline
$e_{i,m}$& 161.40 & 83.53&67.60&	10.94&31.86&	5.60&7.88&	0.28&4.08&	0.03&2.53&	0.02\\
$v_{i,m}$& 15.03	& 3.02&	12.68&	3.02&9.36& 3.01&1.14&	0.01&1.07&	0.01&0.85&	0.01\\
\hline
\hline
\end{tabular}
\end{table}

\begin{figure}[t!]\caption{\small Normalized eigenvalues of the zero-frequency spectral densities of simulated data for $n=200$ and  $T=1000$. Blue crosses: levels, $\mbf Y_{n,m}$;
red circles: log-volatilities, $\mbf h_{n,m}$.}\label{fig:eval_dgp}
\centering \smallskip\noindent 
\setlength{\tabcolsep}{.01\textwidth}
\begin{tabular}{@{}c}
 \includegraphics[width=.7\textwidth,trim=3.5cm 1.6cm 2cm 0cm,clip]{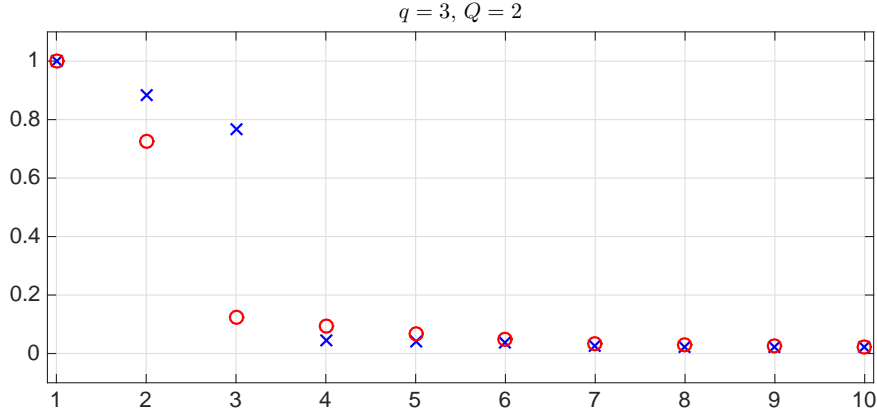} \\
\end{tabular}
\end{figure}

Furthermore, notice that  $\mbf e_{n,m}$, by construction, is a singular vector (as it should be)  and has the role of a common level innovation. Moreover, the elements of $\mbf v_{n,m}$,  in general,  are cross-sectionally dependent. As a consequence,  both $\mbf Y_{n,m}$ and $\mbf h_{n,m}$ have an approximate dynamic factor structure. In Figure \ref{fig:eval_dgp} we show scree-plots with the ten largest eigenvalues of the zero-frequency sample spectral density matrices of $\mbf Y_{n,m}$ (blue crosses), and $\mbf h_{n,m}$ (red circles), normalized by the largest zero-frequency eigenvalue, averaged over all $\mathcal M$ realisations, when $n=200$, $T=1000$, $q=3$, and $Q=2$.

\subsection{Results}
\begin{figure}[t!]\caption{\small Simulation results. True (blue) and estimated (red) common components of levels, $\widehat{X}_{it,m}$, and of volatilties, $\exp(\widehat{\chi}_{it,m})$, when $n=200$, $T=1000$, $q=1$, $Q=1$, and $\kappa_T=0.2$. One series and one realisation.}\label{fig:sim1014}
\centering \smallskip\noindent 
\setlength{\tabcolsep}{.01\textwidth}
\begin{tabular}{@{}cc}
\includegraphics[width=.5\textwidth,trim=3.5cm 1.6cm 2cm 0cm,clip]{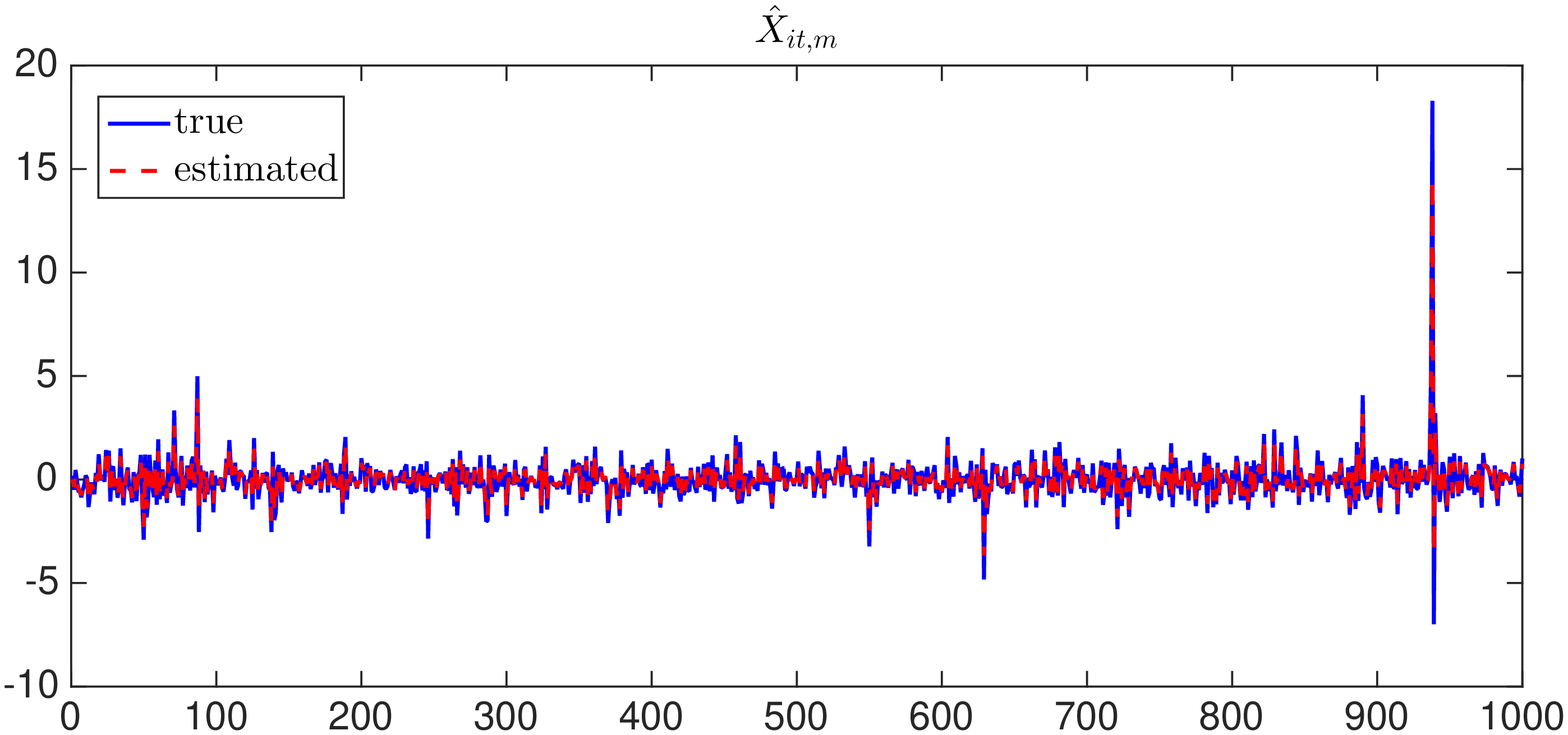}&
  \includegraphics[width=.5\textwidth,trim=3.5cm 1.6cm 2cm 0cm,clip]{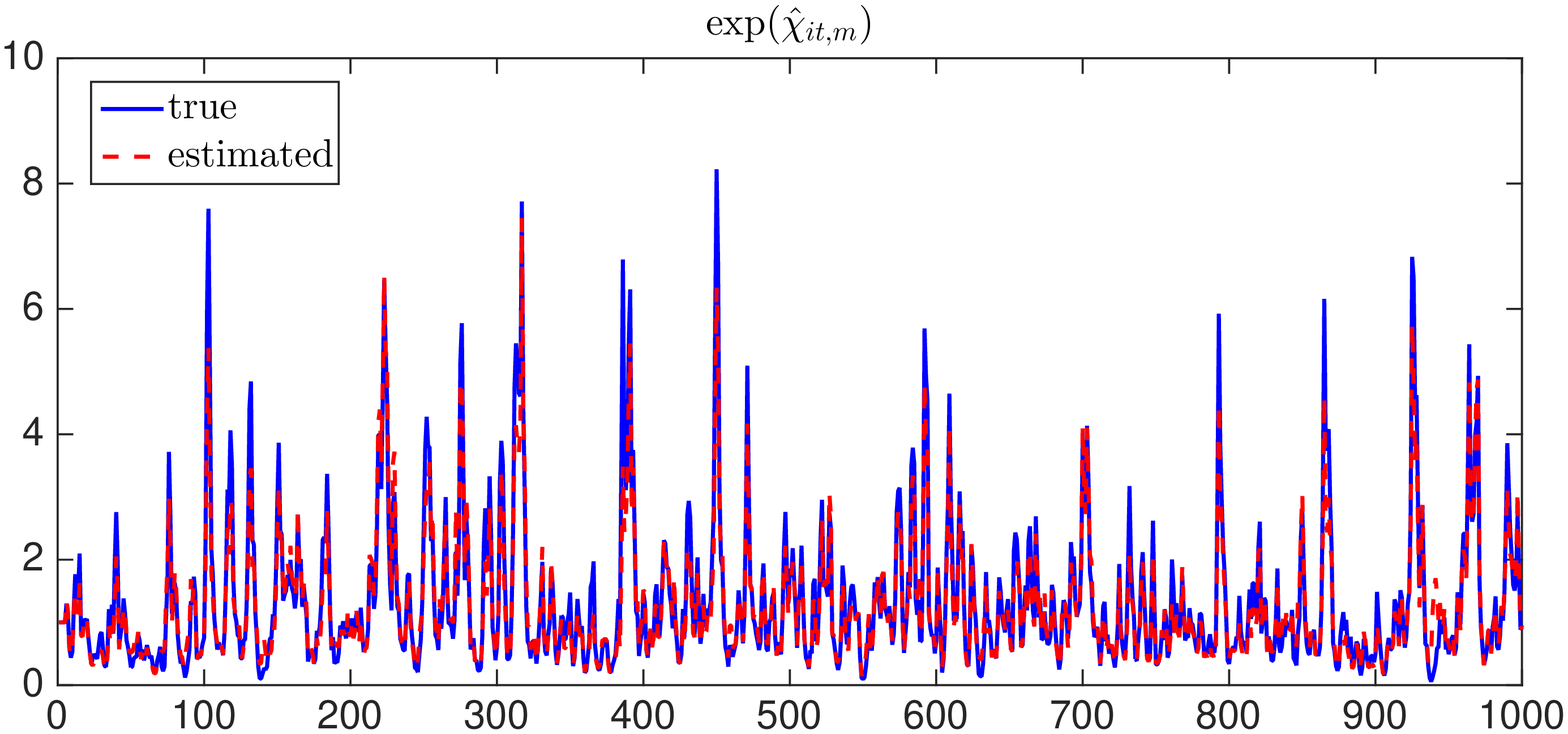} \\
\end{tabular}
\end{figure}

For each replication, we estimate the model as described in Section \ref{sec:est}. The capping constants $\kappa_T$ and the bandwidths $B_T$ and $M_T$ involved in the estimation of the spectral density are chosen as in the empirical analysis of the next section. Specifically, we let $\kappa_T\in\{0, 0.2,0.4\}$, while the bandwidths values  are~$B_T=2$ and $M_T=10$ for $T=200$,  $B_T=2$ and $M_T=15$ for $T=500$,  $B_T=2$ and $M_T=20$ for~$T=1000$ (see   Appendix~\ref{app:sim} for results based on other values). Once we obtain estimated common components $\widehat{X}_{i,m}$ for the levels  and $\widehat{\chi}_{i,m}$ for the log-volatilities, we  compute the  global error measures
\begin{align}
&MSE^X =\frac1{\mathcal MnT}\sum_{m=1}^{\mathcal M}{\sum_{i=1}^n\sum_{t=1}^T ({X}_{it,m}-\widehat{X}_{it,m})^2}, \quad
MSE^\chi=\frac1{\mathcal MnT}\sum_{m=1}^{\mathcal M}\sum_{i=1}^n\sum_{t=1}^T ({\chi}_{it,m}-\widehat{\chi}_{it,m})^2,\nn\\
&MAD^X =\frac1{\mathcal MnT}\sum_{m=1}^{\mathcal M}{\sum_{i=1}^n\sum_{t=1}^T \vert{X}_{it,m}-\widehat{X}_{it,m}\vert}, \quad
MAD^\chi=\frac1{\mathcal MnT}\sum_{m=1}^{\mathcal M}\sum_{i=1}^n\sum_{t=1}^T \vert{\chi}_{it,m}-\widehat{\chi}_{it,m}\vert\nn
\end{align}
and  the maximal errors over all realizations:
\begin{align}
&MAX^X ={ \max_{i=1,\ldots, n}\max_{t=1,\ldots, T}\max_{m=1,\ldots, \mathcal M} \vert{X}_{it,m}-\widehat{X}_{it,m}\vert}, \nn\\
&MAX^\chi={ \max_{i=1,\ldots, n}\max_{t=1,\ldots, T}\max_{m=1,\ldots,\mathcal M} \vert{\chi}_{it,m}-\widehat{\chi}_{it,m}\vert}.\nn
\end{align}
Notice that the error in the estimation of the common component  $X_{i,m}$ of the levels (first step of the estimation procedure) has already been studied in \citet{FHLZ17} and \citet{FGLS18}. We therefore  consider it as the benchmark error with respect to which   the performance of the second estimation step, which is the novelty of this paper, is to be compared. Results are provided  in Table~\ref{tab:sim1}. We note that MSE and MAD in the second step tend to be about 1.5 times higher than in the first step, which is not unexpected as first- and second- step errors typically cumulate in a   two-stage   procedure. However, when turning to MAX,  this is no longer  the case, since  levels in our data-generating process display heavier tails than log-volatilities---in line with the typical behavior of  daily stock returns and their volatilities. Increasing $n$ and $T$ improves the performance of all estimators; the role of $n$, in that respect, seems to be the main one---a manifestation of the ``blessing of dimensionality". On the other hand increasing $Q$ the number of common log-volatility shocks, tends to make estimation of the second step harder, but still results are in line with the case $Q=1$. Capping has an effect in controlling the maximum error but does not affect the MSE and MAD results much. 
 To illustrate the good performances of our method, in Figure~\ref{fig:sim1014} we show, for one replication, the estimated (in red) and simulated (in blue) common components of levels, and of volatilities, respectively, for $n=200$, $T=1000$, $q=1$, and $Q=1$ (which is the case exhibiting the heaviest tails),   setting $\kappa_T=0.2$.  The  choice of bandwidths adopted  seems to work quite well, and, comparing to alternative  choices considered in Appendix \ref{app:sim}, it can be shown that $M_T$ must be large enough to capture the persistence in  log-volatilities, while lower values of $B_T$ are enough for levels and do not affect much the second step of estimation.

\begin{table}[t!]
\centering
\caption{\small Simulation results. MSEs and MADs for common components. Bandwidths are $B_T=2$ and $M_T=10$ for~$T=200$; $B_T=2$ and $M_T=15$ for $T=500$; $B_T=2$ and $M_T=20$ for $T=1000$. }\label{tab:sim1}
\vskip .2cm
\footnotesize
\begin{tabular}{l l | cc | cc | cc }
\multicolumn{8}{c}{$q=1$, $Q=1$}\\
\hline\hline
&& \multicolumn{2}{|c|}{$T=200$}& \multicolumn{2}{|c|}{$T=500$}& \multicolumn{2}{|c}{$T=1000$}\\
&& $n=100$ & $n=200$& $n=100$ & $n=200$& $n=100$ & $n=200$\\
\hline\hline
$MSE^X$&&	0.215	&	0.219	&	0.168	&	0.164	&	0.125	&	0.154	\\
$MSE^\chi$&$\kappa_T=0$&	0.368	&	0.321	&	0.302	&	0.247	&	0.251	&	0.241	\\
$MSE^\chi$&$\kappa_T=0.2$	&	0.369	&	0.324	&	0.276	&	0.247	&	0.240	&	0.230	\\
$MSE^\chi$	&	$\kappa_T=0.4$	&	0.377	&	0.334	&	0.279	&	0.238	&	0.238	&	0.230	\\
\hline
$MAD^X$&&	0.278	&	0.245	&	0.255	&	0.226	&	0.234	&	0.228	\\
$MAD^\chi$&$\kappa_T=0$&	0.442	&	0.401	&	0.394	&	0.346	&	0.360	&	0.342	\\
$MAD^\chi$&$\kappa_T=0.2$	&	0.436	&	0.395	&	0.367	&	0.346	&	0.344	&	0.323	\\
$MAD^\chi$	&	$\kappa_T=0.4$	&	0.437	&	0.397	&	0.364	&	0.324	&	0.337	&	0.317	\\
\hline
$MAX^X$&&	10.797	&	13.560	&	17.113	&	16.352	&	14.405	&	19.757	\\
$MAX^\chi$&$\kappa_T=0$&	6.587	&	6.709	&	8.996	&	6.270	&	7.462	&	7.912	\\
$MAX^\chi$&$\kappa_T=0.2$	&	8.259	&	7.247	&	7.967	&	6.270	&	9.328	&	8.431	\\
$MAX^\chi$	&	$\kappa_T=0.4$	&	8.987	&	8.402	&	8.295	&	10.223	&	9.821	&	8.960	\\
\hline\hline
\\
\multicolumn{8}{c}{$q=3$, $Q=2$}\\
\hline\hline
&& \multicolumn{2}{|c|}{$T=200$}& \multicolumn{2}{|c|}{$T=500$}& \multicolumn{2}{|c}{$T=1000$}\\
&& $n=100$ & $n=200$& $n=100$ & $n=200$& $n=100$ & $n=200$\\
\hline\hline
$MSE^X$	&&	0.143	&	0.155	&	0.101	&	0.112	&	0.086	&	0.085	\\
$MSE^\chi$&$\kappa_T=0$	&	0.291	&	0.284	&	0.237	&	0.216	&	0.209	&	0.185	\\
$MSE^\chi$	&	$\kappa_T=0.2$	&	0.262	&	0.261	&	0.197	&	0.199	&	0.179	&	0.163	\\
$MSE^\chi$	&	$\kappa_T=0.4$	&	0.250	&	0.252	&	0.182	&	0.176	&	0.162	&	0.141	\\
\hline
$MAD^X$	&&	0.265	&	0.261	&	0.227	&	0.228	&	0.210	&	0.205	\\
$MAD^\chi$&$\kappa_T=0$	&	0.412	&	0.402	&	0.370	&	0.348	&	0.346	&	0.322	\\
$MAD^\chi$	&	$\kappa_T=0.2$	&	0.389	&	0.381	&	0.336	&	0.327	&	0.317	&	0.299	\\
$MAD^\chi$	&	$\kappa_T=0.4$	&	0.378	&	0.372	&	0.321	&	0.307	&	0.300	&	0.275	\\
\hline
$MAX^X$	&&	6.634	&	8.693	&	9.682	&	12.606	&	10.943	&	19.277	\\
$MAX^\chi$&$\kappa_T=0$	&	5.532	&	5.624	&	5.128	&	6.725	&	4.488	&	5.209	\\
$MAX^\chi$	&	$\kappa_T=0.2$	&	5.688	&	5.800	&	5.561	&	6.204	&	4.813	&	4.675	\\
$MAX^\chi$	&	$\kappa_T=0.4$	&	5.713	&	6.180	&	5.630	&	7.317	&	4.978	&	5.235	\\
\hline\hline
\\
\multicolumn{8}{c}{$q=2$, $Q=3$}\\
\hline\hline
&& \multicolumn{2}{|c|}{$T=200$}& \multicolumn{2}{|c|}{$T=500$}& \multicolumn{2}{|c}{$T=1000$}\\
&& $n=100$ & $n=200$& $n=100$ & $n=200$& $n=100$ & $n=200$\\
\hline\hline
$MSE^X$	&		&	0.151	&	0.161	&	0.112	&	0.129	&	0.082	&	0.085	\\
$MSE^\chi$	&	$\kappa_T=0$	&	0.356	&	0.323	&	0.290	&	0.277	&	0.247	&	0.224	\\
$MSE^\chi$	&	$\kappa_T=0.2$	&	0.324	&	0.299	&	0.259	&	0.247	&	0.210	&	0.190	\\
$MSE^\chi$	&	$\kappa_T=0.4$	&	0.311	&	0.287	&	0.248	&	0.231	&	0.193	&	0.173	\\
\hline
$MAD^X$	&		&	0.279	&	0.271	&	0.241	&	0.248	&	0.212	&	0.213	\\
$MAD^\chi$	&	$\kappa_T=0$	&	0.461	&	0.434	&	0.414	&	0.398	&	0.382	&	0.360	\\
$MAD^\chi$	&	$\kappa_T=0.2$	&	0.437	&	0.415	&	0.387	&	0.372	&	0.350	&	0.327	\\
$MAD^\chi$	&	$\kappa_T=0.4$	&	0.426	&	0.404	&	0.376	&	0.356	&	0.334	&	0.311	\\
\hline
$MAX^X$	&		&	5.395	&	9.180	&	5.900	&	8.936	&	6.210	&	11.411	\\
$MAX^\chi$	&	$\kappa_T=0$	&	4.833	&	5.066	&	5.009	&	5.654	&	5.111	&	5.571	\\
$MAX^\chi$	&	$\kappa_T=0.2$	&	4.988	&	5.325	&	5.269	&	5.998	&	5.355	&	5.654	\\
$MAX^\chi$	&	$\kappa_T=0.4$	&	5.058	&	5.748	&	5.660	&	6.208	&	5.599	&	5.413	\\
\hline\hline
\end{tabular}
\end{table}

Finally, for  $T=1000$, we estimated the model using the first 900 observations, then  ran a recursive pseudo-out-of-sample forecasting exercise constructing one-step-ahead prediction intervals for the remaining $100$ observations (from $901$ to $1000$), as described in Section~\ref{eq:int}. The $\alpha/2$-upper and $\alpha/2$-lower bounds $\widehat{\mathcal U}_{i,\tau+1|\tau,m}(\alpha/2)$ and~$\widehat{\mathcal L}_{i,\tau+1|\tau,m}(\alpha/2)$ of  prediction intervals  with coverage probability $(1 - \alpha)$ are then computed for each series and replication and each out-of-sample observation. From the latter,  we compute the observed coverage frequencies across all series and replications
\[
C(\alpha) := \frac 1 {\mathcal Mn100}\sum_{m=1}^\mathcal M\sum_{i=1}^n \sum_{\tau=900}^{999} 
\mathbb I\Big(\widehat{\mathcal L}_{i,\tau+1|\tau,m}(\alpha/2)\le Y_{i,\tau+1,m}\! \le \widehat{\mathcal U}_{i,\tau+1|\tau,m}(\alpha/2)\Big)\nn
\]
and the proportions of coverage violations in the upper and lower tails,
\begin{align}
&V_{+}(\alpha/2) := \frac 1 {\mathcal Mn100}\sum_{m=1}^\mathcal M\sum_{i=1}^n \sum_{\tau=900}^{999} \mathbb I\Big(Y_{i,\tau+1,m}\! > \widehat{\mathcal U}_{i,\tau+1|\tau,m}(\alpha/2)\Big),\nn
\end{align}
and 
\begin{align}
&V_{-}(\alpha/2)  :=\frac 1 {\mathcal Mn100}\sum_{m=1}^\mathcal M\sum_{i=1}^n \sum_{\tau=900}^{999}  \mathbb I\Big(Y_{i,\tau+1,m}\! < \widehat{\mathcal L}_{i,\tau+1|\tau,m}(\alpha/2)\Big),\nn
\end{align}
respectively. Results are shown in Table \ref{tab:covsim}.  Overall performances  look reasonably good---the larger  $n$ and $T$, the better. We note that capping has a clear  effect on the empirical coverage; too much capping seems to affect mostly the cases in which $\alpha=0.32$ and $0.2$. No capping at all   works quite well in practice, despite the fact that  theoretical results require  $\kappa_T>0$. Moreover, the same comments apply to   empirical coverage as for  the choice of bandwidths,  with the additional finding that  higher values of $B_T$ yield more reliable prediction performances (see Appendix \ref{app:sim}).  

\begin{table}[t!]
\centering
\caption{
\small Simulation results. Empirical coverage and  frequencies of prediction  bounds violations, averaged over all $n$ series and all $\mathcal M$ replications, for $T=1000$ and $\mathcal M=200$.  Bandwidths values:  $B_T=2$ and $M_T=10$ for~$T=200$; $B_T=2$ and $M_T=15$ for~$T=500$; $B_T=2$ and $M_T=20$ for~$T=1000$.}\label{tab:covsim}
\vskip .2cm
\footnotesize
\begin{tabular}{l l | ccccc | ccccc }
\multicolumn{12}{c}{$q=1$, $Q=1$}\\
\hline
\hline
&&\multicolumn{5}{c|}{$n=100$}&\multicolumn{5}{c}{$n=200$}\\
\hline
&&\multicolumn{5}{c|}{$\alpha$}&\multicolumn{5}{c}{$\alpha$}\\
&&0.32&0.2&0.1&0.05&0.01&0.32&0.2&0.1&0.05&0.01\\
\hline
$C(\alpha)$& $\kappa_T=0$&0.6409&	0.7637&	0.8667&	0.9312&	0.9869&0.6765&	0.7992&	0.9082&	0.9573&	0.9926 \\
$V_+(\alpha/2)$&&0.1810&	0.1195&	0.0674&	0.0342&	0.0057&0.1635&	0.1026&	0.0470&	0.0221&	0.0040\\	
$V_-(\alpha/2)$&&0.1781&	0.1168&	0.0659&	0.0346&	0.0074&0.1601&	0.0983&	0.0449&	0.0206&	0.0035\\	
\hline
$C(\alpha)$& $\kappa_T=0.2$&0.6691&	0.7769&	0.8685&	0.9197&	0.9681 &0.7201&	0.8285&	0.9226	&0.9628&	0.9934\\
$V_+(\alpha/2)$&&0.1636&	0.1124&	0.0682&	0.0403&	0.0153&0.1422&	0.0876&	0.0392&	0.0194&	0.0034\\
$V_-(\alpha/2)$&&0.1673&	0.1107&	0.0633&	0.0400&	0.0166&	0.1378&	0.0840&	0.0383&	0.0179&	0.0033\\	
\hline
$C(\alpha)$& $\kappa_T=0.4$& 0.7072&	0.7987&	0.8799	&0.9238	&0.9699&0.7119&	0.7957&	0.8763&	0.9257&	0.9703\\
$V_+(\alpha/2)$&&0.1453&	0.1023&	0.0617&	0.0384&	0.0145&0.1429&	0.1007&	0.0600&	0.0360&	0.0150\\	
$V_-(\alpha/2)$&&0.1475&	0.0990&	0.0584&	0.0378&	0.0156&0.1453&	0.1037&	0.0638&	0.0383&	0.0147\\	
\hline
\hline
\\
\multicolumn{12}{c}{$q=3$, $Q=2$}\\
\hline
\hline
&&\multicolumn{5}{c|}{$n=100$}&\multicolumn{5}{c}{$n=200$}\\
\hline
&&\multicolumn{5}{c|}{$\alpha$}&\multicolumn{5}{c}{$\alpha$}\\
&&0.32&0.2&0.1&0.05&0.01&0.32&0.2&0.1&0.05&0.01\\
\hline
$C(\alpha)$& $\kappa_T=0$& 0.6350&	0.7523&	0.8543&	0.9131&	0.9688&0.6718&	0.7917&	0.8929&	0.9457&	0.9895\\
$V_+(\alpha/2)$&&0.1810&	0.1252&	0.0749&	0.0449&	0.0157&0.1619&	0.1037&	0.0530&	0.0277&	0.0053\\
$V_-(\alpha/2)$&&0.1840&	0.1225&	0.0708&	0.0420&	0.0155&	0.1664&	0.1047&	0.0542&	0.0266&	0.0053\\
\hline
$C(\alpha)$& $\kappa_T=0.2$&0.6767&	0.7775&	0.8665&	0.9206&	0.9701&0.7031&	0.8081&	0.8986&	0.9469&	0.9916\\
$V_+(\alpha/2)$&&0.1601&	0.1145&	0.0691&	0.0422&	0.0162&0.1458&0.0935&	0.0503&	0.0250&	0.0039\\
$V_-(\alpha/2)$&&0.1632&	0.1080&	0.0644&	0.0372&	0.0137&	0.1512&	0.0985&	0.0512&	0.0282&	0.0046\\	
\hline
$C(\alpha)$& $\kappa_T=0.4$&0.7129&	0.7993&	0.8803&	0.9267&	0.9724&0.7565&	0.8447&	0.9222&	0.9610&	0.9923 \\
$V_+(\alpha/2)$&&0.1445&	0.1033&	0.0614&	0.0384&	0.0147&0.1209&	0.0780&	0.0382&	0.0196&	0.0043\\	
$V_-(\alpha/2)$&&0.1426&	0.09740&	0.0583&	0.0349&	0.0129&0.1227&	0.0774&	0.0397&	0.0195&	0.0035\\	
\hline
\hline
\\
\multicolumn{12}{c}{$q=2$, $Q=3$}\\
\hline
\hline
&&\multicolumn{5}{c|}{$n=100$}&\multicolumn{5}{c}{$n=200$}\\
\hline
&&\multicolumn{5}{c|}{$\alpha$}&\multicolumn{5}{c}{$\alpha$}\\
&&0.32&0.2&0.1&0.05&0.01&0.32&0.2&0.1&0.05&0.01\\
\hline
$C(\alpha)$& $\kappa_T=0$&0.6888&	0.8045&	0.8981&	0.9500&	0.9874&0.6391&	0.7563&	0.8623&	0.9215&	0.9784 \\
$V_+(\alpha/2)$&&0.1568&	0.0995&	0.0527&	0.0258&	0.0065&0.1786&	0.1212&	0.0678&	0.0387&	0.0107\\	
$V_-(\alpha/2)$&&	0.1544&	0.0960&	0.0492&	0.0242&	0.0061&0.1824&	0.1226&	0.0700&	0.0399&	0.0110\\
\hline
$C(\alpha)$& $\kappa_T=0.2$&0.7335&0.8290&	0.9105&	0.9539&	0.9890&0.6770&	0.7814&	0.8752&	0.9277&	0.9791\\
$V_+(\alpha/2)$&&0.1345&	0.0863&	0.0459&	0.0239&	0.0056&0.1602&	0.1077&	0.0613&	0.0355&	0.0103\\	
$V_-(\alpha/2)$&&0.1320&	0.0847&	0.0436&	0.0222&	0.0054&0.1629&	0.1110&	0.0636&	0.0368&	0.0106\\	
\hline
$C(\alpha)$& $\kappa_T=0.4$&0.7733&0.8539&	0.9213&	0.9595&	0.9898&0.7167&	0.8066&	0.8879&	0.9352&	0.9809 \\
$V_+(\alpha/2)$&&0.1154&	0.0747&	0.0408&	0.0207&	0.0055&0.1395&	0.0953&	0.0551&	0.0326&	0.0093\\	
$V_-(\alpha/2)$&&0.1113&	0.0714&	0.0379&	0.0198&	0.0047&0.1439&	0.0981&	0.0570&	0.0322&	0.0099\\	
\hline
\hline
\end{tabular}
\end{table}

\setcounter{equation}{0}
\section{Interval prediction for S\&P100 returns }\label{sec:emp}

In this section, we apply our methodology to a panel of $n=90$ daily returns of stocks from the Standard \& Poor's 100 Index. Data are observed from January 4, 2000 through September 30, 2013, for a total of~$T=3456$ observations. We run a pseudo-out-of-sample forecasting exercise by estimating the model using data over the period $t=1,\ldots, \tau$, with $\tau=(T-M),\ldots, (T-1)$ and $M=~\!1948$,   corresponding to an evaluation period running from January 3, 2006 through  September 27, 2013. For each value of $\tau$, we estimate the $n=90$ one-step-ahead prediction intervals as defined in \eqref{eq:Itt1hat}. The data cover the following sectors (in parentheses, the number of series in each sector): Consumer Discretionary~(11), Consumer Staples (10), Energy (12), Financials (13), Health Care (11), Industrials~(14), Information Technology~(12), Materials (3), Telecommunications Services (2), Utilities (2) (see Appendix \ref{app:data} for the names of individual stocks).

Although we should,  in principle,  fully re-estimate the whole model  at each of the $M$ iterations, some quantities were kept fixed throughout  the exercise. In particular, 
when applied to the full~$n\times~\!T$ panel,  the \citet{hallinliska07} criterion returns $\widehat q=3$ common factors for the level panel and~$\widehat Q=~\!2$ common factors for log-volatility panel: those values are used in all subsequent analyzes. We also choose the bandwidths by minimizing, over a grid of possible bandwidth values, the mean-squared errors 
\begin{align}
\frac 1 {nT} \sum_{i=1}^n \sum_{t=1}^T(Y_{it}-\widehat X_{it|t-1})^2 \quad\text{and}\quad \frac 1 {nT} \sum_{i=1}^n \sum_{t=1}^T(\widehat h_{it}-\widehat \chi_{it|t-1})^2,\nn
\end{align}
respectively, leading to possibly  distinct bandwidths for $\widehat X_{it|t-1}$ and $\widehat \chi_{it|t-1}$. More precisely, we first determine $B_T$ and then  determine $M_T$ using the chosen $B_T$ to compute $\widehat h_{it}$. As a result we throughout use $B_T=2$ and~$M_T=17$. The VAR orders and the  orders of their truncated inverse MA representations needed to compute impulse responses  are set as follows:
\begin{inparaenum}[(i)]
\item $\text{deg}[\mbf A_n(L)]=1$, with inverse MA truncated at lag $\bar k_1=20$; 
\item $\text{deg}[\mbf C_n(L)]=1$, with inverse MA truncated at lag $\bar k_2=20$; 
\item $\text{deg}[\mbf M_n(L)]=5$, with inverse  MA truncated at lag~$\bar k_1^*=~\!100$;
\item $\text{deg}[\mbf P_n(L)]=1$, with inverse  MA truncated at lag $\bar k_2^*=100$.
\end{inparaenum}
The estimation of the GDFM is based on 10 cross-sectional permutations, as explained at the end of Section~\ref{sec:est_summary}. Finally, regarding the choice of the capping constant $\kappa_T$, we choose $\kappa_T\in\{0,\, 0.1,\, 0.25,\, 0.5\}$ irrespective of~$i$; note that, with reference to Assumption (R), we have $\log^{-1} T=  0.12$. Also note that, on  the average across the~$M$ iterations, 6\%, out of the total $n\tau$ observations, are capped when $\kappa_T=0.1$, 14\%   when~$\kappa_T=~\!0.25$, and 27\%   when $\kappa_T=0.5$.

For any given sample size $\tau$, we compute the quantiles of $\widehat{w}_i$ using $(\widehat w_{i,\tau-\ell+1},\ldots,\widehat w_{i,\tau})$, where we\linebreak  set~$\ell\in\{126, 252, 504, \tau\}$, hence using either the past six months, one year, or two years of available data, or using all available past observations. Denoting by $\widehat {\mbf w}^{(\ell)}_{i}$ the vector  of  the most recent $\ell$ observations (so that $\widehat {\mbf w}^{(\tau)}_{i}$ coincides with $\widehat {\mbf w}_{i}$), for  levels  $\alpha\in\{0.32,0.2,0.1,0.05,0.01\}$ and window sizes $\ell$, and for $\tau=(T-M),\ldots, (T-1)$, we obtain the estimates
\begin{align}
&\widehat{\mathcal U}_{i,\tau+1|\tau}^{(\ell)}(\alpha):=\widehat Y_{i,\tau+1|\tau} + \widehat s_{i,\tau+1|\tau}\, \widehat w^{(\ell)}_{i(\lceil \ell(1-\alpha)\rceil)},&\quad \quad \quad 
\widehat{\mathcal L}_{i,\tau+1|\tau}^{(\ell)}(\alpha):=\widehat Y_{i,\tau+1|\tau} + \widehat s_{i,\tau+1|\tau}\, \widehat w^{(\ell)}_{i(\lceil \ell\alpha\rceil)},\nn\\ 
&\widehat{\mathcal I}_{i,\tau+1|\tau}^{(\ell)}(\alpha):=\big[\widehat{\mathcal L}^{(\ell)}_{i,\tau+1|\tau}(\alpha^-),\widehat{\mathcal U}^{(\ell)}_{i,\tau+1|\tau}(\alpha^+)\big], \quad\text{  and}\!\!\!\!\!\!\!\!\!\!\!\!\!\!\!\!\!\!&\widehat{\mathcal H}_{i,\tau+1|\tau}^{(\ell)}(\alpha):=\mathbb I\big(Y_{i,\tau+1}\in \widehat{\mathcal I}_{i,\tau+1|\tau}^{(\ell)}(\alpha)\big).\ \ \  \ \,\nn 
\end{align}
\subsection{Coverage performance: qualitative analysis}
For each of the $n=90$ series considered we compute the coverage frequency
\[
C_i^{(\ell)}(\alpha) := \frac 1 {M} \sum_{\tau=T-M}^{T-1} \widehat{\mathcal H}_{i,\tau+1|\tau}^{(\ell)}(\alpha)=\frac 1M\sum_{\tau=T-M}^{T-1} \mathbb I\Big(\widehat{\mathcal L}_{i,\tau+1|\tau}^{(\ell)}(\alpha^-)\le Y_{i,\tau+1}\! \le \widehat{\mathcal U}_{i,\tau+1|\tau}^{(\ell)}(\alpha^+)\Big),
\]
the proportions 
\begin{align}
V^{(\ell)}_{i,+}(\alpha^+) \! := \frac 1 {M}\!\! \sum_{\tau=T-M}^{T-1}\! \!  \!\!\mathbb I\Big(Y_{i,\tau+1}\! > \widehat{\mathcal U}_{i,\tau+1|\tau}^{(\ell)}(\alpha^+)\Big)
  \text{ and }  
V^{(\ell)}_{i,-}(\alpha^-)\!  := \frac 1 {M}\!\! \sum_{\tau=T-M}^{T-1} \! \! \!\!\mathbb I\Big(Y_{i,\tau+1}\! < \widehat{\mathcal L}_{i,\tau+1|\tau}^{(\ell)}(\alpha^-)\Big)\nn
\end{align}
of coverage violations in the upper and lower tails, and the average interval length
\[
L_{i}^{(\ell)}(\alpha):=\frac 1{M} \sum_{\tau=T-M}^{T-1} \l( \widehat{\mathcal U}_{i,\tau+1|\tau}^{(\ell)}(\alpha)-\widehat{\mathcal L}_{i,\tau+1|\tau}^{(\ell)}(\alpha)\r).
\]
Table \ref{tab:viol} reports, for $\alpha^+=\alpha^-=\alpha/2$ with $\alpha\in\{0.32,0.2,0.1,0.05,0.01\}$ (corresponding to  coverage levels 68\%, 80\%, 90\%, 95\% and 99\%) and $\kappa_T\in\{0,0.1,0.25,0.5\}$, the cross-sectional average $C^{(\ell)}(\alpha)$ of the empirical coverage frequencies $C_i^{(\ell)}(\alpha)$, the cross-sectional averages $V^{(\ell)}_+(\alpha/2)$ and $V^{(\ell)}_-(\alpha/2)$ of the proportions   of coverage violations $V^{(\ell)}_{i,+}(\alpha^+)$ and $V^{(\ell)}_{i,-}(\alpha^-)$,  and the cross-sectional average ${L}^{(\ell)}(\alpha)$ of the average interval lengths~$L_{i}^{(\ell)}(\alpha)$.

\begin{table}[t!]
\centering
\caption{\small Standard \& Poor's 100 Index data ($n=90$ daily returns). Empirical coverage, frequency of prediction bounds violations, and average length of prediction intervals for  GDFM, averaged over the cross-section.}\label{tab:viol}
\vskip .2cm
\footnotesize
\begin{tabular}{l | ccccc | ccccc}
\hline
\hline
&\multicolumn{5}{c|}{$\kappa_T=0$}&\multicolumn{5}{c}{$\kappa_T=0.1$}\\
\hline
&\multicolumn{5}{c|}{$\alpha$}&\multicolumn{5}{c}{$\alpha$}\\
&0.32&0.2&0.1&0.05&0.01 &0.32&0.2&0.1&0.05&0.01 \\
\hline
$C^{(126)}(\alpha)$  &0.6709	&	0.7894	&	0.8887	&	0.9400	&	0.9812 &0.6874 &   0.7985&    0.8931&    0.9416&    0.9813	\\
$V^{(126)}_+(\alpha/2)$&0.1641	&	0.1048	&	0.0552	&	0.0299	&	0.0094&0.1559&    0.1002&    0.0533&    0.0291&    0.0095	\\
$V^{(126)}_-(\alpha/2)$&0.1650	&	0.1058	&	0.0561	&	0.0301	&	0.0094&0.1566&    0.1013&    0.0536&    0.0292&    0.0091	\\
${L}^{(126)}(\alpha)$&3.3934 &   4.5156  &  6.1305 &   7.7681  & 12.4174&3.4726  &  4.5726  &  6.1553  &  7.7698  & 12.3130\\
\hline
$C^{(252)}(\alpha)$  &0.6708	&	0.7903	&	0.8902	&	0.9415	&	0.9848 &0.6882&    0.7999&    0.8940&    0.9424&    0.9846	\\
$V^{(252)}_+(\alpha/2)$&0.1647	&	0.1044	&	0.0544	&	0.0289	&	0.0077&0.1560&    0.0998&    0.0526&    0.0287&    0.0078	\\
$V^{(252)}_-(\alpha/2)$&0.1644	&	0.1053	&	0.0554	&	0.0296	&	0.0075&0.1558&    0.1003&    0.0534&    0.0289&    0.0076	\\
${L}^{(252)}(\alpha)$&3.3621  &  4.4794  &  6.0949  &  7.7240  & 12.5008&3.4351 &   4.5290  &  6.1078   & 7.7074   &12.3767\\
\hline
$C^{(504)}(\alpha)$  &0.6711	&	0.7895	&	0.8895	&	0.9412	&	0.9846 &0.6886&    0.7995&    0.8929&    0.9419&    0.9843	\\
$V^{(504)}_+(\alpha/2)$&0.1651	&	0.1057	&	0.0551	&	0.0290	&	0.0078&0.1561&    0.1005&    0.0536&    0.0288&    0.0081 	\\
$V^{(504)}_-(\alpha/2)$&0.1638	&	0.1047	&	0.0554	&	0.0298	&	0.0076&0.1553&    0.1000&    0.0535&    0.0292&    0.0076	\\
${L}^{(504)}(\alpha)$&3.3034 &   4.4179  &  6.0266  &  7.6643 &  12.1190&3.3786 &   4.4708   & 6.0439 &   7.6539 &  12.0462\\
\hline
$C^{(\tau)}(\alpha)$&   0.7010	&	0.8142	&	0.9049	&	0.9506	&	0.9881&0.7187&    0.8244&    0.9096&    0.9523 &   0.9881	 \\
$V^{(\tau)}_+(\alpha/2)$&0.1516	&	0.0933	&	0.0474	&	0.0247	&	0.0061&0.1424&    0.0879&    0.0452&    0.0237&    0.0062	\\
$V^{(\tau)}_-(\alpha/2)$&0.1474	&	0.0925	&	0.0477	&	0.0248	&	0.0058&0.1389&    0.0877&    0.0452&    0.0239&    0.0057	\\
${L}^{(\tau)}(\alpha)$&3.4523 &   4.6632   & 6.4305 &   8.2802 &  13.3895&3.5562 &   4.7560  &  6.5201  &  8.3747 &  13.5115\\
\hline
\hline
&\multicolumn{5}{c|}{$\kappa_T=0.25$}&\multicolumn{5}{c}{$\kappa_T=0.5$}\\
\hline
&\multicolumn{5}{c|}{$\alpha$}&\multicolumn{5}{c}{$\alpha$}\\
&0.32&0.2&0.1&0.05&0.01 &0.32&0.2&0.1&0.05&0.01 \\
\hline
$C^{(126)}(\alpha)$  &0.7126  &  0.8141  &  0.8997  &  0.9452  &  0.9821&0.7552&    0.8391&    0.9119&    0.9507&    0.9836	\\
$V^{(126)}_+(\alpha/2)$&0.1435 &   0.0926  &  0.0500 &   0.0274 &   0.0091&0.1222&    0.0800&    0.0436&    0.0243&    0.0082  	\\
$V^{(126)}_-(\alpha/2)$&0.1439  &  0.0932 &   0.0504  &  0.0274 &   0.0088&0.1226&    0.0809&    0.0446&    0.0251&    0.0081	\\
${L}^{(126)}(\alpha)$&3.6203 &   4.6949 &   6.2419 &   7.8371  & 12.3547&3.9076&    4.9426 &   6.4443  &  8.0189 &  12.5330\\
\hline
$C^{(252)}(\alpha)$  &	0.7138  &  0.8143 &   0.9009 &   0.9452 &   0.9851&0.7556 &   0.8398 &   0.9127&    0.9512&    0.9866\\
$V^{(252)}_+(\alpha/2)$&0.1433 &   0.0928&    0.0491&    0.0271  &  0.0077&0.1224&    0.0798 &   0.0432&    0.0242&    0.0069	\\
$V^{(252)}_-(\alpha/2)$&0.1428 &   0.0929 &   0.0500  &  0.0277  &  0.0072&0.1219&    0.0804&    0.0440&    0.0246&    0.0065	\\
${L}^{(252)}(\alpha)$&3.5796&    4.6500&    6.1923&    7.7700&   12.4405&3.8737&    4.9105 &   6.4108 &   7.9706 &  12.7086\\
\hline
$C^{(504)}(\alpha)$  &0.7149 &   0.8150 &   0.9002  &  0.9449 &   0.9846&0.7588&    0.8422 &   0.9132&    0.9514&    0.9861	\\
$V^{(504)}_+(\alpha/2)$& 0.1430  &  0.0927 &   0.0495 &   0.0274  &  0.0080&0.1204 &   0.0782&    0.0426&    0.0236&    0.0072	\\
$V^{(504)}_-(\alpha/2)$&0.1420 &   0.0923 &   0.0502 &   0.0277  &  0.0074&0.1208&    0.0795&    0.0442 &   0.0250&    0.0067	\\
${L}^{(504)}(\alpha)$&3.5336 &   4.6035 &   6.1513 &   7.7434  & 12.1898&3.8584  &  4.9070  &  6.4371  &  8.0357 &  12.6302\\
\hline
$C^{(\tau)}(\alpha)$&0.7430 &   0.8387  &  0.9162  &  0.9551  &  0.9886& 0.7824&    0.8633 &   0.9283 &   0.9613  &  0.9900   \\
$V^{(\tau)}_+(\alpha/2)$&0.1301  &  0.0808 &   0.0415  &  0.0221  &  0.0061&0.1091&    0.0680&    0.0351  &  0.0189&    0.0053 	\\
$V^{(\tau)}_-(\alpha/2)$&0.1269 &   0.0805 &   0.0422  &  0.0228  &  0.0054&0.1085&    0.0687 &   0.0366  &  0.0198 &   0.0047\\
${L}^{(\tau)}(\alpha)$&3.7420 &   4.9317 &   6.6982  &  8.5677  & 13.7991&4.1045 &   5.2913&    7.0734&    8.9901&   14.4295\\
\hline
\hline
\end{tabular}
\end{table}

Inspection of the table reveals that $C^{(\ell)}(\alpha) \simeq(1-\alpha)$ and $V_+^{(\ell)}(\alpha/2) \simeq V_-^{(\ell)}(\alpha/2) \simeq \alpha/2$, which is a qualitative confirmation of the validity of our methodology (see Section~\ref{backtestSec} for  more formal validation). Three remarks emerge from these results. 
First, regarding the sensitivity of our procedure to capping, lower values of $\kappa_T$, in general,  provide better results when $\alpha$ is higher, while larger values of~$\kappa_T$ provide better results for lower values of $\alpha$;  in all cases,    $\kappa_T=0.5$  yields a mostly  conservative coverage frequency higher than $(1-\alpha)$. In particular, note  that the choice of $\kappa_T=0$ (no capping at all), although  ruled out by Assumption~(R),  still provides very good results. 
Second, setting $\ell=\tau$, that is, considering the entire past history to compute quantiles   apparently is not the best strategy, and shorter horizons $\ell$ seem   preferable. This finding is possibly related to some time variation in the distribution of the innovations of  log-volatilities at horizons longer than one year. Third, for any given~$\alpha$,  shorter intervals are obtained when setting $\ell=252$ or $504$ regardless of the choice of $\kappa_T$. Overall, choosing~$\kappa_T=0.1$ and~$\ell=252$ or $504$ works best for $\alpha=0.32$ and $0.2$, while~$\kappa_T=0.25$ and $\ell=126$ or $252$ works best for~$\alpha=0.1, 0.05$, and $0.01$.

\begin{figure}[t!]\caption{\small One-step-ahead 90\% conditional prediction intervals (in red; $\ell=252$): America International Group (AIG), Bank of America (BAC), Citigroup (C), Goldman Sachs (GS), JPMorgan Chase (JPM), Morgan Stanley (MS).}\label{fig:stocks}
\centering \smallskip\noindent 
\setlength{\tabcolsep}{.01\textwidth}
\begin{tabular}{@{}cc}
 \includegraphics[width=.45\textwidth,trim=3.5cm 1.6cm 2cm 0cm,clip]{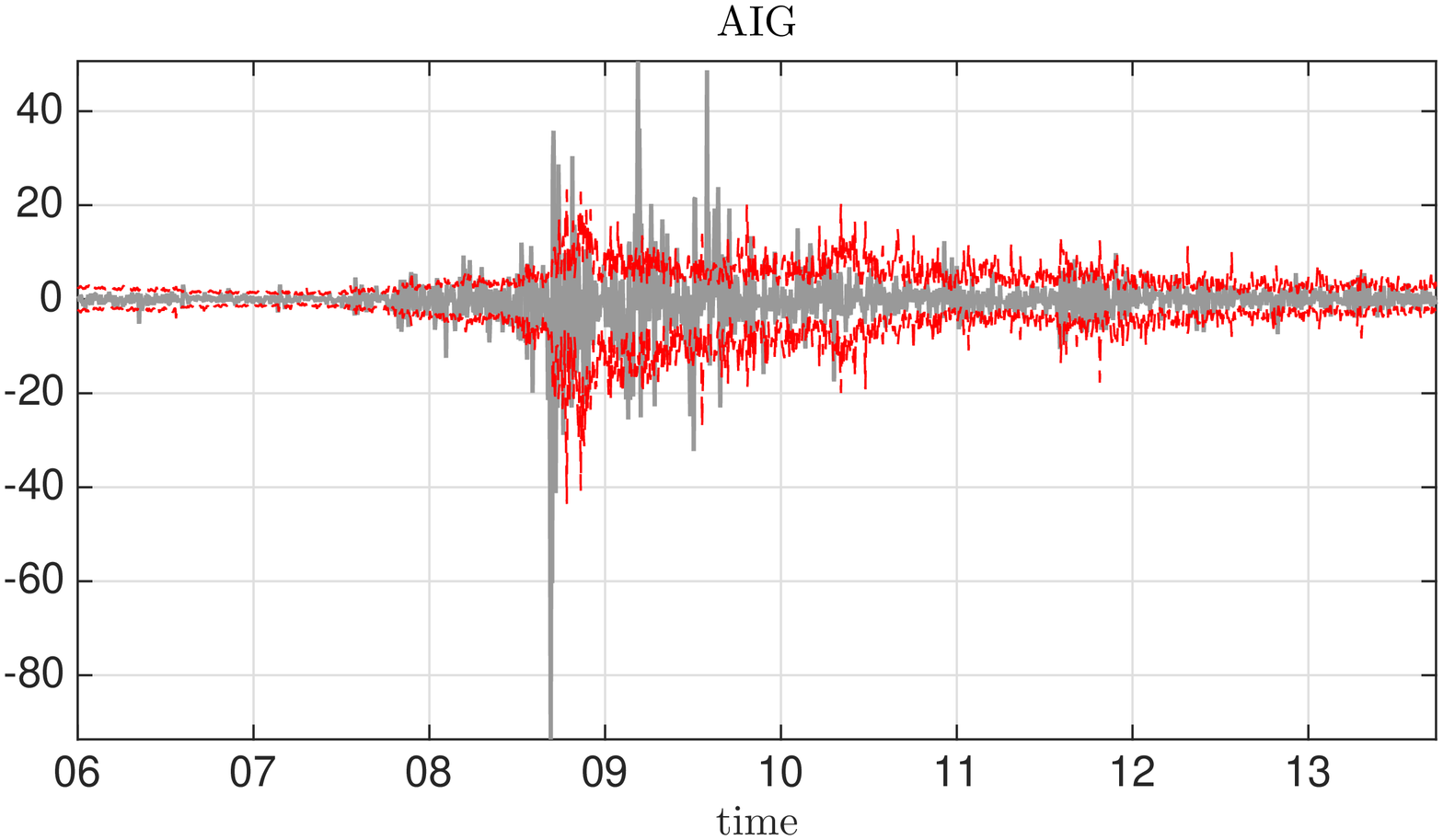}&
 \includegraphics[width=.45\textwidth,trim=3.5cm 1.6cm 2cm 0cm,clip]{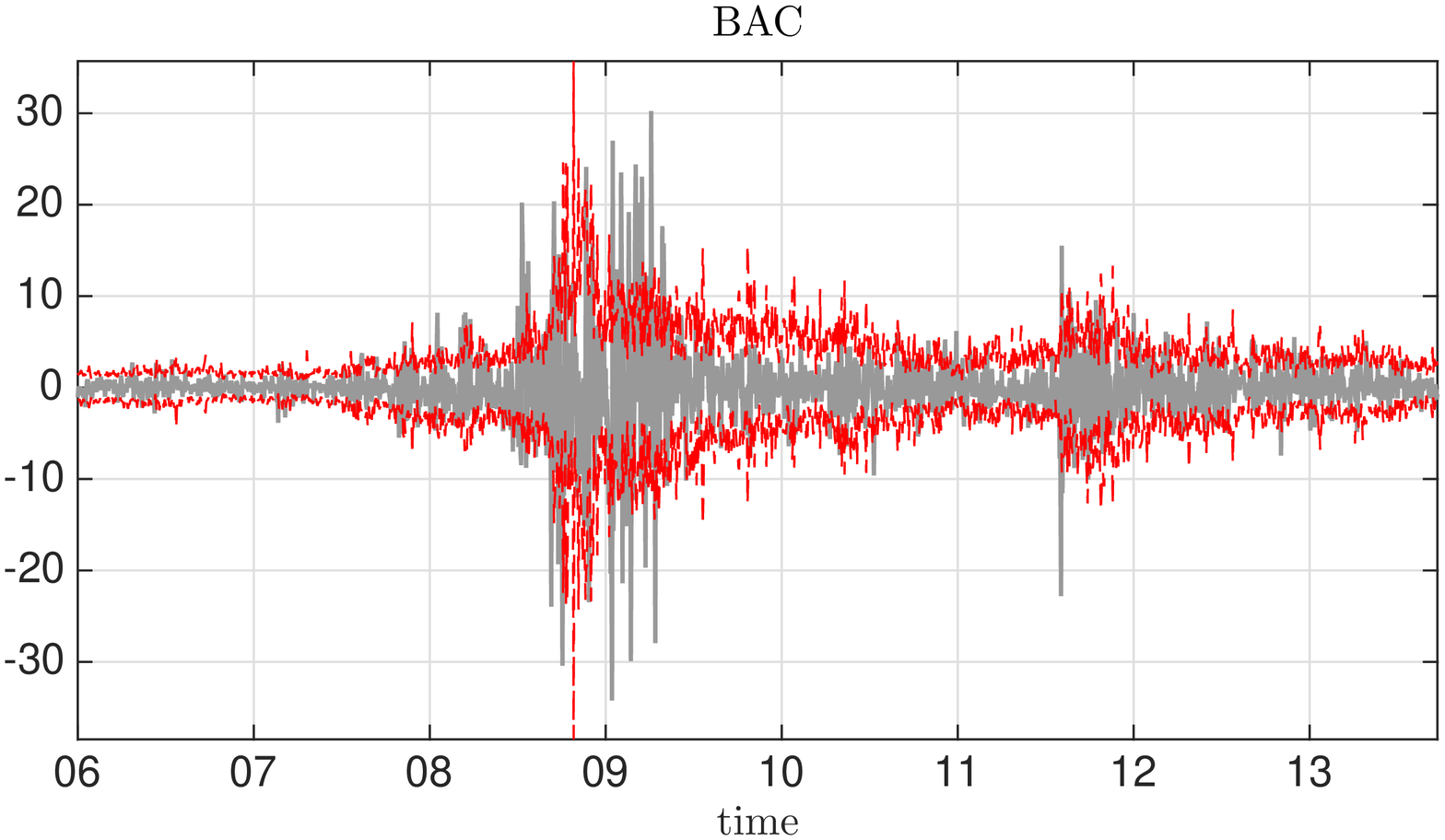} \\
  \includegraphics[width=.45\textwidth,trim=3.5cm 1.6cm 2cm 0cm,clip]{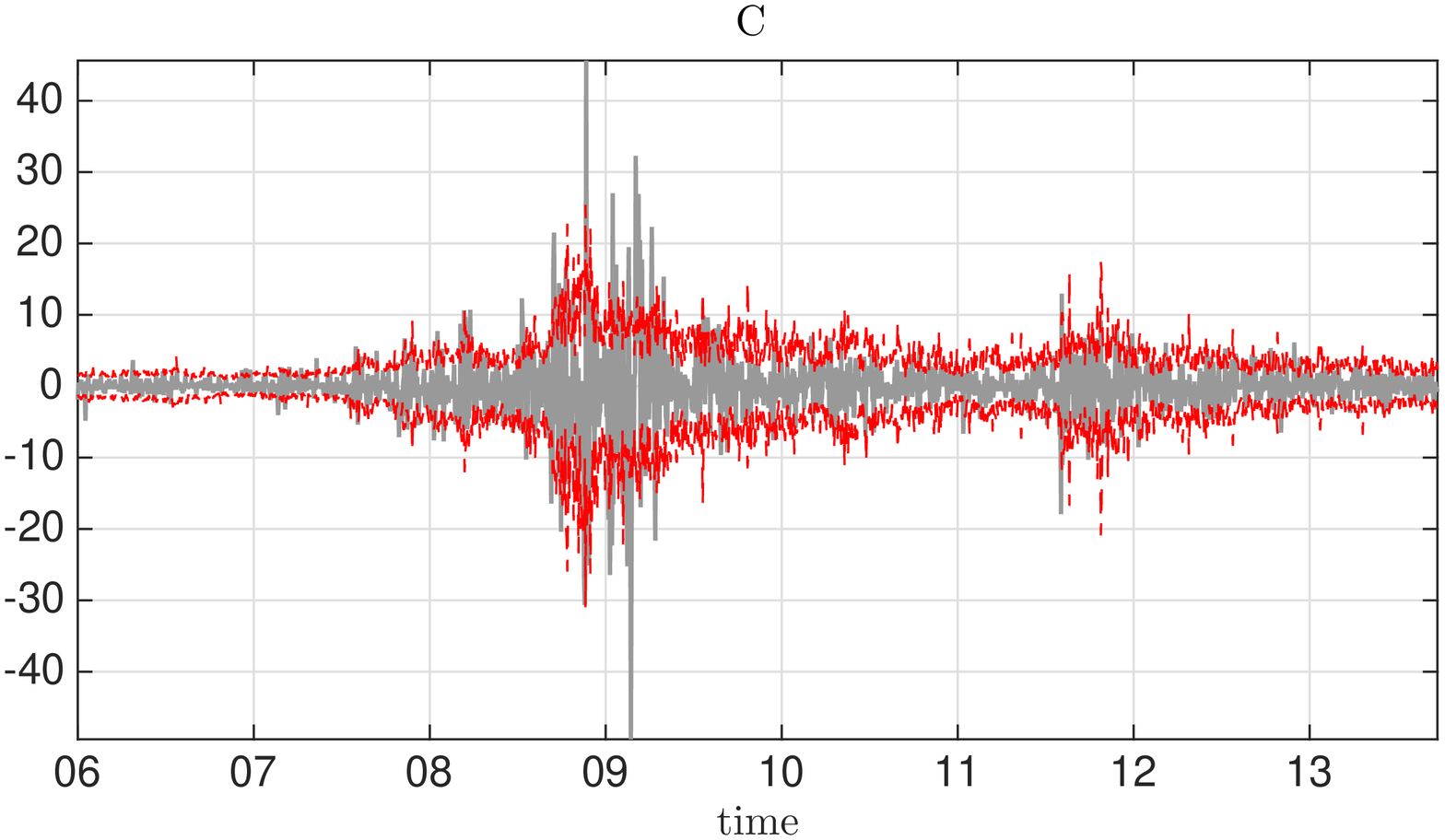}&
 \includegraphics[width=.45\textwidth,trim=3.5cm 1.6cm 2cm 0cm,clip]{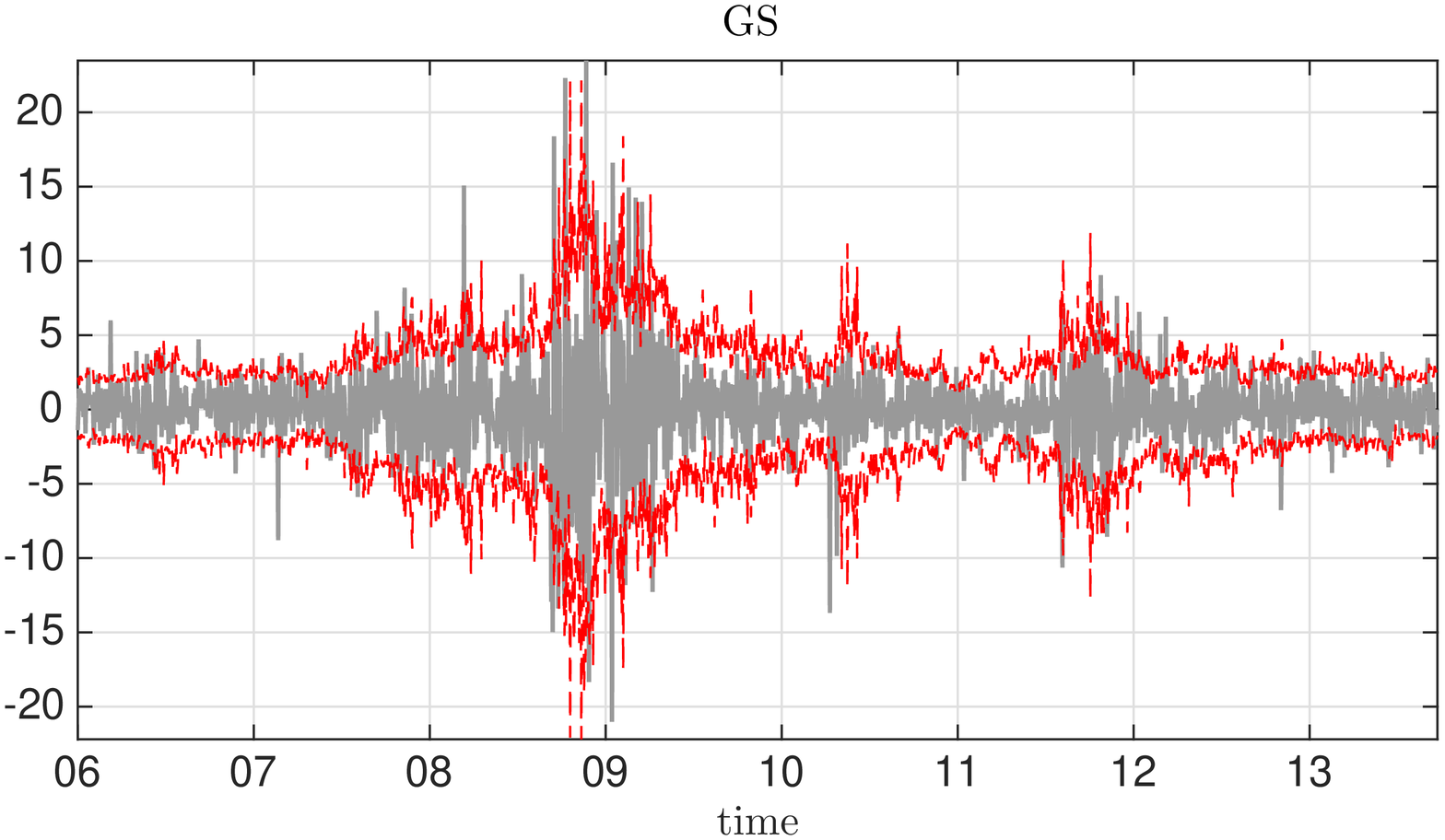}\\
  \includegraphics[width=.45\textwidth,trim=3.5cm 1.6cm 2cm 0cm,clip]{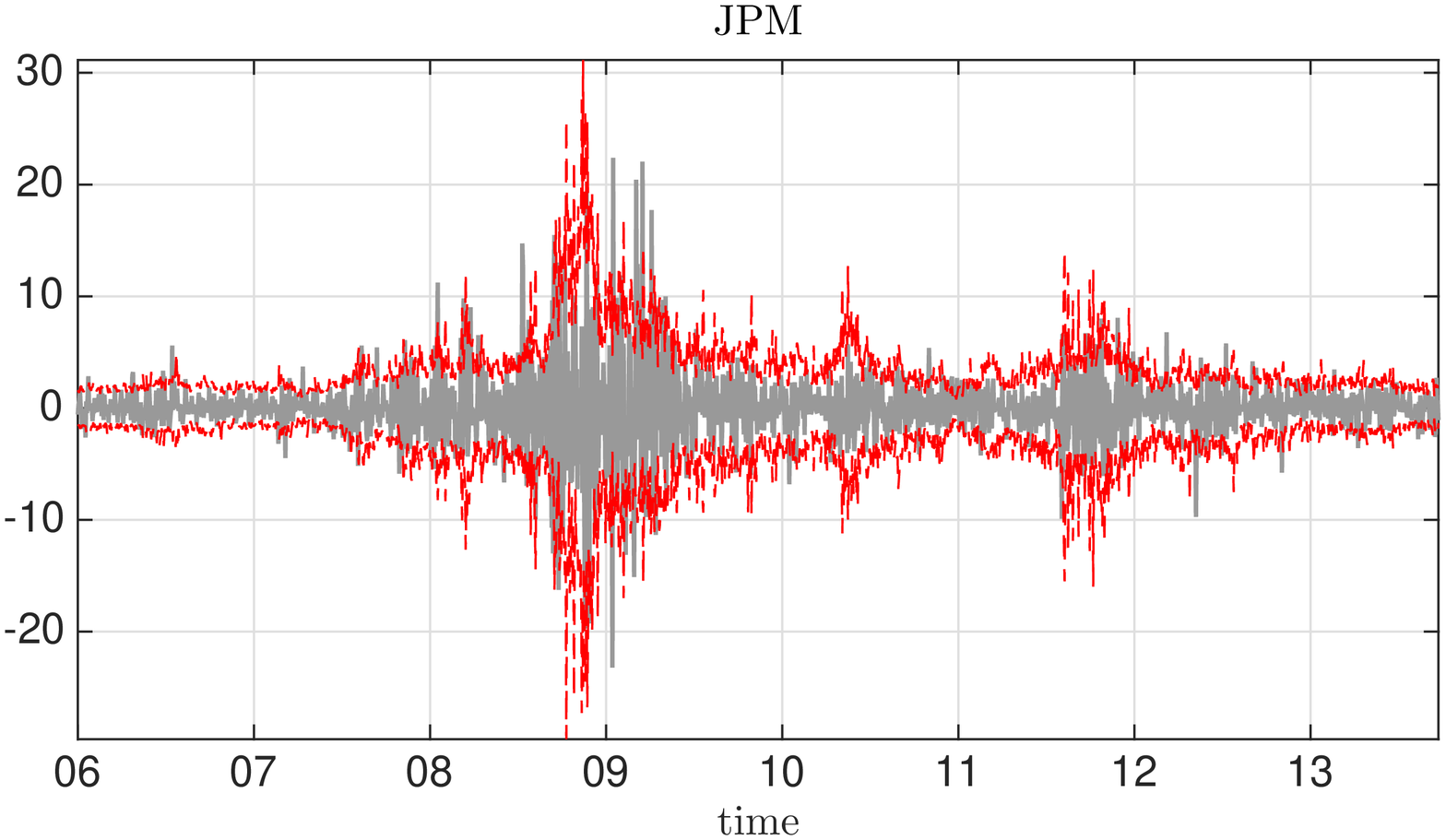}&
  \includegraphics[width=.45\textwidth,trim=3.5cm 1.6cm 2cm 0cm,clip]{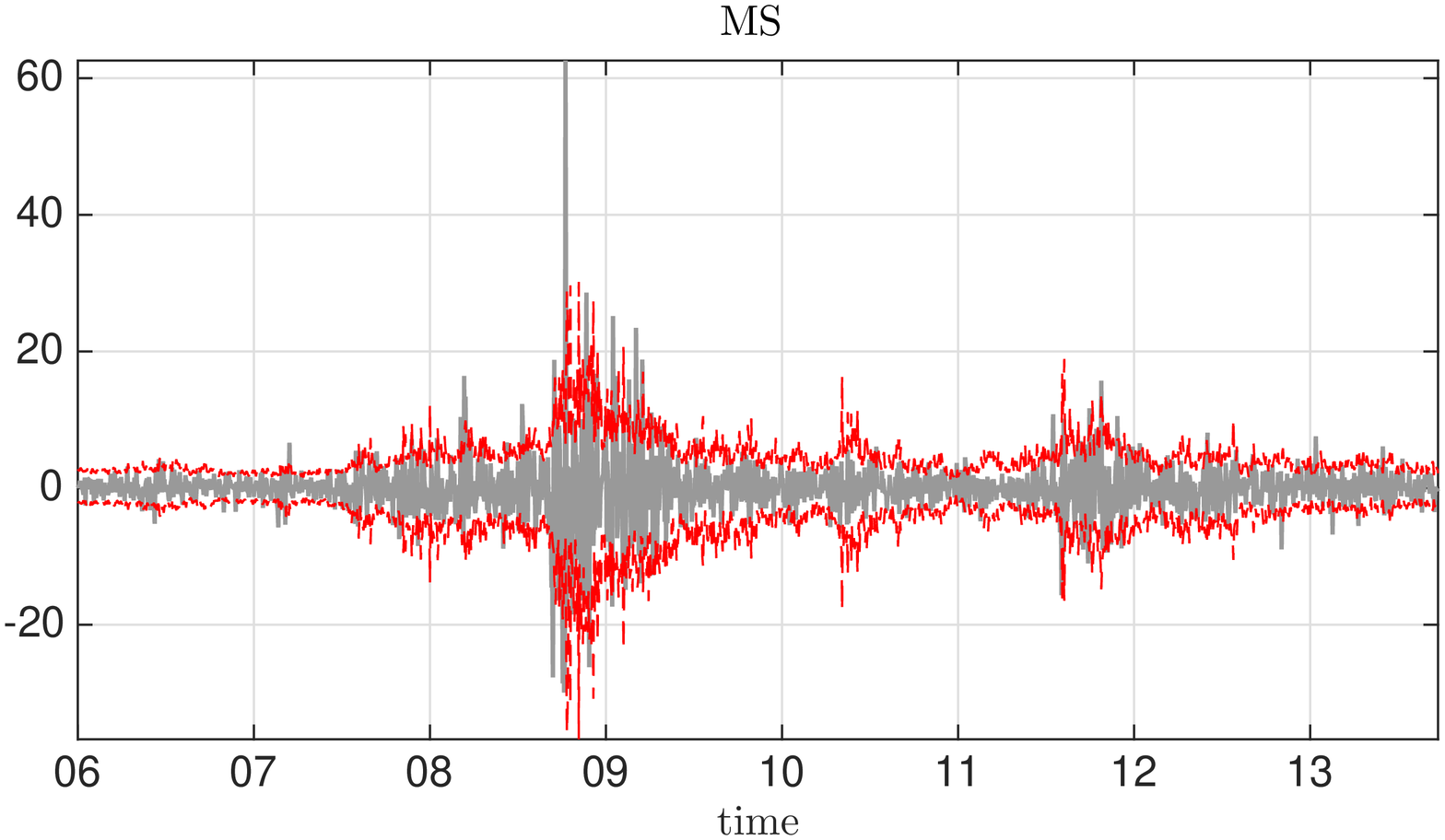}\\
 \end{tabular}
\end{figure}

\begin{figure}[t!]\caption{\small One-step-ahead 90\% conditional prediction intervals (in red; $\ell=252$): Apple (AAPL), Microsoft (MSFT), Amazon (AMZN), Wallgreens (WAG), Exxon Mobil (XOM), Johnson \& Johnson (JNJ), Boeing (BA), General Electric (GE).} \label{fig:stocks2}
\centering \smallskip\noindent 
\setlength{\tabcolsep}{.01\textwidth}
\begin{tabular}{@{}cc}
     \includegraphics[width=.45\textwidth,trim=3.5cm 1.6cm 2cm 0cm,clip]{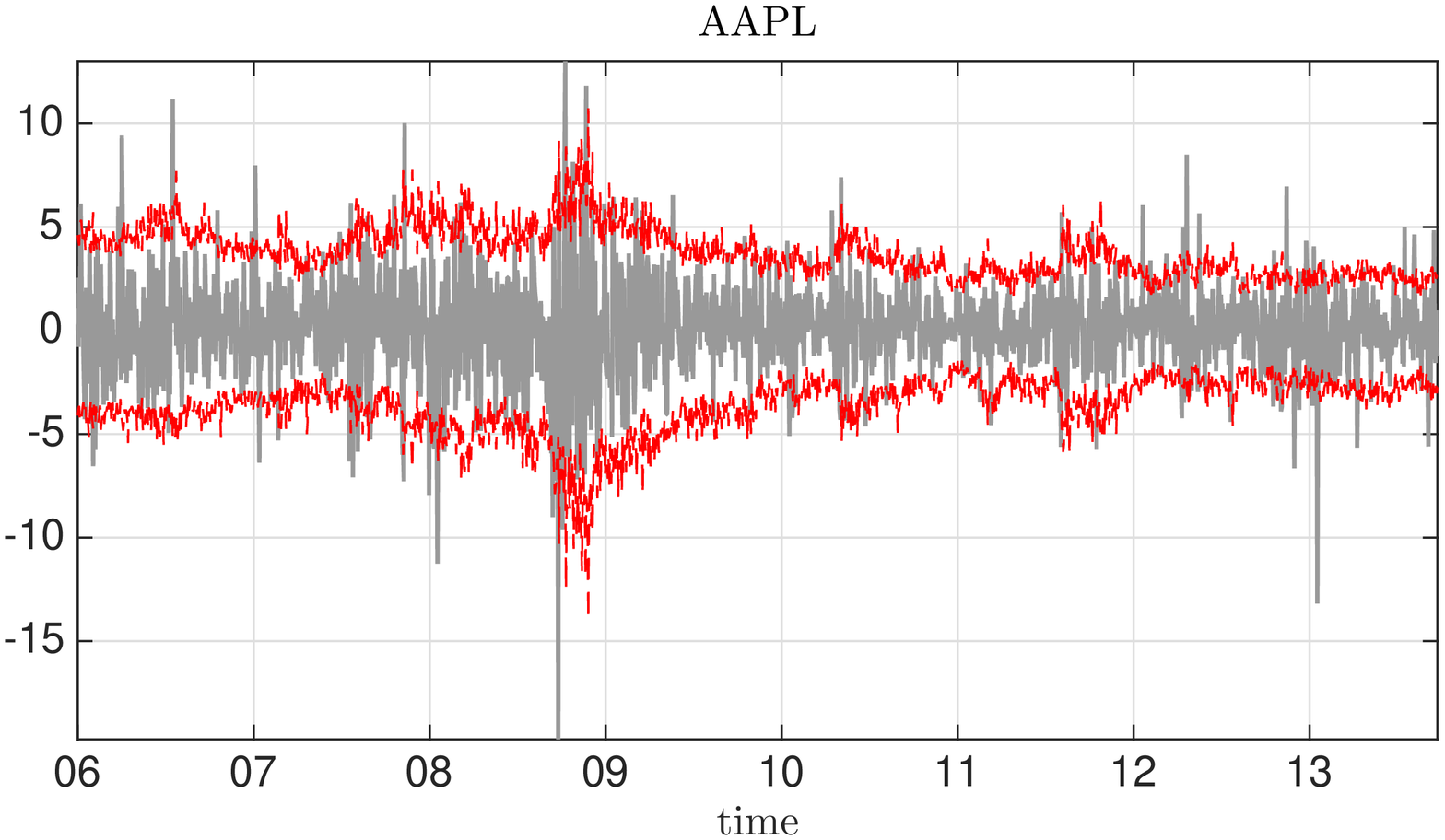}&
  \includegraphics[width=.45\textwidth,trim=3.5cm 1.6cm 2cm 0cm,clip]{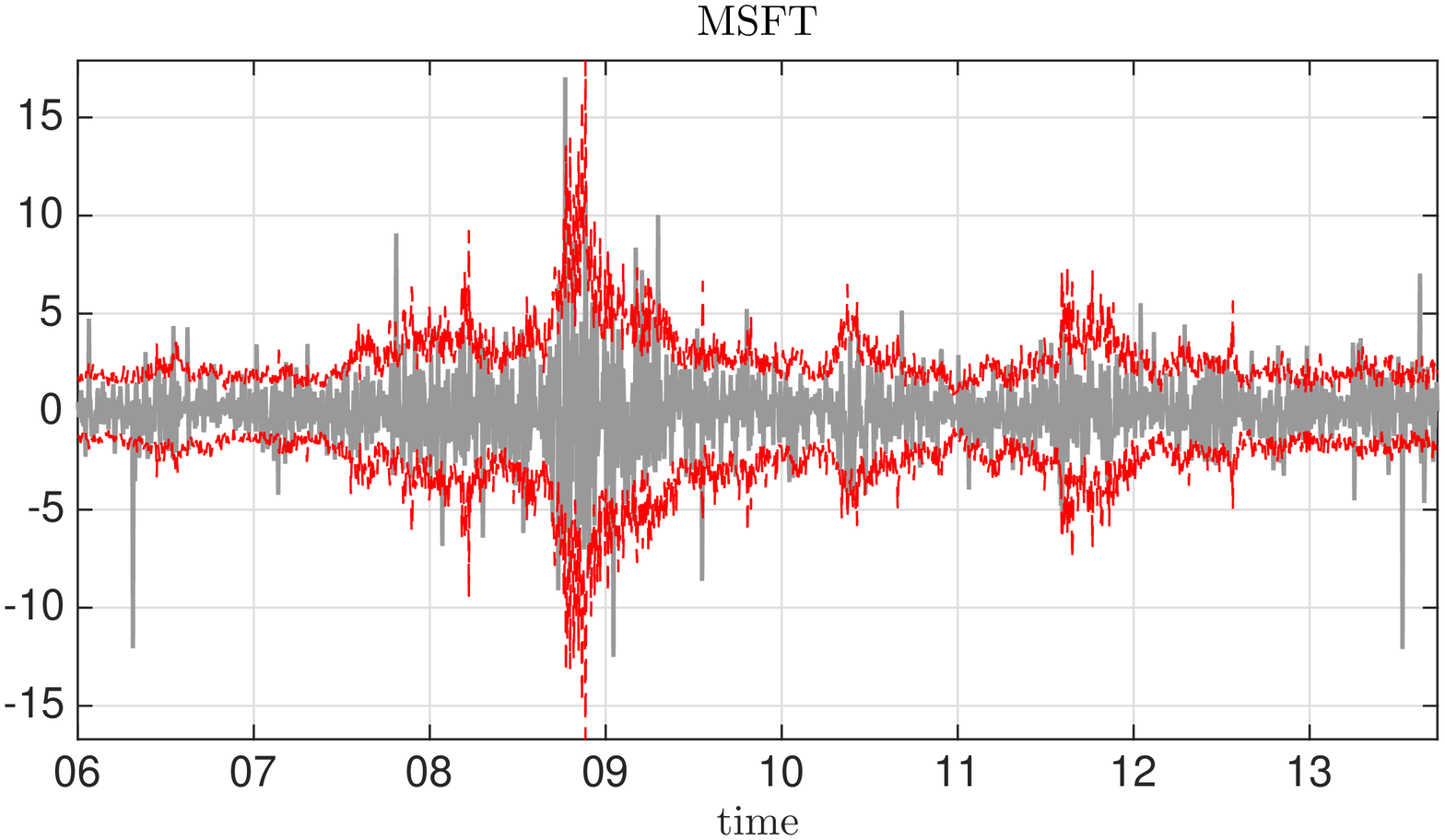}\\
 \includegraphics[width=.45\textwidth,trim=3.5cm 1.6cm 2cm 0cm,clip]{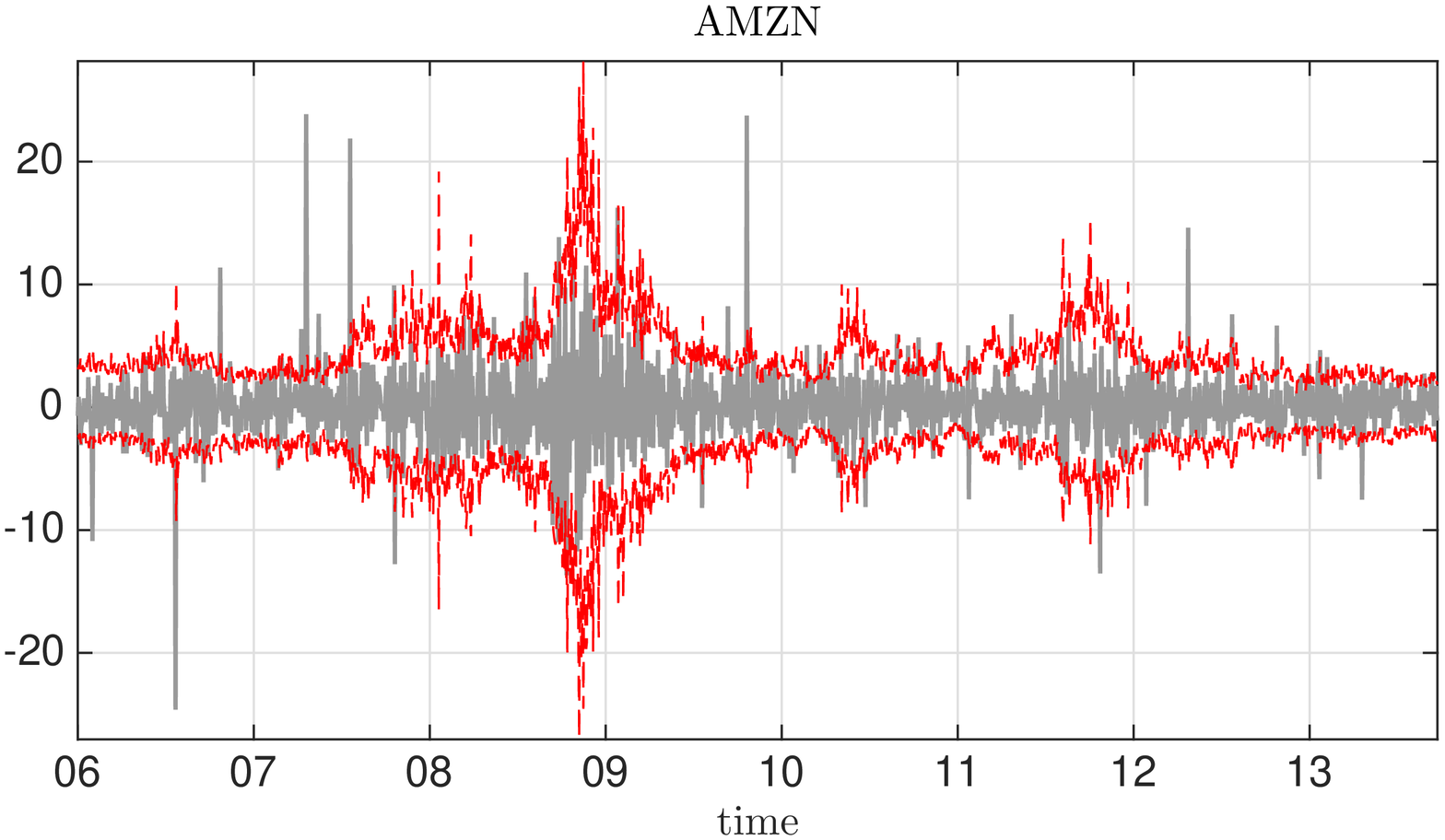}&
 \includegraphics[width=.45\textwidth,trim=3.5cm 1.6cm 2cm 0cm,clip]{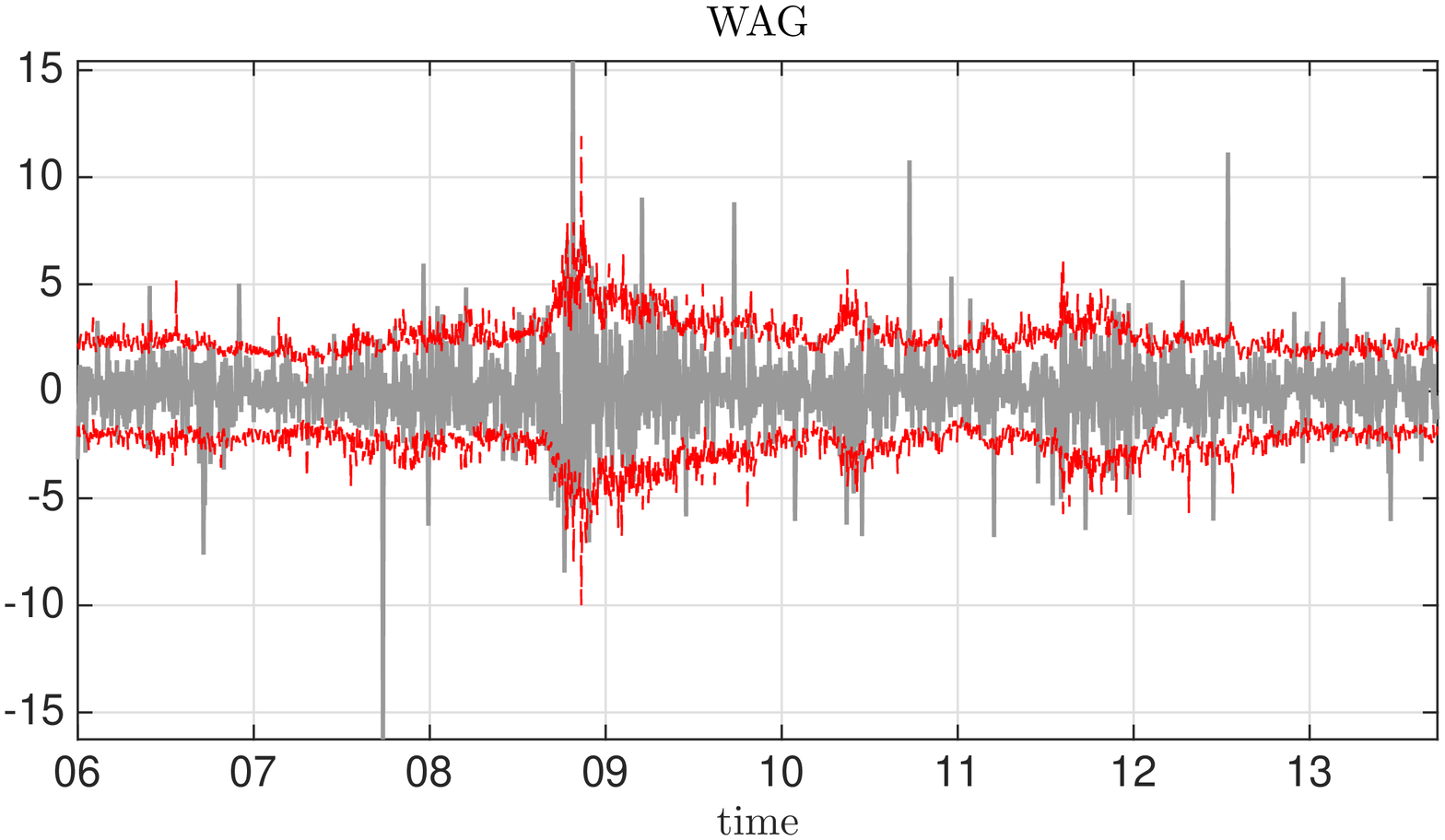} \\
  \includegraphics[width=.45\textwidth,trim=3.5cm 1.6cm 2cm 0cm,clip]{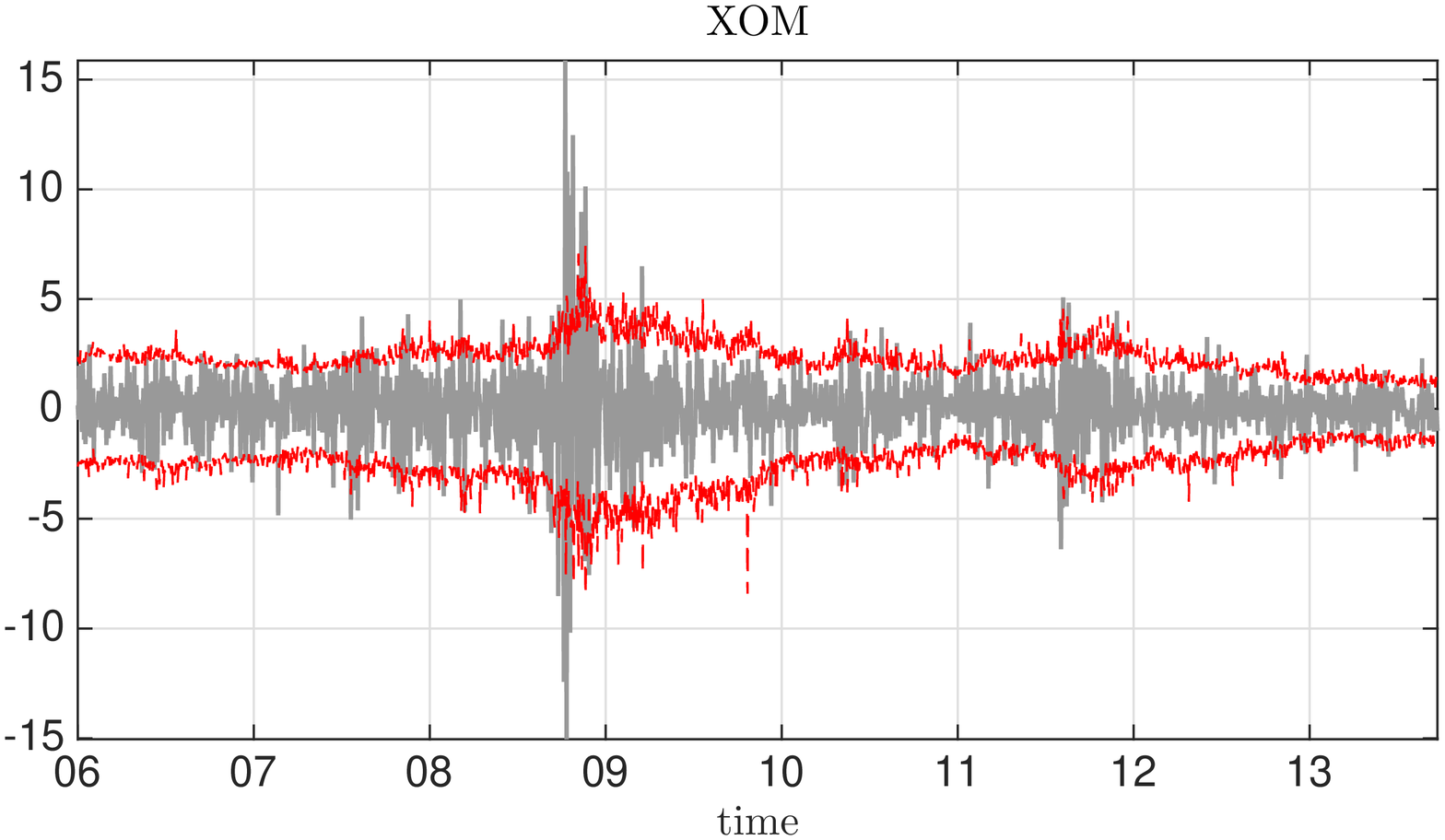}&
 \includegraphics[width=.45\textwidth,trim=3.5cm 1.6cm 2cm 0cm,clip]{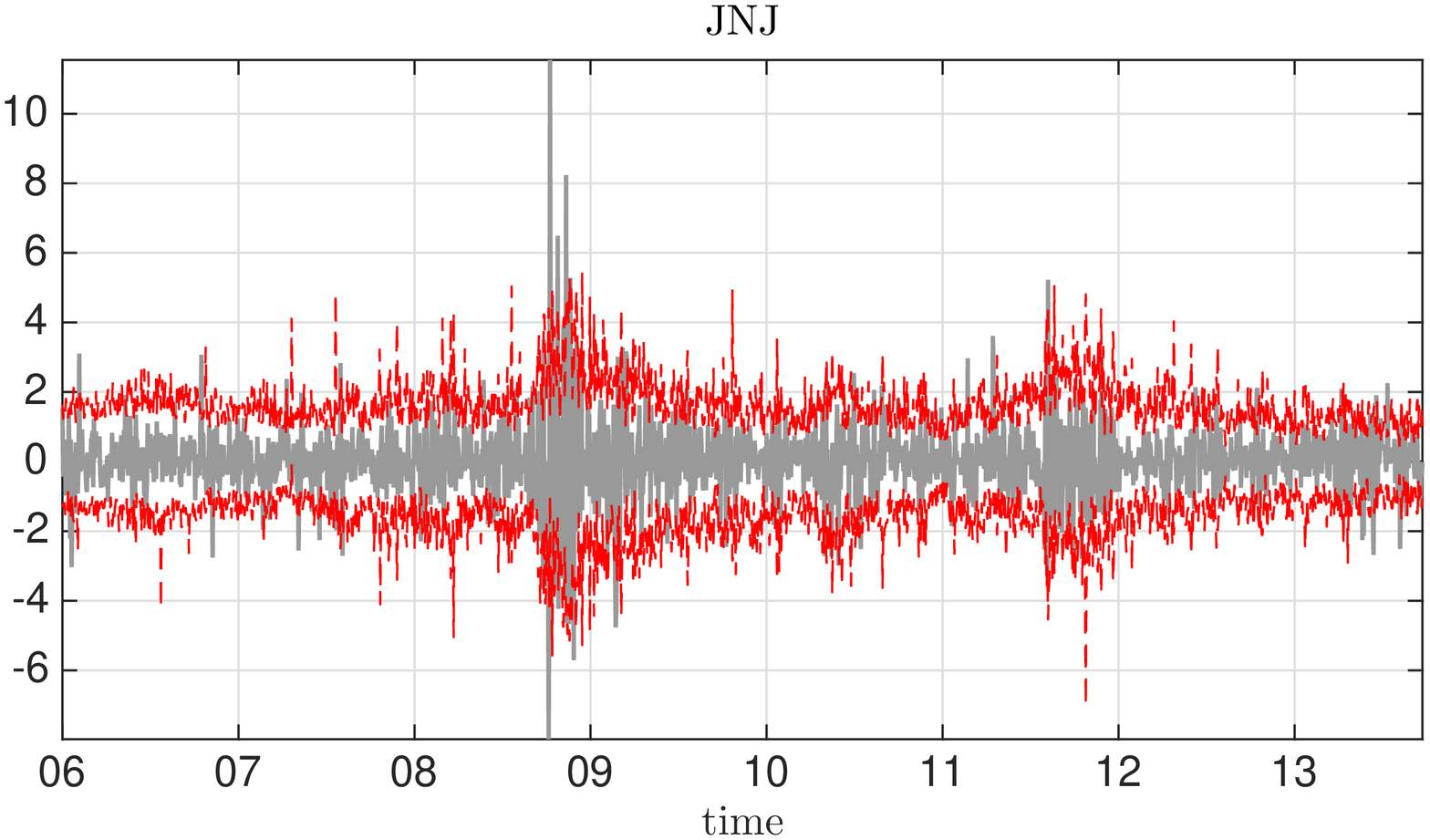}\\
  \includegraphics[width=.45\textwidth,trim=3.5cm 1.6cm 2cm 0cm,clip]{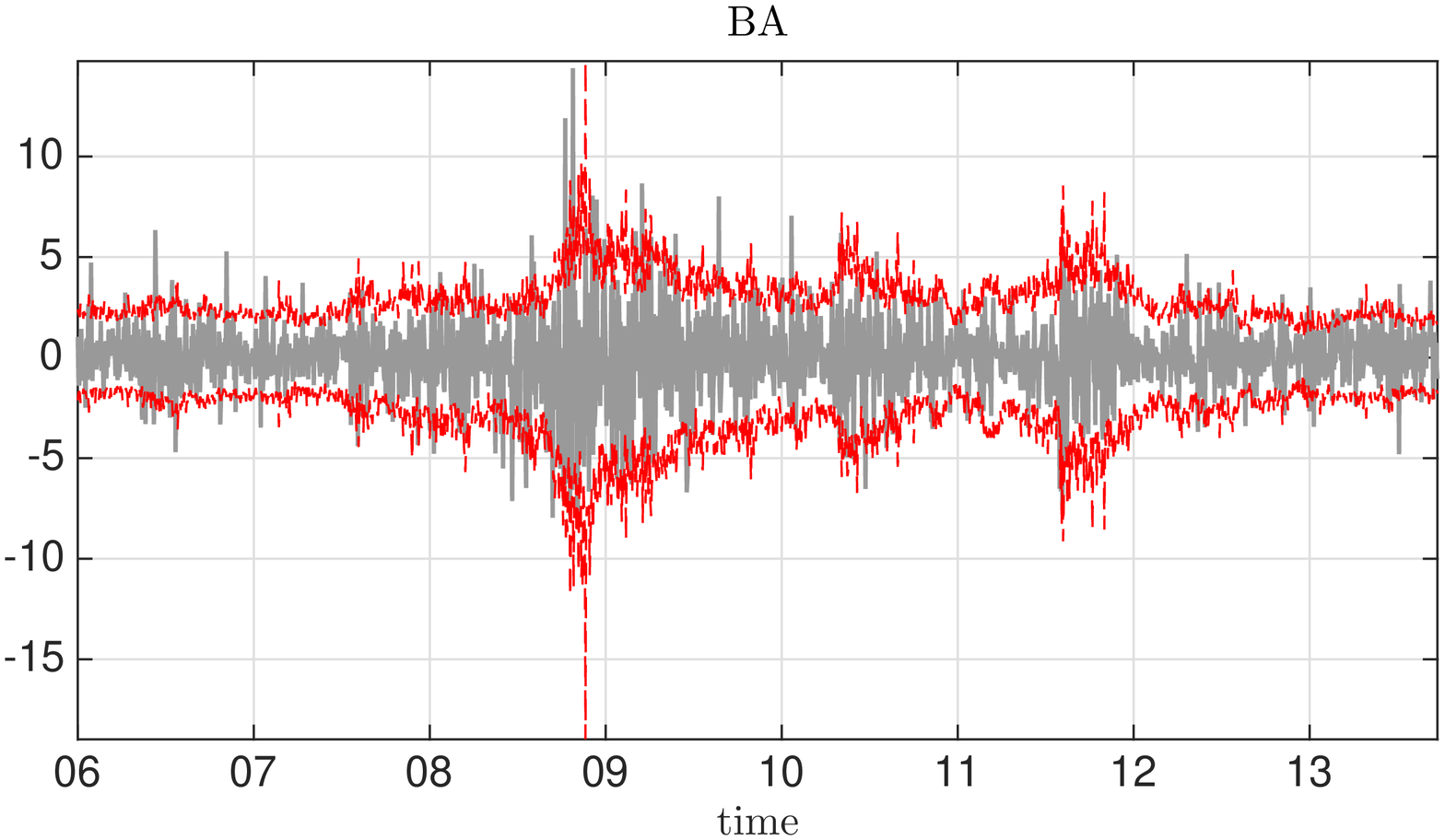}&
  \includegraphics[width=.45\textwidth,trim=3.5cm 1.6cm 2cm 0cm,clip]{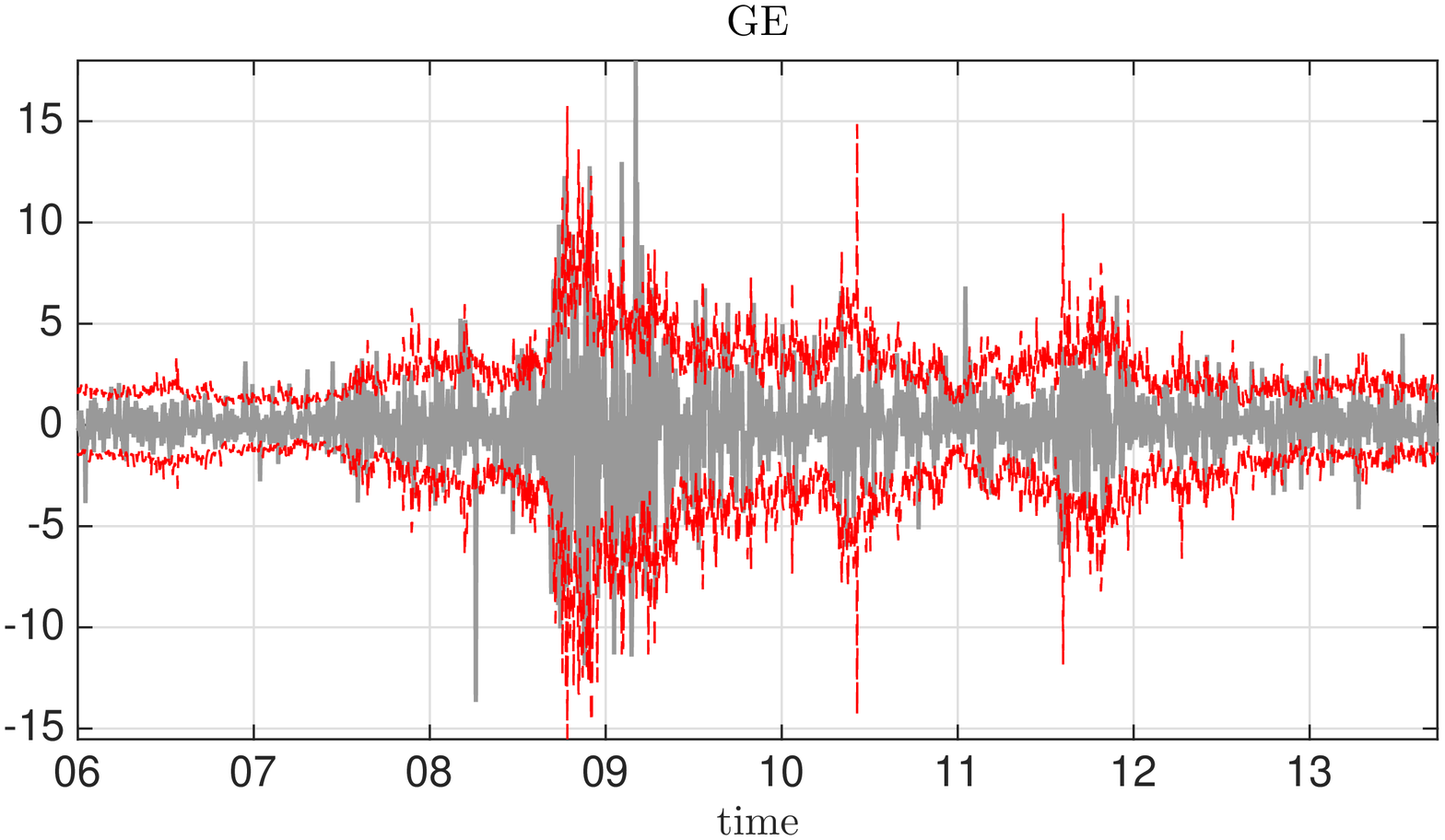}\\
 \end{tabular}
\end{figure}

In Figures \ref{fig:stocks} and \ref{fig:stocks2}, we set $\kappa_T=0.25$ and $\ell=252$ and we show (in grey) $Y_{i,\tau+1}$ for some selected individual stocks, together with (in red) the estimated upper and lower bounds of the 90\% one-step-ahead prediction interval, i.e. $\widehat{\mathcal U}^{(252)}_{i,\tau+1|\tau}(0.05)$ and $\widehat{\mathcal L}^{(252)}_{i,\tau+1|\tau}(0.05)$, respectively.  Figure \ref{fig:stocks}  shows results for six of the most volatiles stocks in our dataset, all belonging to the financial sector: America International Group (AIG), Bank of America (BAC), Citigroup (C), Goldman Sachs (GS), JPMorgan Chase (JPM), Morgan Stanley (MS). Figure \ref{fig:stocks2} provides the same results for eight  relevant  non-financial stocks: Apple (AAPL), Microsoft (MSFT), Amazon (AMZN), Wallgreens (WAG), Exxon Mobil (XOM), Johnson \& Johnson (JNJ), Boeing (BA), General Electric (GE). Volatilities, in those series, which were the most seriously affected by the great financial crisis, are notoriously hard to predict.

\subsection{Coverage: comparison with GARCH}
The novelty of our prediction intervals is that they are exploiting the information contained in the available cross-section of $n=90$ stocks. This is in sharp contrast with the usual GARCH approach, which is strongly univariate, and disregards cross-sectional information by analyzing the $n$ series one by one. Moreover, estimating 90 univariate GARCH models requires much more computing time than estimating our model. GARCH nevertheless constitute the more common practice in this context, and serves as a natural benchmark.
 
We therefore compare  our prediction intervals with those obtained by fitting, {via quasi-maximum likelihood,} univariate GARCH(1,1) models to all series in our panel. Specifically, for each series $i$,  we estimate the model
\begin{align}
Y_{it} &= \E[y_{it}]+ \sigma_{it}\epsilon_{it}, \quad \epsilon_{it}\stackrel{iid}{\sim} (0,1),\qquad t=1,\ldots, \tau,\nn\\
\sigma_{it}^2&=\omega_i +\gamma_i  Y_{it-1}^2+ \beta_i\sigma_{it-1}^2, \quad \omega_i>0,\ \gamma_i,\beta_i\ge 0,\ \gamma_i+\beta_i<1.\nn
\end{align}
 For  given $\tau=(T-M),\ldots, (T-1)$,  we obtain estimated parameters $\widehat{\omega}_i, \widehat{\gamma}_i$ and $\widehat{\beta}_i$, from which we compute the estimated volatilities $\widehat{\sigma}_{it}^2$ and the innovation values~$\widehat{\epsilon}_{i t}=Y_{it}/\widehat{\sigma}_{it}$, $t=1,\ldots, \tau$. 
 Innovation quantiles are computed from $(\widehat \epsilon_{i,\tau-\ell+1},\ldots,\widehat \epsilon_{i,\tau})$, where as before we set $\ell\in\{126,\,  252,\,  504,\,  \tau\}$. Then, for any given level $\alpha$ and window size $\ell$, and for $\tau=(T-M),\ldots, (T-1)$, given the one-step-ahead volatility pre- \linebreak
 dictor~$\widehat{\sigma}_{i,\tau+1|\tau}^2 = \widehat{\omega}_i+\widehat{\gamma}_i Y_{i,\tau}^2+\widehat{\beta}_i \widehat{\sigma}_{i,\tau}^2$, we compute the the upper and lower confidence bounds 
\[ 
\widehat{\mathcal U}_{i,\tau+1|\tau}^{(\ell)\text{\tiny GARCH}}(\alpha):=\bar Y_{i} + \widehat \sigma_{i,\tau+1|\tau}\, \widehat \epsilon^{(\ell)}_{i(\lceil \ell(1-\alpha)\rceil)}  \quad \text{and}\quad 
\widehat{\mathcal L}_{i,\tau+1|\tau}^{(\ell)\text{\tiny GARCH}}(\alpha):=\bar Y_{i} + \widehat \sigma_{i,\tau+1|\tau}\, \widehat \epsilon^{(\ell)}_{i(\lceil \ell\alpha\rceil)},\nn
\] 
yielding the one-step-ahead prediction intervals 
\[
\widehat{\mathcal I}_{i,\tau+1|\tau}^{(\ell)\text{\tiny GARCH}}(\alpha):=\big[\widehat{\mathcal L}^{(\ell)\text{\tiny GARCH}}_{i,\tau+1|\tau}(\alpha/2),\widehat{\mathcal U}^{(\ell)\text{\tiny GARCH}}_{i,\tau+1|\tau}(\alpha/2)\big]
\] 
and the indicators of correct interval prediction $\widehat{\mathcal H}_{i,\tau+1|\tau}^{(\ell)\text{\tiny GARCH}}(\alpha):=\mathbb I(Y_{i,\tau+1}\in \widehat{\mathcal I}_{i,\tau+1|\tau}^{(\ell)\text{\tiny GARCH}}(\alpha))$.  Based on these quantities,   we then compute, for $\alpha\in\{0.32,\, 0.2,\, 0.1,\, 0.05,\, 0.01\}$, 
the empirical coverage frequency, denoted as $C_i^{(\ell)\text{\tiny{GARCH}}}(\alpha)$, the 
proportions of coverage violations in the upper and lower tail, denoted as $V_{i,+}^{(\ell)\text{\tiny{GARCH}}}(\alpha/2)$ and~$V_{i,-}^{(\ell)\text{\tiny{GARCH}}}(\alpha/2)$, respectively,  and the average interval length, denoted as  ${L}_i^{(\ell)\text{\tiny{GARCH}}}(\alpha)$. 
Averages of these quantities over the $n$ series under study are shown in Table \ref{tab:garch}. Inspection of this table reveals that the GDFM performances are slightly better than the GARCH ones in terms of coverage frequencies, based on similar interval lengths. This, however, is mainly a descriptive  and, due to cross-sectional dependence, somewhat misleading assessment, which ideally should be  reinforced into a more formal testing analysis.

\begin{table}[t!]
\centering
\caption{\small  Standard \& Poor's 100 Index data ($n=90$ daily returns). Empirical coverage, frequency of prediction bounds  violations, and average length of prediction intervals for  GARCH, averaged over the cross-section. }\label{tab:garch}
\vskip .2cm
\footnotesize
\begin{tabular}{l | ccccc }
\hline
\hline
&\multicolumn{5}{c}{$\alpha$}\\
&0.32&0.2&0.1&0.05&0.01 \\
\hline
$C^{(126)\text{\tiny{GARCH}}}(\alpha)$&0.6755 &   0.7947  &  0.8933  &  0.9429  &  0.9834\\
$V_{+}^{(126)\text{\tiny{GARCH}}}(\alpha/2)$&0.1576 &   0.0991 &   0.0507 &   0.0267  &  0.0076\\
$V_{-}^{(126)\text{\tiny{GARCH}}}(\alpha/2)$&0.1669  &  0.1062&    0.0560 &   0.0304 &   0.0090\\
$L^{(126)\text{\tiny{GARCH}}}(\alpha)$&3.4401 &   4.5562  &  6.1207 &   7.7282   & 12.3986\\
\hline
$C^{(252)\text{\tiny{GARCH}}}(\alpha)$&0.6786  &  0.7981 &   0.8968  &  0.9460 &   0.9871\\
$V_{+}^{(252)\text{\tiny{GARCH}}}(\alpha/2)$&0.1567  &  0.0978 &   0.0491  &  0.0255  &  0.0060\\
$V_{-}^{(252)\text{\tiny{GARCH}}}(\alpha/2)$& 0.1647  &  0.1041  &  0.0541 &   0.0285  &  0.0069\\
$L^{(252)\text{\tiny{GARCH}}}(\alpha)$&3.4142  &  4.5235 &   6.0755  &  7.6329 &  12.2536\\
\hline
$C^{(504)\text{\tiny{GARCH}}}(\alpha)$&0.6807  &  0.7994  &  0.8983  &  0.9479  &  0.9878\\
$V_{+}^{(504)\text{\tiny{GARCH}}}(\alpha/2)$&0.1560 &   0.0975  &  0.0488   & 0.0248 &   0.0056\\
$V_{-}^{(504)\text{\tiny{GARCH}}}(\alpha/2)$&0.1633  &  0.1031 &   0.0529  &  0.0274  &  0.0066\\
$L^{(504)\text{\tiny{GARCH}}}(\alpha)$&3.3822   & 4.4801  & 6.0220  &  7.5581 &  11.7469\\
\hline
$C^{(\tau)\text{\tiny{GARCH}}}(\alpha)$&0.6920 &   0.8077 &   0.9036 &   0.9510  &  0.9897\\
$V_{+}^{(\tau)\text{\tiny{GARCH}}}(\alpha/2)$&0.1520  &  0.0935  &  0.0458   & 0.0228   & 0.0048 \\
$V_{-}^{(\tau)\text{\tiny{GARCH}}}(\alpha/2)$& 0.1560   & 0.0988  &  0.0505 &   0.0262 &   0.0055\\
$L^{(\tau)\text{\tiny{GARCH}}}(\alpha)$&3.4139  &  4.4942 &   6.0156  &  7.5369 &  11.6268\\
\hline
\hline
\end{tabular}
\end{table}

A formal comparison between the GDFM and GARCH(1,1) coverage performances should take into account the fact that the coverage results of the two methods, for given $i$ and $\tau$, are not independent. The situation is quite similar to that of comparing paired proportions,  where tests are to  be carried out on the basis of the traditional \cite{McN47}   test. For  given $\alpha$ and $\ell$, consider, for all $i$,  the events (discordant GDFM and GARCH coverage results)
\begin{align}
\mathcal A_{i,\tau+1|\tau}^{(\ell)}(\alpha)&:=\l\{Y_{i,\tau+1}\in \widehat{\mathcal I}_{i,\tau+1|\tau}^{(\ell)}(\alpha) \cap Y_{i,\tau+1}\notin \widehat{\mathcal I}_{i,\tau+1|\tau}^{(\ell)\text{\tiny GARCH}}(\alpha)\r\}\nn\\
\mathcal B_{i,\tau+1|\tau}^{(\ell)}(\alpha)&:=\l\{Y_{i,\tau+1}\notin \widehat{\mathcal I}_{i,\tau+1|\tau}^{(\ell)}(\alpha) \cap Y_{i,\tau+1}\in \widehat{\mathcal I}_{i,\tau+1|\tau}^{(\ell)\text{\tiny GARCH}}(\alpha)\r\},\nn
\end{align}
and define 
\[
n_{12i}^{(\ell)}(\alpha) := \sum_{\tau=T-M}^{T-1} \mathbb I\l(\mathcal A_{i,\tau+1|\tau}^{(\ell)}(\alpha) \r) \quad\text{ and }\quad 
n_{21i}^{(\ell)}(\alpha) := \sum_{\tau=T-M}^{T-1} \mathbb I\l(\mathcal B_{i,\tau+1|\tau}^{(\ell)}(\alpha)\r).
\]
Consider the null  hypothesis under which the indicators of a successful interval prediction in both methods are i.i.d.~Bernoulli,  with identical (but otherwise unspecified) coverage probabilities. 
The McNemar test of that hypothesis is conditioning on the sum $n_{\text{disc},i}^{(\ell)}(\alpha):=n_{12i}^{(\ell)}(\alpha) + n_{21i}^{(\ell)}(\alpha)$ of discordant coverage results: concordant results indeed carry no information on a difference between coverage probabilities. Conditional on $n_{\text{disc},i}^{(\ell)}(\alpha)$,  the null distribution of  $n_{12i}^{(\ell)}(\alpha)$  is binomial  Bin$(n_{\text{disc},i}^{(\ell)}(\alpha),\,  0.5)$. At probability level~$\delta$, the test rejects in favour of a better GDFM coverage for ``large values'' of $n_{12i}^{(\ell)}(\alpha)$, in favour of a better GARCH coverage for ``small values'' of the same (equivalently, ``large values'' of~$n^{(\ell)}_{21i}$), with critical values the $(1-\delta)$ and $\delta$ binomial quantiles, respectively.

Table \ref{tab:comparetest1} reports the McNemar empirical rejection frequencies (over the $n=90$ series)---in favour of a better GDFM coverage in the left-hand panel, in favour of a better GARCH coverage in the right-hand one. We consider the cases in which $\alpha=0.1$ or $0.05$, $\ell=126$ or $252$, $\kappa_T=0.25$ (for the GDFM); testing was performed at significance levels  $\delta=0.1$, $0.05$, and $0.01$. Irrespective of $\ell$ and $\alpha$, the GDFM approach appears to outperform, quite consistently and significantly,  the GARCH one.

\begin{table}[t!]
\centering
\caption{\small Standard \& Poor's 100 Index data ($n=90$ daily returns). Proportions of McNemar rejections in favour of a better GDFM coverage (left-hand panel), in favour of a better GARCH coverage (right-hand panel).
 }\label{tab:comparetest1}
\vskip .2cm
\footnotesize
\begin{tabular}{l  | ccc | ccc }
\hline
\hline
&\multicolumn{3}{|c}{better GDFM coverage}&\multicolumn{3}{|c}{better GARCH coverage}\\
\hline
$\alpha=0.1$ & $\delta=0.1$ & $\delta=0.05$ & $\delta=0.01$ &$\delta=0.1$ & $\delta=0.05$ & $\delta=0.01$\\
\hline
$\ell=126$&0.6000  &  0.5444  &  0.3667 & 0.1000&    0.0778 &   0.0667\\
$\ell=252$&0.5333  &  0.4556   & 0.2889 & 0.1333&    0.1222 &   0.0889\\
\hline
\hline
$\alpha=0.05$ & $\delta=0.1$ & $\delta=0.05$ & $\delta=0.01$ &$\delta=0.1$ & $\delta=0.05$ & $\delta=0.01$\\
\hline
$\ell=126$&0.4111 &   0.2778  &  0.1556 &0.1111 &   0.0778 &   0.0556\\
$\ell=252$&0.2556 &   0.2000  &  0.0556&0.1778  &  0.1222  &  0.1111\\
\hline
\hline
\end{tabular}
\end{table}

\subsection{Coverage:  backtesting}\label{backtestSec}
As explained in Section~\ref{eq:int}, a   formal  assessment of the validity of our approach can be based on  the backtesting procedure proposed by \citet{christoffersen1998}. The idea consists in testing the null hypothesis \eqref{eq:null} under which the $\widehat{\mathcal H}_{i,\tau+1|\tau}(\alpha)$'s (the indicators of a successful interval prediction) are i.i.d.~Bernoulli$ (1-\alpha)$.  Depending on the objectives, several alternatives can be considered. One can be interested (Section \ref{521}) in the validity of interval prediction or    the  sharpness of the nominal coverage level. Else, one may consider (Section \ref{522}) alternatives of serial dependence. Or, those two issues can be combined (Section \ref{523}) by merging the corresponding  alternatives. 

Irrespective of the alternative, however, it should be insisted that all those tests---one for each cross-sectional item---are intrinsically univariate. When simultaneously  performing several or all of them, one should be extremely cautious with the interpretation of the results. The tables we are providing below are reporting   empirical   rejection frequencies (over the $n=90$ series). Those $n$ tests, however, are not functionally interrelated (as they would be if the prediction intervals were based on the quantiles of common shocks only); hence, they are not about  testing the validity of {\it joint prediction intervals} with global asymptotic coverage level $(1-\alpha)$. Neither are they mildly interrelated (as they would be if the prediction intervals were exclusively based on idiosyncratic quantiles), providing  joint prediction intervals with global asymptotic coverage level of the order of $(1-\alpha)^n$. High rejection frequencies across the $n$ series thus do  not imply  bad forecasting properties, but can result from   complex cross-sectional dependencies. A standard attitude would consist  in adopting a  Bonferroni or a  \v Sid\' ak  correction; for $n=90$, and for a global  testing level of $1\%$, this would lead to implementing the~$n=90$ individual tests at an overly conservative level $\delta \approx 0.0001 = 10^{-4}$---a level at which none of the null hypotheses under study is rejected.   

All  tests below are  performed for $\kappa_T=0.25$, $\alpha=0.1$ or $0.05$, $\ell=126$ or $252$; testing significance levels are~$\delta=0.1$, $0.05$, and $0.01$.

\subsubsection{Testing for valid or sharp conditional coverage probabilities}\label{521}
If we are interested in the validity of interval prediction, the relevant testing problems are (one-sided)
\beq\label{eq:testUConesided}
H_{0i}: \E[\widehat{\mathcal H}_{i,\tau+1|\tau}^{(\ell)}(\alpha)] \geq (1-\alpha)\quad\text{ versus }\quad H_{1i}: \E[\widehat{\mathcal H}^{(\ell)}_{i,\tau+1|\tau}(\alpha)]< (1-\alpha).
\eeq
If instead we are interested in testing whether $(1-\alpha)$, as a nominal confidence level, is sharp, the testing problems are   (still one-sided) 
\beq\label{eq:testUConesided2}
H_{0i}: \E[\widehat{\mathcal H}_{i,\tau+1|\tau}^{(\ell)}(\alpha)] \leq (1-\alpha)\quad\text{ versus }\quad H_{1i}: \E[\widehat{\mathcal H}^{(\ell)}_{i,\tau+1|\tau}(\alpha)]> (1-\alpha).
\eeq
Both testing problems \eqref{eq:testUConesided} and  \eqref{eq:testUConesided2},
 admit a level-$\delta$ uniformly most powerful solution, rejecting $H_{0i}$ whenever the test statistic 
$$
n_{1i}^{(\ell)}(\alpha) :=\sum_{\tau=T-M}^{T-1}\widehat{\mathcal H}_{i,\tau+1|\tau}^{(\ell)}(\alpha)
$$
falls below the binomial Bin$(M, 1-\alpha)$ quantile of order~$\delta$ when testing \eqref{eq:testUConesided}, or above  the \linebreak Bin$(M, 1-\alpha)$ quantile of order~$(1-\delta)$ when testing \eqref{eq:testUConesided2}. Since $M$ is large, the same tests are well approximated by rejecting $H_{0i}$ whenever the proportion  $n_{1i}^{(\ell)}(\alpha) /M$ of correct coverage is smaller than~$(1-\alpha) - z_{\delta}\sqrt{\alpha(1-\alpha)}$ when testing \eqref{eq:testUConesided}, or larger than $(1-\alpha) + z_{\delta}\sqrt{\alpha(1-\alpha)}$ when testing~\eqref{eq:testUConesided2}, where~$z_{\delta}$ stands for the $(1-\delta)$ standard normal quantile. A two-sided coverage test can also be computed
\beq\label{eq:2sided}
LR_{\text{cover},i}^{(\ell)}(\alpha):= \big(n_{1i}^{(\ell)}(\alpha) -M(1-\alpha)\big)^2/M\alpha(1-\alpha),
\eeq
with asymptotic $\chi^2_{(1)}$ null distribution (as $M\to\infty$).\footnote{It is easily seen that $LR_{\text{cover},i}^{(\ell)}(\alpha)$ is equivalent, up to a constant term, to the so-called ``unconditional coverage'' likelihood ratio test statistic proposed in  Section 3.1 of \citet{christoffersen1998}  which therefore yields the same results.} 

Table \ref{tab:testUC_valid)}  reports the empirical   rejection frequencies (over $n=90$ series) when testing \eqref{eq:testUConesided} (left-hand panel) and~\eqref{eq:testUConesided2} (right-hand panel), respectively and using the normal approximation of the binomial. The general comments above apply when interpreting those tables: the only valid global conclusions are those resulting from Bonferroni or \v Sid\' ak corrections, which do not lead to any rejections.

\begin{table}[t!]
\centering
\caption{\small Standard \& Poor's 100 Index data ($n=90$ daily returns). Proportion of rejections when testing for valid nominal coverage~\eqref{eq:testUConesided} (left-hand panel) and for sharp nominal coverage \eqref{eq:testUConesided2} (middle panel), and when considering the two-sided test~\eqref{eq:2sided} (right-hand panel)}\label{tab:testUC_valid)}
\vskip .2cm
\footnotesize
\begin{tabular}{l  | ccc | ccc | ccc}
\hline
\hline
&\multicolumn{3}{|c}{valid nominal coverage test}&\multicolumn{3}{|c}{sharp nominal coverage test}&\multicolumn{3}{|c}{two-sided coverage test}\\
\hline
$\alpha=0.1$ & $\delta=0.1$ & $\delta=0.05$ & $\delta=0.01$& $\delta=0.1$ & $\delta=0.05$ & $\delta=0.01$& $\delta=0.1$ & $\delta=0.05$ & $\delta=0.01$\\
\hline
$\ell=126$&0.1444 &   0.1222  &  0.0889&0.2444&    0.1444&    0.0333&0.2667  &  0.1889  &  0.0778\\
$\ell=252$&0.1556 &   0.1333  &  0.0778&0.3111&    0.2111&    0.0889&0.3444 &   0.2556 &   0.1333\\
\hline
\hline
$\alpha=0.05$ & $\delta=0.1$ & $\delta=0.05$ & $\delta=0.01$& $\delta=0.1$ & $\delta=0.05$ & $\delta=0.01$& $\delta=0.1$ & $\delta=0.05$ & $\delta=0.01$\\
\hline
$\ell=126$&0.2889&    0.1889&    0.1333&0.0111&         0.0000&         0.0000&0.1889&    0.1556&    0.1000\\
$\ell=252$&0.3000&    0.2000&    0.1444&0.0556&    0.0111&         0.0000&0.2111 &   0.1556  &  0.1333\\
\hline
\hline
\end{tabular}
\end{table}

\subsubsection{Testing against serial dependence}\label{522}
If the alternative of interest is serial dependence among coverage indicators, we propose considering, for each individual stock  $i$,   alternatives of  binary first-order Markov dependence. More precisely,  defining the transition probabilities 
$$p_{hk,i}(\alpha)=\mathrm{P}\Big(\widehat{\mathcal H}^{(\ell)}_{i,\tau+1|\tau}(\alpha)=k\,\Big\vert\, \widehat{\mathcal H}^{(\ell)}_{i,\tau|\tau-1}(\alpha)=h\Big),\quad h,k=1,0,$$
we consider the testing problem (with unspecified unconditional probability $p_i(\alpha)$ of correct coverage)
\beq\label{eq:testID}
H_{0i}:p_{01,i}(\alpha)=p_{11,i}(\alpha)=:p_i(\alpha) \;\quad\text{ versus }\quad H_{1i}:p_{01,i}(\alpha)\neq p_{11,i}(\alpha);
\eeq
 note that $p_{01,i}(\alpha)=p_{11,i}(\alpha)$ automatically 
 implies $p_{00,i}(\alpha)=p_{10,i}(\alpha)$. 
  Defining 
  \begin{align}
n_{11i}^{(\ell)}(\alpha) := \sum_{\tau=T-M+1}^{T-1}\widehat{\mathcal H}_{i,\tau+1|\tau}^{(\ell)}(\alpha)\widehat{\mathcal H}_{i,\tau|\tau-1}^{(\ell)}(\alpha),
\qquad  & n_{10i}^{(\ell)}(\alpha) := n_{1i}^{(\ell)}(\alpha)-n_{11i}^{(\ell)}(\alpha),\nn\\
n_{01i}^{(\ell)}(\alpha) := \sum_{\tau=T-M+1}^{T-1}\widehat{\mathcal H}_{i,\tau+1|\tau}^{(\ell)}(\alpha)\big(1-\widehat{\mathcal H}_{i,\tau|\tau-1}^{(\ell)}(\alpha)\big),\qquad & n_{00i}^{(\ell)}(\alpha) := n_{0i}^{(\ell)}(\alpha)-n_{01i}^{(\ell)}(\alpha),\nn
\end{align}
the statistics
\[
\pi_i^{(\ell)}(\alpha):=\big( {n_{01i}^{(\ell)}(\alpha)+n_{11i}^{(\ell)}(\alpha)}\big)/
M
\]
are estimators of the $p_i(\alpha)$'s under the null, while 
\begin{align}
{\pi}_{11i}^{(\ell)}(\alpha) &:= {n_{11i}^{(\ell)}(\alpha)}/{n_{1i}^{(\ell)}(\alpha)}, \qquad\qquad \qquad{\pi}_{10i}^{(\ell)}(\alpha):=1-{\pi}_{11i}^{(\ell)}(\alpha),\nn\\
{\pi}_{01i}^{(\ell)}(\alpha) &:=\ {n_{01i}^{(\ell)}(\alpha)}/\big( M-{n_{1i}^{(\ell)}(\alpha)}\big), \ \ \text{and}\quad {\pi}_{00i}^{(\ell)}(\alpha):=1-{\pi}_{01i}^{(\ell)}(\alpha)\nn
\end{align}
are estimating the transition probabilities  $p_{hk,i}(\alpha)$ under the alternative. Log-likelihoods under the null and the alternative   are  
\begin{align}
L_{0i}(\alpha)=&\, (n_{00i}^{(\ell)}(\alpha)+n_{10i}^{(\ell)}(\alpha))\log[1-\pi_i^{(\ell)}(\alpha)]+n_{01i}^{(\ell)}(\alpha)+n_{11i}^{(\ell)}(\alpha)\log [\pi_i^{(\ell)}(\alpha)], \nn
\end{align}
and
\begin{align}
L_{1i}^{(\ell)}(\alpha)=& \,n_{00i}^{(\ell)}(\alpha)\log [1-{\pi}_{01i}^{(\ell)}(\alpha)]+n_{01i}^{(\ell)}(\alpha)\log[{\pi}_{01i}^{(\ell)}(\alpha)] \nn\\
&\hspace{33mm}+n_{10i}^{(\ell)}(\alpha)\log [1-{\pi}_{11i}^{(\ell)}(\alpha)]+n_{11i}^{(\ell)}(\alpha)\log [{\pi}_{11i}^{(\ell)}(\alpha)],\nn
\end{align}
respectively. For any given $i$, $\alpha$ and $\ell$, thus, we can construct a  likelihood-ratio   test for  \eqref{eq:testID}, based on the asymptotically $ \chi^2_{(1)}$ null distribution (as $M\to\infty$) of $LR_{\text{{ind}},i}^{(\ell)}(\alpha) := 2\big[L_{1i}^{(\ell)}(\alpha)-L_{0i}(\alpha)\big]$ (see also Section 3.2 in \citealp{christoffersen1998}). More general alternatives, involving higher-order serial dependencies, could be considered as well, based on the tests proposed by \cite{DHM98}.

In Table \ref{tab:testID} (left-hand panel), we report the proportions of rejections (over the~$n$ series) when testing~\eqref{eq:testID}. The same remarks apply as in the interpretation of Table \ref{tab:testUC_valid)}.

\subsubsection{Combined test}\label{523}

Combining the above tests, a likelihood ratio test (given $i$,  $\alpha$, and $\ell$) for 
\beq\label{eq:testCC}
H_{0i}:p_{01,i}(\alpha)=p_{11,i}(\alpha)=(1-\alpha) \ \text{ versus }\ H_{1i}:p_{01,i}(\alpha)\neq p_{11,i}(\alpha)\ \text{ or }\  p_{01,i}(\alpha)=p_{11,i}(\alpha)\neq (1-\alpha)
\eeq
can be based on the asymptotically $\chi^2_{(2)}$ (as $M\to\infty$) null distribution of 
\[
LR_i^{(\ell)}(\alpha) = LR_{\text{cover},i}^{(\ell)}(\alpha)+LR_{\text{{ind}},i}^{(\ell)}(\alpha)
\]
(see also Section 3.3 in \citealp{christoffersen1998}). The fraction of rejections (over $n$ series) when testing~\eqref{eq:testCC} is reported in Table \ref{tab:testID} (right-hand panel). The same remarks as in Table \ref{tab:testUC_valid)} still apply. 

\begin{table}[t!]
\centering
\caption{\small Standard \& Poor's 100 Index data ($n=90$ daily returns). Proportion of rejections when testing against serial  dependence \eqref{eq:testID} (left-hand panel) and in the combined problem \eqref{eq:testCC}  (right-hand panel). }\label{tab:testID}
\vskip .2cm
\footnotesize
\begin{tabular}{l  | ccc | ccc}
\hline
\hline
$\alpha=0.1$ & $\delta=0.1$ & $\delta=0.05$ & $\delta=0.01$&  $\delta=0.1$ & $\delta=0.05$ & $\delta=0.01$\\
\hline
$\ell=126$&0.3222&    0.2222&    0.0778&0.3667&    0.2222  &  0.1222\\
$\ell=252$&0.4000&    0.3556 &   0.1889&0.4889&    0.4222   & 0.2556\\
\hline
\hline
$\alpha=0.05$ & $\delta=0.1$ & $\delta=0.05$ & $\delta=0.01$& $\delta=0.1$ & $\delta=0.05$ & $\delta=0.01$\\
\hline
$\ell=126$&0.2778&    0.2000&    0.0556&0.2778&    0.2111&   0.1222\\
$\ell=252$&0.3667 &   0.2667 &   0.1667&0.3556&    0.2667&    0.2222\\
\hline
\hline
\end{tabular}
\end{table}

\subsection{Discussion}
In Table \ref{tab:sel_ser}, we report (four panels, according to the values of $\alpha$ and $\ell$) the ten individual series for which the four  tests above return the most significant rejections. Rejecting  in \eqref{eq:testUConesided} the null hypothesis of a valid coverage (``small''  values of $n_{1i}^{(\ell)}/M$)  means that the approximations we are making in the construction of the intervals lead to a loss of prediction accuracy for that specific series:  the intervals for that series are not wide enough---equivalently, their actual coverage probability  is less than the nominal $(1-\alpha)$ level. The   series listed in the first column of each panel  thus   are ``hardest to predict''.  Among them are stocks belonging to the Financial sector, as America International Group (AIG), Bank of America (BAC), and Citigroup (C). These series, in particular, were among those mostly affected by the great financial crisis. Rejecting  in \eqref{eq:testUConesided2} the null hypothesis of a sharp coverage (``large'' values of~$n_{1i}^{(\ell)}/M$) also means that  the approximations we are making in the construction of the intervals   lead to a loss of prediction accuracy for that specific series, now  in the sense that we could do better: the intervals for that series are too wide---their actual coverage probability   is more than the nominal $(1-\alpha)$ level. The   series listed in the second column of each panel thus   are ``easiest to predict''. Among them, stocks belonging to the Energy and Consumers sectors, as Exxon Mobil (XOM), Cisco Systems (CSCO), and McDonalds (MCD).

When testing against serial dependence, rejection (``large''  values of $LR_{\text{{ind}},i}^{(\ell)}(\alpha)$)  indicates that the predictive information available in past observations has not been fully exploited in the construction of the prediction intervals.  This could be the case, for example,  if some informative   idiosyncratic cross-correlation is available:  idiosyncratic cross-correlations indeed are not   captured by our univariate autoregressive modelling of idiosyncratic components. 
Alternative multivariate models for idiosyncratic components, such as sparse VAR, are likely to improve on this  (see e.g. the approach proposed in \citealp{barigozzihallin15c}), and could be incorporated into our two-step GDFM approach. We do not explore this any further in this paper, though. Such dependencies  could be related  to sectoral co-movements which, being specific to some restricted sector, are not captured by the market-wide factors. This seems to be the case especially for Financial and Energy stocks.
 A symptom of that phenomenon is the fact that the explained variance of the common component of the Financial stock returns is about 30\% less than the variance explained by the common component of all other stock returns.  The   importance of this   idiosyncratic variation, which is not accounted for by our approach, may explain why combined tests of correct coverage and independence exhibit, for Financial stock returns,   high rejection frequencies.   

\begin{sidewaystable}[h!]
\centering
\caption{\small Standard \& Poor's 100 Index data ($n=90$ daily returns).  Series tickers for which the   null hypotheses considered in Section~5.3  are rejected most significantly.  }\label{tab:sel_ser}
\vskip .3cm
\footnotesize
\begin{tabular}{l | llll | l | llll}
\hline
\hline
$\alpha = 0.1$ 	&  smallest  	&  	largest	 	&	largest						&	largest 					 &$\alpha = 0.05$ 	&  	smallest	&	largest&largest&largest\\
			& $n_{1i}^{(\ell)}(\alpha)/M$ 	&	$n_{1i}^{(\ell)}(\alpha)/M$	&	$LR_{\text{{ind}},i}^{(\ell)}(\alpha)$	& 	$LR_{i}^{(\ell)}(\alpha)$&& $n_{1i}^{(\ell)}(\alpha)/M$ 	&	$n_{1i}^{(\ell)}(\alpha)/M$	&	$LR_{\text{{ind}},i}^{(\ell)}(\alpha)$	& 	$LR_{i}^{(\ell)}(\alpha)$\\
\hline
$\ell=126$	&	BAC	&	MCD	&	AIG	&	BAC	&	$\ell=126$	&	SPG	&	COST	&	AIG	&	AIG	\\
	&	SPG	&	CSCO	&	BRK.B	&	SPG	&		&	BAC	&	MCD	&	BRK.B	&	BAC	\\
	&	C	&	CVX	&	AMGN	&	AIG	&		&	C	&	TGT	&	AMGN	&	SPG	\\
	&	AIG	&	GILD	&	SPG	&	C	&		&	AIG	&	EMC	&	DVN	&	C	\\
	&	WFC	&	MO	&	BAC	&	WFC	&		&	WFC	&	WMT	&	MRK	&	WFC	\\
	&	USB	&	TXN	&	COP	&	BRK.B	&		&	BRK.B	&	CVX	&	BAC	&	COP	\\
	&	JPM	&	WMT	&	SO	&	AMGN	&		&	SO	&	GILD	&	XOM	&	BRK.B	\\
	&	COF	&	XOM	&	AAPL	&	USB	&		&	USB	&	T	&	CVS	&	DVN	\\
	&	BRK.B	&	EMC	&	APC	&	COP	&		&	MS	&	COP	&	COP	&	APC	\\
	&	MS	&	SLB	&	JNJ	&	JPM	&		&	COF	&	CSCO	&	EXC	&	USB	\\
\hline
\hline
$\alpha = 0.1$ 	&  smallest  	&  	largest	 	&	largest						&	largest 					 &$\alpha = 0.05$ 	&  	smallest	&	largest&largest&largest\\
			& $n_{1i}^{(\ell)}(\alpha)/M$ 	&	$n_{1i}^{(\ell)}(\alpha)/M$	&	$LR_{\text{{ind}},i}^{(\ell)}(\alpha)$	& 	$LR_{i}^{(\ell)}(\alpha)$&& $n_{1i}^{(\ell)}(\alpha)/M$ 	&	$n_{1i}^{(\ell)}(\alpha)/M$	&	$LR_{\text{{ind}},i}^{(\ell)}(\alpha)$	& 	$LR_{i}^{(\ell)}(\alpha)$\\
\hline																						
$\ell=252$	&	BAC	&	GILD	&	AIG	&	BAC	&	$\ell=252$	&	SPG	&	MCD	&	AIG	&	AIG	\\
	&	SPG	&	MCD	&	SPG	&	SPG	&		&	BAC	&	GILD	&	BRK.B	&	BAC	\\
	&	C	&	XOM	&	COP	&	AIG	&		&	C	&	ORCL	&	COP	&	SPG	\\
	&	AIG	&	TXN	&	BRK.B	&	C	&		&	AIG	&	QCOM	&	DVN	&	C	\\
	&	WFC	&	CSCO	&	DVN	&	WFC	&		&	WFC	&	CVX	&	EXC	&	WFC	\\
	&	USB	&	CVX	&	APC	&	BRK.B	&		&	BRK.B	&	CSCO	&	OXY	&	BRK.B	\\
	&	MS	&	EBAY	&	LLY	&	USB	&		&	JPM	&	WMT	&	ALL	&	COP	\\
	&	BRK.B	&	EMC	&	BAC	&	MS	&		&	USB	&	COST	&	BAC	&	SO	\\
	&	JPM	&	TGT	&	UNH	&	SO	&		&	FCX	&	EMC	&	SPG	&	OXY	\\
	&	COF	&	MO	&	C	&	AMGN	&		&	TWX	&	TGT	&	KO	&	ALL	\\
\hline
\hline
\end{tabular}
\end{sidewaystable}

\section{Conclusions}\label{sec:conc}

In this paper, we consider a two-step GDFM approach for jointly modelling stock returns and their volatilities in order to build conditional prediction intervals. A careful study of the consistency  properties (as the cross-sectional dimension $n$ and the sample size $T$ both tend to infinity) of the resulting estimators is conducted. Those  results   are the theoretical foundation  of   (\citealp{barigozzihallin15a,barigozzihallin15b,barigozzihallin15c}, and \citealp{BHS18}); here, we are using them in the construction of one-step-ahead prediction intervals. 

We then apply our methodology to a panel of 90 daily returns of stocks listed in the S\&P100. Through a recursive exercise, we show that we are able to obtain one-step-ahead prediction intervals which are in general more accurate than univariate GARCH methods.   

Many  extensions of this work are possible, which are left for future research. First, our empirical results indicate that, by exploiting also the cross-sectional lagged dependencies among idiosyncratic components, we could achieve better coverage especially for those series belonging to the Financial sector, which remains strongly interconnected even after controlling for common factors. This could be achieved  by computing predictions of idiosyncratic components by fitting multivariate models such as  sparse VARs. Second, our methodology immediately allows us to consider bivariate or multivariate prediction intervals.  Third, asymmetric prediction intervals can also be considered. In particular, Value-at-Risk indicators are readily computable;  moreover, by considering many values of the coverage, we can approximate the whole conditional distribution of  returns. Last, another possible application consists in the construction od prediction intervals for macroeconomic variables as GDP or inflation taking into account,  in a way similar to \citet{jurado2015},  the uncertainty related to the business cycle. 

\small
\bibliography{BH_biblio} 

\begin{thebibliography}{}

\bibitem[A{\"\i}t-Sahalia and Xiu, 2017]{ait2017}
A{\"\i}t-Sahalia, Y. and Xiu, D. (2017).
\newblock Using principal component analysis to estimate a high dimensional
  factor model with high-frequency data.
\newblock {\em Journal of Econometrics}, 201:384--399.

\bibitem[Alizadeh et~al., 2002]{ABD02}
Alizadeh, S., Brandt, M.~W., and Diebold, F.~X. (2002).
\newblock Range-based estimation of stochastic volatility models.
\newblock {\em The Journal of Finance}, 57:1047--1091.

\bibitem[Anderson and Deistler, 2008]{andersondeistler08}
Anderson, B.~D. and Deistler, M. (2008).
\newblock Generalized linear dynamic factor models. {A} structure theory.
\newblock In {\em 47th IEEE Conference on Decision and Control}.

\bibitem[Asai et~al., 2006]{AMY06}
Asai, M., McAleer, M., and Yu, J. (2006).
\newblock Multivariate stochastic volatility: {A} review.
\newblock {\em Econometric Reviews}, 25:145--175.

\bibitem[Bai and Ng, 2002]{baing02}
Bai, J. and Ng, S. (2002).
\newblock Determining the number of factors in approximate factor models.
\newblock {\em Econometrica}, 70:191--221.

\bibitem[Bai and Ng, 2008]{baing08JoE}
Bai, J. and Ng, S. (2008).
\newblock Forecasting economic time series using targeted predictors.
\newblock {\em Journal of Econometrics}, 146:304--317.

\bibitem[Barigozzi and Hallin, 2016]{barigozzihallin15a}
Barigozzi, M. and Hallin, M. (2016).
\newblock General dynamic factors and volatilities: Recovering the market
  volatility shocks.
\newblock {\em The Econometrics Journal}, 19:C33--C60.

\bibitem[Barigozzi and Hallin, 2017a]{barigozzihallin15b}
Barigozzi, M. and Hallin, M. (2017a).
\newblock General dynamic factors and volatilities: Estimation and forecasting.
\newblock {\em Journal of Econometrics}, 201:307--321.

\bibitem[Barigozzi and Hallin, 2017b]{barigozzihallin15c}
Barigozzi, M. and Hallin, M. (2017b).
\newblock Networks, dynamic factors, and the volatility analysis of
  high-dimensional financial series.
\newblock {\em Journal of the Royal Statistical Society, Series C},
  66:581--605.

\bibitem[Barigozzi et~al., 2018]{BHS18}
Barigozzi, M., Hallin, M., and Soccorsi, S. (2018).
\newblock Identification of global and local shocks in international financial
  markets via general dynamic factor models.
\newblock {\em Journal of Financial Econometrics}.
\newblock available online.

\bibitem[Bauwens et~al., 2006]{BLR06}
Bauwens, L., Laurent, S., and Rombouts, J. V.~K. (2006).
\newblock Multivariate {GARCH} models: {A} survey.
\newblock {\em Journal of the Applied Econometrics}, 21:79--109.

\bibitem[Boivin and Ng, 2006]{boivinng06}
Boivin, J. and Ng, S. (2006).
\newblock Are more data always better for factor analysis?
\newblock {\em Journal of Econometrics}, 127:169--194.

\bibitem[Borovkov, 2000]{borovkov00}
Borovkov, A.~A. (2000).
\newblock Large deviation probabilities for random walks with semiexponential
  distributions.
\newblock {\em Siberian Mathematical Journal}, 41:1061--1093.

\bibitem[Brillinger, 2001]{brillinger2001}
Brillinger, D. (2001).
\newblock {\em Time Series: Data Analysis and Theory}.
\newblock Classics in Applied Mathematics. Society for Industrial and Applied
  Mathematics.

\bibitem[Chamberlain and Rothschild, 1983]{chamberlainrotshild83}
Chamberlain, G. and Rothschild, M. (1983).
\newblock Arbitrage, factor structure, and mean--variance analysis on large
  asset markets.
\newblock {\em Econometrica}, 51:1281--304.

\bibitem[Chicheportiche and Bouchaud, 2015]{CB15}
Chicheportiche, R. and Bouchaud, J.-P. (2015).
\newblock A nested factor model for non-linear dependencies in stock returns.
\newblock {\em Quantitative Finance}, 15:1789--1804.

\bibitem[Christoffersen, 1998]{christoffersen1998}
Christoffersen, P.~F. (1998).
\newblock Evaluating interval forecasts.
\newblock {\em International Economic Review}, 39:841--862.

\bibitem[Connor and Korajczyk, 1993]{CK93}
Connor, G. and Korajczyk, R.~A. (1993).
\newblock A test for the number of factors in an approximate factor model.
\newblock {\em the Journal of Finance}, 48:1263--1291.

\bibitem[Connor et~al., 2006]{CKL06}
Connor, G., Korajczyk, R.~A., and Linton, O. (2006).
\newblock The common and specific components of dynamic volatility.
\newblock {\em Journal of Econometrics}, 132:231--255.

\bibitem[Davidson, 1994]{davidson1994}
Davidson, J. (1994).
\newblock {\em Stochastic Limit Theory: An Introduction for Econometricians}.
\newblock Oxford University Press.

\bibitem[den Haan and Levin, 1997]{denhaan97}
den Haan, W.~J. and Levin, A.~T. (1997).
\newblock A practitioner's guide to robust covariance matrix estimation.
\newblock In {\em Robust Inference}, volume~15 of {\em Handbook of Statistics}.
  Elsevier.

\bibitem[Diebold and Nerlove, 1989]{DN89}
Diebold, F.~X. and Nerlove, M. (1989).
\newblock The dynamics of exchange rate volatility: a multivariate latent
  factor {ARCH} model.
\newblock {\em Journal of Applied Econometrics}, 4:1--21.

\bibitem[Dufour et~al., 1998]{DHM98}
Dufour, J., Hallin, M., and Mizera, I. (1998).
\newblock Generalized run tests for heteroscedastic time series.
\newblock {\em Journal of Nonparametric Statistics}, 9:39--86.

\bibitem[Engle and Marcucci, 2006]{EM06}
Engle, R.~F. and Marcucci, J. (2006).
\newblock A long--run pure variance common features model for the common
  volatilities of the {D}ow {J}ones.
\newblock {\em Journal of Econometrics}, 132:7--42.

\bibitem[Fan et~al., 2013]{FLM13}
Fan, J., Liao, Y., and Mincheva, M. (2013).
\newblock Large covariance estimation by thresholding principal orthogonal
  complements.
\newblock {\em Journal of the Royal Statistical Society, Series B},
  75:603--680.

\bibitem[Fan et~al., 2015]{fan15}
Fan, J., Liao, Y., and Shi, X. (2015).
\newblock Risks of large portfolios.
\newblock {\em Journal of Econometrics}, 186:367--387.

\bibitem[Forni et~al., 2009]{FGLR09}
Forni, M., Giannone, D., Lippi, M., and Reichlin, L. (2009).
\newblock Opening the black box: Structural factor models versus structural
  {VAR}s.
\newblock {\em Econometric Theory}, 25:1319--1347.

\bibitem[Forni et~al., 2018]{FGLS18}
Forni, M., Giovannelli, A., Lippi, M., and Soccorsi, S. (2018).
\newblock Dynamic factor model with infinite-dimensional factor space:
  Forecasting.
\newblock {\em Journal of Applied Econometrics}, 33:625--642.

\bibitem[Forni et~al., 2000]{FHLR00}
Forni, M., Hallin, M., Lippi, M., and Reichlin, L. (2000).
\newblock The generalized dynamic factor model: Identification and estimation.
\newblock {\em The Review of Economics and Statistics}, 82:540--554.

\bibitem[Forni et~al., 2015]{FHLZ15}
Forni, M., Hallin, M., Lippi, M., and Zaffaroni, P. (2015).
\newblock Dynamic factor models with infinite-dimensional factor spaces:
  One-sided representations.
\newblock {\em Journal of Econometrics}, 185:359--371.

\bibitem[Forni et~al., 2017]{FHLZ17}
Forni, M., Hallin, M., Lippi, M., and Zaffaroni, P. (2017).
\newblock Dynamic factor models with infinite dimensional factor space:
  Asymptotic analysis.
\newblock {\em Journal of Econometrics}, 199:74--92.

\bibitem[Forni and Lippi, 2001]{fornilippi01}
Forni, M. and Lippi, M. (2001).
\newblock The generalized dynamic factor model: {R}epresentation theory.
\newblock {\em Econometric Theory}, 17:1113--1141.

\bibitem[Forni and Lippi, 2011]{fornilippi11}
Forni, M. and Lippi, M. (2011).
\newblock The unrestricted dynamic factor model: One-sided representation
  results.
\newblock {\em Journal of Econometrics}, 163:23--28.

\bibitem[Francq and Zakoian, 2011]{FZ11}
Francq, C. and Zakoian, J.-M. (2011).
\newblock {\em {GARCH} Models: Structure, Statistical Inference and Financial
  Applications}.
\newblock John Wiley \& Sons.

\bibitem[Francq and Zakoian, 2019]{FZ19}
Francq, C. and Zakoian, J.-M. (2019).
\newblock Virtual {H}istorical {S}imulation for estimating the conditional
  {VaR} of large portfolios.
\newblock mimeo.

\bibitem[Hallin and Lippi, 2013]{hallinlippi13}
Hallin, M. and Lippi, M. (2013).
\newblock Factor models in high--dimensional time series. {A} time-domain
  approach.
\newblock {\em Stochastic Processes and their Applications}, 123:2678--2695.

\bibitem[Hallin and Li{\v s}ka, 2007]{hallinliska07}
Hallin, M. and Li{\v s}ka, R. (2007).
\newblock Determining the number of factors in the general dynamic factor
  model.
\newblock {\em Journal of the American Statistical Association}, 102:603--617.

\bibitem[Hamilton, 1994]{hamilton1994}
Hamilton, J.~D. (1994).
\newblock {\em Time series analysis}.
\newblock Princeton University Press.

\bibitem[Hannan, 1970]{hannan1970}
Hannan, E.~J. (1970).
\newblock {\em Multiple Time Series}.
\newblock John Wiley \& Sons.

\bibitem[Harvey et~al., 1992]{HRS92}
Harvey, A., Ruiz, E., and Sentana, E. (1992).
\newblock Unobserved component time series models with arch disturbances.
\newblock {\em Journal of Econometrics}, 52:129--157.

\bibitem[Jurado et~al., 2015]{jurado2015}
Jurado, K., Ludvigson, S.~C., and Ng, S. (2015).
\newblock Measuring uncertainty.
\newblock {\em American Economic Review}, 105:1177--1216.

\bibitem[Kuchibhotla and Chakrabortty, 2018]{KC18}
Kuchibhotla, A.~K. and Chakrabortty, A. (2018).
\newblock Moving beyond sub-gaussianity in high dimensional statistics:
  {A}pplications in covariance matrix estimation and linear regressions.
\newblock arXiv preprint arXiv:1804.02605.

\bibitem[L{\"u}tkepohl, 2005]{lutkepohl2005}
L{\"u}tkepohl, H. (2005).
\newblock {\em New Introduction to Multiple Time Series Analysis}.
\newblock Springer Science \& Business Media.

\bibitem[McNemar, 1947]{McN47}
McNemar, Q.~M. (1947).
\newblock Note on the sampling error of the difference between correlated
  proportions or percentages.
\newblock {\em Psychometrika}, 12:153--157.

\bibitem[Merlev{\`e}de et~al., 2011]{MPR11}
Merlev{\`e}de, F., Peligrad, M., and Rio, E. (2011).
\newblock A bernstein type inequality and moderate deviations for weakly
  dependent sequences.
\newblock {\em Probability Theory and Related Fields}, 151:435--474.

\bibitem[Ng et~al., 1992]{ENR92}
Ng, V., Engle, R.~F., and Rothschild, M. (1992).
\newblock A multi-dynamic-factor model for stock returns.
\newblock {\em Journal of Econometrics}, 52:245--266.

\bibitem[Parzen, 1957]{parzen57}
Parzen, E. (1957).
\newblock On consistent estimates of the spectrum of a stationary time series.
\newblock {\em The Annals of Mathematical Statistics}, 28:329--348.

\bibitem[Pe\~na and Box, 1987]{pena1987identifying}
Pe\~na, D. and Box, G.~E. (1987).
\newblock Identifying a simplifying structure in time series.
\newblock {\em Journal of the American statistical Association}, 82:836--843.

\bibitem[Priestley, 2001]{priestley01}
Priestley, M.~B. (2001).
\newblock {\em Spectral Analysis and Time Series}.
\newblock Elsevier Academic Press.

\bibitem[Sentana et~al., 2008]{SCF08}
Sentana, E., Calzolari, G., and Fiorentini, G. (2008).
\newblock Indirect estimation of large conditionally heteroskedastic factor
  models, with an application to the {D}ow 30 stocks.
\newblock {\em Journal of Econometrics}, 146:10--25.

\bibitem[Stock and Watson, 2002]{stockwatson2002}
Stock, J.~H. and Watson, M.~W. (2002).
\newblock Forecasting using principal components from a large number of
  predictors.
\newblock {\em Journal of the American Statistical Association}, 97:1167--1179.

\bibitem[Tiao, 1972]{tiao1972asymptotic}
Tiao, G.~C. (1972).
\newblock Asymptotic behaviour of temporal aggregates of time series.
\newblock {\em Biometrika}, 59:525--531.

\bibitem[Tiao and Box, 1981]{tiao1981modeling}
Tiao, G.~C. and Box, G.~E. (1981).
\newblock Modeling multiple time series with applications.
\newblock {\em Journal of the American Statistical Association}, 76:802--816.

\bibitem[Tiao and Guttman, 1980]{tiao1980forecasting}
Tiao, G.~C. and Guttman, I. (1980).
\newblock Forecasting contemporal aggregates of multiple time series.
\newblock {\em Journal of Econometrics}, 12:219--230.

\bibitem[Tiao and Hillmer, 1978]{tiao1978some}
Tiao, G.~C. and Hillmer, S.~C. (1978).
\newblock Some consideration of decomposition of a time series.
\newblock {\em Biometrika}, 65:497--502.

\bibitem[Tiao and Tsay, 1989]{tiao1989model}
Tiao, G.~C. and Tsay, R.~S. (1989).
\newblock Model specification in multivariate time series.
\newblock {\em Journal of the Royal Statistical Society: Series B
  (Methodological)}, 51:157--195.

\bibitem[Truc\'{\i}os et~al., 2019]{Trucios19}
Truc\'{\i}os, .~C., Mazzeu, J., Hallin, M., Zevallos, M., Hotta, L., and
  Pereira, P.~V. (2019).
\newblock Forecasting conditional covariance matrices in high-dimensional time
  series with application to dynamic portfolio optimization: a general dynamic
  factor approach.
\newblock Technical Report DOI: 10.13140/RG.2.2.23950.82241, ECARES.

\bibitem[Tsay and Tiao, 1985]{tsay1985use}
Tsay, R.~S. and Tiao, G.~C. (1985).
\newblock Use of canonical analysis in time series model identification.
\newblock {\em Biometrika}, 72:299--315.

\bibitem[Uchaikin and Zolotarev, 2011]{UZ11}
Uchaikin, V.~V. and Zolotarev, V.~M. (2011).
\newblock {\em Chance and Stability: Stable Distributions and their
  Applications}.
\newblock Walter de Gruyter.

\bibitem[Vershynin, 2012]{vershynin12}
Vershynin, R. (2012).
\newblock Introduction to the non-asymptotic analysis of random matrices.
\newblock In Eldar, Y. and Kutyniok, G., editors, {\em Compressed Sensing.
  Theory and Applications.} Cambridge University Press.

\bibitem[Vladimirova and Arbel, 2019]{VA19}
Vladimirova, M. and Arbel, J. (2019).
\newblock Sub-{W}eibull distributions: generalizing sub-{G}aussian and
  sub-{E}xponential properties to heavier-tailed distributions.
\newblock arXiv preprint arXiv:1905.04955.

\bibitem[Weyl, 1912]{Weyl}
Weyl, H. (1912).
\newblock Das asymptotische {V}erteilungsgesetz der {E}igenwerte linearer
  partieller {D}ifferentialgleichungen.
\newblock {\em Mathematische Annalen}, 71:441--479.

\bibitem[Wu and Zaffaroni, 2018]{WZ18}
Wu, W.~B. and Zaffaroni, P. (2018).
\newblock Asymptotic theory for spectral density estimates of general
  multivariate time series.
\newblock {\em Econometric Theory}, 34:1--22.

\end{thebibliography}
\bibliographystyle{apalike}

  \setcounter{section}{0}%
\setcounter{subsection}{-1}
\setcounter{equation}{0}
\setcounter{lemma}{0}
\renewcommand{\thelemma}{A\arabic{lemma}}
\renewcommand{\theequation}{A\arabic{equation}}

\appendix
\small
\section{Technical Appendix}\label{sec:app_proof}
\subsection{Proofs of Lemmas \ref{lem:dyn_eval}, \ref{lem:stat_eval}, and \ref{lem:eval_vol}}
\textsc{Proof of Lemma \ref{lem:dyn_eval}.} From Assumption L1{\it (v)}, for any $i=1,\ldots,n$ and any $\theta\in[-\pi,\pi]$,
\[
\vert d_i(e^{-i\theta})\vert \le \sum_{k=0}^\infty \vert d_{ik} e^{-i\theta k}\vert\le \sum_{k=0}^\infty \vert d_{ik} \vert\le M_2.
\] 
Let $\sigma_{ij}^Z(\theta)$ stand for    entry $(i,j)$ of $\bm\Sigma_n^Z(\theta)$. From Assumption~L1{\it (iv)}, for all $n>\bar n$, we have
\begin{align}
\sup_{\theta\in[-\pi,\pi]} \lambda_{n1}^Z(\theta) &\le\sup_{\theta\in[-\pi,\pi]}\max_{i=1,\ldots, n} \sum_{j=1}^n \vert \sigma_{ij}^Z(\theta)\vert\nn\\
&=\sup_{\theta\in[-\pi,\pi]}\max_{i=1,\ldots, n} \frac 1 {2\pi}\sum_{j=1}^n \vert d_i(e^{-i\theta})\Cov(v_{it},v_{jt}) d_j(e^{i\theta})\vert\le  {M_2^2C^v}/{2\pi}.\nn
\end{align}
where we used the fact that $\lambda_{n1}^Z(\theta)=\Vert\bm\Sigma_n^Z(\theta)\Vert\le \Vert\bm\Sigma_n^Z(\theta)\Vert_1=\max_{i=1,\ldots, n} \sum_{j=1}^n \vert \sigma_{ij}^Z(\theta)\vert$. This proves part~{\it (i)}. Parts {\it (ii)} and {\it (iii)} are consequences of Assumption L3, part~{\it (i)} above, and Weyl's inequality (\citealp{Weyl}).~\hfill$\square$\bigskip

\noindent
\textsc{Proof of Lemma \ref{lem:stat_eval}.}  Part {\it (i)} follows from Proposition 4 in \citet{FHLZ17}. Parts {\it (ii)} and {\it (iii)} are consequences of Assumption L5, part {\it (i)} above, and Weyl's inequality. \hfill$\square$\bigskip

\noindent
\textsc{Proof of Lemma \ref{lem:eval_vol}.} Parts {\it (i)}-{\it (iii)} follow as in Lemma \ref{lem:dyn_eval},   parts {\it (iv)}-{\it (vi)}  as in Lemma \ref{lem:stat_eval}.\hfill$\square$\bigskip

\subsection{Estimation of spectral densities}
\begin{lemma} \label{lem:spettro_level} Let $\sigma_{ij}^Y(\theta)$ and $\widehat{\sigma}_{ij}^Y(\theta)$ stand for the $(i,j)$  entries of $\bm\Sigma_n^Y(\theta)$ and $\widehat{\bm\Sigma}_n^Y(\theta)$, respectively. Then, 
\begin{compactenum}[{\it (i)}]
\item letting  $\theta_h:= {\pi h}/{B_T}$ with $\vert h\vert \le B_T$,  under Assumptions (L1)-(L2),
\[
\E\l[\max_{|h|\le B_T} \big\vert\widehat{\sigma}_{ij}^Y(\theta_h)-{\sigma}_{ij}^Y(\theta_h)\big\vert^2\r] \le \frac{C_1B_T^2}{T} + \frac{C_2} {B_T^2},
\]
where $C_1>0$ and $C_2>0$ are finite and independent of $i$ and $j$;
\item letting $\theta_\ell:= {\pi \ell}/{M_T}$ with $\vert \ell\vert \le M_T$,  under Assumptions (V1)-(V2), 
\[
\E\l[\max_{|\ell| \le M_T} \big\vert\widehat{\sigma}_{ij}^h(\theta_\ell)-{\sigma}_{ij}^h(\theta_\ell)\big\vert^2\r] \le  \frac{C_3M_T^2}{T} + \frac{C_4} {M_T^2},
\]
 where $C_3>0$ and $C_4>0$ are finite and independent of $i$ and $j$.
\end{compactenum}
\end{lemma}
\noindent
\textsc{Proof of Lemma \ref{lem:spettro_level}.} For any given $i,j$, we have
\begin{align}
&\E\l[\max_{|h|\le B_T} \big\vert\widehat{\sigma}_{ij}^Y(\theta_h)-{\sigma}_{ij}^Y(\theta_h)\big\vert^2\r] =\E\l[\max_{|h|\le B_T} \big\vert\widehat{\sigma}_{ij}^Y(\theta_h)-\E[\widehat{\sigma}_{ij}^Y(\theta_h)]+\E[\widehat{\sigma}_{ij}^Y(\theta_h)]-{\sigma}_{ij}^Y(\theta_h)\big\vert^2\r] \nn\\
&\le 2 \E\l[\max_{|h|\le B_T} \big\vert\widehat{\sigma}_{ij}^Y(\theta_h)-\E[\widehat{\sigma}_{ij}^Y(\theta_h)]\big\vert^2\r] + 2\E\l[\max_{|h|\le B_T} \big\vert\E[\widehat{\sigma}_{ij}^Y(\theta_h)]-{\sigma}_{ij}^Y(\theta_h)\big\vert^2\r] =2( I +I\mspace{-2mu}I),\text{ say}.
\end{align}
Considering the first term $I$, under Assumptions (L1{\it ii}), (L1{\it v}), (L1{\it vii}), and (L1{\it viii}) (finite fourth-order innovation moments  and summability of common and  idiosyncratic coefficients), we have that the variance of the lag-window estimator is such that
\beq\label{eq:var_sp}
 \max_{|h|\le B_T}\E\Big[\big\vert\widehat{\sigma}_{ij}^Y(\theta_h)-\E[\widehat{\sigma}_{ij}^Y(\theta_h)]\big\vert^2\Big] \le {C_1^* B_T}/{T},
\eeq
for some   finite  $C_1^*>0$ independent of $i$ and $j$ (see also the first term on the right-hand side of equation~(5) in \citealp{hallinliska07}). This is a classical result which is proved, for example, in Theorem 5A of  \citet{parzen57}. Then, 
\begin{align}\label{eq:bon_var}
I&=\E\l[\max_{|h|\le B_T} \big\vert\widehat{\sigma}_{ij}^Y(\theta_h)-\E[\widehat{\sigma}_{ij}^Y(\theta_h)]\big\vert^2\r] \le \sum_{|h|\le B_T} \E\l[\big\vert\widehat{\sigma}_{ij}^Y(\theta_h)-\E[\widehat{\sigma}_{ij}^Y(\theta_h)]\big\vert^2\r]\nn\\
&\le (2B_T+1)   \max_{|h|\le B_T}\E\Big[\big\vert\widehat{\sigma}_{ij}^Y(\theta_h)-\E[\widehat{\sigma}_{ij}^Y(\theta_h)]\big\vert^2\Big] \le {C_1 B^2_T}/{T},
\end{align}
where $C_1>0$ is finite and independent of $i$ and $j$ 
(see also Chapter 6 by \citealp{priestley01}).\medskip

Turning to $I\mspace{-2mu}I$,  (see also Proposition 6 in \citealp{FHLZ17})
\begin{align}
2\pi \big\vert&\E[\widehat{\sigma}_{ij}^Y(\theta_h)]-{\sigma}_{ij}^Y(\theta_h)\big\vert = \bigg\vert\sum_{k=-T+1}^{T-1}\mathrm K\l({k}/{B_T}\r)\E[\widehat{\gamma}_{ijk}^Y] e^{-ik\theta_h} - \sum_{k=-\infty}^{\infty} \gamma_{ijk}^Ye^{-ik\theta_h}\bigg\vert\nn\\
&\le\bigg\vert\sum_{k=-T+1}^{T-1}\l(\mathrm K\l({k}/{B_T}\r)-1\r) \gamma_{ijk}^Ye^{-ik\theta_h}\bigg\vert+\bigg\vert\sum_{k=-T+1}^{T-1}\mathrm K\l({k}/{B_T}\r)\frac{|k|}{T}\gamma_{ijk}^Ye^{-ik\theta_h}\bigg\vert+\bigg\vert\sum_{|k|\ge T}\gamma_{ijk}^Ye^{-ik\theta_h}\bigg\vert\nn\\
&= I\mspace{-2mu}I\mspace{-2mu}I + IV + V, \text{ say,}\label{eq:bias_spectra}
\end{align}
owing to the fact that $\E[\widehat{\gamma}_{ijk}^Y]=\gamma_{ijk}^Y \l(1-\frac{|k|}{T}\r)$. In order to bound each term of \eqref{eq:bias_spectra}, note that, because of Assumption (L2), there exists a finite constant $D>0$ and a constant~$\phi\in (0,1)$, both independent of $i$ and $j$, such that
\beq\label{eq:D}
\vert \gamma_{ijk}^Y\vert \le \vert \gamma_{ijk}^X\vert + \vert \gamma_{ijk}^Z\vert \le D\phi^{|k|}.
\eeq
For term $I\mspace{-2mu}I\mspace{-2mu}I$ in \eqref{eq:bias_spectra}, using \eqref{eq:D} and the Bartlett kernel, $\mathrm K\l( k/{B_T}\r)=\l(1- {\vert k\vert}/{B_T}\r)$, we have
\begin{align}\label{eq:bias_spectra3}
I\mspace{-2mu}I\mspace{-2mu}I\le D\sum_{k=-\infty}^{\infty}\phi^{|k|} \frac{\vert k\vert}{B_T}\le  {2D\phi}/{(1-\phi^2) B_T},
\end{align}
irrespective of   $i$, $j$, and $\theta_h$. Similarly, for terms $IV$ and $V$, 
\beq\label{eq:bias_spectra4}
IV\le D\sum_{k=-\infty}^{\infty}\phi^{|k|}  {\vert k\vert}/{T}\le  {2D\phi}/{(1-\phi^2) T}\quad\text{and}\quad V\le D\sum_{|k|\ge T}\phi^{|k|} {\vert k\vert}/{T}\le {2D\phi}/{(1-\phi^2) T},
\eeq
irrespective  of $i$, $j$, and $\theta_h$,  and since ${\vert k\vert}/{T}>1$ when $|k|\ge T$. 
 By substituting \eqref{eq:bias_spectra3} and \eqref{eq:bias_spectra4}  into \eqref{eq:bias_spectra}, we obtain that~$I\mspace{-2mu}I\le (C_2/B_T^2)$ with $C_2>0$ finite and independent of $i$ and $j$. This proves part {\it (i)}. Part {\it (ii)} follows along the same lines.~\hfill$\square$\bigskip


\subsection{Proof of Proposition \ref{prop:level}}\label{app:prop1}
From Lemma \ref{lem:spettro_level}{\it (i)}, we have the following (see also Lemma~1 in \citealp{FHLZ17}) 
\begin{align}
\E&\l[\max_{|h|\le B_T} \frac 1{n^2}\Big\Vert\widehat{\bm\Sigma}_n^Y(\theta_h)-{\bm\Sigma}_n^Y(\theta_h)\Big\Vert^2\r]\le 
\E\l[\max_{|h|\le B_T} \frac 1{n^2}\mbox{\rm tr}\Big\{\Big(\widehat{\bm\Sigma}_n^Y(\theta_h)-{\bm\Sigma}_n^Y(\theta_h)\Big)\Big(\widehat{\bm\Sigma}_n^Y(\theta_h)-{\bm\Sigma}_n^Y(\theta_h)\Big)\Big\}\r]\label{eq:spettriuniformi_larnacaFF}\\
&=\frac 1 {n^2} \E\l[\max_{|h|\le B_T} \sum_{i=1}^n\sum_{j=1}^n \big\vert\widehat{\sigma}_{ij}^Y(\theta_h)-{\sigma}_{ij}^Y(\theta_h)\big\vert^2\r]
\le\frac 1 {n^2} \sum_{i=1}^n\sum_{j=1}^n\E\l[\max_{|h|\le B_T}  \big\vert\widehat{\sigma}_{ij}^Y(\theta_h)-{\sigma}_{ij}^Y(\theta_h)\big\vert^2\r]
\le {C_1B_T^2}/T+ {C_2}/{B_T^2}.\nn
\end{align}
Let $\bm{\ell}_i$ denote the $n$-dimensional vector with 1 in entry $i$ and 0 elsewhere. Then, since $\bm{\ell}_i$ is non-random,
\begin{align}
\E&\l[\max_{i=1,\ldots, n} \max_{|h|\le B_T} \frac{1}{n}\Big\Vert\bm{\ell}^\prime_i\Big(\widehat{\bm\Sigma}_n^Y(\theta_h)-{\bm\Sigma}_n^Y(\theta_h)\Big)\Big\Vert^2\r]=\E\l[\max_{i=1,\ldots, n} \max_{|h|\le B_T}\frac 1 n\bm{\ell}^\prime_i\Big(\widehat{\bm\Sigma}_n^Y(\theta_h)-{\bm\Sigma}_n^Y(\theta_h)\Big)\Big(\widehat{\bm\Sigma}_n^Y(\theta_h)-{\bm\Sigma}_n^Y(\theta_h)\Big)^\prime\bm{\ell}_i\r]\nn\\
&=\max_{i=1,\ldots, n}\E\l[\max_{|h|\le B_T}\frac 1 n\bm{\ell}^\prime_i\Big(\widehat{\bm\Sigma}_n^Y(\theta_h)-{\bm\Sigma}_n^Y(\theta_h)\Big)\Big(\widehat{\bm\Sigma}_n^Y(\theta_h)-{\bm\Sigma}_n^Y(\theta_h)\Big)\bm{\ell}_i\r]=\max_{i=1,\ldots, n}\frac 1n  \E\l[ \max_{|h|\le B_T}\sum_{j=1}^n\big\vert\widehat{\sigma}_{ij}^Y(\theta_h)-{\sigma}_{ij}^Y(\theta_h)\big\vert^2\r]\nn\\
&\le \max_{i=1,\ldots, n}\frac 1n  \sum_{j=1}^n\E\l[ \max_{|h|\le B_T}\big\vert\widehat{\sigma}_{ij}^Y(\theta_h)-{\sigma}_{ij}^Y(\theta_h)\big\vert^2\r]\le {C_1B_T^2}/T+ {C_2}/{B_T^2}.\label{eq:spettriuniformi_larnaca2}
\end{align}
Hence, by \eqref{eq:spettriuniformi_larnacaFF}, \eqref{eq:spettriuniformi_larnaca2}, and Chebychev's inequality
\begin{align}
&\max_{|h|\le B_T} \frac 1n\Big\Vert\widehat{\bm\Sigma}_n^Y(\theta_h)-{\bm\Sigma}_n^Y(\theta_h)\Big\Vert= O_{\rm P}\l(\max\l({B_T}/{\sqrt T},1/{B_T}\r)\r),\label{eq:spettriuniformi_larnaca}
\end{align}
and
\beq\label{eq:spettriuniformi_larnaca3}
\max_{i=1,\ldots, n} \max_{|h|\le B_T} \frac{1}{\sqrt n}\Big\Vert\bm{\ell}^\prime_i\Big(\widehat{\bm\Sigma}_n^Y(\theta_h)-{\bm\Sigma}_n^Y(\theta_h)\Big)\Big\Vert= O_{\rm P}\l(\max\l({B_T}/{\sqrt T},1/{B_T}\r)\r).
\eeq
Note that \eqref{eq:spettriuniformi_larnaca} and \eqref{eq:spettriuniformi_larnaca3} hold independently of $n$.  Moreover, using \eqref{eq:spettriuniformi_larnaca} and Lemma \ref{lem:dyn_eval}{\it (i)}, we have
\begin{align}
\max_{|h|\le B_T} \frac 1n\Big\Vert\widehat{\bm\Sigma}_n^Y(\theta_h)-{\bm\Sigma}_n^X(\theta_h)\Big\Vert&\le  
\max_{|h|\le B_T} \frac 1n\Big\Vert\widehat{\bm\Sigma}_n^Y(\theta_h)-{\bm\Sigma}_n^Y(\theta_h)\Big\Vert+ \max_{|h|\le B_T} \frac 1n\Big\Vert\widehat{\bm\Sigma}_n^Z(\theta_h)\Big\Vert\nn\\
&= O_{\rm P}\l(\max\l({B_T}/{\sqrt T},1/{B_T},1/{n}\r)\r).\label{eq:sonotroppe}
\end{align}
Then, note that, because of Lemma \ref{lem:dyn_eval}{\it (i)} and since $\Vert\bm\ell_i\Vert=1$,
\begin{align}\label{eq:spettriuniformi_larnaca4}
\max_{i=1,\ldots, n}  \max_{|h|\le B_T} \frac 1{n}\Big\Vert\bm{\ell}^\prime_i\widehat{\bm\Sigma}_n^Z(\theta_h)\Big\Vert^2&
\le \max_{\bm w : \Vert\bm w\Vert=1}  \max_{|h|\le B_T} \frac 1{n}\bm w^\prime\widehat{\bm\Sigma}_n^Z(\theta_h)\widehat{\bm\Sigma}_n^{Z}(\theta_h)\bm w = \frac 1 n \Big\Vert \widehat{\bm\Sigma}_n^Z(\theta_h)\Big\Vert^2 = O(1/n).
\end{align}
Hence, from \eqref{eq:spettriuniformi_larnaca2} and \eqref{eq:spettriuniformi_larnaca4}, following the same approach as in \eqref{eq:sonotroppe}, it follows that
\begin{align}
&\max_{i=1,\ldots, n} \max_{|h|\le B_T} \frac{1}{\sqrt n}\Big\Vert\bm{\ell}^\prime_i\Big(\widehat{\bm\Sigma}_n^Y(\theta_h)-{\bm\Sigma}_n^X(\theta_h)\Big)\Big\Vert= O_{\rm P}\l(\max\l({B_T}/{\sqrt T},1/{B_T},1/{\sqrt n}\r)\r).\label{eq:spettriuniformi_larnaca5}
\end{align}
It follows from \eqref{eq:sonotroppe} that, for all $j=1,\ldots, q$ (see also Lemma 2{\it (i)} in \citealp{FHLZ17})
\beq\label{eq:cons_dyn_eval_level}
\max_{|h|\le B_T}\frac 1 n\Big\vert \widehat{\lambda}_{nj}^Y(\theta_h)-{\lambda}_{nj}^X(\theta_h)\Big\vert\le \max_{|h|\le B_T} \frac 1n\Big\Vert\widehat{\bm\Sigma}_n^Y(\theta_h)-{\bm\Sigma}_n^X(\theta_h)\Big\Vert=O_{\rm P}\l(\max\l({B_T}/{\sqrt T},1/{B_T},1/{n}\r)\r).
\eeq
Let $\widehat{\bm\Lambda}^Y_{n} (\theta_h)$ and $\bm\Lambda^X_{n} (\theta_h)$ be the $q\times q$ diagonal matrices with the $q$ largest eigenvalues of $\widehat{\bm\Sigma}_n^Y(\theta_h)$ and $\bm\Sigma_n^X(\theta_h)$, respectively. Then, from \eqref{eq:cons_dyn_eval_level},
\beq\label{eq:cons_dyn_eval_level1more}
\max_{|h|\le B_T}\frac 1 {n}\Big \Vert\widehat{\bm\Lambda}^Y_{n} (\theta_h)-\bm\Lambda^X_{n} (\theta_h)
\Big\Vert\le \frac 1 {n}\sum_{j=1}^q \max_{|h|\le B_T}\Big\vert\widehat{\lambda}_{nj}^Y(\theta_h)-{\lambda}_{nj}^X(\theta_h)\Big\vert=O_{\rm P}\l(\max\l({B_T}/{\sqrt T},1/{B_T},1/{n}\r)\r),
\eeq
and, from Assumption (L3) and Lemma \ref{lem:dyn_eval}{\it (ii)} (see also Lemma 2{\it (ii)} in \citealp{FHLZ17}),
\beq\label{eq:cons_dyn_eval_level2}
\max_{|h|\le B_T} n \big\Vert (\bm\Lambda^X_{n} (\theta_h))^{-1}\big\Vert = O(1), \qquad \max_{|h|\le B_T} n \big\Vert (\widehat{\bm\Lambda}^Y_{n} (\theta_h))^{-1}\big\Vert = O_{\rm P}(1).
\eeq
Moreover, using \eqref{eq:sonotroppe} and following Lemma 3 in \citet{FHLZ17}, it can be shown that there exist  $q\times q$ complex diagonal matrices $\bm{\mathcal J}(\theta_h)$ with entries having unit modulus, such that
\beq\label{eq:cons_dyn_evec_level}
\max_{|h|\le B_T} \Big\Vert \widehat{\mbf P}_n^{Y\dag}(\theta_h){\mbf P}_n^X(\theta_h)-\bm{\mathcal J}(\theta_h)\Big\Vert =O_{\rm P}\l(\max\l({B_T}/{\sqrt T},1/{B_T},1/{n}\r)\r).
\eeq
Now, note that, because of Assumption (L1{\it i}), (L1{\it ii}), (L1{\it iii}), and (L1{\it v}), $\sigma_{ii}^X(\theta_h)=\sum_{j=1}^q\lambda_{nj}^X(\theta_h) \,|p_{ij}^X(\theta_h)|^2\le M$,
for some $M>0$ finite and independent of $i$ and $\theta_h$. Therefore, from Assumption (L3) we get (see also equation (B5) in \citealp{FHLZ17})
\beq\label{eq:nonloso}
\max_{i=1,\ldots,n}\max_{|h|\le B_T}\sqrt n \Big\Vert\bm{\ell}^\prime_i  {\mbf P}_n^{X}(\theta_h)\Big\Vert\le M^{*},
\eeq
for some $M^*>0$ finite and independent of $n$. Therefore, 
from \eqref{eq:cons_dyn_evec_level}, \eqref{eq:nonloso} and using \eqref{eq:spettriuniformi_larnaca5}, it is possible to prove that (see the proof of Lemma 4 in \citealp{FHLZ17} for details)
\beq
\max_{i=1,\ldots,n}\max_{|h|\le B_T}\sqrt n \Big\Vert\bm{\ell}^\prime_i \Big( {\mbf P}_n^{X}(\theta_h)\bm{\mathcal J}(\theta_h)-\widehat{\mbf P}_n^{Y}(\theta_h)\Big)\Big\Vert=O_{\rm P}\l(\max\l({B_T}/{\sqrt T},1/{B_T},1/\sqrt{n}\r)\r),\label{eq:cons_dyn_evec_level2}
\eeq
and, from \eqref{eq:cons_dyn_eval_level1more}, \eqref{eq:cons_dyn_eval_level2}, and \eqref{eq:cons_dyn_evec_level2}, we have that (see the proof of Lemma 4 in \citealp{FHLZ17} for details)
\beq
\max_{i=1,\ldots,n}\max_{|h|\le B_T}\Big\Vert\bm{\ell}^\prime_i \Big( {\mbf P}_n^{X}(\theta_h)(\bm\Lambda^X_{n} (\theta_h))^{1/2}\bm{\mathcal J}(\theta_h)-\widehat{\mbf P}_n^{Y}(\theta_h)(\widehat{\bm\Lambda}^Y_{n} (\theta_h))^{1/2}\Big)\Big\Vert=O_{\rm P}\l(\max\l({B_T}/{\sqrt T},1/{B_T},1/\sqrt{n}\r)\r).\label{eq:cons_dyn_evec_level3}
\eeq
The estimator of the spectral density matrix of $\mbf X_n$ is defined as $\widehat{\bm\Sigma}_{n}^X(\theta_h):=\widehat{\bf P}^Y_{n} (\theta_h) \widehat{\bm \Lambda}^Y_{n} (\theta_h) \widehat{\bf P}^{Y\dag}_{n} (\theta_h)$, with entries $\widehat{\sigma}_{ij}^X(\theta_h)$. Then, \eqref{eq:cons_dyn_evec_level3} implies
(see also Proposition 7 in \citealp{FHLZ17})
\begin{align}\label{eq:cons_common_spectra_level}
\max_{i,j=1,\ldots, n} \max_{|h|\le B_T} \Big\vert\widehat{\sigma}_{ij}^X(\theta_h)-{\sigma}_{ij}^X(\theta_h)\Big\vert=\max_{i,j=1,\ldots, n} \max_{|h|\le B_T}\Big\Vert\bm\ell_i^\prime(\widehat{\bm\Sigma}_{n}^X(\theta_h)-{\bm\Sigma}_{n}^X(\theta_h))\bm\ell_j\Big\Vert = O_{\rm P}(\rho_{nT}),
\end{align}
where   $\rho_{nT}:=\max\big({B_T}/{\sqrt T},1/{B_T},1/{\sqrt n}\big)$.\\

Moving to the autocovariances of the common component,   the $(i,j)$ entry ${\gamma}^X_{ijk}$ of ${\bm\Gamma}^X_{nk}$ is obtained as  the inverse Fourier transform
\[
{\gamma}^X_{ijk} = \int_{-\pi}^\pi e^{ik\theta} \sigma_{ij}^X(\theta)\mathrm d\theta.
\]
Denoting by $\widehat{\gamma}^X_{ijk}$ the entries of 
the estimated autocovariances  $\widehat{\bm\Gamma}^X_{nk}$, we have (see also Proposition 8 in \citealp{FHLZ17})
\begin{align}
\vert \widehat{\gamma}^X_{ijk}- {\gamma}^X_{ijk}\vert &\le \frac{\pi}{B_T}\sum_{|h|\le B_T}  \vert\widehat{\sigma}_{ij}^X(\theta_h)-{\sigma}_{ij}^X(\theta_h)\vert+\frac{\pi}{B_T}\sum_{|h|\le B_T} \max_{\theta_{h-1}\le\theta\le \theta_h} \vert e^{i k\theta_h}\sigma_{ij}^X(\theta_h)-e^{i k\theta}\sigma_{ij}^X(\theta)\vert\nn\\
&\le \pi \max_{|h|\le B_T} \Big\vert\widehat{\sigma}_{ij}^X(\theta_h)-{\sigma}_{ij}^X(\theta_h)\Big\vert + \frac {C_k}{B_T}= O_{\rm P}(\rho_{nT}),\label{eq:cons_autocov_level}
\end{align}
where we used \eqref{eq:cons_common_spectra_level}, the fact that the functions $\theta\mapsto e^{ik\theta}$ and $\theta\mapsto \sigma_{ij}^X(\theta)$ are of bounded variation, and Assumption~(K). Moreover, in view of \eqref{eq:cons_common_spectra_level},  \eqref{eq:cons_autocov_level} holds uniformly in $i$ and $j$:
\beq\label{eq:cons_autocov_level_unif}
\max_{i,j=1,\ldots, n}\vert \widehat{\gamma}^X_{ijk}- {\gamma}^X_{ijk}\vert= O_{\rm P}(\rho_{nT}).
\eeq 
 Notice, however, that \eqref{eq:cons_autocov_level} does not hold uniformly in $k$, which poses no problem since we always consider~$k\le S$, with~$S<\infty$   because of Assumption (L4).

Hereafter, for simplicity of notation and without loss of generality, we assume that $n=m(q+1)$ (with $m\in\mathbb{N}$) and all VARs are of order one;  namely, $\mbf A^{(\ell)}(L)=(\mbf I_{q+1}-\mbf A_1^{(\ell)}L)$ for   $\ell=1,\ldots, m$. Consider the traditional Yule-Walker esti\-mator~$
\widehat{\mbf A}_1^{(\ell)}:= \widehat{\bm\Gamma}_1^{X^{(\ell)}} \big[\widehat{\bm\Gamma}_0^{X^{(\ell)}}\big]^{-1}
$ 
of $\mbf A_1^{(\ell)}$ (see also \eqref{eq:yw_pop}). Then, the $n\times n$ block-diagonal VAR ope\-rator~${\mbf A}_n(L)=(\mbf I_n-{\mbf A}_{n1}L)$ with diagonal blocks $\mbf I_{q+1}-\mbf A_1^{(1)},\ldots,\mbf I_{q+1}- \mbf A_1^{(m)}$ has  block-diagonal  estimator~$\widehat{\mbf A}_n(L)=(\mbf I_n-\widehat{\mbf A}_{n1}L)$,   with  diagonal blocks   $\mbf I_{q+1}-\widehat{\mbf A}_1^{(1)},\ldots, \mbf I_{q+1}- \widehat{\mbf A}_1^{(m)}$. As a consequence of~\eqref{eq:cons_autocov_level}, we have (see also Proposition 9 \citealp{FHLZ17})
\begin{align}
\label{eq:cons_var_level}
&\max_{\ell=1,\ldots, m}\Vert \widehat{\mbf A}_1^{(\ell)}-{\mbf A}_1^{(\ell)}\Vert= O_{\rm P}(\rho_{nT}). 
\end{align}
Let ${\mbf a}_{i}'$ and   $\widehat{\mbf a}_{i}'$ denote the $i$-th rows of ${\mbf A}_{n1}$ and $\widehat{\mbf A}_{n1}$, respectively.
Since  $\mbf A_{n1}$ has only $n(q+1)^2$ non-zero entries, and since each of its $n$ rows has only $(q+1)$ non-zero entries,  we also have
\beq
\label{eq:cons_var_level_row}
\max_{i=1,\ldots, n}\Vert \widehat{\mbf a}_{i}'-{\mbf a}_{i}'\Vert= O_{\rm P}(\rho_{nT})\quad\text{and}\quad \frac 1 {\sqrt n}\Vert \widehat{\mbf A}_{n1}-{\mbf A}_{n1}\Vert= O_{\rm P}(\rho_{nT}),
\eeq 
where uniformity over $i$ is a consequence of \eqref{eq:cons_autocov_level_unif}.\medskip

Turning to  $\mbf H_n$ and $\widehat{\mbf H}_n$,  with $i$-th rows $\bm h_i^\prime$ and $\widehat{\bm h}_i^\prime$, respectively,  we have 
\beq\label{eq:cons_H}
 \max_{i=1,\ldots, n}\Vert \widehat{\bm h}_i'-\bm h_i'\mbf J\Vert= O_{\rm P}(\rho_{nT}) \quad\text{and}\quad \frac 1 {\sqrt n}\Vert \widehat{\mbf H}_n-{\mbf H}_n\mbf J\Vert= O_{\rm P}(\rho_{nT}),
\eeq
where $\mbf J$ is some $q\times q$ diagonal matrix with entries $\pm 1$   (see also Proposition 10 in \citealp{FHLZ17} which since $\widehat{\mbf H}_n$ is a matrix of eigenvectors is based on steps similar to those leading to \eqref{eq:cons_dyn_evec_level2}). 
Because~${\mbf A}_n(L)$  and~$\widehat{\mbf A}_n(L)$ are block-diagonal, so are ${\mbf B}_n(L)=[{\mbf A}_n(L)]^{-1}$ and $\widehat{\mbf B}_n(L)=[\widehat{\mbf A}_n(L)]^{-1}$. Hence, all rows of ${\mbf B}_n(L)$ and~$\widehat{\mbf B}_n(L)$, irrespective of $n$, have at most $(q+1)$ non-zero entries. It thus follows from~\eqref{eq:cons_var_level_row} and \eqref{eq:cons_H} that, for  any~$k\ge 0$,
\beq\label{eq:irf_cons_level}
 \max_{i=1,\ldots, n}\Vert \widehat{\mbf b}_{ik}'-\mbf b_{ik}'\mbf J\Vert  
 =O_{\rm P}(\rho_{nT}).
\eeq
This completes the proof of part {\it (a)} of the proposition.\medskip

The estimator $\widehat{\mbf A}_n(L)$ provides an estimator $\widehat{\mbf Y}^*_n:=\widehat{\mbf A}_n(L)\mbf Y_n$ for the filtered process ${\mbf Y}^*_n$. Consider   the estimated factors
\begin{align}
\widehat{\mbf u}_t=\frac{1}{n}{\widehat{\mbf H}_n'\widehat{\mbf Y}^*_{nt}}&=
\frac{1}{n}{\big(\widehat{\mbf H}_n'\widehat{\mbf A}_n(L)-\mbf J\mbf H_n'{\mbf A}_n(L)\big)\mbf Y_{nt}}+
\frac{1}{n}{\mbf J\mbf H_n'\mbf A_n(L)\mbf Y_{nt}}\nn\\
&=\frac{1}{n}{\big(\widehat{\mbf H}_n'\widehat{\mbf A}_n(L)-\mbf J\mbf H_n'{\mbf A}_n(L)\big)\mbf Y_{nt}}+
\frac{1}{n}{\mbf J\mbf H_n'\mbf A_n(L)\mbf X_{nt}}+\frac{1}{n}{\mbf J\mbf H_n'\mbf A_n(L)\mbf Z_{nt}}\nn\\
&=\frac{1}{n}{\big(\widehat{\mbf H}_n'\widehat{\mbf A}_n(L)-\mbf J\mbf H_n'{\mbf A}_n(L)\big)\mbf Y_{nt}}+
\frac{1}{n}{\mbf J\mbf H_n'\mbf H_n\mbf u_t}+\frac{1}{n}{\mbf J\mbf H_n'\mbf A_n(L)\mbf Z_{nt}}\nn\\
&=\frac{1}{n}{\big(\widehat{\mbf H}_n'\widehat{\mbf A}_n(L)-\mbf J\mbf H_n'{\mbf A}_n(L)\big)\mbf Y_{nt}}+
\mbf J\mbf u_t+\frac{1}{n}{\mbf J\mbf H_n'\mbf A_n(L)\mbf Z_{nt}},\nn
\end{align}
where we used the identification constraints in Assumption~(I{\it i}). Then, 
\begin{align}\label{eq:error_level}
\max_{t=1,\ldots, T} \Vert \widehat{\mbf u}_t-\mbf J\mbf u_t\Vert&\le \max_{t=1,\ldots, T}\frac 1n \big\Vert{\big[\widehat{\mbf H}_n'\widehat{\mbf A}_n(L)-\mbf J\mbf H_n'{\mbf A}_n(L)\big]\mbf Y_{nt}}\big\Vert + \max_{t=1,\ldots, T}\frac 1 n\big\Vert\mbf J\mbf H_n'\mbf A_n(L)\mbf Z_{nt}\big\Vert\nn\\
&=A+B,\ \text{say.}
\end{align}
Term $A$ in \eqref{eq:error_level} is such that
\begin{align}
 \max_{t=1,\ldots, T}\frac 1n &\big\Vert{\big(\widehat{\mbf H}_n'\widehat{\mbf A}_n(L)-\mbf J\mbf H_n'{\mbf A}_n(L)\big)\mbf Y_{nt}}\big\Vert\label{eq:error_level_1}\\
& \le 
 \frac1{\sqrt n} \Big[\big\Vert\widehat{\mbf H}_n-\mbf H_n\mbf J\big\Vert+\big\Vert\widehat{\mbf H}_n'\widehat{\mbf A}_{n1}-\mbf J{\mbf H}_n'{\mbf A}_{n1}\big\Vert\Big]\,\max_{t=1,\ldots, T}\big\Vert {\mbf Y_{nt}}/{\sqrt n}\big\Vert \\ 
 &=O_{\rm P}(\rho_{nT})\max_{t=1,\ldots, T}\big\Vert {\mbf Y_{nt}}/{\sqrt n}\big\Vert,\nn
\end{align}
because of \eqref{eq:cons_var_level_row} and \eqref{eq:cons_H}. Moreover, 
\begin{align}
\max_{t=1,\ldots, T}\big\Vert {\mbf Y_{nt}}/{\sqrt n}\big\Vert&\le \max_{t=1,\ldots, T}\big\Vert{\mbf X_{nt}}/{\sqrt n}\big\Vert + \max_{t=1,\ldots, T}\big\Vert{\mbf Z_{nt}}/{\sqrt n}\big\Vert\nn\\
&\le\frac{1}{\sqrt{n}}\max_{t=1,\ldots, T} \big\Vert  \sum_{k=0}^{\infty}{\mbf B_{nk}\mbf u_{t-k}}\big\Vert  +  \max_{t=1,\ldots, T}\max_{i=1,\ldots ,n} \vert Z_{it}\vert\nn\\
&\le \max_{t=1,\ldots, T} \max_{i=1,\ldots, n}  \sum_{k=0}^{\infty} \Vert \mbf b_{ik}' \Vert\, \Vert\mbf u_{t}\Vert+  \max_{t=1,\ldots, T}\max_{i=1,\ldots ,n} \vert Z_{it}\vert\nn\\
&\le M_1 \sqrt q \max_{t=1,\ldots, T}\max_{j=1,\ldots ,q} \vert u_{jt}\vert+  \max_{t=1,\ldots, T}\max_{i=1,\ldots ,n} \vert Z_{it}\vert
= AI + AI\mspace{-2mu}I,\ \text{say}.\label{eq:unif_Y}
\end{align}
In Assumption (T{\it i}) and (T{\it iii}) we can set $K_u=1$ and $K_Z=1$ by replacing $u_{jt}$  and $Z_{it}$ with $u_{jt}/\Vert u_{jt}\Vert_{\psi_1}$   and $Z_{it}/\Vert Z_{it}\Vert_{\psi_1}$, respectively;  since the sub-exponential norms are assumed to be finite, there is no loss of generality in this choice. Now, using Assumption~(T{\it i}), and since, by Assumption (L1{\it i}), $\E[u_{jt}]=0$ for all~$j$, we have,   for all $\lambda$ such that $|\lambda|\le  1/{e}$  (see also Lemma 5.15 in \citealp{vershynin12}),
\begin{align}
\max_{j=1,\ldots, q}\E[\exp(\lambda u_{jt})] &= \max_{j=1,\ldots, q}
\bigg\{1 + \lambda \E[u_{jt}]+\sum_{p=2}^{\infty}\frac{\lambda ^p\E[(u_{jt})^p]}{p!}\bigg\}\le \bigg\{1 + \sum_{p=2}^{\infty}\frac{\vert\lambda\vert^p p^p}{p!}\bigg\}\nn\\
&\le \bigg\{1 + \sum_{p=2}^{\infty}\vert\lambda\vert^p  e^p\bigg\}= 1+ e^2\lambda^2\le  \exp(\lambda^2 e^{2}),\label{eq:mgfexpu}
\end{align}
where we used the fact that $p!\ge \l(p /e\r)^p$. Then, for any $\epsilon>0$ and  $|\lambda|\le  1/{e}$, we have from \eqref{eq:mgfexpu} that
\begin{align}
\mathrm{P}(u_{jt}>\epsilon)&=\mathrm{P}\l(\exp\l(u_{jt}\lambda\r)> \exp\l(\epsilon \lambda\r)\r)
\le  \E\l[\exp\l(u_{jt}\lambda\r)\r] \exp\l(- \epsilon \lambda\r)\le \exp\l(\lambda^2 e^2- \epsilon \lambda\r).\label{eq:tailssubexp}
\end{align}
Similarly, we have $\mathrm{P}(u_{jt}<-\epsilon)\le \exp\l(\lambda^2 e^2-\epsilon\lambda\r)$. Without loss of generality, we may  set $\lambda =   1/{3}$, which \linebreak yields $\mathrm{P}(|u_{jt}|>\epsilon)\le K_u^* \exp\l(- {\epsilon}/{3}\r)$ for some finite $K_u^*>0$. By Bonferroni inequality, we then obtain 
\beq
\mathrm{P}\big(\max_{t=1,\ldots, T}\max_{j=1,\ldots ,q} \vert u_{jt}\vert>\epsilon\big)\le TK_u^* \exp\l(- {\epsilon}/3\r).\label{eq:bonf_u}
\eeq
Therefore, term $AI$ on the right-hand side  of \eqref{eq:unif_Y} is  $O_{\rm P}(\log T)$. Turning to $AI\mspace{-2mu}I$ on the right-hand side  of \eqref{eq:unif_Y}, notice that, since $\Vert\bm\ell_i\Vert=1$, then
\[
\max_{i=1,\ldots ,n} \Vert Z_{it}\Vert_{\psi_1} = \max_{i=1,\ldots ,n} \Vert \bm\ell_i^\prime\mbf Z_{t}\Vert_{\psi_1} \le \sup_{\bm w_n : \Vert \bm w_n\Vert=1 }\Vert \bm w_n^\prime\mbf Z_{t}\Vert_{\psi_1} \le K_Z,
\] for all $n\in\mathbb N$
by Assumption (T1{\it iii}). Therefore, using Bonferroni inequality, we obtain
\begin{align}
\mathrm{P} \Big(\max_{t=1,\ldots, T}\max_{i=1,\ldots ,n} \vert Z_{it}\vert>\epsilon\Big)\le T K_Z^* \exp\l(-{\epsilon}/{3}\r)\label{eq:bonf_Z}.
\end{align}
Hence, $AI\mspace{-2mu}I$ on the right-hand side  of \eqref{eq:unif_Y} is~$O_{\rm P}(\log T)$. 
By substituting  \eqref{eq:unif_Y} into \eqref{eq:error_level_1}, we conclude that term $A$ in \eqref{eq:error_level} is $O_{\rm P}(\rho_{nT}\log T)$. 
\medskip 

Turning to term $B$ in \eqref{eq:error_level}, we have (note that $\Vert \mbf J\Vert=1$)
\begin{align}\label{eq:error_level_2}
\max_{t=1,\ldots, T}\frac 1 n\big\Vert\mbf J\mbf H_n'\mbf A_n(L)\mbf Z_{nt}\big\Vert
&\le \max_{t=1,\ldots, T}\frac{1}{n}\big\Vert{\mbf H'_n\mbf Z_{nt}}\big\Vert
+\max_{t=1,\ldots, T}\frac{1}{n}\big\Vert{\mbf H'_n\mbf A_{n1}\mbf Z_{nt-1}}\big\Vert\nn\\
&\le \max_{t=1,\ldots, T}\frac{1}{{\sqrt n}}\big\Vert{{\mbf P^{X^*}_n}'\mbf Z_{nt}}\big\Vert + \max_{t=1,\ldots, T}\frac{1}{{\sqrt n}}\big\Vert {{\mbf P^{X^*}_n}'\mbf A_{n1}\mbf Z_{nt-1}}\big\Vert\nn\\
&\le \sqrt{\frac{q}{n}}\Big( \max_{t=1,\ldots, T} \max_{j=1,\ldots, q}\big\vert{\mbf p_{nj}^{X^*}}'\mbf Z_{nt}\big\vert +  \max_{t=1,\ldots, T} \max_{j=1,\ldots, q}\big\vert{\mbf p_{nj}^{X^*}}'\mbf A_{n1}\mbf Z_{nt-1}\big\vert\Big)\nn\\
&= BI\  + BI\mspace{-2mu}I,\ \text{say},
\end{align}
where we used the identification constraints of Assumption (I{\it i}). Repeating the same arguments as above, we can show that, if~$|\lambda|\le  1 /e$, then
\begin{align}
\sup_{\bm w_n : \Vert\bm w_n\Vert= 1}\E[\exp(\lambda \bm w_n'\mbf Z_{nt})] 
\le  \exp(\lambda^2 e^{2}).\label{eq:mgfexp_vec}
\end{align}
Without loss of generality we can set  $\lambda =   1/3$ in \eqref{eq:mgfexp_vec}, and, since $\Vert\mbf p_{nj}^{X^*}\Vert=1$, using the same reasoning as in~\eqref{eq:tailssubexp} and the Bonferroni inequality, for any $\epsilon>0$, we obtain, for some finite $K_Z^*>0$, 
 \begin{align}
\mathrm{P} \Big( \max_{t=1,\ldots, T}\max_{j=1,\ldots ,q}\Big\vert{\mbf p_{nj}^{X^*}}'\mbf Z_{nt}\Big\vert> \epsilon) \le K_Z^* Tq \exp\l(- {\epsilon}/3\r).\nn
\end{align}
 Therefore, $BI$ on the right-hand side  of \eqref{eq:error_level_2} is such that
\beq
BI = \sqrt{q/{ n}}\max_{j=1,\ldots ,q} \max_{t=1,\ldots, T}\big\vert{\mbf p_{nj}^{X^*}}'\mbf Z_{nt}\big\vert = O_{\rm P}\l({\log T}/{\sqrt n}\r).\label{eq:PZ}
\eeq
Last, because of stationarity in Assumption (L4{\it iii}), $\Vert\mbf A_{n1}\Vert<1$  thus $\Vert\mbf p_{nj}^{X^*}\mbf A_{n1}\Vert\le 1$,  and, since if Assumption~(T{\it iii}) holds for~$\sup_{\bm w_n : \Vert\bm w_n\Vert=1}$  it also holds  for $\sup_{\bm w_n : \Vert\bm w_n\Vert\le1}$, the same reasoning as for \eqref{eq:PZ} yields 
\beq
BI\mspace{-2mu}I = \sqrt{q/{ n}}\max_{j=1,\ldots ,q} \max_{t=1,\ldots, T}\big\vert{\mbf p_{nj}^{X^*}}'\mbf A_{n1}\mbf Z_{nt-1}\big\vert = O_{\rm P}\l({\log T}/{\sqrt n}\r).\label{eq:PAZ}
\eeq
Substituting \eqref{eq:PZ} and \eqref{eq:PAZ} in \eqref{eq:error_level_2}, we conclude that term $B$  is $O_{\rm P}(\rho_{nT}\log T/\sqrt n)$. Therefore,  term~$A$, which\linebreak  is~$O_{\rm P}(\rho_{nT}\log T)$, 
dominates in \eqref{eq:error_level};  part {\it (b)} of the proposition follows.\medskip 

From parts {\it (a)} and {\it (b)} and \eqref{eq:cons_var_level}, it immediately follows that
\begin{align}
&\max_{i=1,\ldots, n}\,\max_{t=1,\ldots, T} \vert\widehat{e}_{it}-{e}_{it}\vert=\max_{i=1,\ldots, n}\max_{t=1,\ldots, T} \vert\widehat{\bm h}_i'\widehat{\mbf u}_{t}-\bm h_i'\mbf u_{t}\vert=O_{\rm P}(\rho_{nT}\log T). \label{eq:consistency_eit}
\end{align}
Now, for some finite $M>0$ that does not depend on  $i$,
\[
\max_{i=1,\ldots, n}\bigg\Vert \sum_{k=0}^{\bar k_1}\widehat{\mbf b}_{ik}'-\sum_{k=0}^{\infty}\mbf b_{ik}'\mbf J\bigg\Vert \le \max_{i=1,\ldots, n} \bigg\Vert \sum_{k=0}^{\bar k_1}(\widehat{\mbf b}_{ik}'-\mbf b_{ik}'\mbf J)\bigg\Vert + \max_{i=1,\ldots, n} \sum_{|k|\ge \bar k_1} \Vert \mbf b_{ik}'\Vert \frac {|k|}{\bar k_1} \le O_{\rm P}(\rho_{nT})+ \frac M{\bar k_1}; 
\]
the bound on the first term on the right-hand side follows from \eqref{eq:irf_cons_level} and Proposition 3.6 in~\citet{lutkepohl2005}, which holds for any sequence $\bar k_1\to\infty$, while  for the  second term we used Assumption (L1{\it ii}). By taking $\bar k_1$ large enough ($\bar k_1\simeq B_T^{-1}$, say), the second term can be made smaller than the first one. Since~$X_{it}=\mbf b_i'(L)\mbf u_t$ and $\widehat{X}_{it}:=\widehat{\mbf b}_i'(L)\widehat{\mbf u}_t$, it follows that 
\begin{align}
&\max_{i=1,\ldots, n}\,\max_{t=1,\ldots, T} \vert\widehat{X}_{it}-{X}_{it}\vert=O_{\rm P}(\rho_{nT}\log T)\nn
\end{align}
and
\begin{align}&\max_{i=1,\ldots, n}\,\max_{t=1,\ldots, T} \vert\widehat{Z}_{it}-{Z}_{it}\vert=\max_{i=1,\ldots, n}\,\max_{t=1,\ldots, T} \vert Y_{it}-\widehat{X}_{it}-Y_{it}+{X}_{it}\vert=O_{\rm P}(\rho_{nT}\log T).\label{eq:cons_idio_level}
\end{align}

For simplicity of notation and  without loss of generality as far as this proof is concerned, let us assume that
$$[d_i(L)]^{-1}=:c_i(L)=(1-c_{i1}L)\quad\text{and}\quad  \widehat{c}_i(L)=(1-\widehat{ c}_{i1}L).$$
 Then, for any given $i=1,\ldots,n$, we define the estimator
\beq\label{eq:def_c}
\widehat{ c}_{i1} :=  {\sum_{t=2}^T \widehat{Z}_{it}\widehat{Z}_{i,t-1}}\l({\sum_{t=2}^T \widehat{Z}_{i,t-1}^2}\r)^{-1}.
\eeq
For the numerator of \eqref{eq:def_c}, we have
\begin{align}\label{eq:cons_var_idio_level}
\bigg\vert\frac 1 T\sum_{t=2}^T \Big(\widehat{Z}_{it}\widehat{Z}_{i,t-1}- {Z}_{it}{Z}_{i,t-1}\Big)\bigg\vert\le& \bigg\vert\frac 1 T\sum_{t=2}^T \Big(\widehat{Z}_{it}- {Z}_{it}\Big){Z}_{i,t-1}\bigg\vert+\bigg\vert\frac 1 T\sum_{t=2}^T \Big(\widehat{Z}_{it}- {Z}_{it}\Big){Z}_{it}\bigg\vert \\
&+\bigg\vert\frac 1 T\sum_{t=2}^T \Big(\widehat{Z}_{it}- {Z}_{it}\Big)\Big(\widehat{Z}_{i,t-1}-{Z}_{i,t-1}\Big)\bigg\vert = CI + CI\mspace{-2mu}I + CI\mspace{-2mu}I\mspace{-2mu}I,\ \text{say}.\nn
\end{align}
First consider term $CI$:
 \begin{align}
CI&\le 
 \max_{t=1,\ldots,T}\vert\widehat{Z}_{it}- {Z}_{it}\vert\; \bigg\vert\frac 1 T\sum_{t=2}^T{Z}_{i,t-1}\bigg\vert
\le \Big(\max_{t=1,\ldots, T}\vert \widehat{Z}_{it}- {Z}_{it}\vert\Big) \Big(\max_{t=1,\ldots, T}\vert{Z}_{i,t-1}\vert
\Big)
= O_{\rm P}(\rho_{nT}\log^2T),\nn
\end{align}
uniformly over $i$ because of \eqref{eq:cons_idio_level} and \eqref{eq:bonf_Z}.
%
%
 A similar reasoning  shows that  $CI\mspace{-2mu}I$  in \eqref{eq:cons_var_idio_level}  also is~$O_{\rm P}(\rho_{nT}\log^2 T)$ uniformly in $i$, while $CI\mspace{-2mu}I\mspace{-2mu}I=O_{\rm P}(\rho_{nT}\log T)$ uniformly over $i$. 
Turning to the denominator of \eqref{eq:def_c}, we can show that, uniformly in~$i$,
\beq\label{eq:cons_var_idio_level2}
\bigg\vert\frac 1 T\sum_{t=2}^T \Big(\widehat{Z}_{i,t-1}^2- {Z}_{it}^2\Big)\bigg\vert = O_{\rm P}(\rho_{nT}\log T).
\eeq
Consider then the (infeasible) oracle estimator 
 $
\widetilde{ c}_{i1} :=  {\sum_{t=2}^T {Z}_{it}{Z}_{i,t-1}}/{\sum_{t=2}^T {Z}_{i,t-1}^2}
$ 
we would construct if the idiosyncratic components were observed. That oracle is 
such that
\beq
\widetilde{ c}_{i1} -c_{i1} = {\sum_{t=2}^T {Z}_{i,t-1} v_{it}}\l({\sum_{t=2}^T {Z}_{i,t-1}^2}\r)^{-1}.\label{eq:cols}
\eeq
Because of Assumption (L1{\it viii}), $\E[Z_{i,t-1}^2]$ and $\E[(Z_{i,t-1}v_{it})^2]$ are finite, and  
$$\E[Z_{i,t-1}v_{it}]=\E[\E[Z_{i,t-1}v_{it}| Z_{i,t-1}]]=\E[Z_{i,t-1}\E[v_{it}| Z_{i,t-1}]]=0,\quad i\in\mathbb N.$$  
The  summability, uniform over $i$,  of MA coefficients implies the summability, uniform over $i$, of the autocovariances of the~$Z_{it}$'s, hence the ergodicity, for all $i$, of $\{Z_{it}\vert t\in\mathbb Z\}$. Therefore,  the denominator of~\eqref{eq:cols} is such that
\beq\label{eq:cols_D}
\bigg\vert\frac 1 T \sum_{t=2}^T {Z}_{i,t-1}^2-\E[Z_{i,t-1}^2]\bigg\vert = o_{\rm P}(1).
\eeq
Turning to  the numerator,  note that  $\E[{Z}_{i,t-1} v_{it}|v_{i,t-1}]=0$  for all $i$, so that $\{{Z}_{i,t-1} v_{it}\}$ is a martingale difference sequence;  moreover, because of finite fourth moments and summability of the MA coefficients, it is uniformly integrable (see Proposition 7.7 in \citealp{hamilton1994}). Therefore, by Theorem 19.8 in \citet{davidson1994}, we have ergodicity and therefore, for all~$i$, 
\[
\bigg\vert\frac 1 T \sum_{t=2}^T (Z_{i,t-1}v_t)^2 - \E[(Z_{i,t-1}v_t)^2] \bigg\vert= o_{\rm P}(1).
\]
This, along with weak stationarity, implies that all conditions for the central limit theorem for martingale differences, as stated, for instance, in Theorem 24.3 of \citet{davidson1994}, hold, yielding, for all $i$,
\beq\label{eq:cols_N}
\bigg\vert\frac 1 T \sum_{t=2}^T {Z}_{i,t-1} v_{it} -\E[Z_{i,t-1}v_{it}]\bigg\vert= O_{\rm P}\l(1 /{\sqrt T}\r).
\eeq
Going back to~ \eqref{eq:cols},   \eqref{eq:cols_D} and \eqref{eq:cols_N} entail  $\max_{i=1,\ldots, n}|\widetilde{ c}_{i1} -c_{i1}|=O_{\rm P}\big( 1/ {\sqrt T}\big)$;  therefore, from \eqref{eq:def_c}, \eqref{eq:cons_var_idio_level}, \eqref{eq:cons_var_idio_level2}, and \eqref{eq:cols_N},
\beq\label{eq:cons_var_c}
\max_{i=1,\ldots, n}\vert\widehat{ c}_{i1} -c_{i1} \vert \le  \max_{i=1,\ldots, n}\vert\widehat{ c}_{i1} -\widetilde c_{i1} \vert+\max_{i=1,\ldots, n}\vert\widetilde { c}_{i1} -c_{i1} \vert = O_{\rm P}(\rho_{nT}\log^2 T),
\eeq
which in turn implies part {\it (c)} of the proposition.  
 Last, defining 
 $
\widehat{v}_{it}:=\widehat{Z}_{it}-\widehat{ c}_{i1}\widehat{Z}_{i,t-1}
$, 
we have, in view of~\eqref{eq:cons_idio_level} and~\eqref{eq:cons_var_c}, 
\beq\label{eq:adessobasta}
\max_{i=1,\ldots, n} \max_{t=1,\ldots, T} \vert\widehat{v}_{it}-{v}_{it}\vert = O_{\rm P}(\rho_{nT}\log^2 T).
\eeq
  Part {\it (d)} of the proposition follows.\hfill $\square$


\subsection{Proof of Proposition \ref{prop:vol}}\label{app:prop2}
It follows from  Proposition \ref{prop:level} parts {\it (b)} (see also \eqref{eq:consistency_eit}) and {\it (d)}  (see also \eqref{eq:adessobasta}) that
\beq\label{eq:cons_inn2}
\max_{i=1,\ldots, n}\max_{t=1,\ldots, T} \vert (\widehat{e}_{it}+\widehat{v}_{it})-(e_{it}+v_{it})\vert=\max_{i=1,\ldots, n}\max_{t=1,\ldots, T} \vert \widehat{s}_{it}-s_{it}\vert = O_{\rm P}(\rho_{nT}\log^2 T).
\eeq
 Assumption~(R) implies that, for any $i$ and any $t\in{\mathcal T}^c_{i;nT}:=\{1,\ldots ,T\}\setminus {\mathcal T}_{i;nT}$,  \begin{align}
\vert \widehat{h}_{it}-h_{it}\vert &= \vert \log \widehat{s}_{it}^{\, 2}-\log s_{it}^2\vert = 2\vert \log\vert\widehat{s}_{it}\vert -\log\vert s_{it}\vert\vert \le \frac 2 {\kappa_T} \vert \widehat{s}_{it}- s_{it}\vert.\label{eq:cont_map}
\end{align}
From \eqref{eq:cons_inn2} and \eqref{eq:cont_map} we obtain
\beq\label{eq:cons_h}
\max_{i=1,\ldots, n}\max_{t\in{\mathcal T}_{i;nT}^c} \vert \widehat{h}_{it}-h_{it}\vert= O_{\rm P}\l({\rho_{nT}\log^2 T}/{\kappa_T}\r).
\eeq 

Hereafter, let $\mathbb T_{ij;nT}:=\mathcal T_{i;nT}\cup\mathcal T_{j,nT}$. Denoting by $\widehat{\gamma}_{ijk}^{h}$ the oracle estimators (computed from the unavailable $\mbf h_n$ values) of $\mbf h_n$'s lag $k$ cross-covariances and by $\widehat{\gamma}_{ijk}^{\widehat h}$ the estimator obtained by plugging in the estimated values $\widehat{\mbf h}_n$ of $\mbf h_n$ for the actual ones, we have, 
  for any $i,j,k$,
\begin{align}
\vert\widehat{\gamma}_{ijk}^{\widehat h}-\widehat{\gamma}_{ijk}^{h}\vert\le &
\Bigg\vert\frac 1 T\sum_{\substack{t=k+1\\t,(t-k)\in\mathbb T_{ij;nT}^c}}^T\Big(\widehat{h}_{it}\widehat{h}_{jt-k}-h_{it}h_{jt-k}\Big)\Bigg\vert+
\Bigg\vert\frac 1 T\sum_{\substack{t=k+1\\ t,(t-k)\in\mathbb T_{ij;nT}}}^T\Big(\widehat{h}_{it}\widehat{h}_{jt-k}-h_{it}h_{jt-k}\Big)\Bigg\vert\nn\\
&+\Bigg\vert\frac 1 T\sum_{\substack{t=k+1\\ t\in\mathbb T_{ij;nT}\\(t-k)\in\mathbb T_{ij;nT}^c}}^T\Big(\widehat{h}_{it}\widehat{h}_{jt-k}-h_{it}h_{jt-k}\Big)
\Bigg\vert+\Bigg\vert\frac 1 T\sum_{\substack{t=k+1\\ t\in\mathbb T_{ij;nT}^c\\(t-k)\in\mathbb T_{ij;nT}}}^T\Big(\widehat{h}_{it}\widehat{h}_{jt-k}-h_{it}h_{jt-k}\Big)\Bigg\vert \nn\\
=& DI+DI\mspace{-2mu}I+DI\mspace{-2mu}I\mspace{-2mu}I+DIV,\ \text{ say.}\label{eq:bias}
\end{align}

Considering term $DI$ first, 
\begin{align}
DI \le&\bigg\vert\frac 1 T\sum_{\substack{t=k+1\\ t,(t-k)\mathbb T_{ij;nT}^c}}^T\Big(\widehat{h}_{it}-h_{it}\Big)h_{jt-k}\Big)\bigg\vert + \bigg\vert\frac 1 T\sum_{\substack{t=k+1\\ t,(t-k)\in\mathbb T_{ij;nT}^c}}^T\Big(\widehat{h}_{jt-k}-h_{jt-k}\Big)h_{it}\bigg\vert\nn\\
&+\bigg\vert\frac 1 T\sum_{\substack{t=k+1\\ t,(t-k)\in\mathbb T_{ij;nT}^c}}^T\Big(\widehat{h}_{it}-h_{it}\Big)\Big(\widehat{h}_{jt-k}-h_{jt-k}\Big)\bigg\vert
=DI_A + DI_B+ DI_C,\ \text{ say.}\label{eq:gamma_hat_h}
\end{align}
For   $DI_A$  we have
(recall that $\mathbb T_{ij;nT}^c=\mathcal T_{i;nT}^c\cap\mathcal T_{j,nT}^c$)
\begin{align}
DI_A&\le 
\max_{t\in\mathbb T_{ij;nT}^c}|\widehat{h}_{it}- {h}_{it}|
\;\bigg\vert
\frac 1 T\sum_{\substack{t=k+1\\(t-k)\in\mathbb T_{ij;nT}^c}}^T {h}_{jt-k}
\bigg\vert
\le  \max_{t=1,\ldots,T}|\widehat{h}_{it}- {h}_{it}| \max_{t=1,\ldots,T} \vert h_{jt-k}\vert
,\label{eq:I_A}
\end{align}
since $ {|\mathbb T_{ij;nT}^c|}/{T}\le 1$. Note that \eqref{eq:I_A} holds independently of all $k$ such that $|k|\le M_T$. Then, since $\Vert\bm\ell_j\Vert=1$, by Assumption (T{\it iv}) we have that
\[
\max_{j=1,\ldots ,n} \Vert \xi_{jt}\Vert_{\psi_1} = \max_{j=1,\ldots ,n} \Vert \bm\ell_j^\prime\bm\xi_{t}\Vert_{\psi_1} \le \sup_{\bm w_n : \Vert \bm w_n\Vert=1 }\Vert \bm w_n^\prime\bm\xi_{t}\Vert_{\psi_1} \le K_\xi,
\] 
for all $n\in\mathbb N$. Therefore, using also Assumptions (V1{\it ii}) and (T{\it ii}), following the same steps leading to \eqref{eq:bonf_u} and \eqref{eq:bonf_Z}, we have
\begin{align}
\max_{j=1,\ldots, n}\max_{t=1,\ldots, T}\vert{h}_{jt}\vert \le M_5 \max_{j=1,\ldots Q} \max_{t=1,\ldots, T} \vert\varepsilon_{jt}\vert + 
\max_{j=1,\ldots, n}\max_{t=1,\ldots, T} \vert \xi_{jt}\vert = O_{\rm P}(\log T).\label{eq:bonf_h}
\end{align} 
Substituting \eqref{eq:cons_h} and \eqref{eq:bonf_h} into \eqref{eq:I_A} we have $DI_A=O_{\rm P}\l({\rho_{nT}\log^3 T}/{\kappa_T}\r)$, uniformly over $i,j,k$.
%
Terms $DI_B$ and $DI_C$  can be treated similarly, and therefore $DI=O_{\rm P}\l({\rho_{nT}\log^3 T}/{\kappa_T}\r)$, uniformly over $i,j,k$. Turning to   $DI\mspace{-2mu}I$, first notice that, for $t\in {\mathcal T}_{i;nT}$, we have~$\widehat h_{it}=\log (\kappa_T^2)$ for all $i$. Then,  $DI\mspace{-2mu}I$ is bounded from above by  $DI\mspace{-2mu}I_A+DI\mspace{-2mu}I_B+DI\mspace{-2mu}I_C,
$ 
where, as in \eqref{eq:I_A}, because of Assumption (R), we have 
\begin{align}
DI\mspace{-2mu}I_A&\le  \bigg[\frac{\vert\mathbb T_{ij;nT}\vert}T\Big( \max_{t\in\mathbb T_{ij;nT}} |\widehat{h}_{it}- {h}_{it} |\Big)\bigg]\Big(
\max_{t=1,\ldots, T}\vert{h}_{jt}\vert
\Big)\nn\\
&\le  \bigg[\frac{2\max_{i=1,\ldots,n}\vert\mathcal T_{i;nT}\vert}T\Big( |\log (\kappa_T^2)| +\max_{t\in\mathbb T_{ij;nT}} \vert{h}_{it} |\Big)\bigg]\Big(
\max_{t=1,\ldots, T}\vert{h}_{jt}\vert
\Big)\nn\\
&\le  \bigg[o_{\rm P}\l(\frac 1 {\sqrt T}\r)\Big( |\log (\kappa_T^2)| +\max_{t\in\mathbb T_{ij;nT}} \vert{h}_{it} |\Big)\bigg] \Big(
\max_{t=1,\ldots, T}\vert{h}_{jt}\vert
\Big)\nn\\
&= o_{\rm P}\l(\frac{\vert\log\kappa_T^2\vert\log T}{\sqrt T}\r)+ o_{\rm P}\l(\frac{\log^2 T}{\sqrt T}\r) = o_{\rm P}\l(\frac{\vert\log\kappa_T^2\vert\log T}{\sqrt T}\r)+ o_{\rm P}(\rho_{nT}\log^3 T),\label{eq:termIIA}
\end{align}
uniformly over $i,j,k$, because of \eqref{eq:bonf_h}. Terms $DI\mspace{-2mu}I_B$ and $DI\mspace{-2mu}I_C$   are  analogous to terms $DI_B$ and $DI_C$, and can be treated similarly. 
It follows that $I\mspace{-2mu}I=o_{\rm P}\l({\vert\log\kappa_T^2\vert\log T}/{\sqrt T}\r)+o_{\rm P}(\rho_{nT}\log^3 T)$. The same result can be obtained along similar lines for terms $DI\mspace{-2mu}I\mspace{-2mu}I$ and $DIV$. 

Therefore, from \eqref{eq:bias},  and since $\kappa_T$ is of order $\log^{-\varphi} T$ by Assumption (R), we obtain 
\begin{align}
\max_{i,j=1,\ldots,n}\max_{\vert k\vert \le M_T} \big\vert\widehat{\gamma}_{ijk}^{\widehat h}-\widehat{\gamma}_{ijk}^{h}\big\vert &= O_{\rm P}\l({\rho_{nT}\log^3 T}/{\kappa_T}\r)+ o_{\rm P}\l({\vert\log\kappa_T^2\vert\log T}/{\sqrt T}\r)+o_{\rm P}(\rho_{nT}\log^3 T)\nn \\
&=O_{\rm P}(\rho_{nT}\log^{3+\varphi} T)+ o_{\rm P}\l({\log\log T\log T} /{\sqrt T}\r)=O_{\rm P}(\rho_{nT}\log^{3+\varphi} T).\label{eq:gammaijk_h} 
\end{align}
Now let $\widehat{\sigma}_{ij}^{\widehat h}(\theta_\ell)$ be the $(i,j)$-th entry of the estimated spectral density computed from  $\widehat{\mbf h}_n$: then,
using \eqref{eq:gammaijk_h} and the definition of the Bartlett kernel, we have
\begin{align}
\max_{i,j=1,\ldots, n}\max_{|\ell|\le M_T}\big|\widehat{\sigma}_{ij}^{\widehat h}(\theta_\ell)-\widehat{\sigma}_{ij}^{ h}(\theta_\ell)\big|&=\max_{i,j=1,\ldots, n}\max_{|\ell|\le M_T}\bigg\vert\frac 1{ 2\pi }\sum_{k=-T+1}^{T-1}\mathrm K\l(\frac{k}{M_T}\r)e^{-ik\theta_\ell}\big(\widehat{\gamma}_{ijk}^{\widehat h}-\widehat{\gamma}_{ijk}^{h}\big)\bigg\vert\nn\\
&\le  \frac 1{ 2\pi }\sum_{|k|\le M_T} \bigg(1-\frac{k}{M_T}\bigg)  \max_{i,j=1,\ldots, n}\big\vert\widehat{\gamma}_{ijk}^{\widehat h}-\widehat{\gamma}_{ijk}^{h}\big\vert\nn\\
&\le \frac{(2M_T+1)}{ 2\pi } \max_{i,j=1,\ldots, n}\max_{|k|\le M_T}\big\vert\widehat{\gamma}_{ijk}^{\widehat h}-\widehat{\gamma}_{ijk}^{h}\big\vert = O_{\rm P}(M_T \rho_{nT}\log^{3+\varphi} T).\nn
\end{align}
Therefore,
\begin{align}
&\max_{|\ell|\le M_T} \frac 1n\Big\Vert\widehat{\bm\Sigma}_n^{\widehat h}(\theta_\ell)-\widehat{\bm\Sigma}_n^h(\theta_h)\Big\Vert
= O_{\rm P}(M_T \rho_{nT}\log^{3+\varphi} T).\label{eq:spettriuniformi_larnaca_vol2}
\end{align}
Moreover, from Lemma \ref{lem:spettro_level}(ii), we have the following (see also Lemma~1 in \citealp{FHLZ17})
\begin{align}
\E\l[\max_{|\ell|\le M_T} \frac 1{n^2}\Big\Vert\widehat{\bm\Sigma}_n^h(\theta_\ell)-{\bm\Sigma}_n^h(\theta_\ell)\Big\Vert^2\r]&\le\frac 1 {n^2} \sum_{i=1}^n\sum_{j=1}^n\E\l[\max_{|\ell|\le M_T}  \big\vert\widehat{\sigma}_{ij}^h(\theta_\ell)-{\sigma}_{ij}^h(\theta_\ell)\big\vert^2\r]
\le {C_3M_T^2}/T+ {C_4}/{M_T^2},\nn
\end{align}
and, by Chebychev's inequality,
\begin{align}
&\max_{|\ell|\le M_T} \frac 1n\Big\Vert\widehat{\bm\Sigma}_n^h(\theta_\ell)-{\bm\Sigma}_n^h(\theta_\ell)\Big\Vert= O_{\rm P}\l(\max\l({M_T}/{\sqrt T},1/{M_T}\r)\r).\label{eq:spettriuniformi_larnaca_vol}
\end{align}
From \eqref{eq:spettriuniformi_larnaca_vol2}, \eqref{eq:spettriuniformi_larnaca_vol} and Lemma \ref{lem:eval_vol}(i), and, since $M_T \rho_{nT}=\tau_{nT}$,
\begin{align}
\max_{|\ell|\le M_T} \frac 1n\Big\Vert\widehat{\bm\Sigma}_n^{\widehat h}(\theta_\ell)-{\bm\Sigma}_n^\chi(\theta_\ell)\Big\Vert&\le
\max_{|\ell|\le M_T} \frac 1n\Big\Vert\widehat{\bm\Sigma}_n^{\widehat h}(\theta_\ell)-\widehat{\bm\Sigma}_n^h(\theta_\ell)\Big\Vert+
\max_{|\ell|\le M_T} \frac 1n\Big\Vert\widehat{\bm\Sigma}_n^{h}(\theta_\ell)-{\bm\Sigma}_n^h(\theta_\ell)\Big\Vert+
\max_{|\ell|\le M_T} \frac 1n\Big\Vert{\bm\Sigma}_n^\xi(\theta_\ell)\Big\Vert\nn\\
&= O_{\rm P}\l(\max\l(\tau_{nT}\log^{3+\varphi} T,{M_T}/{\sqrt T},1/{M_T},1/n\r)\r)=O_{\rm P}(\tau_{nT}\log^{3+\varphi} T).\nn
\end{align}
and, following the same steps leading to \eqref{eq:cons_common_spectra_level}, we can prove that:
\beq
\max_{i,j=1,\ldots, n}\max_{|\ell|\le M_T} \big\vert \widehat{\sigma}_{ij}^{\chi}(\theta_\ell)-{\sigma}_{ij}^{\chi}(\theta_\ell)\big\vert = O_{\rm P}(\tau_{nT}\log^{3+\varphi} T).\label{eq:lastone}
\eeq
From there on, the proof of Proposition~2 is strictly identical to that of  Proposition \ref{prop:level}. In particular: for part {\it (a)} we have the same rate of convergence as in \eqref{eq:lastone}; for part {(\it b)} we have an additional $\log T$ term, by following the same reasoning leading to the bound of \eqref{eq:error_level}); and for parts {(\it c)} and {\it (d)} we have one more $\log T$ term, by following the same reasoning leading to \eqref{eq:cons_var_c} and \eqref{eq:adessobasta}, respectively.  \hfill$\square$
\clearpage
  \setcounter{figure}{0}
\renewcommand{\thefigure}{B\arabic{figure}}

\section{Assumption (R): empirical evidence }\label{sec:app_R}
In this section, we provide some empirical evidence that Assumption (R) holds in the S\&P100 panel under study. 
From the estimated $nT$ realisations of the estimated panel $\{\widehat {\mbf s}_{t}\}$ obtained in the previous section, we let the sample size vary and we simulate $M$ artificial datasets $\{\widehat {\mbf s}^*_{t}\}$ of size $N\times T_j$, $j=1,\ldots, M$, by uniformly sampling with replacement from $\{\widehat {\mbf s}_{t}\}$. In particular, we consider $M=199$ different sample sizes such that~$T_1=100$, $T_{M}=10000$ and $T_j=T_{j-1}+50$ for~$j=2,\ldots, M$;  for each value of $T_j$, we simulate $N=150$ time series. 

\begin{figure}[h!]\caption{Large sample behaviour of $r(j,\varphi,K,\epsilon)$ as a function of $T_j$}\label{fig:trim}
\centering \noindent 
\setlength{\tabcolsep}{.01\textwidth}
\begin{tabular}{@{}cccc}
 \footnotesize $\varphi=1.6$, $\epsilon=0.01$&\footnotesize $\varphi=1.8$, $\epsilon=0.01$& \footnotesize $\varphi=2$, $\epsilon=0.01$& \footnotesize $\varphi=2.2$, $\epsilon=0.01$\\
 \includegraphics[width=.25\textwidth,trim=0cm .2cm 0cm 0cm,clip]{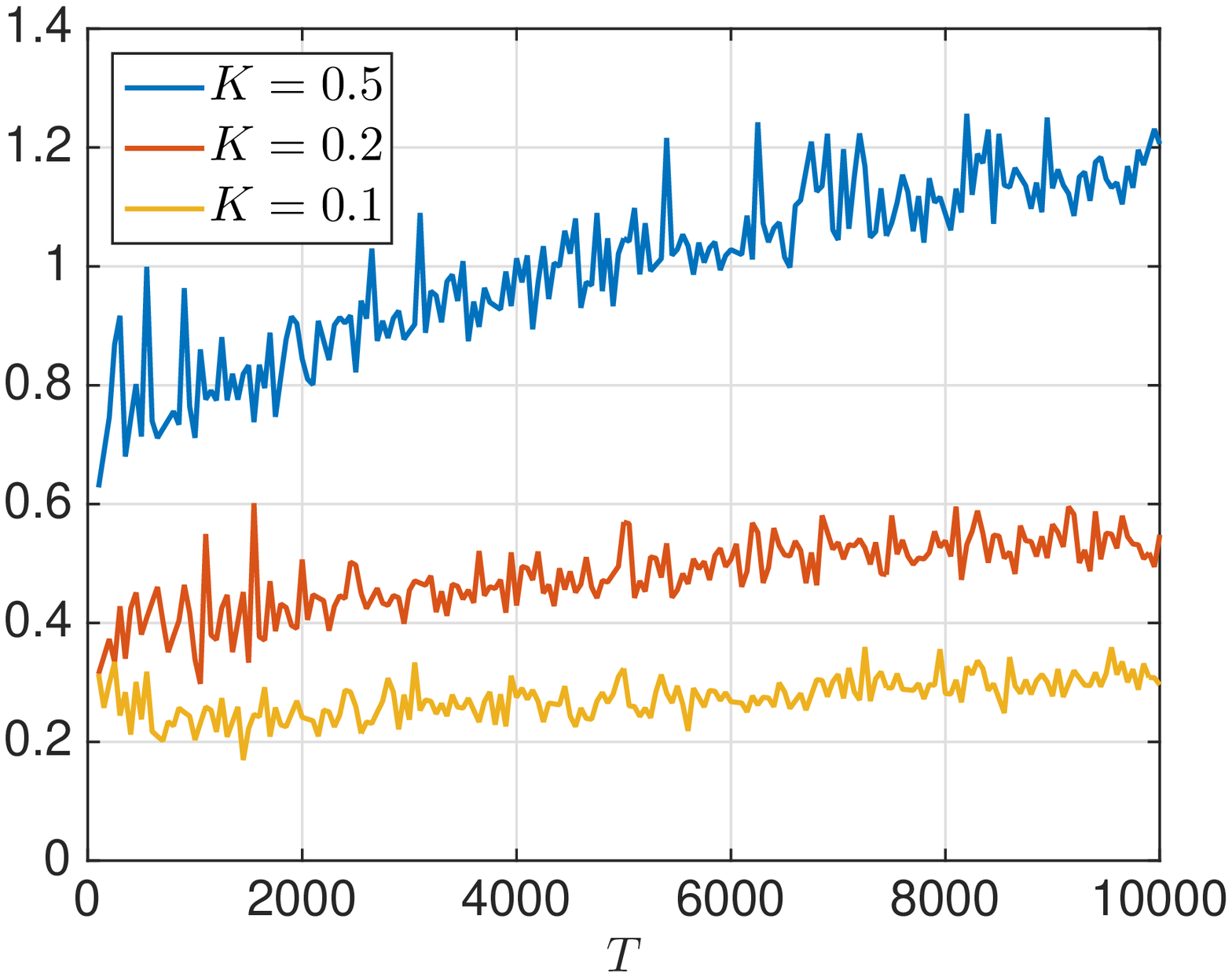}&
 \includegraphics[width=.25\textwidth,trim=0cm .2cm 0cm 0cm,clip]{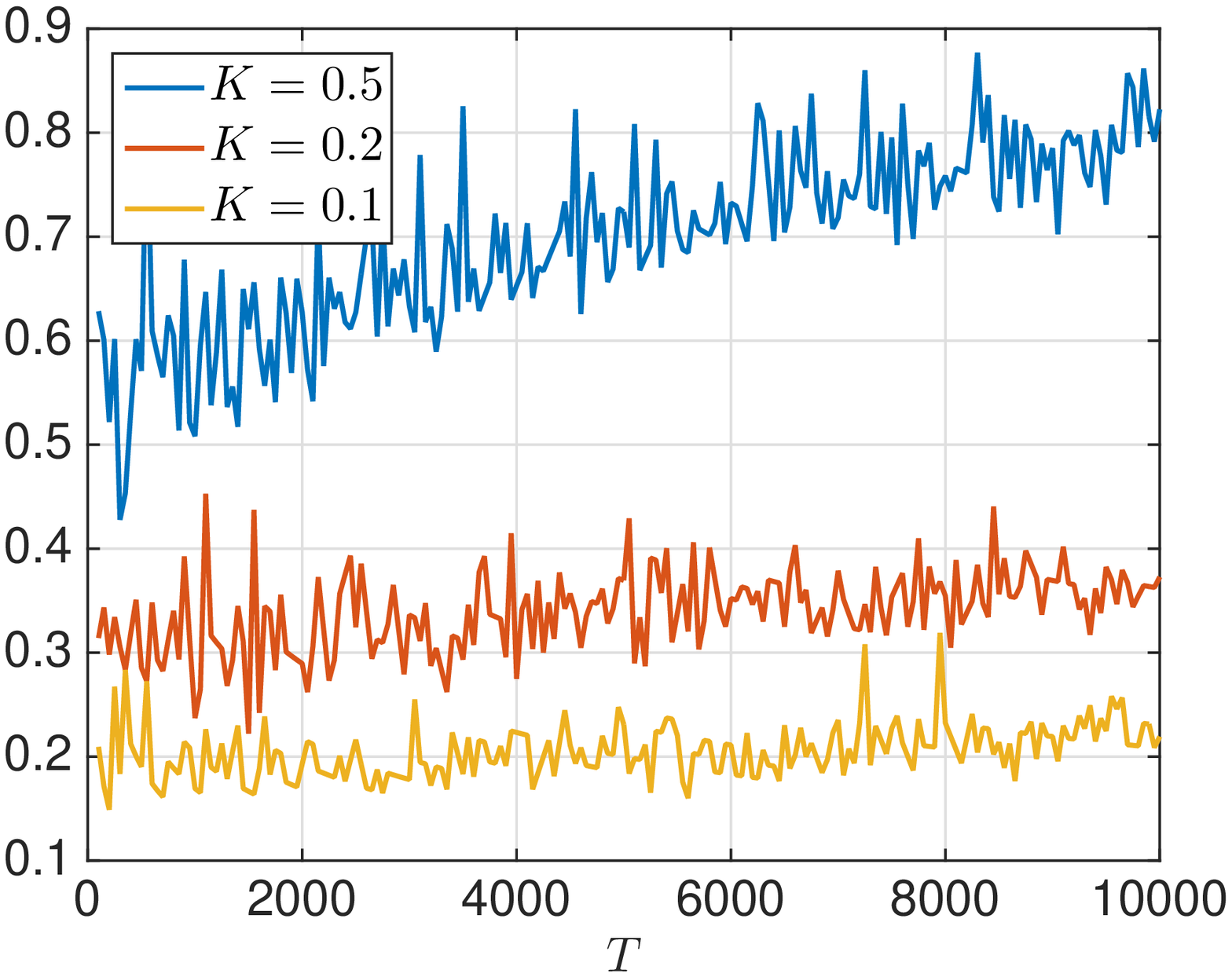}&
 \includegraphics[width=.25\textwidth,trim=0cm .2cm 0cm 0cm,clip]{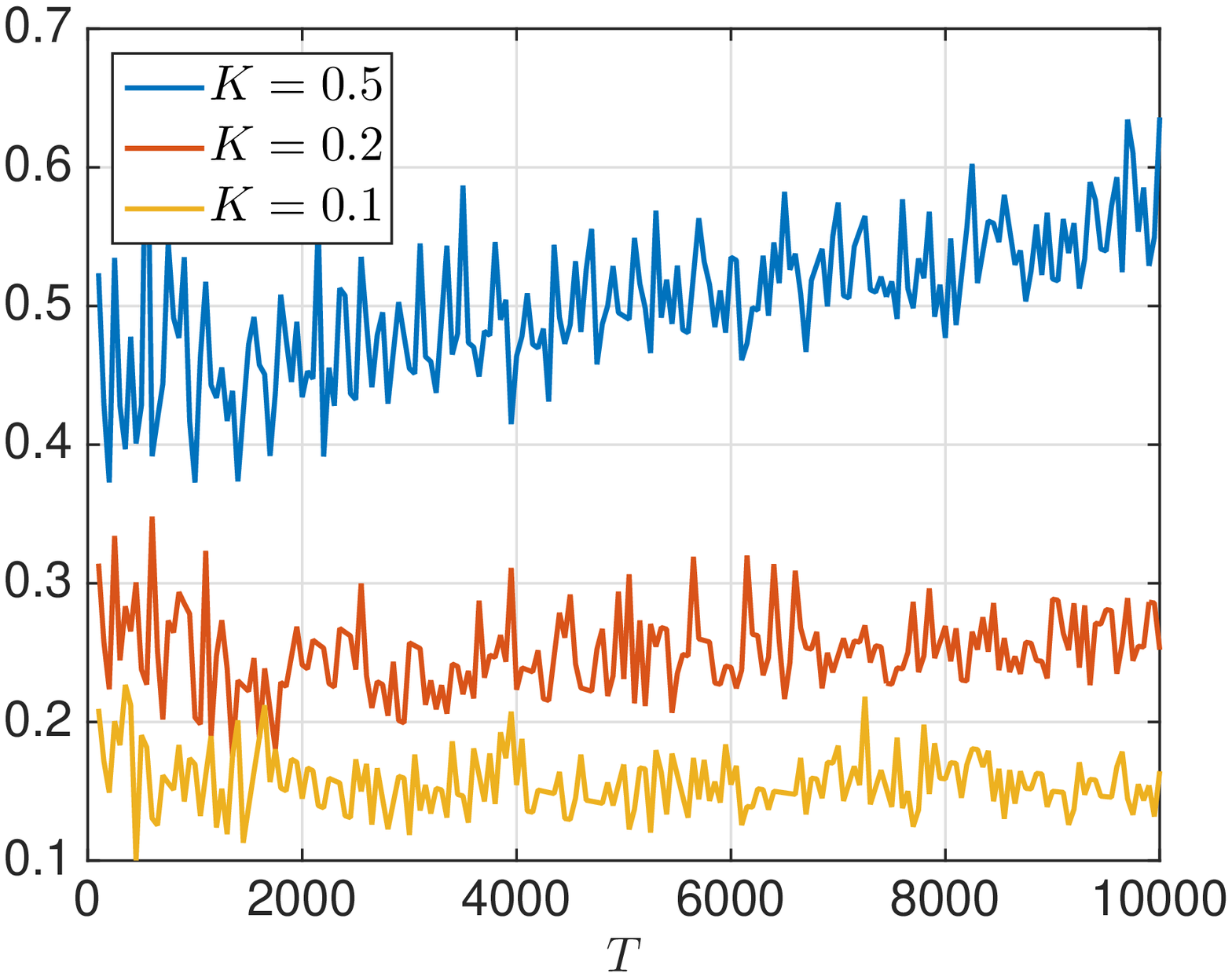}&
  \includegraphics[width=.25\textwidth,trim=0cm .2cm 0cm 0cm,clip]{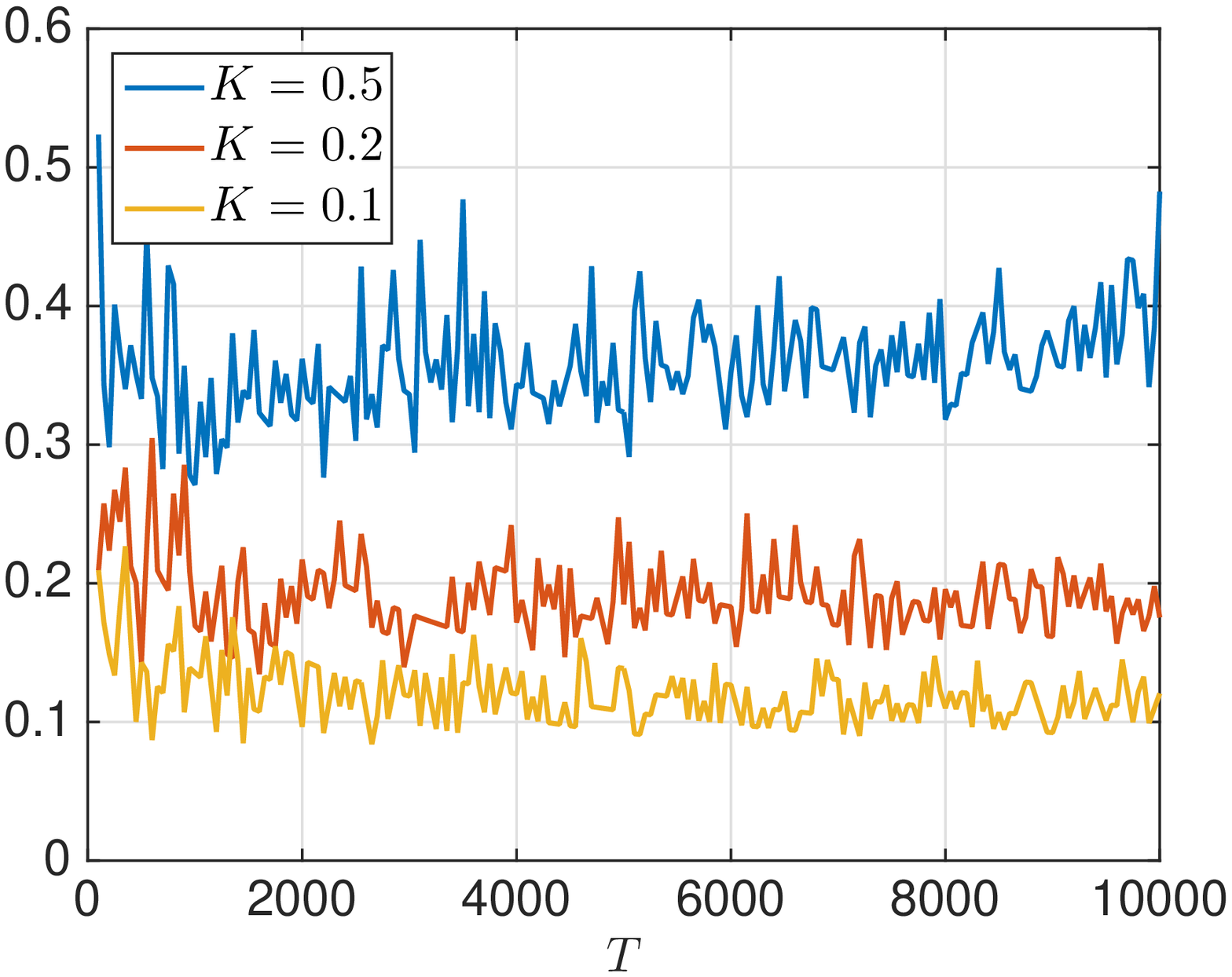}\\
 \footnotesize $\varphi=2.4$, $\epsilon=0.01$&\footnotesize $\varphi=2.6$, $\epsilon=0.01$& \footnotesize $\varphi=2.8$, $\epsilon=0.01$& \footnotesize $\varphi=3$, $\epsilon=0.01$\\
 \includegraphics[width=.25\textwidth,trim=0cm .2cm 0cm 0cm,clip]{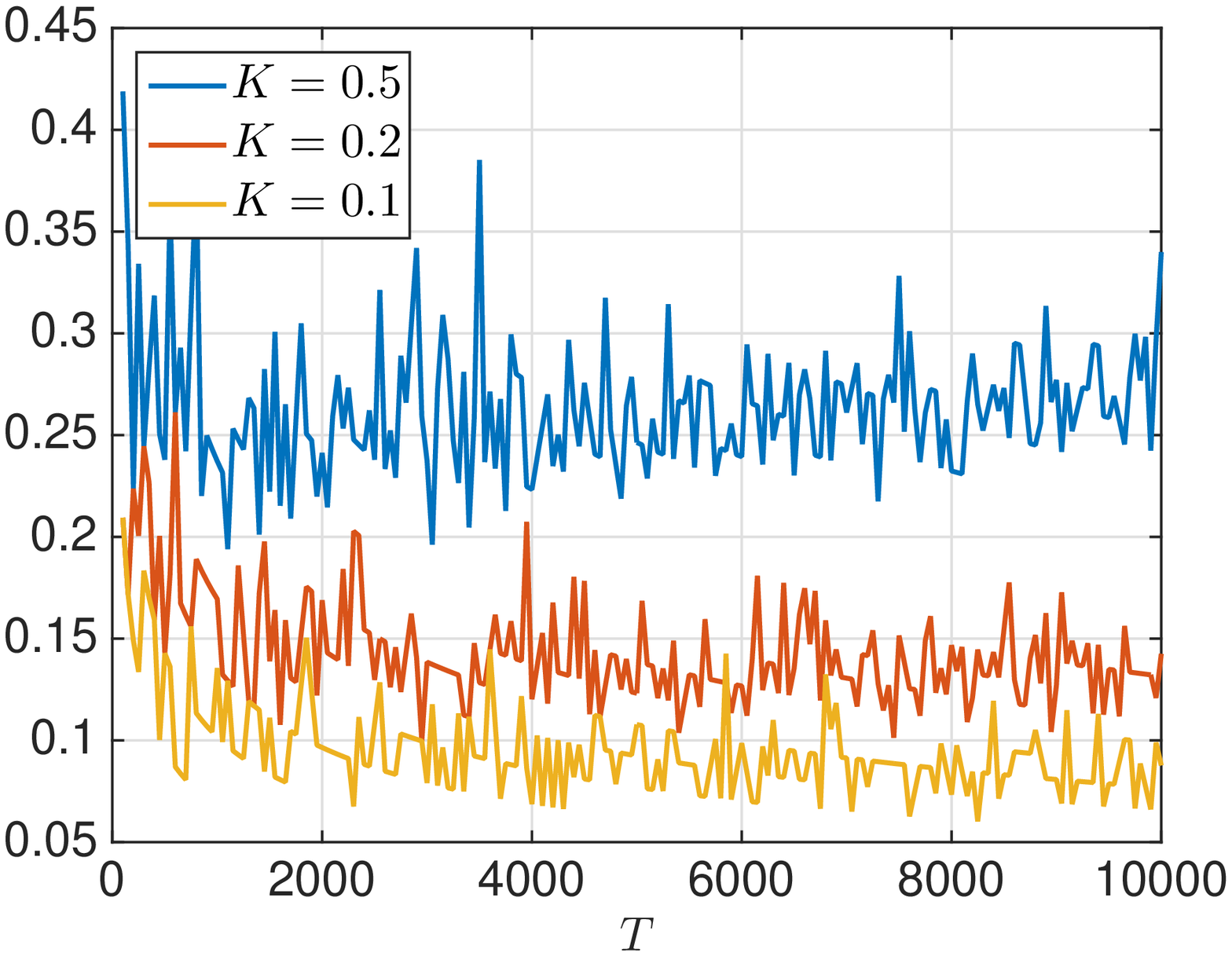}&
 \includegraphics[width=.25\textwidth,trim=0cm .2cm 0cm 0cm,clip]{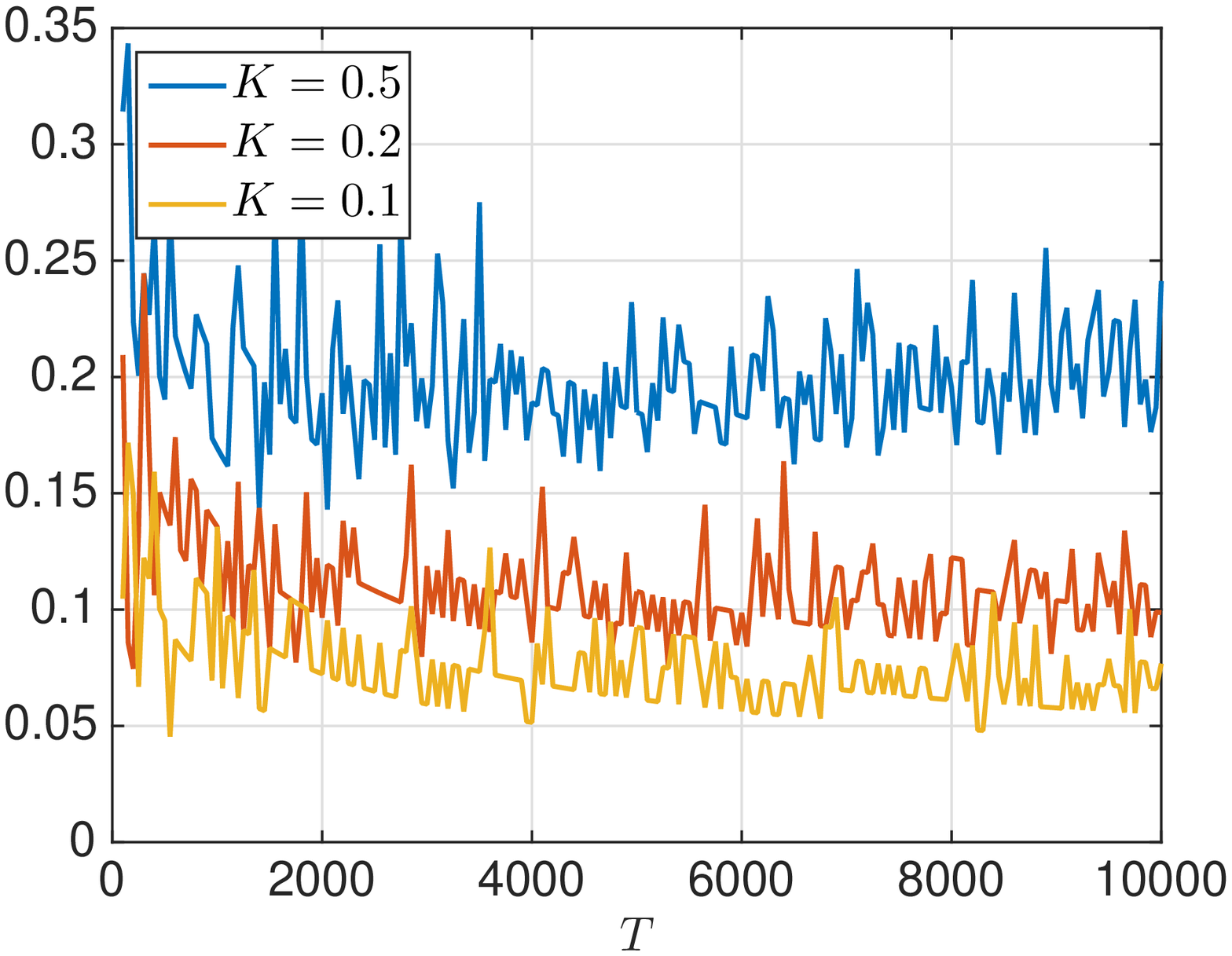}&
 \includegraphics[width=.25\textwidth,trim=0cm .2cm 0cm 0cm,clip]{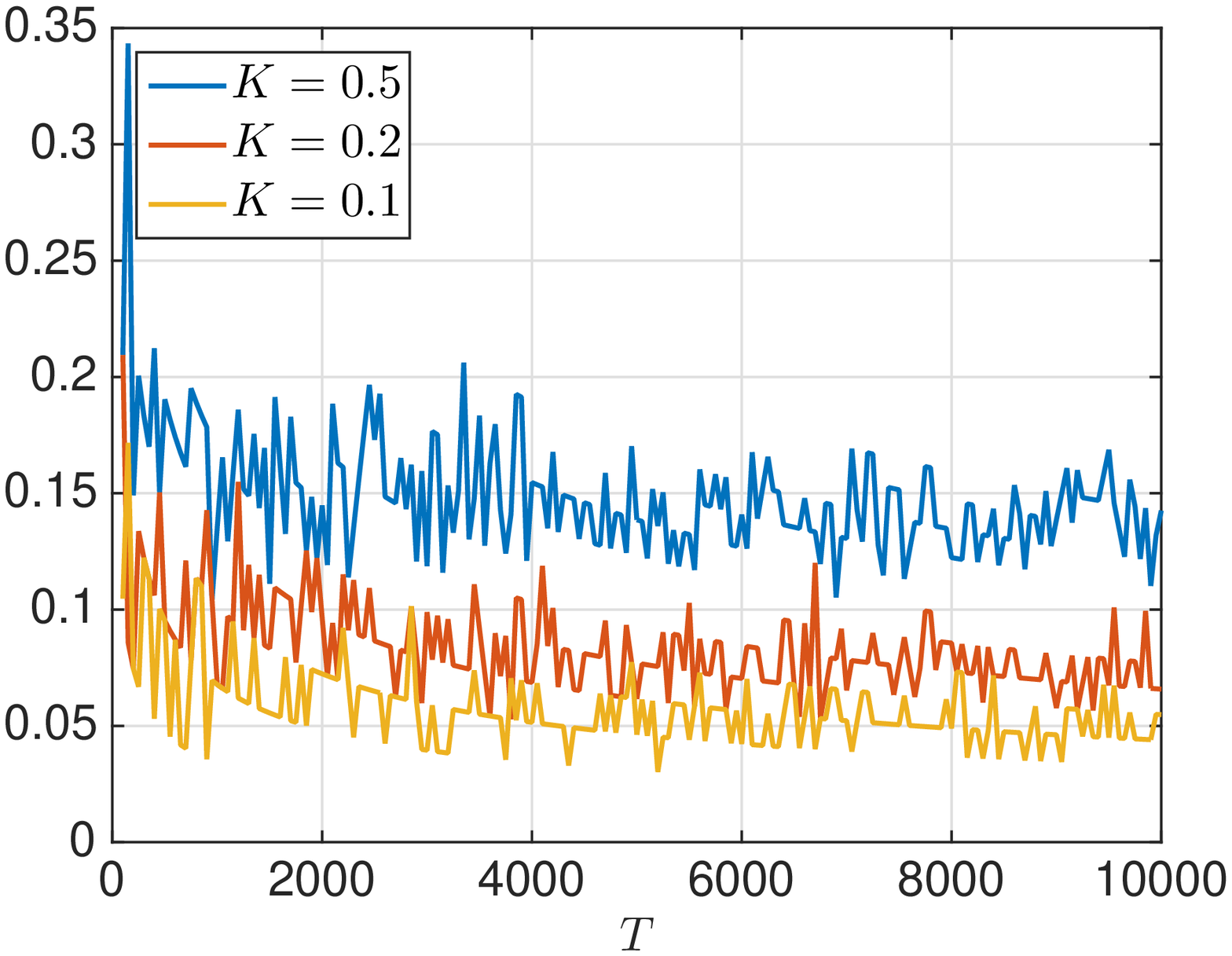}&
  \includegraphics[width=.25\textwidth,trim=0cm .2cm 0cm 0cm,clip]{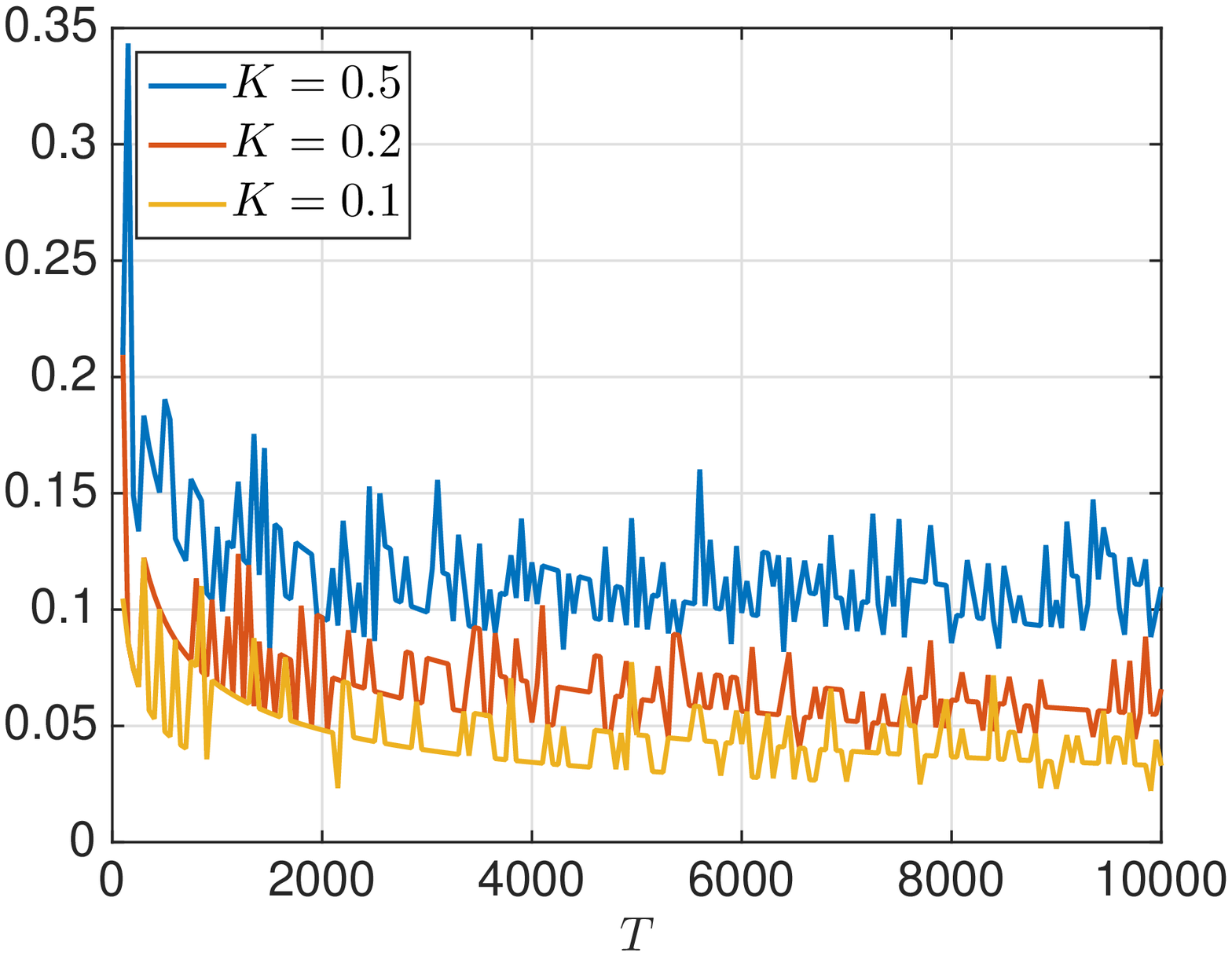}\\
 \footnotesize $\varphi=1.6$, $\epsilon=0.1$&\footnotesize $\varphi=1.8$, $\epsilon=0.1$& \footnotesize $\varphi=2$, $\epsilon=0.1$& \footnotesize $\varphi=2.2$, $\epsilon=0.01$\\
 \includegraphics[width=.25\textwidth,trim=0cm .2cm 0cm 0cm,clip]{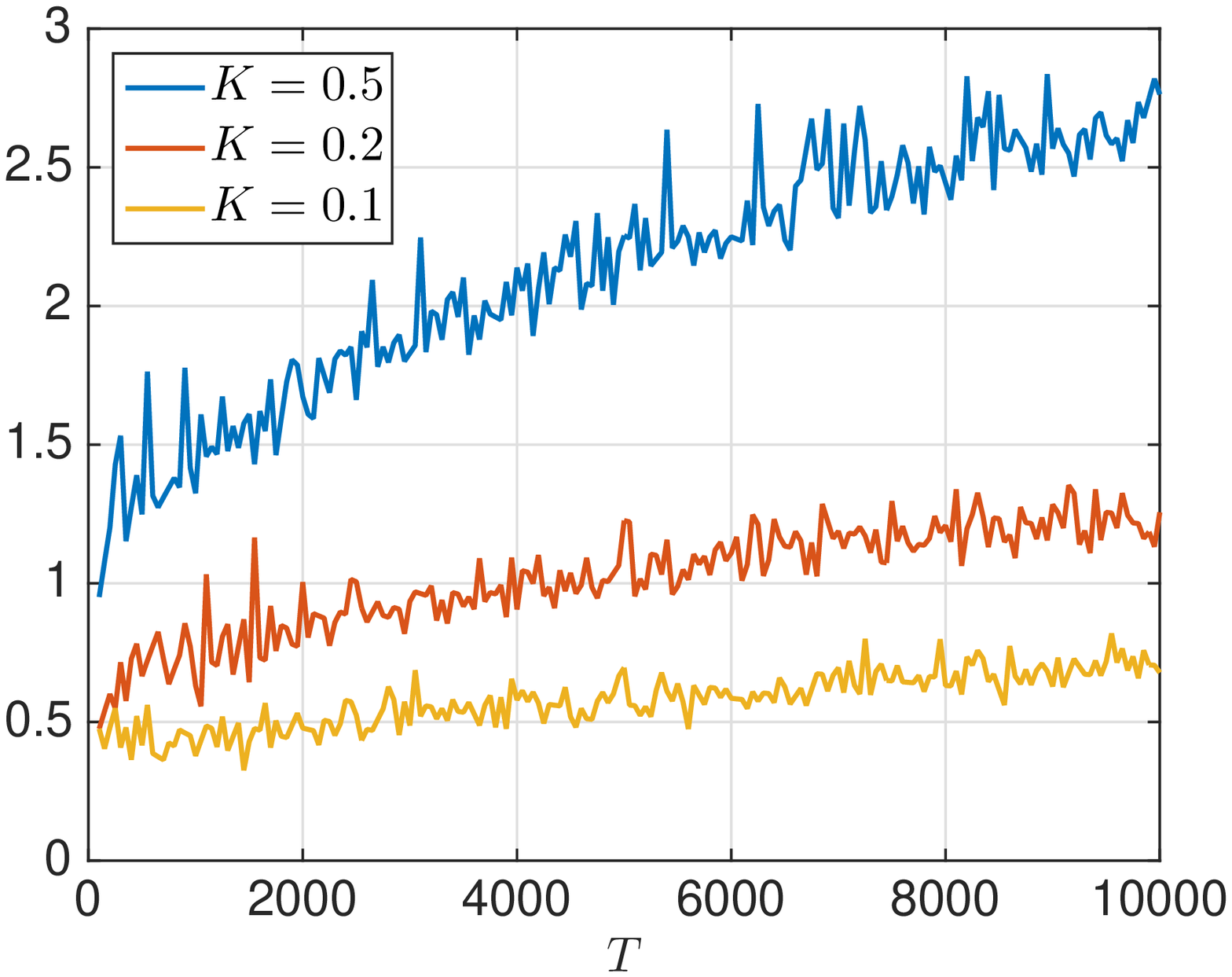}&
 \includegraphics[width=.25\textwidth,trim=0cm .2cm 0cm 0cm,clip]{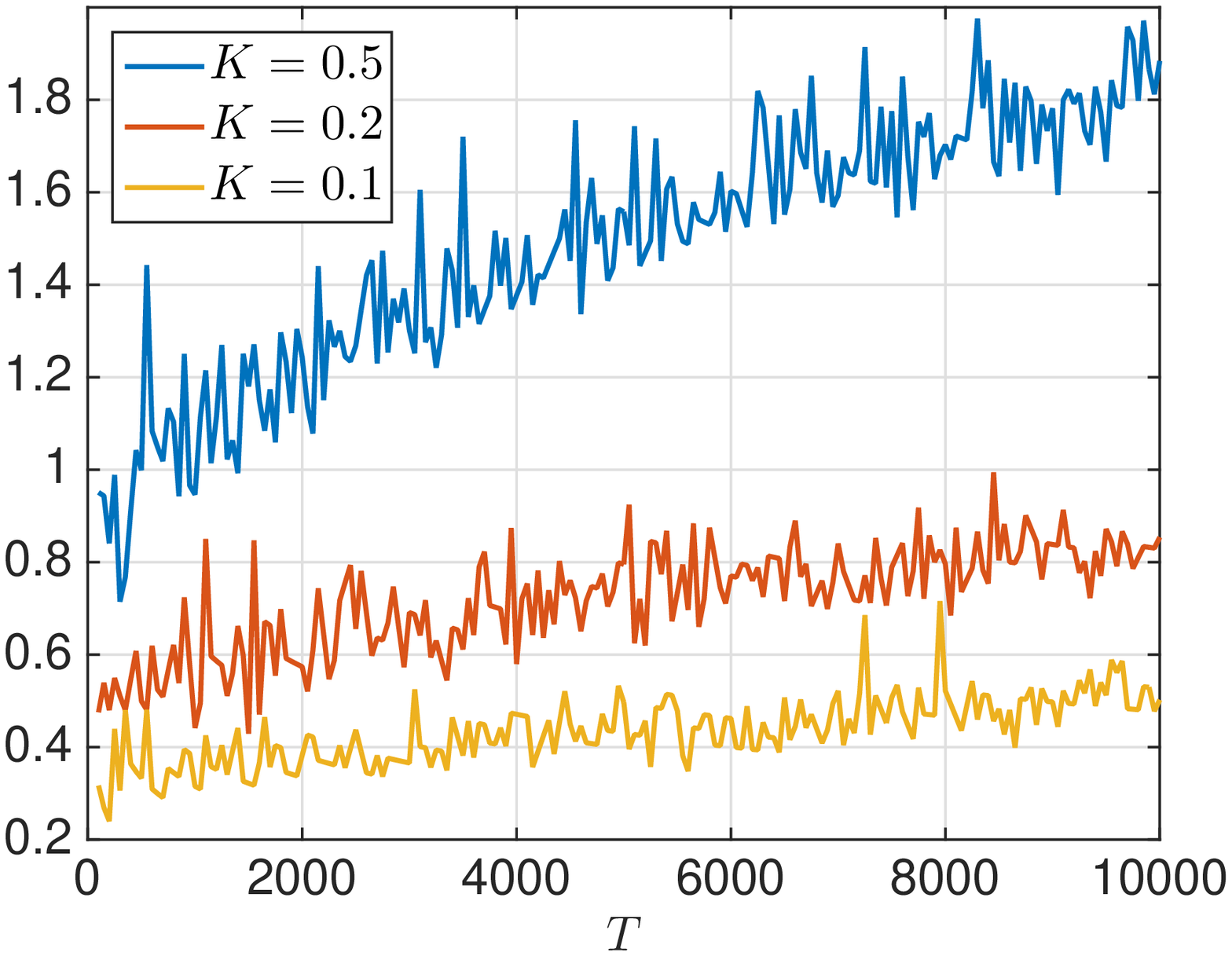}&
 \includegraphics[width=.25\textwidth,trim=0cm .2cm 0cm 0cm,clip]{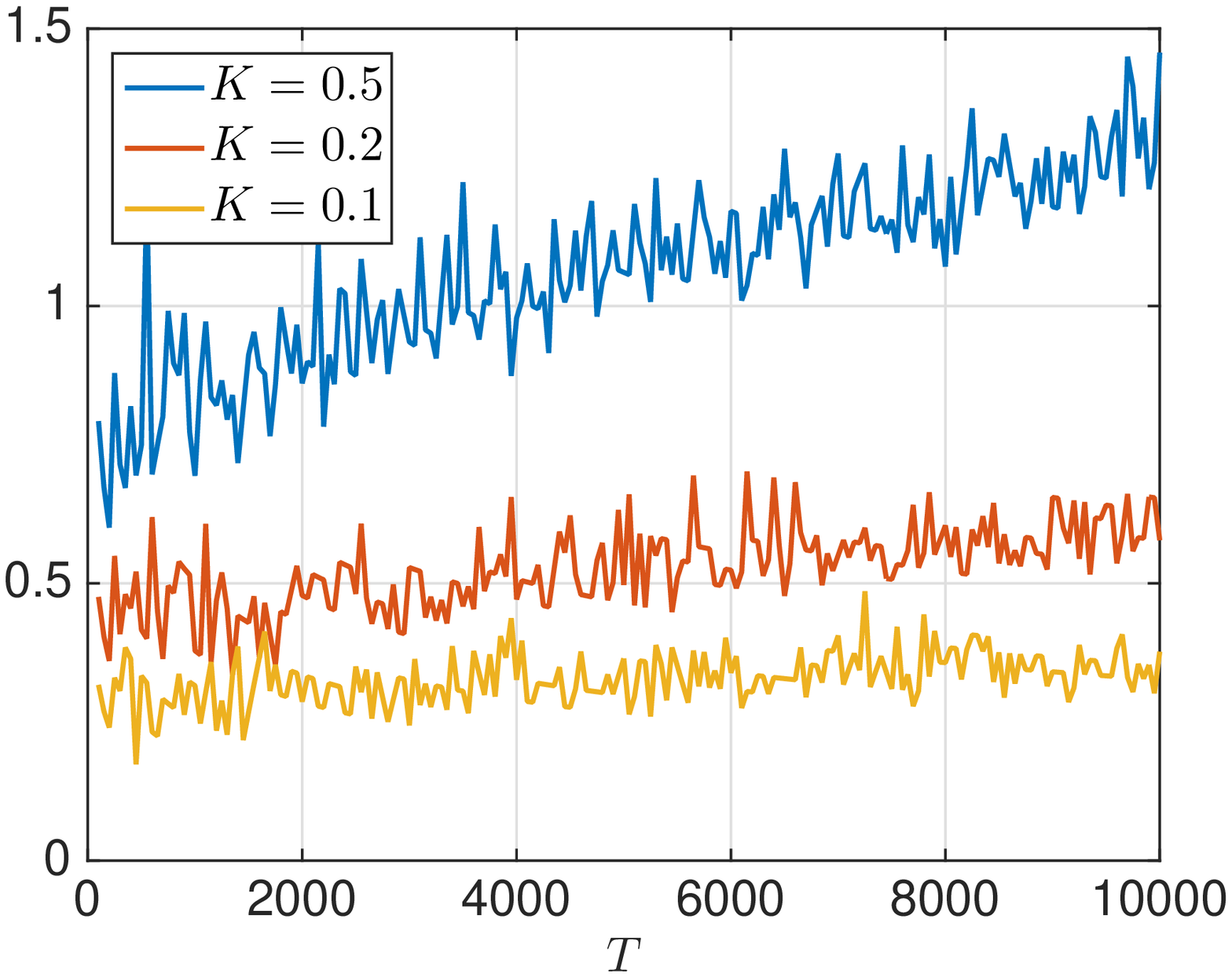}&
  \includegraphics[width=.25\textwidth,trim=0cm .2cm 0cm 0cm,clip]{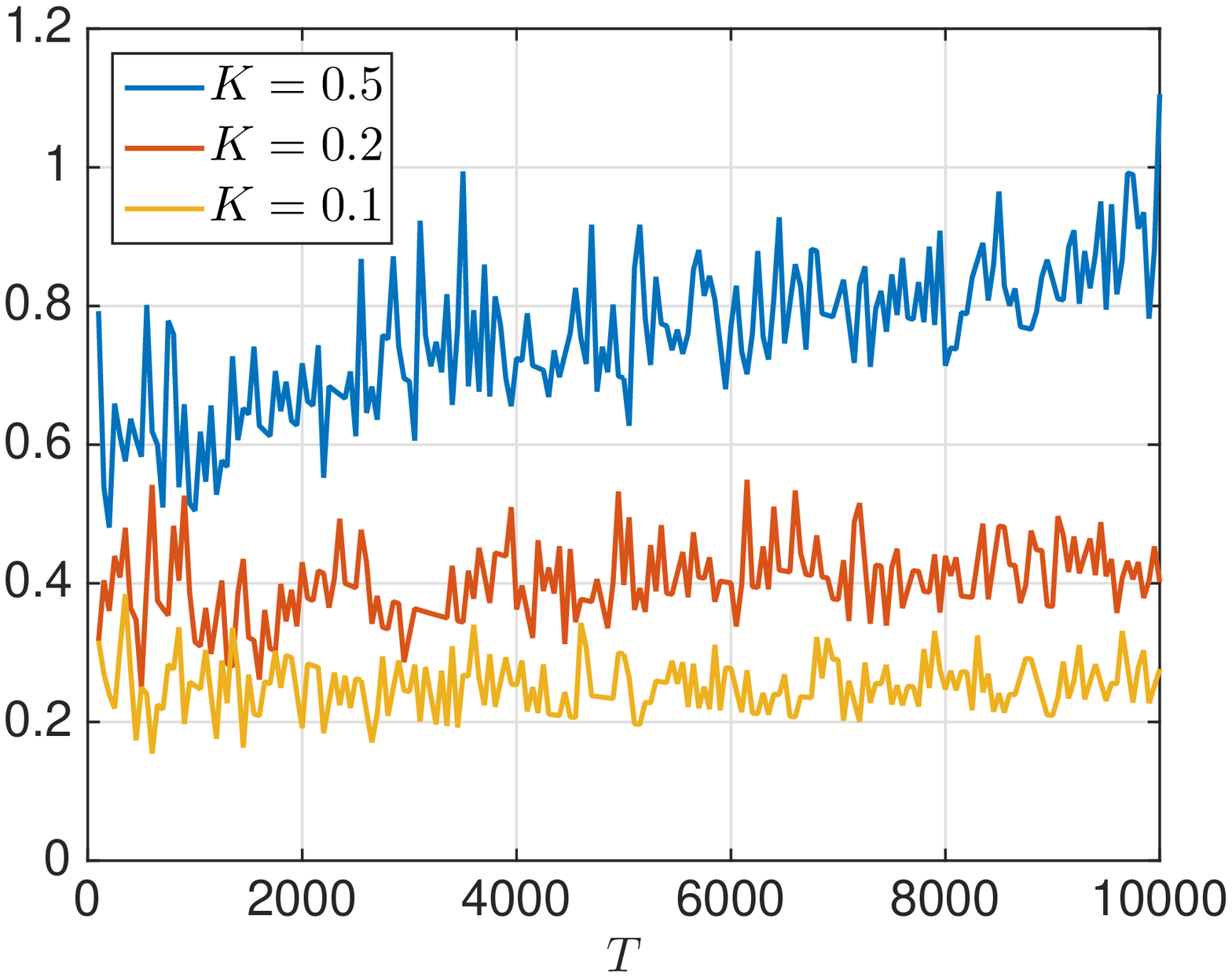}\\
 \footnotesize $\varphi=2.4$, $\epsilon=0.1$&\footnotesize $\varphi=2.6$, $\epsilon=0.1$& \footnotesize $\varphi=2.8$, $\epsilon=0.1$& \footnotesize $\varphi=3$, $\epsilon=0.01$\\
 \includegraphics[width=.25\textwidth,trim=0cm .2cm 0cm 0cm,clip]{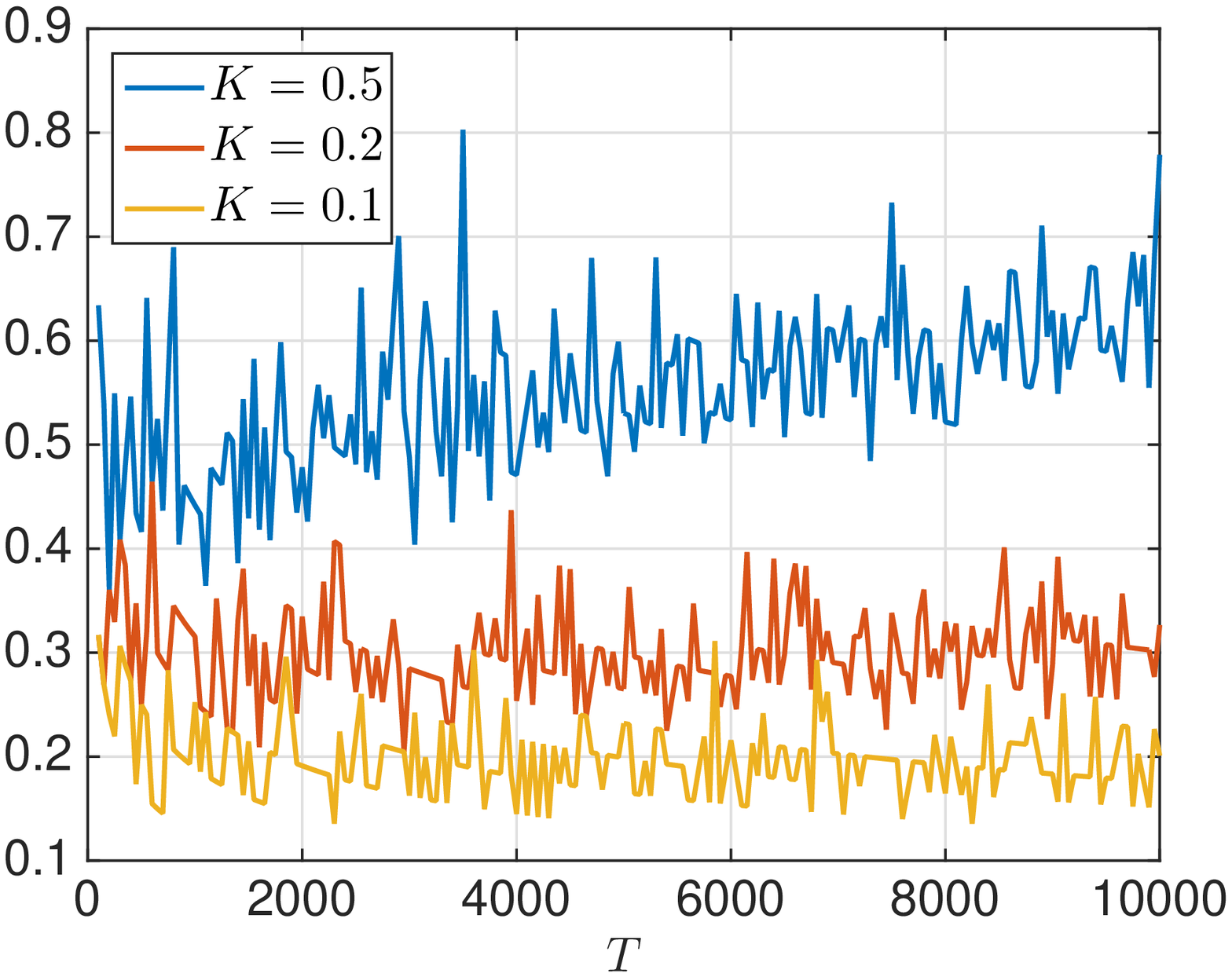}&
 \includegraphics[width=.25\textwidth,trim=0cm .2cm 0cm 0cm,clip]{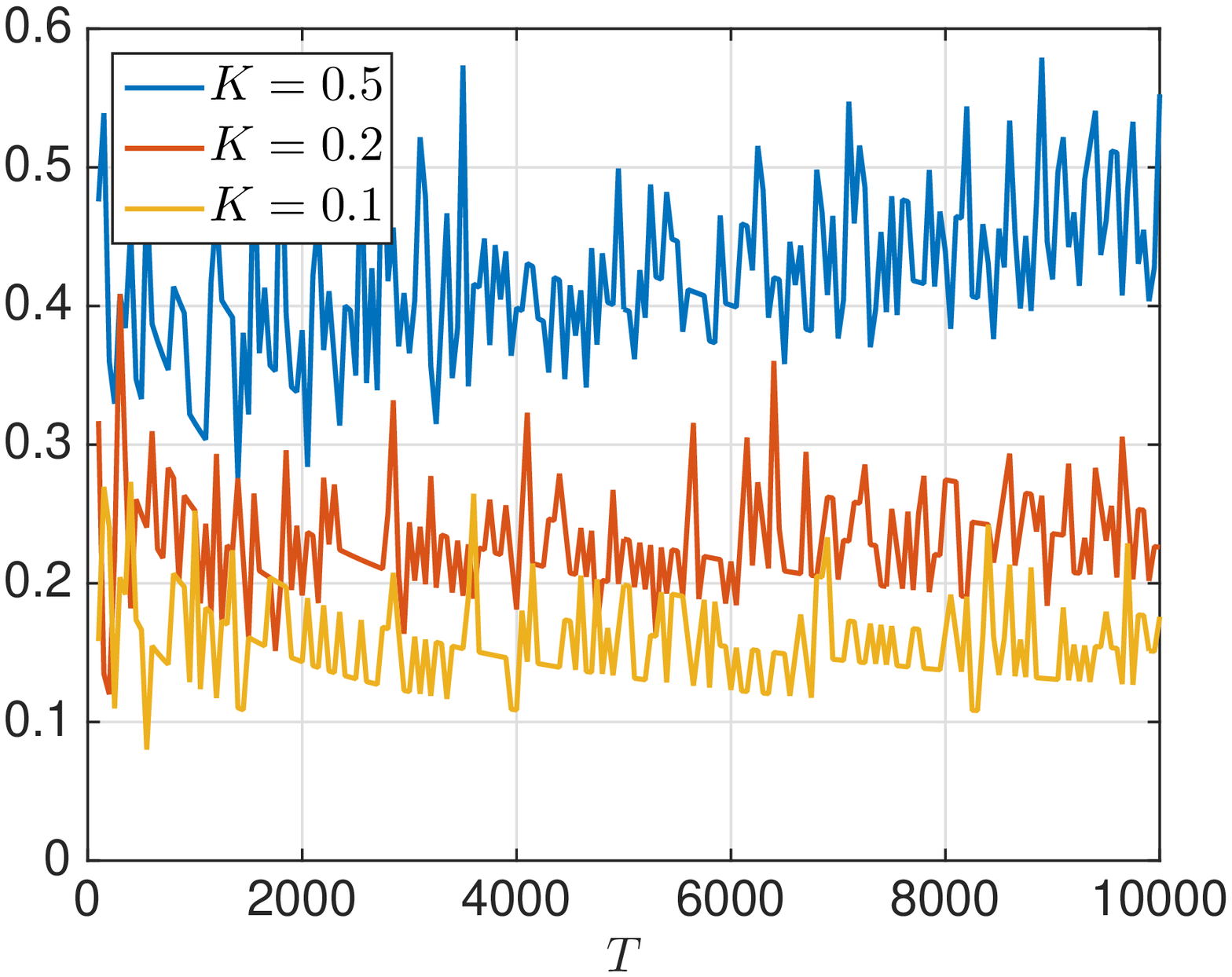}&
 \includegraphics[width=.25\textwidth,trim=0cm .2cm 0cm 0cm,clip]{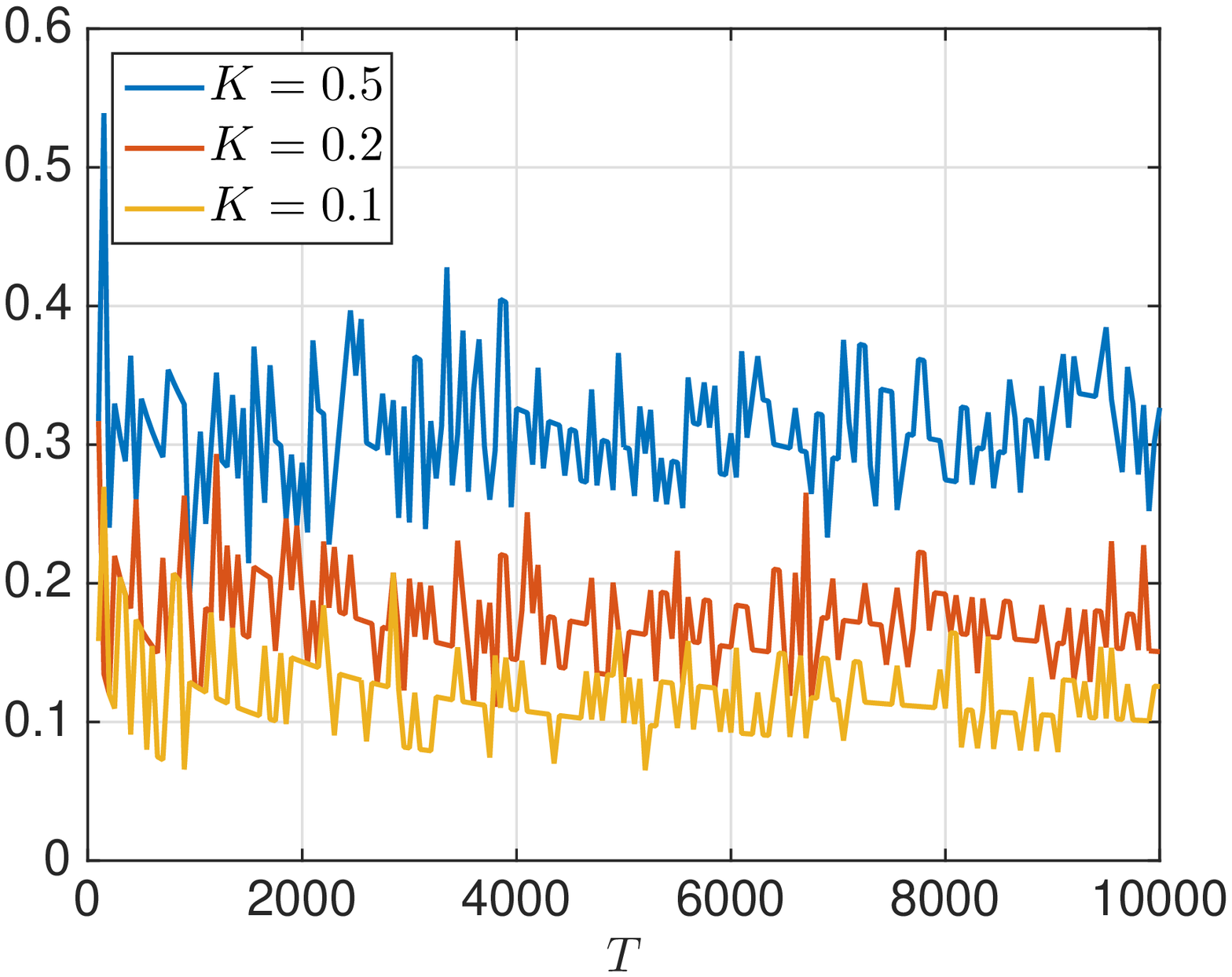}&
  \includegraphics[width=.25\textwidth,trim=0cm .2cm 0cm 0cm,clip]{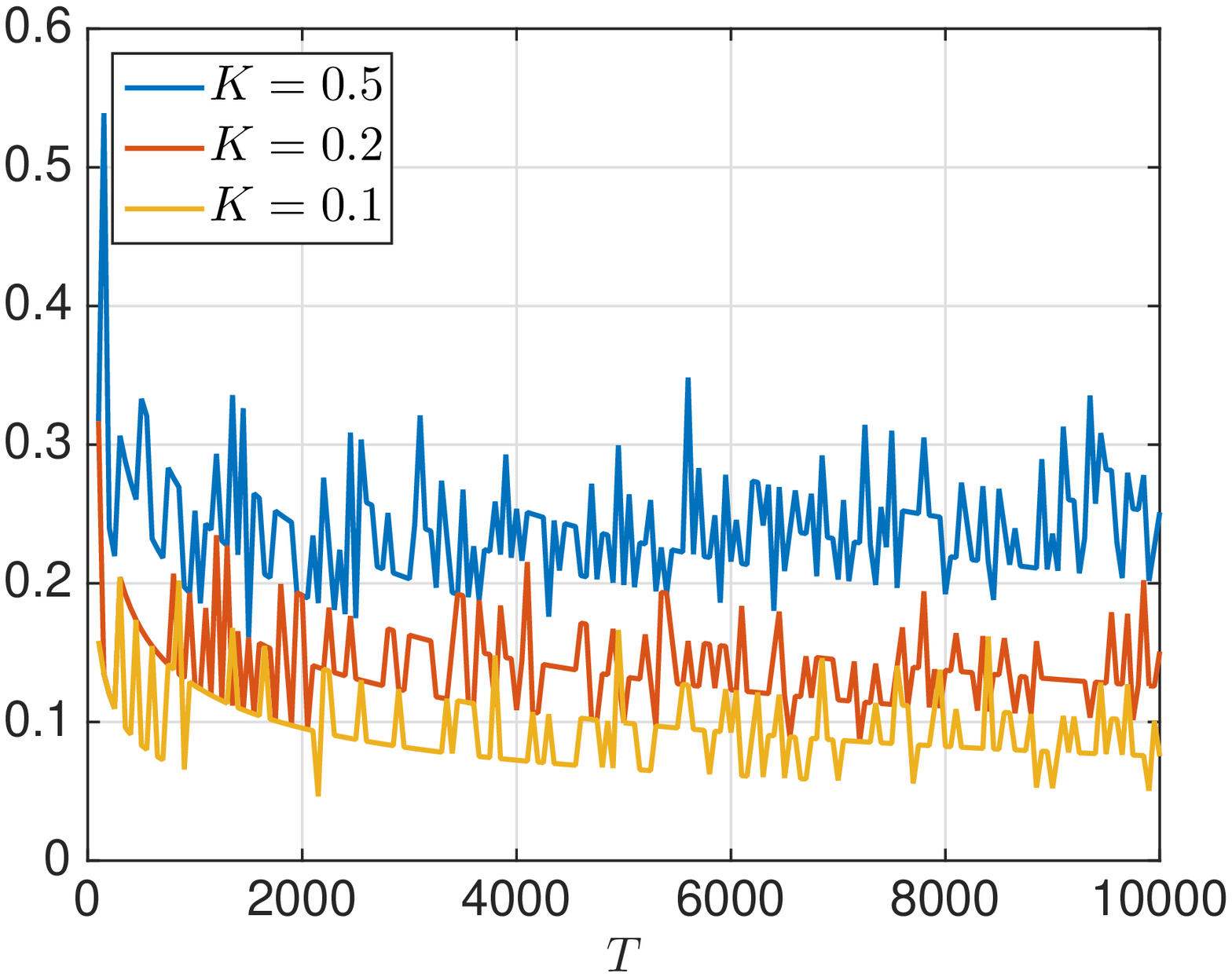}\\
  \end{tabular}
\end{figure}

We set $\kappa_T =   K/{\log^\varphi T}$, and for each given $T_j$ we compute the cardinality of the set $\mathcal T_{i,NT_j}$. If Assumption~(R) holds, then the quantity $r(j,\varphi,K, \epsilon):=\max_{i=1,\ldots, N}T^{\epsilon}{\vert \mathcal T_{i,NT_j}\vert}/\sqrt {T_j}$ should tend to a constant as~$T_j$ grows, for any $\epsilon>0$. In Figure \ref{fig:trim}, we report $r(j,\varphi,K,\epsilon)$, as function of $T_j$, when $\epsilon\in\{0.01,\, 0.1\}$, $\varphi\in\{1.6,\, 1.8,\, 2,\, 2.2,\, 2.4,\, 2.6,\, 2.8,\,  3\}$ and~$K\in\{0.5,\, 0.2,\, 0.1\}$.

\clearpage
  \setcounter{table}{0}%
\renewcommand{\thetable}{C\arabic{table}}
\section{S\&P100 data}\label{app:data}
\begin{table}[h!]
\caption{S\&P100 constituents.}\label{tab:ticker}

\centering
\footnotesize
\begin{tabular}{llll}
&\\
\hline
\hline
Ticker & Name\\
\hline
&\\
AAPL	&	Apple Inc.	&HPQ	&	Hewlett Packard Co.	\\
ABT	&	Abbott Laboratories&IBM	&	International Business Machines	\\
AEP	&	American Electric Power Co.&INTC	&	Intel Corporation	\\
AIG	&	American International Group Inc.&JNJ	&	Johnson \& Johnson Inc.	\\
ALL	&	Allstate Corp.	&JPM	&	JP Morgan Chase \& Co.	\\
AMGN	&	Amgen Inc.	&KO	&	The Coca-Cola Company	\\
AMZN	&	Amazon.com	&LLY	&	Eli Lilly and Company	\\
APA	&	Apache Corp.	&LMT	&	Lockheed-Martin	\\
APC	&	Anadarko Petroleum Corp.&LOW	&	Lowe's	\\
AXP	&	American Express Inc.&MCD	&	McDonald's Corp.	\\
BA	&	Boeing Co.&MDT	&	Medtronic Inc.	\\
BAC	&	Bank of America Corp.	&MMM	&	3M Company	\\
BAX	&	Baxter International Inc.&MO	&	Altria Group		\\
BK	&	Bank of New York	&MRK	&	Merck \& Co.	\\
BMY	&	Bristol-Myers Squibb &MS	&	Morgan Stanley		\\
BRK.B	&	Berkshire Hathaway	&MSFT	&	Microsoft \\
C	&	Citigroup Inc.	&NKE	&	Nike		\\
CAT	&	Caterpillar Inc.&NOV	&	National Oilwell Varco		\\
CL	&	Colgate-Palmolive Co.	&NSC	&	Norfolk Southern Corp.	\\
CMCSA	&	Comcast Corp.&ORCL	&	Oracle Corporation	\\
COF	&	Capital One Financial Corp.	&OXY	&	Occidental Petroleum Corp.	\\
COP	&	ConocoPhillips		&PEP	&	Pepsico Inc.	\\
COST	&	Costco	&PFE	&	Pfizer Inc.	\\
CSCO	&	Cisco Systems		&PG	&	Procter \& Gamble Co.	\\
CVS	&	CVS Caremark&QCOM	&	Qualcomm Inc.	\\
CVX	&	Chevron	&RTN	&	Raytheon Co.	\\
DD	&	DuPont&SBUX	&	Starbucks Corporation	\\
DELL	&	Dell		&SLB	&	Schlumberger	\\
DIS	&	The Walt Disney Company	&SO	&	Southern Company	\\
DOW	&	Dow Chemical	&SPG	&	Simon Property Group, Inc.	\\
DVN	&	Devon Energy&T	&	AT\&T Inc.	\\
EBAY	&	eBay Inc.&TGT	&	Target Corp.		\\
EMC	&	EMC Corporation	&TWX	&	Time Warner Inc.	\\
EMR	&	Emerson Electric Co.	&TXN	&	Texas Instruments	\\
EXC	&	Exelon	&UNH	&	UnitedHealth Group Inc.	\\
F	&	Ford Motor	&UNP	&	Union Pacific Corp.	\\
FCX	&	Freeport-McMoran	&UPS	&	United Parcel Service Inc.	\\
FDX	&	FedEx	&USB	&	US Bancorp	\\
GD	&	General Dynamics	&UTX	&	United Technologies Corp.	\\
GE	&	General Electric Co.	&VZ	&	Verizon Communications Inc.	\\
GILD	&	Gilead Sciences	&WAG	&	Walgreens	\\
GS	&	Goldman Sachs	&WFC	&	Wells Fargo	\\
HAL	&	Halliburton	&WMB	&	Williams Companies	\\
HD	&	Home Depot	&WMT	&	Wal-Mart	\\
HON	&	Honeywell	&XOM	&	Exxon Mobil Corp.	\\
	\\
\hline
\hline
\end{tabular}
\end{table}

\clearpage
  \setcounter{table}{0}%
\renewcommand{\thetable}{D\arabic{table}}
\section{Additional simulation results}\label{app:sim}

\begin{table}[h!]
\centering
\caption{\small Simulation results. Common components. Values of the bandwidths are: $B_T=5$ and $M_T=5$ for all $T$.}\label{tab:sim2app}
\vskip .2cm
\footnotesize
\begin{tabular}{l l | cc | cc | cc }
\multicolumn{8}{c}{$q=1$, $Q=1$}\\
\hline\hline
&& \multicolumn{2}{|c|}{$T=200$}& \multicolumn{2}{|c|}{$T=500$}& \multicolumn{2}{|c}{$T=1000$}\\
&& $n=100$ & $n=200$& $n=100$ & $n=200$& $n=100$ & $n=200$\\
\hline\hline
$MSE^X$	&		&	0.089	&	0.091	&	0.073	&	0.075	&	0.061	&	0.064	\\
$MSE^\chi$	&	$\kappa_T=0$	&	0.450	&	0.412	&	0.630	&	0.607	&	0.793	&	0.734	\\
$MSE^\chi$	&	$\kappa_T=0.2$	&	0.469	&	0.448	&	0.673	&	0.641	&	0.750	&	0.730	\\
$MSE^\chi$	&	$\kappa_T=0.4$	&	0.455	&	0.477	&	0.671	&	0.672	&	0.788	&	0.820	\\
\hline
$MAD^X$	&		&	0.184	&	0.167	&	0.176	&	0.158	&	0.170	&	0.157	\\
$MAD^\chi$	&	$\kappa_T=0$	&	0.486	&	0.458	&	0.575	&	0.555	&	0.650	&	0.619	\\
$MAD^\chi$	&	$\kappa_T=0.2$	&	0.491	&	0.468	&	0.591	&	0.563	&	0.620	&	0.608	\\
$MAD^\chi$	&	$\kappa_T=0.4$	&	0.481	&	0.487	&	0.583	&	0.579	&	0.630	&	0.637	\\
\hline
$MAX^X$	&		&	9.016	&	10.336	&	11.809	&	10.955	&	11.742	&	16.046	\\
$MAX^\chi$	&	$\kappa_T=0$	&	11.385	&	19.027	&	18.554	&	40.420	&	55.108	&	50.038	\\
$MAX^\chi$	&	$\kappa_T=0.2$	&	18.154	&	29.037	&	34.038	&	37.224	&	45.037	&	61.851	\\
$MAX^\chi$	&	$\kappa_T=0.4$	&	20.399	&	23.017	&	33.477	&	44.226	&	65.348	&	51.665	\\
\hline\hline
\\
\multicolumn{8}{c}{$q=3$, $Q=2$}\\
\hline\hline
&& \multicolumn{2}{|c|}{$T=200$}& \multicolumn{2}{|c|}{$T=500$}& \multicolumn{2}{|c}{$T=1000$}\\
&& $n=100$ & $n=200$& $n=100$ & $n=200$& $n=100$ & $n=200$\\
\hline\hline
$MSE^X$	&		&	0.091	&	0.096	&	0.067	&	0.070	&	0.056	&	0.058	\\
$MSE^\chi$	&	$\kappa_T=0$	&	0.361	&	0.277	&	0.373	&	0.259	&	0.420	&	0.257	\\
$MSE^\chi$	&	$\kappa_T=0.2$	&	0.316	&	0.232	&	0.340	&	0.336	&	0.229	&	0.225	\\
$MSE^\chi$	&	$\kappa_T=0.4$	&	0.300	&	0.226	&	0.305	&	0.210	&	0.319	&	0.218	\\
\hline
$MAD^X$	&		&	0.213	&	0.205	&	0.186	&	0.183	&	0.175	&	0.172	\\
$MAD^\chi$	&	$\kappa_T=0$	&	0.458	&	0.395	&	0.461	&	0.380	&	0.489	&	0.376	\\
$MAD^\chi$	&	$\kappa_T=0.2$	&	0.426	&	0.359	&	0.439	&	0.432	&	0.354	&	0.350	\\
$MAD^\chi$	&	$\kappa_T=0.4$	&	0.413	&	0.354	&	0.413	&	0.338	&	0.421	&	0.344	\\
\hline
$MAX^X$	&		&	5.632	&	8.485	&	12.693	&	8.876	&	11.027	&	13.225	\\
$MAX^\chi$	&	$\kappa_T=0$	&	5.151	&	4.585	&	5.254	&	6.031	&	7.377	&	5.736	\\
$MAX^\chi$	&	$\kappa_T=0.2$	&	3.897	&	4.985	&	4.981	&	6.400	&	5.038	&	6.491	\\
$MAX^\chi$	&	$\kappa_T=0.4$	&	5.855	&	4.454	&	7.723	&	7.237	&	5.837	&	5.723	\\
\hline\hline
\\
\multicolumn{8}{c}{$q=2$, $Q=3$}\\
\hline\hline
&& \multicolumn{2}{|c|}{$T=200$}& \multicolumn{2}{|c|}{$T=500$}& \multicolumn{2}{|c}{$T=1000$}\\
&& $n=100$ & $n=200$& $n=100$ & $n=200$& $n=100$ & $n=200$\\
\hline\hline
$MSE^X$	&		&	0.087	&	0.088	&	0.068	&	0.071	&	0.056	&	0.064	\\
$MSE^\chi$	&	$\kappa_T=0$	&	0.355	&	0.300	&	0.384	&	0.274	&	0.425	&	0.280	\\
$MSE^\chi$	&	$\kappa_T=0.2$	&	0.327	&	0.265	&	0.347	&	0.236	&	0.380	&	0.231	\\
$MSE^\chi$	&	$\kappa_T=0.4$	&	0.300	&	0.247	&	0.315	&	0.208	&	0.361	&	0.215	\\
\hline
$MAD^X$	&		&	0.210	&	0.204	&	0.191	&	0.187	&	0.177	&	0.180	\\
$MAD^\chi$	&	$\kappa_T=0$	&	0.459	&	0.419	&	0.475	&	0.396	&	0.496	&	0.399	\\
$MAD^\chi$	&	$\kappa_T=0.2$	&	0.439	&	0.391	&	0.450	&	0.366	&	0.469	&	0.361	\\
$MAD^\chi$	&	$\kappa_T=0.4$	&	0.420	&	0.376	&	0.427	&	0.343	&	0.456	&	0.347	\\
\hline
$MAX^X$	&		&	9.532	&	9.098	&	6.292	&	11.581	&	8.078	&	7.409	\\
$MAX^\chi$	&	$\kappa_T=0$	&	3.978	&	4.204	&	4.184	&	5.321	&	6.107	&	5.417	\\
$MAX^\chi$	&	$\kappa_T=0.2$	&	3.986	&	4.835	&	4.913	&	5.932	&	5.512	&	7.808	\\
$MAX^\chi$	&	$\kappa_T=0.4$	&	4.524	&	4.674	&	4.521	&	6.284	&	5.944	&	5.705	\\
\hline\hline
\end{tabular}
\end{table}

\begin{table}[t!]
\centering
\caption{\small Simulation results. Empirical coverage and  frequency of confidence bound violations averaged over all $n$ series and all $\mathcal M$ replications, when $T=1000$ and $\mathcal M=200$. Values of the bandwidths are: $B_T=5$ and $M_T=5$ for all~$T$.}\label{tab:covsimapp1}
\vskip .2cm
\footnotesize
\begin{tabular}{l l | ccccc | ccccc }
\multicolumn{12}{c}{$q=1$, $Q=1$}\\
\hline
\hline
&&\multicolumn{5}{c|}{$n=100$}&\multicolumn{5}{c}{$n=200$}\\
\hline
&&\multicolumn{5}{c|}{$\alpha$}&\multicolumn{5}{c}{$\alpha$}\\
&&0.32&0.2&0.1&0.05&0.01&0.32&0.2&0.1&0.05&0.01\\
\hline
$C(\alpha)$&$\kappa_T=0$&0.6854&	0.8098&	0.9099&	0.9581&	0.9928&0.6793&	0.7871&	0.8809&	0.9335&	0.9831\\
$V_+(\alpha/2)$&&0.1573&	0.0951&	0.0441&	0.0217&	0.0040&0.1606&	0.1055&	0.0599&	0.0332&	0.0087\\	
$V_-(\alpha/2)$&&	0.1573&	0.0951&	0.0460&	0.0202&	0.0032&0.1601&	0.1074&	0.0593&	0.0333&	0.0083\\
\hline
$C(\alpha)$&$\kappa_T=0.2$&0.7046&	0.7982&	0.8790&	0.9246&	0.9744&0.7126&	0.8207&	0.9070&	0.9490&	0.9841\\
$V_+(\alpha/2)$&&0.1465&	0.0992&	0.0594&	0.0373&	0.0129&0.1439&	0.0891&	0.0461&	0.0247&	0.0081\\	
$V_-(\alpha/2)$&&0.1489&	0.1026&	0.0616&	0.0381&	0.0127&0.1436&	0.0902&	0.0469&	0.0264&	0.0078\\	
\hline
$C(\alpha)$&$\kappa_T=0.4$&0.7531&	0.8440&	0.9197&	0.9543&	0.9845&0.7687&	0.8511&	0.9243&	0.9628&	0.9969\\
$V_+(\alpha/2)$&&0.1217&	0.0759&	0.0401&	0.0229&	0.0084&0.1171&	0.0775&	0.0406&	0.0201&	0.0017\\	
$V_-(\alpha/2)$&&	0.1252& 0.0801&	0.0402&	0.0228&	0.0071&0.1142&	0.0714&	0.0351&	0.0172&	0.0014\\
\hline
\hline
\\
\multicolumn{12}{c}{$q=3$, $Q=2$}\\
\hline
\hline
&&\multicolumn{5}{c|}{$n=100$}&\multicolumn{5}{c}{$n=200$}\\
\hline
&&\multicolumn{5}{c|}{$\alpha$}&\multicolumn{5}{c}{$\alpha$}\\
&&0.32&0.2&0.1&0.05&0.01&0.32&0.2&0.1&0.05&0.01\\
\hline
$C(\alpha)$&$\kappa_T=0$&0.6821&	0.7978&	0.8907&	0.9405&	0.9800&0.6651&	0.7872&	0.8903&	0.9445&	0.9840\\
$V_+(\alpha/2)$&&0.1596&	0.1003&	0.0556&	0.0294&	0.0097&0.1596&	0.1003&	0.0556&	0.0294&	0.0097\\	
$V_-(\alpha/2)$&&	0.1583&	0.1019&	0.0537&	0.0301&	0.0103&0.1583&	0.1019&	0.0537&	0.0301&	0.0103\\
\hline
$C(\alpha)$&$\kappa_T=0.2$&0.6837&	0.7787&	0.8609&	0.9089&	0.9602&0.7277&	0.8290&	0.9124&	0.9552&	0.9910\\
$V_+(\alpha/2)$&&0.1560&	0.1097&	0.0689&	0.0442&	0.0186&0.1387&	0.0865&	0.0442&	0.0220&	0.0037\\	
$V_-(\alpha/2)$&&0.1603&	0.1116&	0.0702&	0.0469&	0.0212&0.1336&	0.0846&	0.0435&	0.0229&	0.0054\\	
\hline
$C(\alpha)$&$\kappa_T=0.4$&0.7500&	0.8369&	0.9141&	0.9527&	0.9881&0.7551&	0.8431&	0.9149&	0.9543&	0.9915\\
$V_+(\alpha/2)$&&0.1217&	0.0794&	0.0419&	0.0233&	0.0059&0.1232&	0.0782&	0.0409&	0.0219&	0.0036\\	
$V_-(\alpha/2)$&&	0.1283&	0.0837&	0.0440&	0.0240&	0.0060&0.1218&	0.0788&	0.0442&	0.0238&	0.0049\\
\hline
\hline
\\
\multicolumn{12}{c}{$q=2$, $Q=3$}\\
\hline
\hline
&&\multicolumn{5}{c|}{$n=100$}&\multicolumn{5}{c}{$n=200$}\\
\hline
&&\multicolumn{5}{c|}{$\alpha$}&\multicolumn{5}{c}{$\alpha$}\\
&&0.32&0.2&0.1&0.05&0.01&0.32&0.2&0.1&0.05&0.01\\
\hline
$C(\alpha)$&$\kappa_T=0$&0.6137&	0.7191&	0.8259&	0.8916&	0.9598&0.6631&	0.7803&	0.8820&	0.9393&	0.9866\\
$V_+(\alpha/2)$&&0.2034&	0.1490&	0.0894&	0.0552&	0.0217&0.1722&	0.1118&	0.0610&	0.0317&	0.0078\\	
$V_-(\alpha/2)$&&0.1829&	0.1319&	0.0847&	0.0532&	0.0185&0.1648&	0.1080&	0.0571&	0.0290&	0.0057\\	
\hline
$C(\alpha)$&$\kappa_T=0.2$&0.7166&	0.8219&	0.9095&	0.9549&	0.9925&0.7259&	0.8332&	0.9262&	0.9660&	0.9941\\
$V_+(\alpha/2)$&&0.1495&	0.0940&	0.0493&	0.0252&	0.0040&0.1371&	0.0826&	0.0372&	0.0176&	0.0029\\	
$V_-(\alpha/2)$&&	0.1339&	0.0841&	0.0412&	0.0199&	0.0035&0.1371&	0.0843&	0.0367&	0.0164&	0.0031\\
\hline
$C(\alpha)$&$\kappa_T=0.4$&0.7715&	0.8569&	0.9282&	0.9671&	0.9950&0.7265	&0.8165&	0.8986&	0.9466&	0.9918\\
$V_+(\alpha/2)$&&0.1146&	0.0728&	0.0361&	0.0149&	0.0019&0.1232&	0.0782&	0.0409&	0.0219&	0.0036\\	
$V_-(\alpha/2)$&&0.1139&	0.0703&	0.0357&	0.0180&	0.0031&0.1218&	0.07878&	0.0442&	0.0238&	0.0049\\
\hline
\hline
\end{tabular}
\end{table}

\begin{table}[h!]
\centering
\caption{\small Simulation results. Common components. Values of the bandwidths are: $B_T=1$ and $M_T=14$ for~$T=200$; $B_T=1$ and $M_T=22$ for~$T=500$; $B_T=1$ and $M_T=31$ for~$T=1000$.}\label{tab:sim3app}
\vskip .2cm
\footnotesize
\begin{tabular}{l l | cc | cc | cc }
\multicolumn{8}{c}{$q=1$, $Q=1$}\\
\hline\hline
&& \multicolumn{2}{|c|}{$T=200$}& \multicolumn{2}{|c|}{$T=500$}& \multicolumn{2}{|c}{$T=1000$}\\
&& $n=100$ & $n=200$& $n=100$ & $n=200$& $n=100$ & $n=200$\\
\hline\hline
$MSE^X$	&		&	0.467	&	4.568	&	0.334	&	0.344	&	0.271	&	0.251	\\
$MSE^\chi$	&	$\kappa_T=0$	&	0.547	&	0.508	&	0.477	&	0.422	&	0.419	&	0.343	\\
$MSE^\chi$	&	$\kappa_T=0.2$	&	0.561	&	0.527	&	0.481	&	0.427	&	0.412	&	0.331	\\
$MSE^\chi$	&	$\kappa_T=0.4$	&	0.593	&	0.487	&	0.405	&	0.560	&	0.458	&	0.350	\\
\hline
$MAD^X$	&		&	0.417	&	0.386	&	0.366	&	0.330	&	0.337	&	0.294	\\
$MAD^\chi$	&	$\kappa_T=0$	&	0.555	&	0.524	&	0.508	&	0.467	&	0.471	&	0.417	\\
$MAD^\chi$	&	$\kappa_T=0.2$	&	0.556	&	0.526	&	0.503	&	0.460	&	0.459	&	0.399	\\
$MAD^\chi$	&	$\kappa_T=0.4$	&	0.573	&	0.502	&	0.448	&	0.541	&	0.472	&	0.406	\\
\hline
$MAX^X$	&		&	36.332	&	826.833	&	20.032	&	39.123	&	31.799	&	46.432	\\
$MAX^\chi$	&	$\kappa_T=0$	&	7.023	&	8.756	&	7.328	&	8.536	&	8.251	&	8.111	\\
$MAX^\chi$	&	$\kappa_T=0.2$	&	8.059	&	8.743	&	7.758	&	9.147	&	9.171	&	8.843	\\
$MAX^\chi$	&	$\kappa_T=0.4$	&	8.727	&	9.109	&	9.846	&	9.310	&	9.387	&	9.895	\\
\hline\hline
\\
\multicolumn{8}{c}{$q=3$, $Q=2$}\\
\hline\hline
&& \multicolumn{2}{|c|}{$T=200$}& \multicolumn{2}{|c|}{$T=500$}& \multicolumn{2}{|c}{$T=1000$}\\
&& $n=100$ & $n=200$& $n=100$ & $n=200$& $n=100$ & $n=200$\\
\hline\hline
$MSE^X$	&		&	0.255	&	0.250	&	0.159	&	0.157	&	0.116	&	0.121	\\
$MSE^\chi$	&	$\kappa_T=0$	&	0.389	&	0.375	&	0.299	&	0.273	&	0.248	&	0.234	\\
$MSE^\chi$	&	$\kappa_T=0.2$	&	0.369	&	0.268	&	0.212	&	0.360	&	0.241	&	0.202	\\
$MSE^\chi$	&	$\kappa_T=0.4$	&	0.358	&	0.358	&	0.250	&	0.252	&	0.191	&	0.176	\\
\hline
$MAD^X$	&		&	0.357	&	0.337	&	0.284	&	0.272	&	0.248	&	0.245	\\
$MAD^\chi$	&	$\kappa_T=0$	&	0.480	&	0.467	&	0.417	&	0.395	&	0.380	&	0.365	\\
$MAD^\chi$	&	$\kappa_T=0.2$	&	0.465	&	0.391	&	0.348	&	0.454	&	0.367	&	0.334	\\
$MAD^\chi$	&	$\kappa_T=0.4$	&	0.457	&	0.450	&	0.376	&	0.372	&	0.328	&	0.311	\\
\hline
$MAX^X$	&		&	8.732	&	10.161	&	10.535	&	11.145	&	12.495	&	13.895	\\
$MAX^\chi$	&	$\kappa_T=0$	&	5.721	&	5.525	&	5.010	&	5.412	&	4.532	&	5.946	\\
$MAX^\chi$	&	$\kappa_T=0.2$	&	6.036	&	5.137	&	4.693	&	5.637	&	5.194	&	6.250	\\
$MAX^\chi$	&	$\kappa_T=0.4$	&	5.332	&	5.653	&	5.659	&	6.776	&	5.285	&	7.061	\\
\hline\hline
\\
\multicolumn{8}{c}{$q=2$, $Q=3$}\\
\hline\hline
&& \multicolumn{2}{|c|}{$T=200$}& \multicolumn{2}{|c|}{$T=500$}& \multicolumn{2}{|c}{$T=1000$}\\
&& $n=100$ & $n=200$& $n=100$ & $n=200$& $n=100$ & $n=200$\\
\hline\hline
$MSE^X$	&		&	0.298	&	0.286	&	0.194	&	0.184	&	0.134	&	0.143	\\
$MSE^\chi$	&	$\kappa_T=0$	&	0.480	&	0.439	&	0.387	&	0.353	&	0.311	&	0.300	\\
$MSE^\chi$	&	$\kappa_T=0.2$	&	0.463	&	0.426	&	0.358	&	0.327	&	0.277	&	0.266	\\
$MSE^\chi$	&	$\kappa_T=0.4$	&	0.464	&	0.453	&	0.352	&	0.341	&	0.271	&	0.247	\\
\hline
$MAD^X$	&		&	0.393	&	0.370	&	0.316	&	0.303	&	0.267	&	0.272	\\
$MAD^\chi$	&	$\kappa_T=0$	&	0.538	&	0.511	&	0.478	&	0.453	&	0.428	&	0.417	\\
$MAD^\chi$	&	$\kappa_T=0.2$	&	0.526	&	0.500	&	0.455	&	0.432	&	0.400	&	0.388	\\
$MAD^\chi$	&	$\kappa_T=0.4$	&	0.525	&	0.515	&	0.449	&	0.438	&	0.392	&	0.372	\\
\hline
$MAX^X$	&		&	9.353	&	10.260	&	8.710	&	14.061	&	9.802	&	12.255	\\
$MAX^\chi$	&	$\kappa_T=0$	&	6.081	&	5.367	&	5.962	&	6.078	&	5.683	&	7.025	\\
$MAX^\chi$	&	$\kappa_T=0.2$	&	5.417	&	5.818	&	6.051	&	6.590	&	6.481	&	7.808	\\
$MAX^\chi$	&	$\kappa_T=0.4$	&	5.905	&	6.056	&	5.951	&	7.419	&	6.519	&	6.513	\\
\hline\hline
\end{tabular}
\end{table}

\begin{table}[t!]
\centering
\caption{\small Simulation results. Empirical coverage and  frequency of confidence bound violations averaged over all $n$ series and all $\mathcal M$ replications, for~$T=1000$ and $\mathcal M=200$. Values of the bandwidths are: $B_T=1$ and $M_T=14$ for~$T=200$; $B_T=1$ and $M_T=22$ for~$T=500$; $B_T=1$ and $M_T=31$ for~$T=1000$.}\label{tab:covsimapp2}
\vskip .2cm
\footnotesize
\begin{tabular}{l l | ccccc | ccccc }
\multicolumn{12}{c}{$q=1$, $Q=1$}\\
\hline
\hline
&&\multicolumn{5}{c|}{$n=100$}&\multicolumn{5}{c}{$n=200$}\\
\hline
&&\multicolumn{5}{c|}{$\alpha$}&\multicolumn{5}{c}{$\alpha$}\\
&&0.32&0.2&0.1&0.05&0.01&0.32&0.2&0.1&0.05&0.01\\
\hline
$C(\alpha)$&$\kappa_T=0$&0.7211&	0.8374&	0.9267&	0.9708&	0.9966&0.6287&	0.7476&	0.8602&	0.9238&	0.9856\\
$V_+(\alpha/2)$&&0.1460&	0.0855&	0.0393&	0.0144&	0.0020&0.1883&	0.1285&	0.0703&	0.0394&	0.0072\\	
$V_-(\alpha/2)$&&	0.1329&	0.0771&	0.0340&	0.0148&	0.0014&0.1831&	0.1240&	0.0696&	0.0368&	0.0073\\
\hline
$C(\alpha)$&$\kappa_T=0.2$&0.7615&	0.8597&	0.9367&	0.9763&	0.9975&0.6841&	0.7864&	0.8846&	0.9346&	0.9882\\
$V_+(\alpha/2)$&&0.1255&	0.0730&	0.0332&	0.0125&	0.0011&0.1603&	0.1090&	0.0585&	0.0335&	0.0057\\	
$V_-(\alpha/2)$&&	0.1130&	0.0673&	0.0301&	0.0112&	0.0014&0.1557&	0.1047&	0.0570&	0.0320&   0.0062\\
\hline
$C(\alpha)$&$\kappa_T=0.4$&0.7747&	0.8623&	0.9368&	0.9708&	0.9966&0.6957&	0.7856&	0.8735&	0.9272&	0.9821\\
$V_+(\alpha/2)$&&0.1135&	0.0678&	0.0291&	0.0133&	0.0012&0.1495&	0.1060&	0.0623&	0.0358&	0.0081\\	
$V_-(\alpha/2)$&&	0.1118&	0.0699&	0.0341&	0.0160&	0.0014&0.1549&	0.1085&	0.0642&	0.0371&	0.0099\\
\hline
\hline
\\
\multicolumn{12}{c}{$q=3$, $Q=2$}\\
\hline
\hline
&&\multicolumn{5}{c|}{$n=100$}&\multicolumn{5}{c}{$n=200$}\\
\hline
&&\multicolumn{5}{c|}{$\alpha$}&\multicolumn{5}{c}{$\alpha$}\\
&&0.32&0.2&0.1&0.05&0.01&0.32&0.2&0.1&0.05&0.01\\
\hline
$C(\alpha)$&$\kappa_T=0$&0.6826&	0.8013&	0.9040&	0.9551&	0.9942&0.6327&	0.7474&	0.8486&	0.9090&	0.9706\\
$V_+(\alpha/2)$&&0.1552&	0.1001&	0.0472&	0.0229&	0.0027&0.1847&	0.1263&	0.0752&	0.0450&	0.0142\\
$V_-(\alpha/2)$&&	0.1622&	0.0986&	0.0488&	0.0220&	0.0031&	0.1826&	0.1264&	0.0763&	0.0461&	0.0153\\
\hline
$C(\alpha)$&$\kappa_T=0.2$&0.7226&	0.8271&	0.9171&	0.9594&	0.9941&0.6749&	0.7732&	0.8639&	0.9157&	0.9720\\
$V_+(\alpha/2)$&&0.1362&	0.0869&	0.0410&	0.0209&	0.0031&0.1636&	0.1134&	0.0672&	0.0415&	0.0135\\
$V_-(\alpha/2)$&&	0.1412&	0.0860&	0.0419&	0.0197&	0.0028&0.1616&	0.1135&	0.0690&	0.0428&	0.0146\\
\hline
$C(\alpha)$&$\kappa_T=0.4$&0.7322&	0.8250&	0.9036&	0.9437&	0.9850&0.7239&	0.8100&	0.8913&	0.9335&	0.9766\\
$V_+(\alpha/2)$&&0.1321&	0.0854&	0.0470&	0.0259&	0.0068&0.1392&	0.0960&	0.0539&	0.0333&	0.0121\\
$V_-(\alpha/2)$&&0.1357&	0.0896&	0.0494&	0.0304&	0.0082&0.1370&	0.0941&	0.0549&	0.0333&	0.0114\\
\hline
\hline
\\
\multicolumn{12}{c}{$q=2$, $Q=3$}\\
\hline
\hline
&&\multicolumn{5}{c|}{$n=100$}&\multicolumn{5}{c}{$n=200$}\\
\hline
&&\multicolumn{5}{c|}{$\alpha$}&\multicolumn{5}{c}{$\alpha$}\\
&&0.32&0.2&0.1&0.05&0.01&0.32&0.2&0.1&0.05&0.01\\
\hline
$C(\alpha)$&$\kappa_T=0$&0.6604&	0.7730&	0.8773&	0.9312&	0.9767&0.6973&	0.8141&	0.9118&	0.9582&	0.9946\\
$V_+(\alpha/2)$&&0.1682&	0.1107&	0.0595&	0.0314&	0.0108&0.1481&	0.0901&	0.0424&	0.0204&	0.0029\\	
$V_-(\alpha/2)$&&0.1714&	0.1163&	0.0632&	0.0374&	0.0125&0.1547&	0.0959&	0.0458&	0.0215&	0.0026\\
\hline
$C(\alpha)$&$\kappa_T=0.2$&0.6979&	0.8006&	0.8899&	0.9367&	0.9794&0.7446&	0.8415&	0.9240&	0.9647&	0.9959\\
$V_+(\alpha/2)$&&0.1507&	0.0973&	0.0521&	0.0288&	0.0100&0.1256&	0.0759&	0.0365&	0.0168&	0.0022\\
$V_-(\alpha/2)$&&	0.1514&	0.1021&	0.0580&	0.0345&	0.0106&0.1298&	0.0827&	0.0396&	0.0186&	0.0020\\
\hline
$C(\alpha)$&$\kappa_T=0.4$&0.7368&	0.8290&	0.9009&	0.9462&	0.9887&0.7557&	0.8420&	0.9160&	0.9566&	0.9902\\
$V_+(\alpha/2)$&&0.1316&	0.0864&	0.0501&	0.0259&	0.0056&0.1273&	0.0834&	0.0434&	0.0219&	0.0047\\	
$V_-(\alpha/2)$&&	0.1316&	0.0846&	0.0490&	0.0279&	0.0057&0.1171&	0.0747&	0.0407&	0.0216&	0.0051\\
\hline
\hline
\end{tabular}
\end{table}

\end{document}